\documentclass[12pt]{article}

\usepackage{graphicx}
\usepackage{epsfig}
\usepackage{color}
\usepackage{subfigure}

\def\MSbar{$\overline{\mathrm{MS}}\ $}
\newcommand{\smartqaq}{q\hskip-7pt\hbox{$^{^{(\!-\!)}}$}}

\newcommand{\smallz}{{\scriptscriptstyle Z}} %
\newcommand{\mz}{M_Z}
\newcommand{\smallw}{{\scriptscriptstyle W}}
\newcommand{\mw}{M_W} 
\newcommand{\oal}{${\cal O}(\alpha)$} 
\newcommand{\oaas}{${\cal O}(\alpha\alpha_s)$}

\newcommand\ba{\begin{eqnarray}}
\newcommand\ea{\end{eqnarray}}
\newcommand{\be}{\begin{equation}}
\newcommand{\ee}{\end{equation}}
\newcommand{\bea}{\begin{eqnarray}}
\newcommand{\eea}{\end{eqnarray}}

\newcommand{\gz}{\Gamma_\smallz}

\newcommand{\dynnlo}{{\tt DYNNLO\,}}
\newcommand{\dynnlops}{{\tt DYNNLOPS\,}}
\newcommand{\fewz}{{\tt FEWZ\,}}
\newcommand{\horace}{{\tt HORACE\,}}

\newcommand{\minlo}{{\tt MiNLO\,}}
\newcommand{\photos}{{\tt PHOTOS\,}}
\newcommand{\powheg}{{\tt POWHEG\,}}
\newcommand{\powhegew}{{\tt POWHEG-(QCD+EW)\,}}

\newcommand{\pythia}{{\tt PYTHIA\,}}
\newcommand{\rady}{{\tt RADY\,}}
\newcommand{\sherpanlo}{{\tt SHERPA\,NLO+PS\,}}
\newcommand{\sherpa}{{\tt SHERPA\,NNLO+PS\,}}
\newcommand{\sherpafo}{{\tt SHERPA-NNLO-FO\,}}
\newcommand{\sanc}{{\tt SANC\,}}
\newcommand{\winhac}{{\tt WINHAC\,}}
\newcommand{\wzgrad}{{\tt WZGRAD\,}}

\def\phist{\ensuremath{\phi^{*}_{\eta}}}
\def\ptl{\ensuremath{p^{l}_\perp\,}}
\def\ptz{\ensuremath{p^{Z}_\perp\,}}
\def\ptv{\ensuremath{p^{V}_\perp}\,}

\begin{document}

\begin{titlepage}

\begin{flushright}
CERN-TH-2016-137\\
CERN-LPCC-2016-002
\end{flushright}

\vspace*{0.2cm}

\begin{center}
{\Large Precision Studies of Observables in $p p \to W \to l\nu_l$  and
$ pp \to \gamma,Z \to l^+ l^-$ processes at the LHC
}

\vspace*{0.2cm}

S.~Alioli$^{a_{1}}$,
A.B.~Arbuzov$^{a_{2},a_{3}}$,
D.Yu.~Bardin$^{a_{2}}$,
L.~Barz\`e$^{a_{4},a_{6}}$,
C.~Bernaciak$^{a_{5}}$,
S.G.~Bondarenko$^{a_{2},a_{3}}$,
C.~Carloni Calame$^{a_{6}}$,
M.~Chiesa$^{a_{6}}$,
S.~Dittmaier$^{a_{7}}$,
G.~Ferrera$^{a_{8}}$,
D.~de Florian$^{a_{9},a_{10}}$,
M.~Grazzini$^{a_{11}}$,
S.~H\"oche$^{a_{12}}$,
A.~Huss$^{a_{13}}$,
S.~Jadach$^{a_{14}}$,
L.V.~Kalinovskaya$^{a_{2}}$,
A.~Karlberg$^{a_{15}}$,
F.~Krauss$^{a_{16}}$,
Y.~Li$^{a_{17}}$,
H.~Martinez$^{a_{4},a_{6}}$,
G.~Montagna$^{a_{4},a_{6}}$,
A.~M\"uck$^{a_{18}}$,
P.~Nason$^{a_{19}}$,
O.~Nicrosini$^{a_{6}}$,
F.~Petriello$^{a_{20},a_{21}}$,
F.~Piccinini$^{a_{6}}$,
W.~P{\l}aczek$^{a_{22}}$,
S.~Prestel$^{a_{12}}$,
E.~Re$^{a_{23}}$,
A.A.~Sapronov$^{a_{2}}$,
M.~Sch\"onherr$^{a_{11}}$,
C.~Schwinn$^{a_{18}}$,
A.~Vicini$^{a_{8}}$,
D.~Wackeroth$^{a_{5},a_{24}}$,
Z.~Was$^{a_{14}}$,
G.~Zanderighi$^{a_{1},a_{15}}$\\
\vskip0.5truecm
\vspace*{2mm}{\sl ${}^{a_{1}}$ TH Division, Physics Department, CERN, CH-1211 Geneva 23, Switzerland}\\
\vspace*{2mm}{\sl ${}^{a_{2}}$
Dzhelepov Laboratory for Nuclear Problems, JINR, Joliot-Curie 6, RU-141980 Dubna, Russia} \\
\vspace*{2mm}{\sl ${}^{a_{3}}$
Bogoliubov Laboratory of Theoretical Physics, JINR, Joliot-Curie 6, RU-141980 Dubna, Russia} \\
\vspace*{2mm}{\sl ${}^{a_{4}}$ Dipartimento di Fisica, Universit\`a di Pavia, 
Via A. Bassi 6, I--27100 Pavia, Italy}\\
\vspace*{2mm}{\sl ${}^{a_{5}}$ Department of Physics, University at Buffalo, The State University of New York, Buffalo, NY 14260, USA}\\
\vspace*{2mm}{\sl ${}^{a_{6}}$ INFN, Sezione di Pavia, Via A. Bassi 6, I--27100 Pavia, Italy}\\
\vspace*{2mm}{\sl ${}^{a_{7}}$ Albert-Ludwigs-University Freiburg, Physikalisches Institut, D-79104 Freiburg, Germany}\\
\vspace*{2mm}{\sl ${}^{a_{8}}$ Tif lab, Dipartimento di Fisica, Universit\`a di Milano and INFN, Sezione di Milano, Via Celoria 16, I--20133 Milano, Italy}\\
\vspace*{2mm}{\sl ${}^{a_{9}}$ Departamento de F\'\i sica and IFIBA, FCEyN, Universidad de Buenos Aires,
(1428) Pabellon 1 Ciudad Universitaria, Capital Federal, Argentina}\\
\vspace*{2mm}{\sl ${}^{a_{10}}$
International Center for Advanced Studies (ICAS), UNSAM, Campus Miguelete, 25 de Mayo y Francia, 
(1650) Buenos Aires, Argentina}\\
\vspace*{2mm}{\sl ${}^{a_{11}}$ 
Physik-Institut, Universit\"at Z\"urich, CH-8057 Z\"urich, Switzerland}\\
\vspace*{2mm}{\sl ${}^{a_{12}}$ SLAC National Accelerator Laboratory, Menlo Park, CA 94025, USA}\\
\vspace*{2mm}{\sl ${}^{a_{13}}$ Institute for Theoretical Physics, ETH, CH-8093 Z\"urich, Switzerland}\\
\vspace*{2mm}{\sl ${}^{a_{14}}$ Institute of Nuclear Physics, Polish Academy of Sciences,\\
  ul.\ Radzikowskiego 152, 31-342 Krakow, Poland}\\
\vspace*{2mm}{\sl ${}^{a_{15}}$  
Rudolf Peierls Centre for Theoretical Physics, 1 Keble Road, University of Oxford, United Kingdom} \\
\vspace*{2mm}{\sl ${}^{a_{16}}$ Institute for Particle Physics Phenomenology, Physics Department,
Durham University, Durham DH1 3LE, United Kingdom}\\
\vspace*{2mm}{\sl ${}^{a_{17}}$ Fermilab, P.O.Box 500, Batavia, IL 60510, USA}\\
\vspace*{2mm}{\sl ${}^{a_{18}}$ Institute for Theoretical Particle Physics and Cosmology, RWTH Aachen University, D-52056 Aachen, Germany}\\
\vspace*{2mm}{\sl ${}^{a_{19}}$ INFN, Sezione di Milano Bicocca, I--20126 Milan, Italy}\\
\vspace*{2mm}{\sl ${}^{a_{20}}$ Department of Physics \& Astronomy, Northwestern University, Evanston, IL 60208, USA}\\
\vspace*{2mm}{\sl ${}^{a_{21}}$ High Energy Physics Division, Argonne National Laboratory, Argonne, IL 60439, USA}\\
\vspace*{2mm}{\sl ${}^{a_{22}}$ Marian Smoluchowski Institute of Physics, Jagiellonian University,\\
         ul.\ {\L}ojasiewicza 11, 30-348 Krakow, Poland}\\
\vspace*{2mm}{\sl ${}^{a_{23}}$ LAPTh, Universit\'e Savoie Mont Blanc, CNRS, B.P.110, \\ Annecy-le-Vieux F-74941, France}\\
\vspace*{2mm}{\sl ${}^{a_{24}}$ Kavli Institute for Theoretical Physics, University of California, Santa Barbara, CA 93106, USA}
\end{center}

\begin{abstract}
\noindent
  This report was prepared in the context of the LPCC {\em Electroweak Precision Measurements at the LHC WG}~\footnote{{\tt https://lpcc.web.cern.ch/lpcc/index.php?page=electroweak\_wg}  } 
  and summarizes the activity of a subgroup dedicated to
  the systematic comparison of public Monte Carlo codes, which describe
  the Drell-Yan processes at hadron colliders, in particular at the
  CERN Large Hadron Collider (LHC).  This work represents an important step towards the
  definition of an accurate simulation framework necessary for very
  high-precision measurements of electroweak (EW) observables such as the
  $W$ boson mass and the weak mixing angle.
  All the codes considered in this report share at least
  next-to-leading-order (NLO) accuracy in the prediction of the total
  cross sections in an expansion either in the strong or in the
  EW coupling constant.  The NLO fixed-order predictions have
  been scrutinized at the technical level, using exactly the same
  inputs, setup and perturbative accuracy, in order to quantify the
  level of agreement of different implementations of the same
  calculation.  A dedicated comparison, again at the technical level,
  of three codes that reach next-to-next-to-leading-order (NNLO) 
  accuracy in quantum chromodynamics (QCD) for the total cross section has also been performed.
  These fixed-order results are a well-defined reference that allows a
  classification of the impact of higher-order sets of radiative
  corrections.  Several examples of higher-order effects due to the
  strong or the EW interaction are discussed in this common
  framework.
  Also the combination of QCD and EW corrections is discussed,
  together with the ambiguities that affect the final result, due to
  the choice of a specific combination recipe.
  All the codes considered in this report have been run by the
  respective authors, and the results presented here constitute a
  benchmark that should be always checked/reproduced before any
  high-precision analysis is conducted based on these codes.
  In order to simplify these benchmarking procedures,
  the codes used in this report, together with the relevant input files and running instructions,
can be found in a repository at 
  {\tt https://twiki.cern.ch/twiki/bin/view/Main/DrellYanComparison}.
\end{abstract}

\end{titlepage}

\tableofcontents

\clearpage

\section{Introduction}
\label{sec:intro}

Precision electroweak (EW) measurements in Drell-Yan-like processes at
the Fermilab Tevatron and CERN Large Hadron Collider (LHC), 
$p p (p \bar p) \to W^\pm \to l^\pm \nu_l$ and 
$ pp (p\bar p)\to \gamma,Z \to l^+ l^-$ ($l=e,\mu$), 
require the development of sophisticated simulation tools 
that should include the best theoretical knowledge available 
(for recent reviews see, e.g., 
\cite{Mangano:2015ejw,Baak:2013fwa,Andersen:2014efa}). 
Several different theoretical effects enter in the accurate evaluation
of total cross sections and kinematic distributions: higher-order QCD
corrections, higher-order EW corrections, the interplay between EW and
QCD effects, matching of fixed-order results with QCD/QED Parton
Showers (PS), tuning of QCD PS to reproduce non-perturbative
low-energy effects, and effects of Parton Distribution Functions (PDF)
and their uncertainties.  The usage of different Monte Carlo (MC)
programs that implement some or all of the above mentioned effects is
not trivial.  

As an explicit example of the need for the best theoretical
predictions, we can consider for instance the measurement of the $W$
boson mass ($\mw$), which is extracted from the transverse mass
distribution of the $l\nu$ pair in $p p (p \bar p) \to W^\pm \to l^\pm
\nu_l$ by means of a template fit to the experimental data.  The
inclusion of different subsets of radiative corrections in the
preparation of the templates modifies the final result of the fit.
Having in mind an accuracy target of ${\cal O}(10$~MeV), it is
important to include the ${\cal O}(\alpha)$ QED final-state radiation
effects which yield a shift of $\mw$ of about 100~MeV-200~MeV (depending
on the precise definition of the final state), but also final-state
multiple photon radiation to all orders, which induces an additional
shift of up to ${\cal O}(-10\%)$ of the \oal ~\cite{CarloniCalame:2003ux}.
One may thus also wonder about the size of the shift in $\mw$ induced
by weak or mixed QCD-EW corrections.  Different subsets of corrections
became available separately in the past years in codes that simulate
purely QCD or purely EW effects.  The combination of QCD and EW
corrections is an important step in the development of the MC programs
that will be used in high-precision measurements and is one of the
main topics of the present report.

The combination of results produced by different
MC simulation codes can be quite difficult and should satisfy some
basic requirements:
\begin{itemize}
\item[1)]
Two codes that have the same perturbative approximation, 
the same input parameters (couplings, masses, PDFs), the same
setup (choice of scales, acceptance cuts), should yield exactly the
same results, within the accuracy of the numerical integration.
\item[2)]
The results of different codes can be meaningfully combined only if
they satisfy the previous point.
\end{itemize}
The size of the mismatches which occur if the first point is not
satisfied may have a larger effect on predictions for EW precision
observables than the anticipated experimental uncertainties. For this
reason it is important to produce a collection of benchmark results for 
total cross sections and kinematic distributions with the most used, publicly
available tools to describe Drell-Yan (DY) processes.  These results should
serve
\begin{itemize}
\item[1)] to verify at any time that a given code works properly according to
what its authors have foreseen, 
\item[2)]
to demonstrate explicitly the level of agreement of different codes
which include identical subsets of radiative corrections, and 
\item[3)]
to expose the impact of different subsets of higher-order corrections and
of differences in their implementations. 
\end{itemize}
In this report, the authors of the MC codes\\ \dynnlo
\cite{Catani:2009sm}, \dynnlops \cite{Karlberg:2014qua}, \fewz
\cite{Gavin:2012sy,Li:2012wna}, \horace \cite{CarloniCalame:2003ux,
  CarloniCalame:2005vc,CarloniCalame:2006zq,CarloniCalame:2007cd},
\photos \cite{Golonka:2005pn}, \powheg \cite{Alioli:2008gx}, \powheg\_{\tt BMNNP}
\cite{Barze:2012tt},
  \powheg\_{\tt BMNNPV} \cite{Barze:2013yca}, \powheg\_{\tt BW} \cite{Bernaciak:2012hj}, \rady
\cite{Dittmaier:2001ay,Dittmaier:2009cr}, \sanc
\cite{Arbuzov:2005dd,Arbuzov:2007db}, \sherpa \cite{Hoeche:2014aia},
\winhac \cite{Placzek:2003zg,Placzek:2009jy,Placzek:2013moa}, and \wzgrad
\cite{Baur:1998kt,Baur:2001ze,Baur:2004ig},\\ 
provide predictions for
a number of observables relevant to the study of charged (CC) and
neutral-current (NC) Drell-Yan processes at the LHC and LHCb~\footnote{For recent $W/Z$ physics results from the LHC see: \\
ATLAS: https://twiki.cern.ch/twiki/bin/view/AtlasPublic/StandardModelPublicResults \\
CMS: https://twiki.cern.ch/twiki/bin/view/CMSPublic/PhysicsResultsSMP \\
LHCb: https://lhcb.web.cern.ch/lhcb/Physics-Results/LHCb-Physics-Results.html}.  Most of
these codes first have been compared, using a common choice of input
parameters, PDFs, renormalization and factorization scales, and
acceptance cuts (tuned comparison), to test the level of technical
agreement 
at leading order (LO), NLO EW and QCD and NNLO QCD, 
before studying the impact of higher-order effects.

The report is structured as follows: 
In Section~\ref{sec:setup} we describe the common setup for the tuned comparison and the observables under study in this report. 
The choice of observables was guided by the relevance to the study of Drell-Yan processes at the LHC, in particular to a precise measurement of the $W$ boson mass.  
In Sections~\ref{sec:tuning_xsec} and \ref{sec:tuning_distri} we present
the results of the tuned comparison at NLO: 
in Section~\ref{sec:tuning_xsec} we show the predictions of NLO-EW and
NLO-QCD total cross sections, and in Section~\ref{sec:tuning_distri}
we show the results at NLO EW and NLO QCD for a sample of kinematic
distributions listed in Section~\ref{sec:setup}.

In Section~\ref{sec:impact} we discuss the impact of higher-order QCD
and EW corrections, i.e. corrections beyond NLO accuracy, on a
selected set of $W$ and $Z$ boson observables.  For each code used in
this study we consider all the subsets of available corrections which
are beyond NLO.  To compute the results presented in this section, we
adopted an EW input scheme, described in Section~\ref{sec:setupbest},
which absorbs known higher-order corrections already in the (N)LO
  predictions, thus minimalizing the impact of neglected orders in
  perturbation theory.
All results obtained in this {\em benchmark} setup can serve as a
benchmark for future studies.  For completeness we provide the results
for the total cross sections at NLO EW and NLO QCD obtained in this
{\em benchmark} setup in Section~\ref{sec:total-xsec-best}.  
In Section~\ref{sec:impact-qcd} we discuss the effects of purely QCD
corrections: 
after a short introduction in Section~\ref{sec:NLO-QCD}
on the impact of the ${\cal O}(\alpha_s)$ corrections on the
observables under study, we consider in Sections~\ref{sec:NNLO-QCD}
and \ref{sec:NNLO-QCD-diff}
exact results at ${\cal O}(\alpha_s^2)$ respectively for the total cross sections and for some differential distributions;
in Section~\ref{sec:QCDallorders} we briefly introduce the problem of matching fixed- and all-order results in perturbation theory; 
we present results of (NLO+PS)-QCD matching in Section \ref{sec:matching-NLOPS-QCD}
and of (NNLO+PS)-QCD matching in Section \ref{sec:nnlops}.
In
Section~\ref{sec:impact-ew} we discuss the effects of purely EW
corrections: after a short introduction in Section~\ref{sec:NLO-EW} on
the role of the ${\cal O}(\alpha)$ corrections on the observables
under study, we compare in Section~\ref{sec:gammainduced} the
predictions for the partonic subprocesses induced by photons, which
are naturally part of the NLO EW results. We discuss different EW
input scheme choices in Section~\ref{sec:renormalization} and the
impact of different gauge boson mass definitions in
Section~\ref{sec:massdef}.  In
Sections~\ref{sec:renormalization-ho-universal},~\ref{sec:renormalization-ho-running},~\ref{sec:QEDshowers},
we describe respectively the impact of higher-order corrections
introduced via the $\rho$ parameter or via the definition of effective
couplings or due to multiple photon radiation described with a QED PS
properly matched to the NLO EW calculation.  The effect of light
fermion-pair emission is discussed in Section~\ref{sec:fermionpair}.
 
In Section~\ref{sec:interplay} we consider the combination of QCD and
EW corrections and discuss some possibilities which are allowed by our
presently incomplete knowledge of the ${\cal O}(\alpha\alpha_s)$
corrections to the DY processes.  
In Section \ref{sec:combi} we compare the results that can be obtained with the codes presently available and discuss the origin of the observed differences. 
In Section~\ref{sec:approximations} the results of a first calculation of ${\cal O}(\alpha\alpha_s)$
corrections in the pole approximation are used to assess the validity
of simple prescriptions for the combination of EW and QCD corrections.

In Appendix~\ref{app:codes} we provide a short description of the MC
codes used in this study.  In Appendix~\ref{app:lhcb} we present a
tuned comparison of the total cross sections at NLO EW and NLO QCD for
$W^\pm$ and $Z$ production with LHCb cuts.

\subsection{Reproducibility of the results: a repository of the codes used in this report}

The goal of this report is to provide a quantitative assessment of the
technical level of agreement of different codes, but also a
classification of the size of higher-order radiative corrections.

The usage of modern MC programs is quite complex and it is not trivial
to judge whether the numerical results ``out-of-the-box'' of a code
are correct.  The numbers presented here, computed by the respective
authors, should be considered as benchmarks of the codes; every user
should thus be able to reproduce them, provided that he/she uses the
same inputs and setup and runs with the appropriate amount of
statistics.

In order to guarantee the reproducibility of the results presented in
this report, we prepared a repository that contains a copy of all the
MC codes used in this study, together with the necessary input files
and the relevant instructions to run them.
The repository can be found at the following URL:\\
{\tt https://twiki.cern.ch/twiki/bin/view/Main/DrellYanComparison}\\
It should be stressed that simulation codes may evolve in time,
because of improvements but also of bug fixes.

\clearpage
\section{Tuned comparison of the codes}
\label{sec:tuned}

\subsection{Setup for the tuned comparison}
\label{sec:setup}
For the numerical evaluation of the cross sections at the LHC
($\sqrt{s}=8$ TeV) we choose the following set of Standard Model input
parameters~\cite{Beringer:1900zz}:
\begin{eqnarray}\label{eq:pars}
G_{\mu} = 1.1663787\times 10^{-5} \; {\rm GeV}^{-2}, 
& \qquad & \alpha= 1/137.035999074, \quad \alpha_s\equiv\alpha_s(M_Z^2)=0.12018
\nonumber \\ 
M_Z = 91.1876 \; {\rm GeV}, & \quad & \Gamma_Z =  2.4952  \; {\rm GeV}
\nonumber  \\ 
M_W = 80.385 \; {\rm GeV}, & \quad & \Gamma_W = 2.085 \; {\rm GeV}
\nonumber  \\
M_H = 125 \; {\rm GeV}, & \quad & 
\nonumber  \\
m_e  = 0.510998928 \; {\rm MeV}, &\quad &m_{\mu}=0.1056583715 \; {\rm GeV},  
\quad m_{\tau}=1.77682 \; {\rm GeV}
\nonumber  \\
m_u=0.06983 \; {\rm GeV}, & \quad & m_c=1.2 \; {\rm GeV}, 
\quad m_t=173.5 \; {\rm GeV}
\nonumber  \\
m_d=0.06984 \; {\rm GeV}, & \quad & m_s=0.15 \; {\rm GeV}, \quad m_b=4.6 \; {\rm GeV} 
\nonumber \\
|V_{ud}| = 0.975, & \quad & |V_{us}| = 0.222 
\nonumber \\
|V_{cd}| = 0.222, & \quad & |V_{cs}| = 0.975 
\nonumber \\
|V_{cb}|=|V_{ts}|=|V_{ub}|& =& |V_{td}|= |V_{tb}|=0  \; .
\end{eqnarray}
We work in the constant width scheme and fix the weak mixing angle by
$c_w=M_W/M_Z$, $s_w^2=1-c_w^2$.  The $Z$ and $W$ boson decay widths
given above are used in the LO, NLO and NNLO evaluations of the cross
sections. The fermion masses only enter through loop contributions to
the vector-boson self energies and as regulators of the collinear
singularities which arise in the calculation of the QED
contribution. The light quark masses are chosen in such a way, that
the value for the hadronic five-flavour contribution to the photon
vacuum polarization, $\Delta
\alpha_{had}^{(5)}(M_Z^2)=0.027572$~\cite{Jegerlehner:2001wq}, is
recovered, which is derived from low-energy $e^+ e^-$ data with the
help of dispersion relations.

To compute the hadronic cross section we use the
MSTW2008~\cite{Martin:2009iq} set of parton distribution functions,
and take the renormalization scale, $\mu_r$, and the QCD factorization
scale, $\mu_{\rm QCD}$, to be the invariant mass of the final-state lepton pair, i.e. 
$\mu_r=\mu_{\rm QCD}=M_{l\nu}$ in the
$W$ boson case and $\mu_r=\mu_{\rm QCD}=M_{l^+l^-}$ in the $Z$ boson
case.

All numerical evaluations of EW corrections require the subtraction of
QED initial-state collinear divergences, which is performed using the
QED DIS scheme.  It is defined analogously to the usual
DIS~\cite{Owens:1992hd} scheme used in QCD calculations, i.e.  by
requiring the same expression for the leading and next-to-leading
order structure function $F_2$ in deep inelastic scattering, which is
given by the sum of the quark distributions. Since $F_2$ data are an
important ingredient in extracting PDFs, the effect of the ${\cal
  O}(\alpha)$ QED corrections on the PDFs should be reduced in the QED
DIS scheme. The QED factorization scale is chosen to be equal to the
QCD factorization scale, $\mu_{QED}=\mu_{QCD}$. The QCD factorization
is performed in the $\overline{{\rm MS}}$ scheme.  The subtraction of
the QED initial state collinear divergences is a necessary step to
obtain a finite partonic cross section. The absence of a QED evolution
in the PDF set MSTW2008 has little phenomenological impact on the
kinematic distributions as discussed in
Section~\ref{sec:gammainduced}. However, to be consistent in the order
of higher order corrections in a {\em best} EW prediction, modern PDFs
which include QED corrections, such as
NNPDF2.3QED~\cite{Ball:2013hta} and CT14QED~\cite{Schmidt:2015zda}, 
should be used.

For NLO EW predictions, we work in the on-shell renormalization scheme
and use the following $Z$ and $W$ mass renormalization constants:
\begin{equation}
\label{renorm}
\delta M_Z^2 = {\cal R}e \Sigma^{Z}(M_Z^2), \quad
\delta M_W^2 = {\cal R}e \Sigma^{W}(M_W^2) \; ,
\end{equation}
where $\Sigma^V$ denotes the transverse part of the
unrenormalized vector-boson self energy.  

For the sake of simplicity and to avoid additional sources of
discrepancies in the tuned comparison we use the fine-structure
constant $\alpha(0)$ throughout in both the calculation of CC and NC
cross sections.  We will discuss different EW input schemes in
Section~\ref{sec:renormalization}.

In the course of the calculation of radiative corrections to $W$ boson
observables the Kobayashi-Maskawa mixing has been neglected, but the
final result for each parton level process has been multiplied with
the square of the corresponding physical matrix element $V_{ij}$.
From a numerical point of view, this procedure does not significantly
differ from a consideration of the Kobayashi-Maskawa matrix in the
renormalisation procedure as it has been pointed out
in~\cite{Denner:1990yz}.

We choose to evaluate the running of the strong coupling constant at
the two-loop level, with five flavours, for LO, NLO and NLO+PS
predictions using as reference value
$\alpha_s^{NLO}(M_{\scriptscriptstyle Z})=0.12018$, which is
consistent with the choice made in the NLO PDF set of MSTW2008. NNLO QCD
predictions use the NNLO PDF set and correspondingly the three-loop
running of $\alpha_s(\mu_r)$, with reference value
$\alpha_s^{NNLO}(M_{\scriptscriptstyle Z})=0.117$.  In
Table~\ref{tab:alphas} we provide $\alpha_s(\mu_r^2)$ for several
choices of the QCD renormalization scale $\mu_r$, which are consistent
with the results provided by the LHAPDF function {\tt
  alphasPDF($\mu_r$)} when called in conjunction with MSTW2008.

\begin{table}
\centering{
\caption{Two-loop and three-loop running of $\alpha_s(\mu_r^2)$. } 
\label{tab:alphas}
\vskip 5.mm
\begin{tabular}{ccc} \hline
$\mu_r$ [GeV]& $\alpha_s$(NLO) & $\alpha_s$(NNLO) \\
 91.1876 & 0.1201789 & 0.1170699 \\
 50          & 0.1324396 & 0.1286845 \\  
100         & 0.1184991 & 0.1154741 \\
200         & 0.1072627 & 0.1047716 \\
500         & 0.0953625 & 0.0933828 \\
\end{tabular}}
\end{table}

The detector acceptance is simulated by imposing the following
transverse momentum ($p_\perp$) and pseudo-rapidity ($\eta$) cuts:
\begin{eqnarray}
{\rm LHC:} &&\, p_\perp^\ell>25~{\rm GeV,} \;~~~~ |\eta(\ell)|<2.5, \; p_\perp^\nu>25~{\rm GeV,} \qquad
\ell=e,\,\mu , 
\nonumber \\
{\rm LHCb:} && \, p_\perp^\ell>20~{\rm GeV,} \; 2<\eta(\ell)<4.5, \; p_\perp^\nu>20~{\rm GeV,} \qquad
\ell=e,\,\mu \; , 
\label{eq:lepcuts}
\end{eqnarray}
where $p_\perp^\nu$ is the missing transverse momentum originating
from the neutrino.  These cuts approximately model the acceptance of
the ATLAS, CMS, and LHCb detectors at the LHC.  In addition to the
separation cuts of Eq.~\ref{eq:lepcuts} we apply a cut on the
invariant mass of the final-state lepton pair of $M_{l^+l^-} > 50$~GeV
and $M(l\nu)>1$~GeV in the case of $\gamma/Z$ production and $W$
production respectively,

Results are provided for the {\em bare} setup, i.e. when only
applying the acceptance cuts of Eq.~\ref{eq:lepcuts}, and the {\em
  calo} setup, which is defined as follows: In addition to the
acceptance cuts, for muons we require that the energy of the photon is
$E_{\gamma}<2$~GeV for $\Delta R(\mu,\gamma)<0.1$.  For electrons we
first recombine the four-momentum vectors of the electron and photon
to an effective electron four-momentum vector when $\Delta
R(e,\gamma)<0.1$ and then apply the acceptance cuts to the recombined
momenta. For both electrons and muons we reject the event for
$E_\gamma > 0.1 \, E_{\mu,e}$ for $0.1 \le \Delta R(e,\gamma) \le
0.4$, where
\[\Delta R(l,\gamma)= \sqrt{(\Phi_l-\Phi_\gamma)^2+(\eta_l-\eta_\gamma)^2} \; .\]
We summarize the lepton identification
requirements in the {\em calo} setup in Table~\ref{tab:th_ewk_c}.

Since we consider predictions inclusive with respect to QCD radiation,
we do not impose any jet definition.

We use the Pythia version 6.4.26, Perugia tune (PYTUNE(320)).  When
producing NLO QCD+EW results with Pythia, the QED showering effects
are switched off by setting {\tt MSTJ(41)=MSTP(61)=MSTP(71)=1}.
\begin{table}
\begin{center}
\begin{tabular}{|c|c|} \hline
\multicolumn{1}{|c|}{electrons} & \multicolumn{1}{|c|}{muons} \\
\hline
combine $e$ and $\gamma$ momentum four vectors, & reject events with 
$E_\gamma>2$~GeV \\ 
if $\Delta R(e,\gamma)<0.1$ & for $\Delta R(\mu,\gamma)<0.1$ \\
\hline
reject events with 
$E_\gamma>0.1~E_e$ & reject events with 
$E_\gamma>0.1~E_\mu$ \\  
for $0.1<\Delta R(e,\gamma)<0.4$ & for $0.1<\Delta R(\mu,\gamma)<0.4$ \\ \hline  
\end{tabular}
\caption{Summary of lepton identification requirements in the {\em calo} setup. } 
\label{tab:th_ewk_c}
\end{center}
\end{table}

In the following we list the observables considered in this study for
charged (CC) and neutral current (NC) processes: $pp \to W^\pm \to
l^\pm \nu_l$ and $pp \to \gamma,Z \to l^+ l^-$ with $l=e,\mu$.

\noindent
\subsubsection{$W$ boson observables}
\begin{itemize}
\item
$\sigma_W$:
total inclusive cross section of $W$ boson production.
\item
$\frac{d\sigma}{dM_\perp(l\nu)}$: transverse mass distribution of the lepton lepton-neutrino pair.
The transverse mass is defined as
\begin{equation}
M_\perp=\sqrt{2p_\perp^\ell p_\perp^\nu (1-\cos\phi^{\ell\nu})} \; ,
\label{eq:mt}
\end{equation}
where $p_\perp^\nu$ is the transverse momentum of the neutrino, and
$\phi^{\ell\nu}$ is the angle between the 
charged lepton and the neutrino in the transverse plane. 
\item
$\frac{d\sigma}{d p_\perp^l}$: charged lepton transverse momentum distribution.
\item
$\frac{d\sigma}{d p_\perp^\nu}$: missing transverse momentum distribution.
\item
$d\sigma_W/d p_\perp^W$:
lepton-pair ($W$) transverse momentum distribution.
\end{itemize}

\noindent
\subsubsection{$Z$ boson observables}
\begin{itemize}
\item
$\sigma_Z$:
total inclusive cross section of $Z$ boson production.
\item
$\frac{d\sigma}{dM_{l^+l^-}}$: invariant mass distribution of the lepton pair.
\item
$\frac{d\sigma}{dp_\perp^l}$: transverse lepton momentum distribution ($l$ is the positively charged lepton).
\item
$d\sigma_Z/dp_\perp^Z$:
lepton-pair ($Z$) transverse momentum distribution.
\end{itemize}

Finally, for the case of $Z$ boson production we add the distribution
in $\phi^*$ to our list of observables. This observable is
defined, e.g., in Ref.~\cite{Banfi:2010cf} as follows:
\[\phi^*=\tan\left(\frac{\pi-\Delta \Phi}{2}\right)\sin(\theta^*_\eta) \; , \]
with $\Delta \Phi=\Phi^- -\Phi^+$ denoting the difference in the
azimuthal angle of the two negatively/positively charged leptons in
the laboratory frame, and
\[\cos(\theta^*_\eta)=\tanh\left(\frac{\eta^- -\eta^+}{2}\right) \; .\]
$\eta^{\pm}$ denote the pseudo rapidity of the negatively/positively
charged lepton.


\clearpage

\subsection{Tuned comparison of total cross sections at NLO EW and NLO QCD with ATLAS/CMS cuts}
\label{sec:tuning_xsec}
In Sections~\ref{sec:wtot} and \ref{sec:ztot} we provide a tuned
comparison of the total cross sections computed at fixed order, namely
LO, NLO EW and NLO QCD, using the setup of Section~\ref{sec:setup} for
the choice of input parameters and ATLAS/CMS acceptance cuts.

All codes can provide LO results, but different codes may include
different sets of higher-order corrections.  We use the symbol
$\times$ in the tables to indicate that a particular correction is not
available in the specified code.  Note that even when working at the
same, fixed order and using the same setup, there can be slight
differences in the implementation of higher-order corrections,
resulting in small numerical differences in the predictions of
different codes.

In Tables 
\ref{tab:xsec-wp-lhc8-atlcms-bare},
\ref{tab:xsec-wm-lhc8-atlcms-bare}, and
\ref{tab:xsec-z-lhc8-atlcms-bare},
we present the results obtained in the {\em bare} treatment of real
photon radiation.  The photon-lepton recombination procedure described
in Section~\ref{sec:setup}, which is only relevant for the codes that include NLO EW
corrections, modifies the total cross section, as shown in Tables
\ref{tab:xsec-wp-lhc8-atlcms-rec},
\ref{tab:xsec-wm-lhc8-atlcms-rec}, and
\ref{tab:xsec-z-lhc8-atlcms-rec}.
 
The total cross section results computed with  
LHCb acceptance cuts can be found in Appendix~\ref{app:lhcb}.

\subsubsection{Results for $W^\pm$ boson production}
\label{sec:wtot}

\begin{table}[h]
\begin{center}
\begin{tabular}{|c|l|l|l|l|}
\hline
       & LO          & NLO       & NLO        &  NLO        \\ 
code   &             & QCD       & EW $\mu$   & EW $e$      \\ 
\hline
HORACE &  2897.38(8) & $\times$          & 2988.2(1)  & 2915.3(1)   \\
\hline
WZGRAD &  2897.33(2) & $\times$           & 2987.94(5) & 2915.39(6)  \\
\hline
RADY   &  2897.35(2) & 2899.2(4) & 2988.01(4) & 2915.38(3)    \\
\hline
SANC   &  2897.30(2) & 2899.9(3) & 2987.77(3) & 2915.00(3)   \\
\hline 
DYNNLO &  2897.32(5) & 2899(1)   & $\times$  & $\times$  \\
\hline
FEWZ       & 2897.2(1)& 2899.4(3) & $\times$ & $\times$  \\
\hline
POWHEG-w & 2897.34(4)  & 2899.41(9)& $\times$ & $\times$ \\
\hline
POWHEG\_BMNNP & 2897.36(5)& 2899.0(1)  & 2988.4(2)  & 2915.7(1) \\
\hline
POWHEG\_BW & 2897.4(1) & 2899.2(3)  & 2987.7(4) &    ($\times$) \\
\hline
\end{tabular}
\caption{\label{tab:xsec-wp-lhc8-atlcms-bare} Tuned comparison of
  total cross sections (in pb) for $p p \to W^+\to l^+ \nu_l+X$ at the
  8 TeV LHC, with ATLAS/CMS cuts and {\em bare} leptons. ($\times$)
  indicates that although {\tt POWHEG\_BW} provides NLO EW results also for
  {\em bare} electrons, due to the smallness of the electron mass
  it would require very high-statistics to obtain per-mille level
  precision. Thus, we recommend to use the {\em bare} setup in {\tt POWHEG\_BW}
  only for muons.  }
\end{center}
\end{table}

\begin{table}[h]
\begin{center}
\begin{tabular}{|c|c|c|c|}
\hline
         & LO         & NLO EW $\mu$ calo & NLO EW $e$ calo  \\ 
code     &            &                                   &  \\ 
\hline
HORACE   & 2897.38(8) & 2899.0(1)  & 3003.5(1) \\
\hline
WZGRAD   & 2897.33(2) & 2898.33(5) & 3003.33(6) \\
\hline
RADY     & 2897.35(2)           & 2898.37(4) & 3003.36(4)  \\
\hline
SANC     & 2897.30(2) & 2898.18(3) & 3003.00(4)  \\
\hline
\end{tabular}
\caption{\label{tab:xsec-wp-lhc8-atlcms-rec} Tuned comparison of total
  cross sections (in pb) for $p p \to W^+\to l^+ \nu_l+X$ at the 8 TeV
  LHC, with ATLAS/CMS cuts and {\em calorimetric} leptons.}
\end{center}
\end{table}

\begin{table}[h]
\begin{center}
\begin{tabular}{|c|l|l|l|l|}
\hline
         & LO              & NLO        & NLO             & NLO   \\ 
code     &                 & QCD        & EW $\mu$        & EW $e$   \\ 
\hline
HORACE   & 2008.84(5) & $\times$          & 2076.48(9) & 2029.15(8) \\
\hline
WZGRAD   & 2008.95(1) & $\times$          & 2076.51(3) & 2029.26(3) \\
\hline
RADY     & 2008.93(1) & 2050.5(2) & 2076.62(2) & 2029.29(2)  \\
\hline
SANC     & 2008.926(8)& 2050.3(3) &2076.56(2)  & 2029.19(3)  \\
\hline 
DYNNLO   & 2008.89(3) & 2050.2(9) & $\times$ & $\times$  \\
\hline
FEWZ     & 2008.9(1)  & 2049.97(8) & $\times$ &  $\times$ \\
\hline
POWHEG-w & 2008.93(3) & 2050.14(5) & $\times$ & $\times$  \\
\hline
POWHEG\_BMNNP &2008.94(3) & 2049.9(1) & 2076.9(1) & 2029.71(6) \\
\hline
POWHEG\_BW & 2009.2(4) & 2050.2(4) & 2076.0(3) &  ($\times$)         \\
\hline
\end{tabular}
\caption{\label{tab:xsec-wm-lhc8-atlcms-bare} Tuned comparison of
  total cross sections (in pb) for $p p \to W^-\to l^- \bar\nu_l+X$ at
  the 8 TeV LHC, with ATLAS/CMS cuts and {\em bare}
  leptons. ($\times$) indicates that although {\tt POWHEG\_BW} provides NLO
  EW results also for {\em bare} electrons, due to the smallness of
  the electron mass it would require very high-statistics to obtain
  per-mille level precision. Thus, we recommend to use the {\em bare}
  setup in {\tt POWHEG\_BW} only for muons. }
\end{center}
\end{table}

\begin{table}[h]
\begin{center}
\begin{tabular}{|c|l|l|l|}
\hline
         & LO         & NLO EW $\mu$ calo & NLO EW $e$ calo  \\ 
code     &            &                   &     \\ 
\hline
HORACE   & 2008.84(5) & 2013.67(7) & 2085.42(8) \\
\hline
WZGRAD   & 2008.95(1) & 2013.42(3) & 2085.26(3)  \\
\hline
RADY     & 2008.93(1)           & 2013.49(2) & 2085.37(2)  \\
\hline 
SANC     & 2008.926(8) & 2013.48(2) & 2085.24(4)  \\
\hline
\end{tabular}
\caption{\label{tab:xsec-wm-lhc8-atlcms-rec}Tuned comparison of total
  cross sections (in pb) for $p p \to W^-\to l^- \bar\nu_l+X$ at the 8
  TeV LHC, with ATLAS/CMS cuts and {\em calorimetric} leptons.}
\end{center}
\end{table}

\clearpage
\subsubsection{Results for $Z$ boson production }
\label{sec:ztot}

\begin{table}[h]
\begin{center}
\begin{tabular}{|c|l|l|l|l|}
\hline
        & LO         & NLO       & NLO      & NLO        \\ 
code    &            & QCD       & EW $\mu$ & EW $e$   \\
\hline
HORACE  & 431.033(9) & $\times$  & 438.74(2)  & 422.08(2) \\
\hline
WZGRAD  & 431.048(7) & $\times$  & 439.166(6) & 422.78(1)   \\
\hline
RADY    & 431.047(4) & 458.16(3) & 438.963(4) & 422.536(5) \\
\hline
SANC    & 431.050(2) & 458.27(3) & 439.004(5) & 422.56(1)  \\
\hline  
DYNNLO  & 431.043(8) & 458.2(2)  &  $\times$          &  $\times$        \\
\hline
FEWZ    & 431.00(1)  & 458.13(2)     & ($\times$)  &  ($\times$) \\
\hline
POWHEG-z &  431.08(4) & 458.19(8)& $\times$ &$\times$  \\
\hline
POWHEG\_BMNNPV & 431.046(9) & 458.16(7) & 438.9(1) & 422.2(2)  \\
\hline
\end{tabular}
\caption{\label{tab:xsec-z-lhc8-atlcms-bare} Tuned comparison of total
  cross sections (in pb) for $p p \to \gamma,Z \to l^- l^++X$ at the 8
  TeV LHC, with ATLAS/CMS cuts and {\em bare} leptons. ($\times$)
  indicates that {\tt FEWZ} provides NLO EW results only in the $G_\mu$
  scheme, and thus no results are available for the setup of the tuned
  comparison (see Section~\ref{sec:setup}).}
\end{center}
\end{table}

\begin{table}[h]
\begin{center}
\begin{tabular}{|c|l|l|l|}
\hline
         & LO             & NLO EW $\mu$ calo & NLO EW $e$ calo  \\ 
code     &                &                   &                  \\ 
\hline
HORACE   & 431.033(9) & 407.67(1)  & 439.68(2) \\
\hline
WZGRAD   &  431.048(7)          & 407.852(7) & 440.29(1) \\
\hline
RADY     & 431.047(4)           & 407.568(6)           & 440.064(5)  \\
\hline
SANC     & 431.050(2)           & 407.687(5) & 440.09(1)  \\
\hline
\end{tabular}
\caption{\label{tab:xsec-z-lhc8-atlcms-rec} Tuned comparison of total
  cross sections (in pb) for $p p \to \gamma , Z\to l^+ l^-+X$ at the
  8 TeV LHC, with ATLAS/CMS cuts and {\em calorimetric} leptons.}

\end{center}
\end{table}

\clearpage

\clearpage

\subsection{Tuned comparison of kinematic distributions at NLO EW and NLO QCD with ATLAS/CMS cuts}
\label{sec:tuning_distri}

In Sections~\ref{sec:wpmcomp},\ref{sec:zcomp} we provide a sample of
the kinematic observables calculated for this report including either
NLO EW or NLO QCD corrections.  The results have been obtained with
different codes, using exactly the same setup as described in
Section~\ref{sec:setup}.  While in earlier
studies~\cite{Buttar:2006zd,Gerber:2007xk}~\footnote{See also a recent study in Ref.~\cite{Badger:2016bpw}.} relative corrections have
been compared, i.e. predictions for NLO/LO ratios of different codes,
we expose here any effects of slight differences in the implementation
of these corrections by comparing the ratios of different NLO EW and
NLO QCD predictions to {\tt HORACE} and {\tt POWHEG}, respectively.  Although
technically the codes under consideration calculate the same quantity,
in practice there are different possible ways to implement these
higher-order corrections in a Monte Carlo integration code, which may
result in ratios slightly different from one.  This tuned comparison
is thus a non-trivial test of these different implementations. The
observed differences can be interpreted as a technical limit of
agreement one can reach, and thus as a lower limit on the theoretical
uncertainty.

The corresponding total cross sections can be found in
Section~\ref{sec:tuning_xsec}.

It is important to note that NLO QCD is not sufficient for the
description of certain observables and kinematic regimes where the
resummation of logarithmic enhanced contributions and/or the inclusion
of NNLO corrections is required, as discussed in detail in
Section~\ref{sec:impact-qcd}. In these cases, the NLO QCD results
presented in this section are only used for technical checks.

\subsubsection{Tuned comparison of $W^\pm$ boson observables}
\label{sec:wpmcomp}

\begin{figure}[h]
\centering
\includegraphics[width=75mm,angle=0]{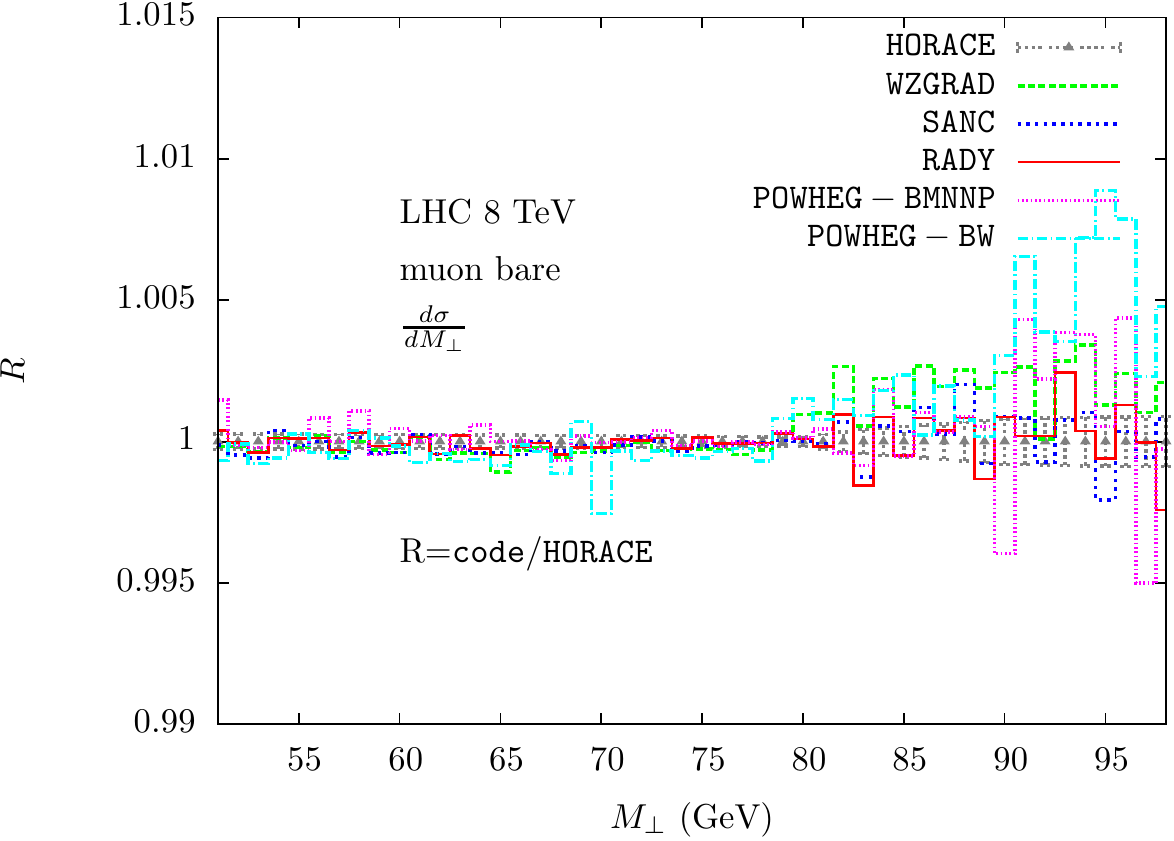}
\includegraphics[width=75mm,angle=0]{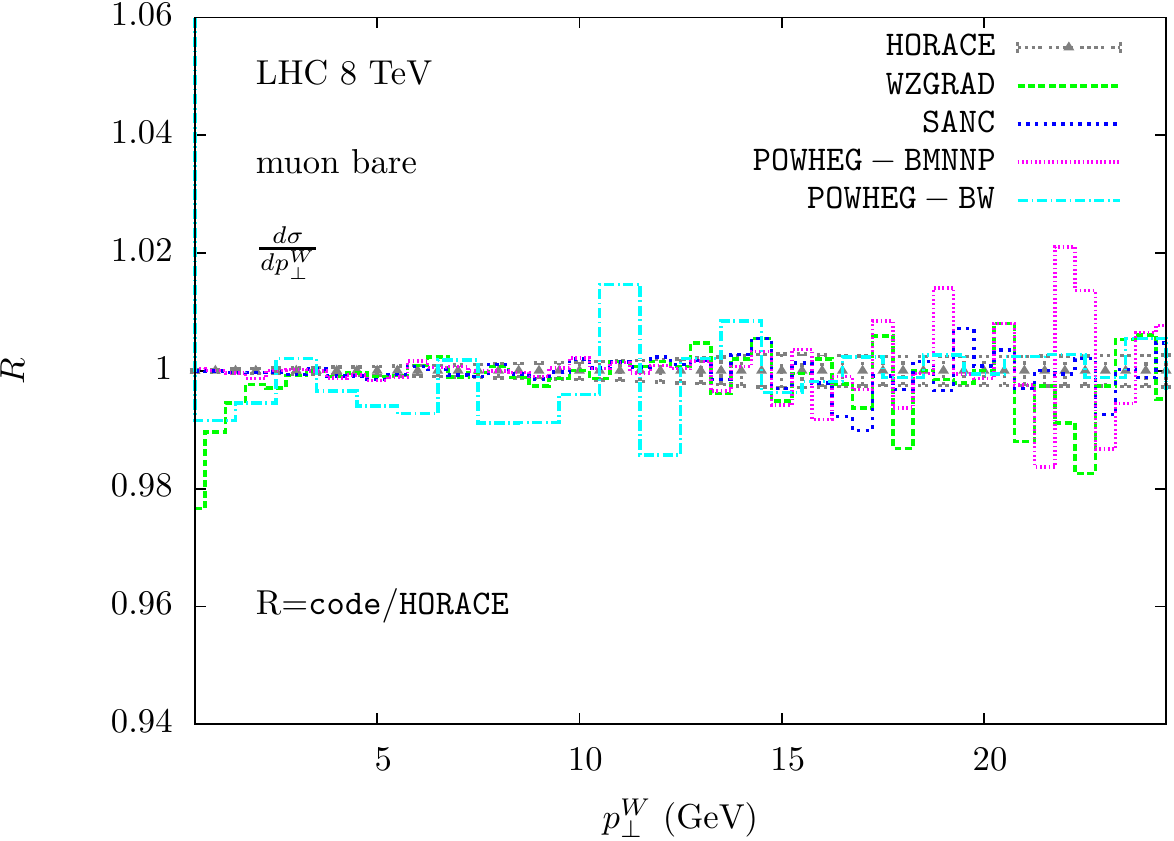}
\caption{Tuned comparison of the lepton-pair transverse mass and
  transverse momentum distributions in $pp\to W^+ \to \mu^+\nu_\mu+X$
  at the 8~TeV LHC with ATLAS/CMS cuts in the {\em bare} setup,
  including NLO EW corrections.}\label{fig:one}
\end{figure}

\begin{figure}[h]
\includegraphics[width=75mm,angle=0]{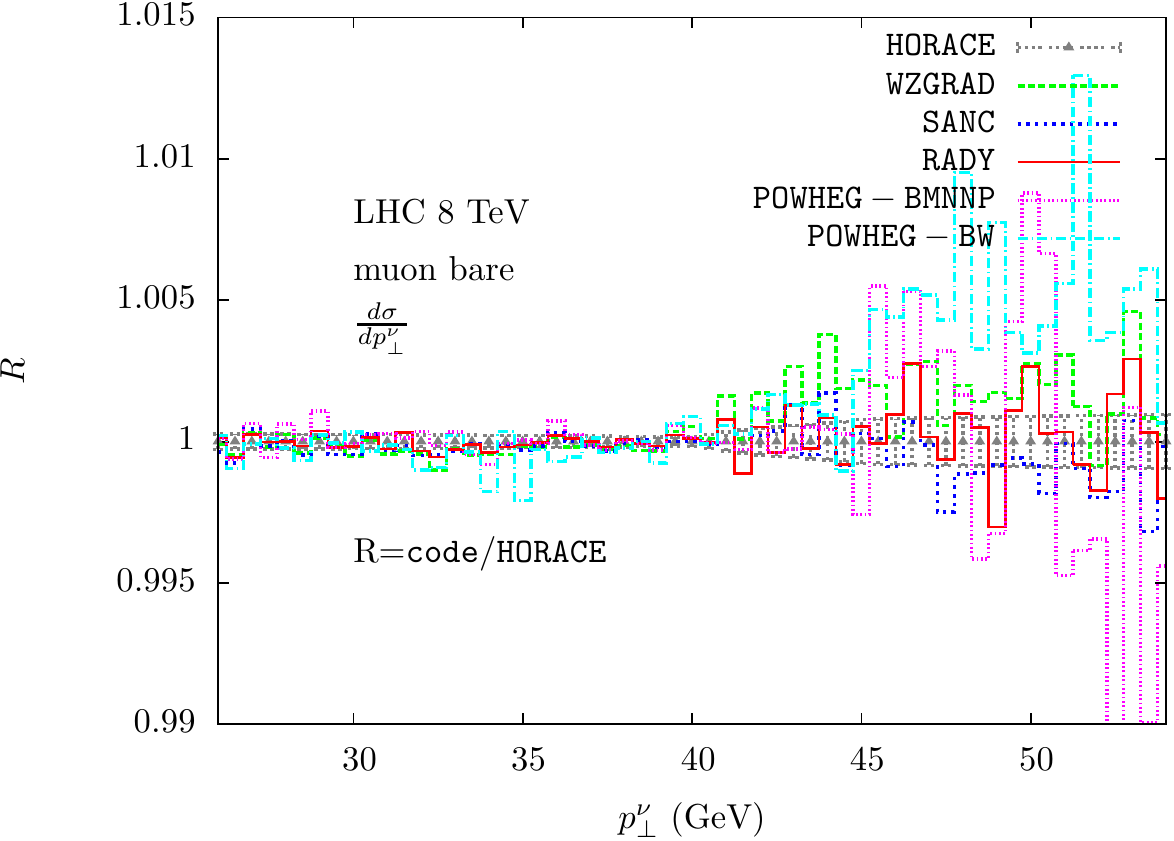}
\includegraphics[width=75mm,angle=0]{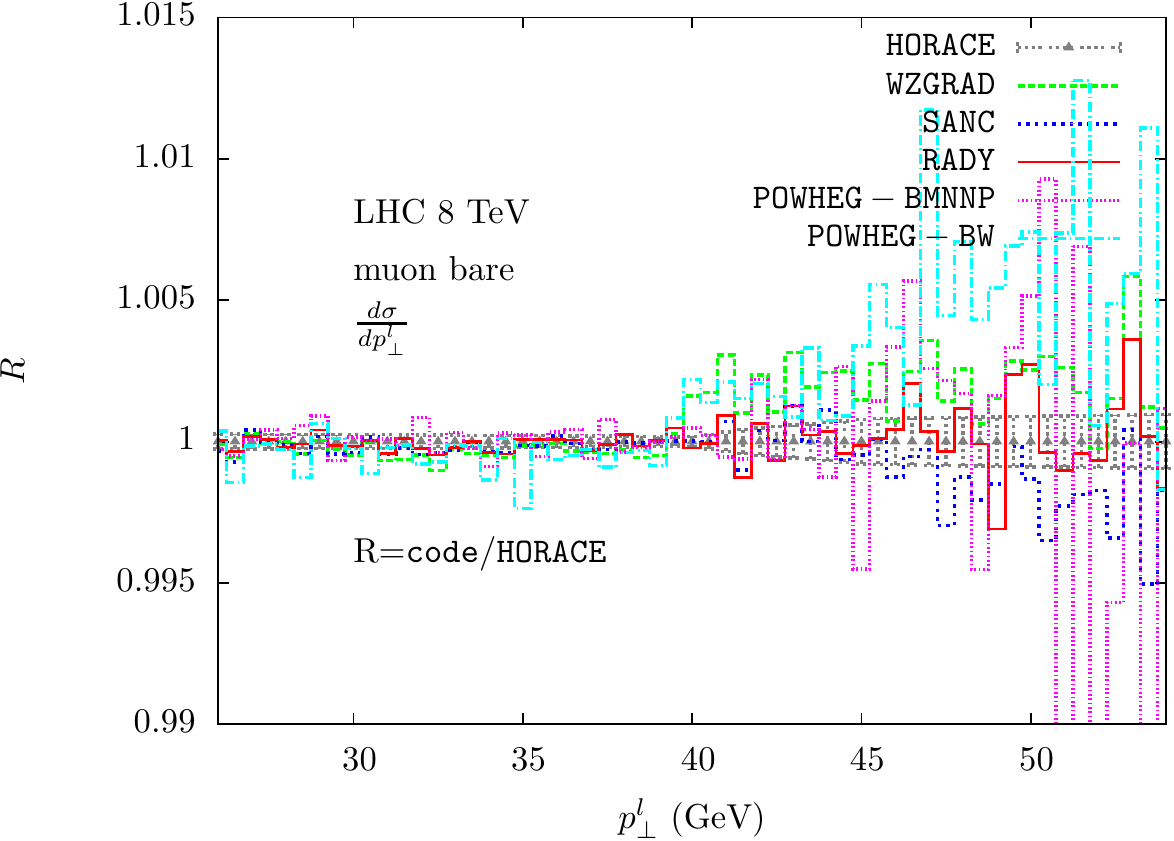}
\caption{Tuned comparison of the muon and muon neutrino transverse
  momentum distributions in $pp\to W^+ \to \mu^+\nu_\mu+X$ at the
  8~TeV LHC with ATLAS/CMS cuts in the {\em bare} setup, including NLO
  EW corrections.}\label{fig:two}
\end{figure}

\begin{figure}[h]
\centering
\includegraphics[width=75mm,angle=0]{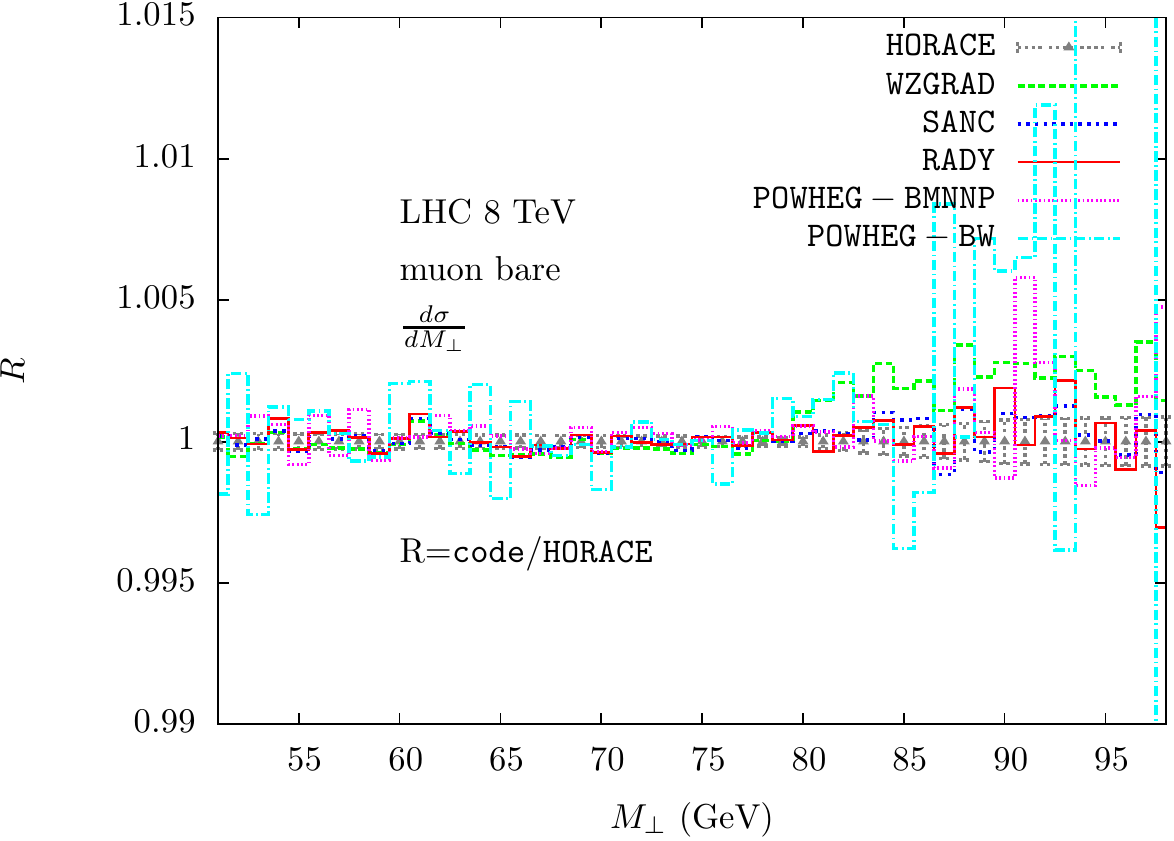}
\includegraphics[width=75mm,angle=0]{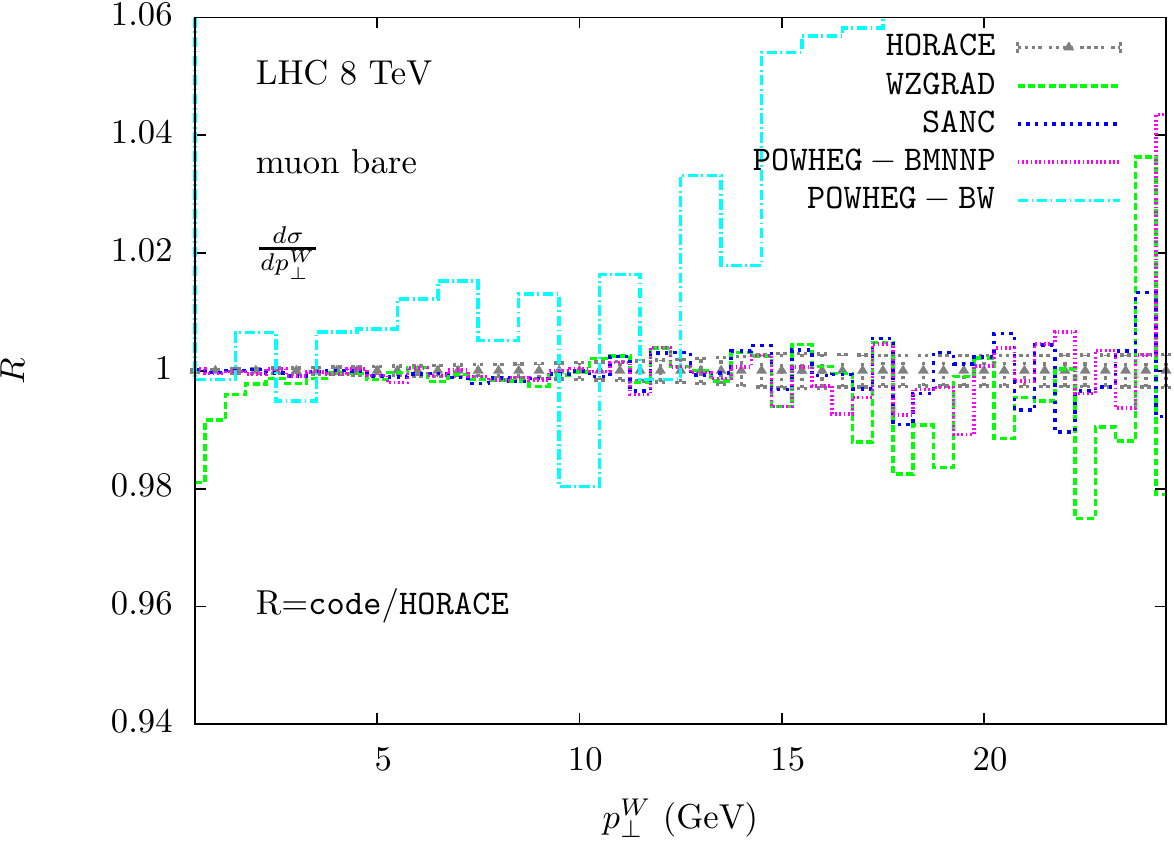}
\caption{Tuned comparison of the lepton-pair transverse mass and
  transverse momentum distributions in $pp\to W^- \to \mu^-\nu_\mu+X$
  at the 8~TeV LHC with ATLAS/CMS cuts in the {\em bare} setup,
  including NLO EW corrections.}\label{fig:three}
\end{figure}

\begin{figure}[h]
\includegraphics[width=75mm,angle=0]{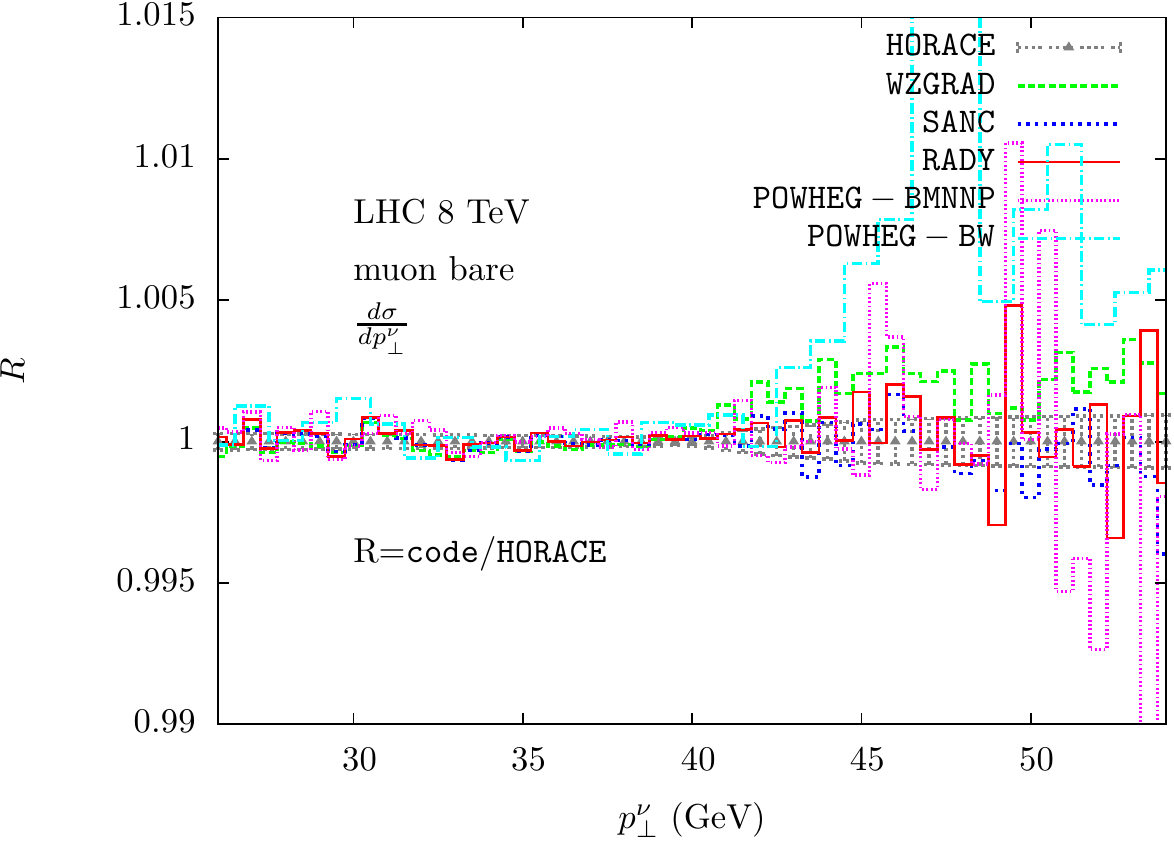}
\includegraphics[width=75mm,angle=0]{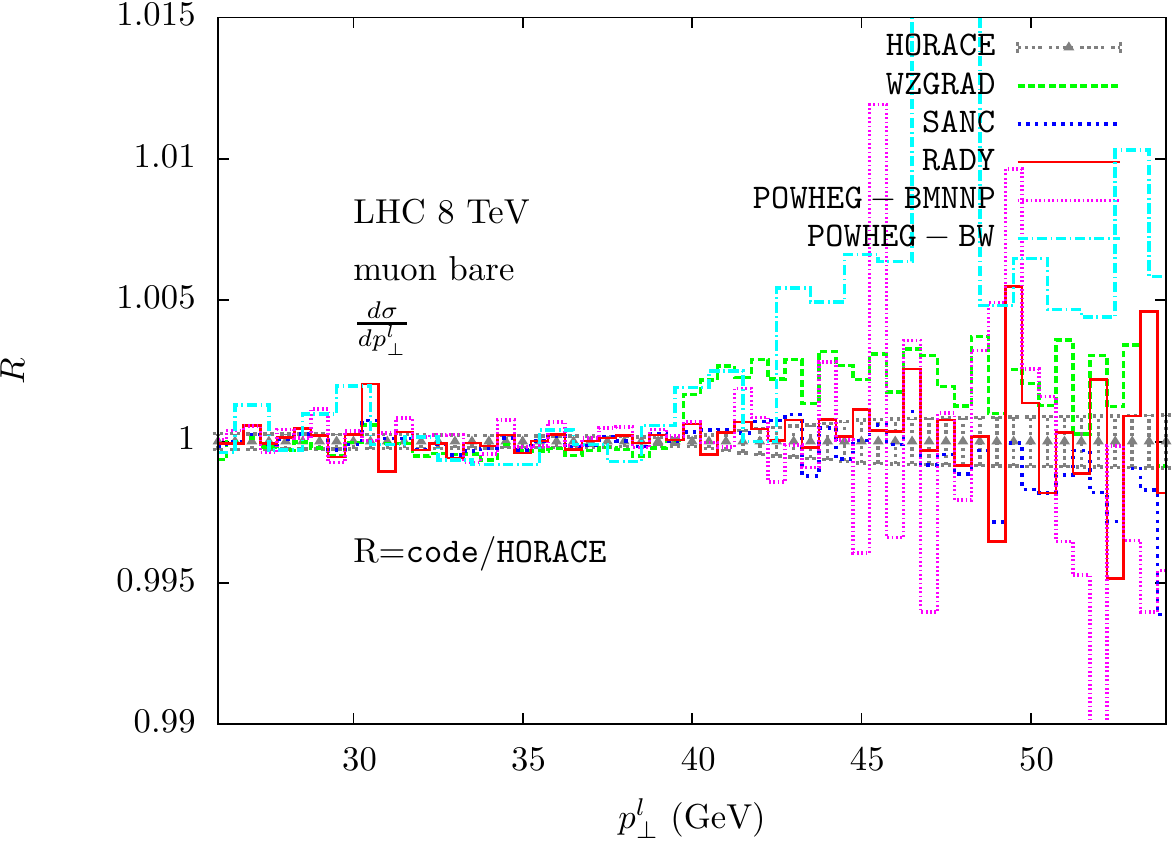}
\caption{Tuned comparison of the muon and muon neutrino transverse
  momentum distributions in $pp\to W^- \to \mu^- \nu_\mu+X$ at the
  8~TeV LHC with ATLAS/CMS cuts in the {\em bare} setup, including NLO
  EW corrections.}\label{fig:four}
\end{figure}

\begin{figure}[h]
\includegraphics[width=75mm,angle=0]{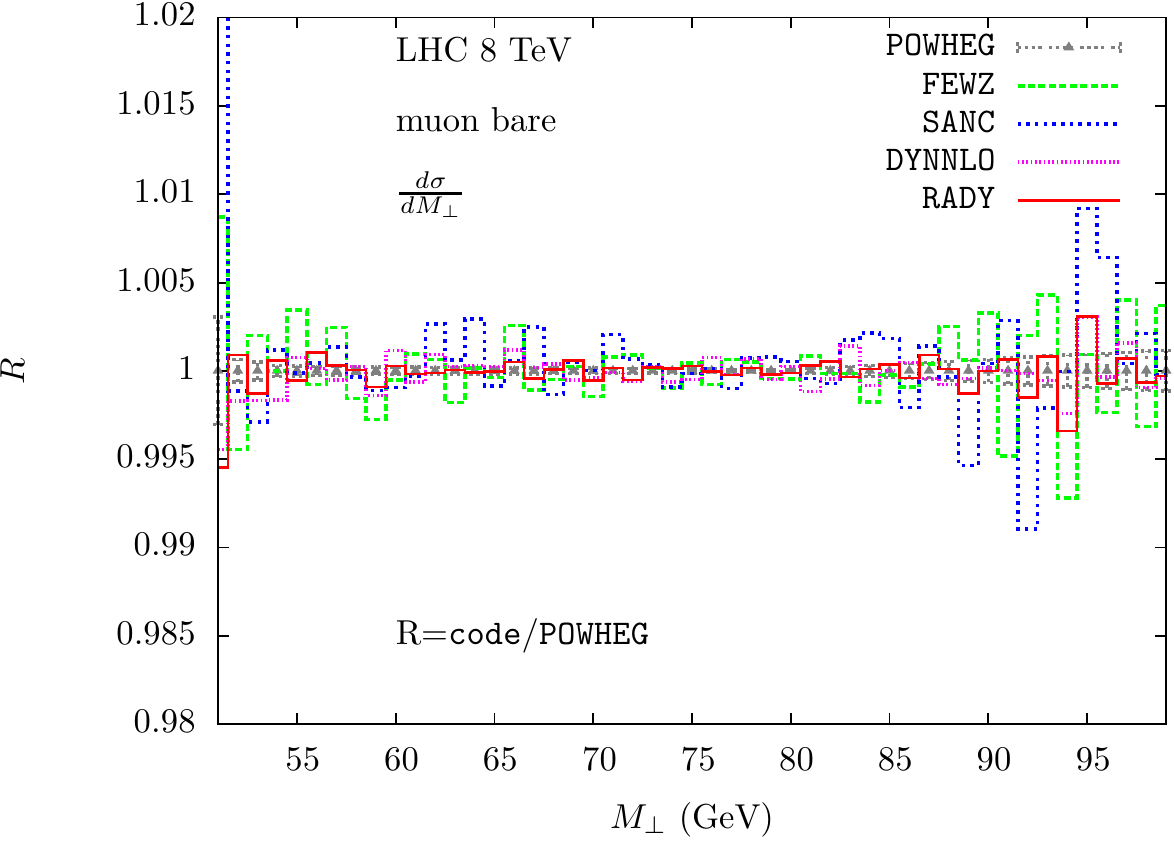}
\includegraphics[width=75mm,angle=0]{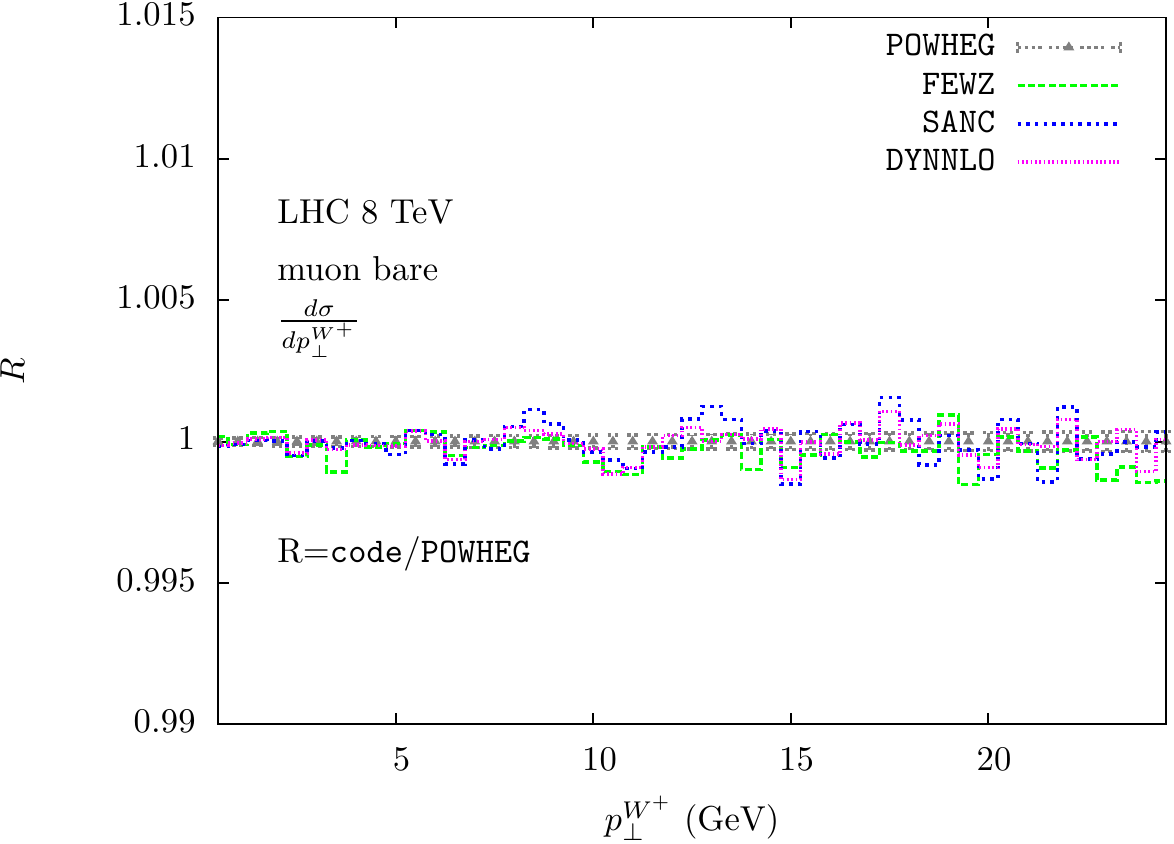}
\caption{Tuned comparison of the lepton-pair transverse mass and
  transverse momentum distribution in $pp\to W^+ \to \mu^+\nu_\mu+X$
  at the 8~TeV LHC with ATLAS/CMS cuts in the {\em bare} setup,
  including NLO QCD corrections.}\label{fig:five}
\end{figure}

\begin{figure}[h]
\includegraphics[width=75mm,angle=0]{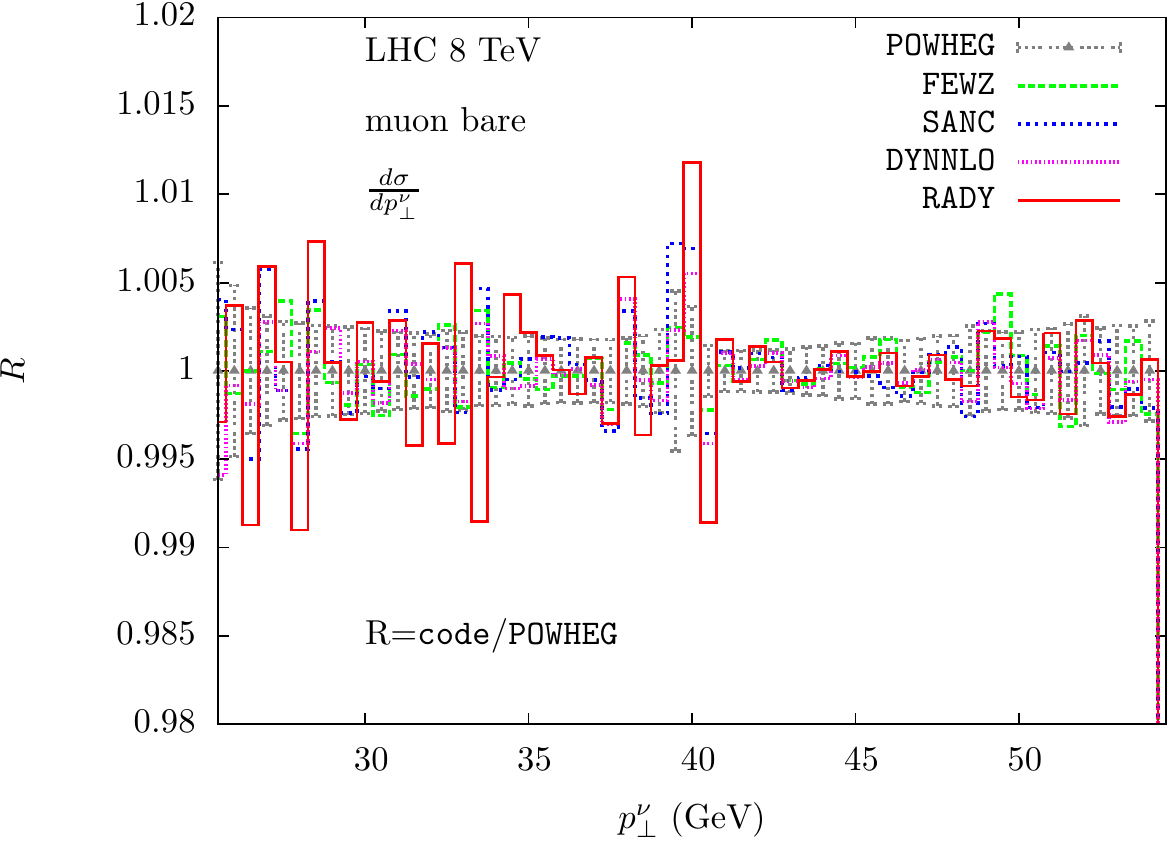}
\includegraphics[width=75mm,angle=0]{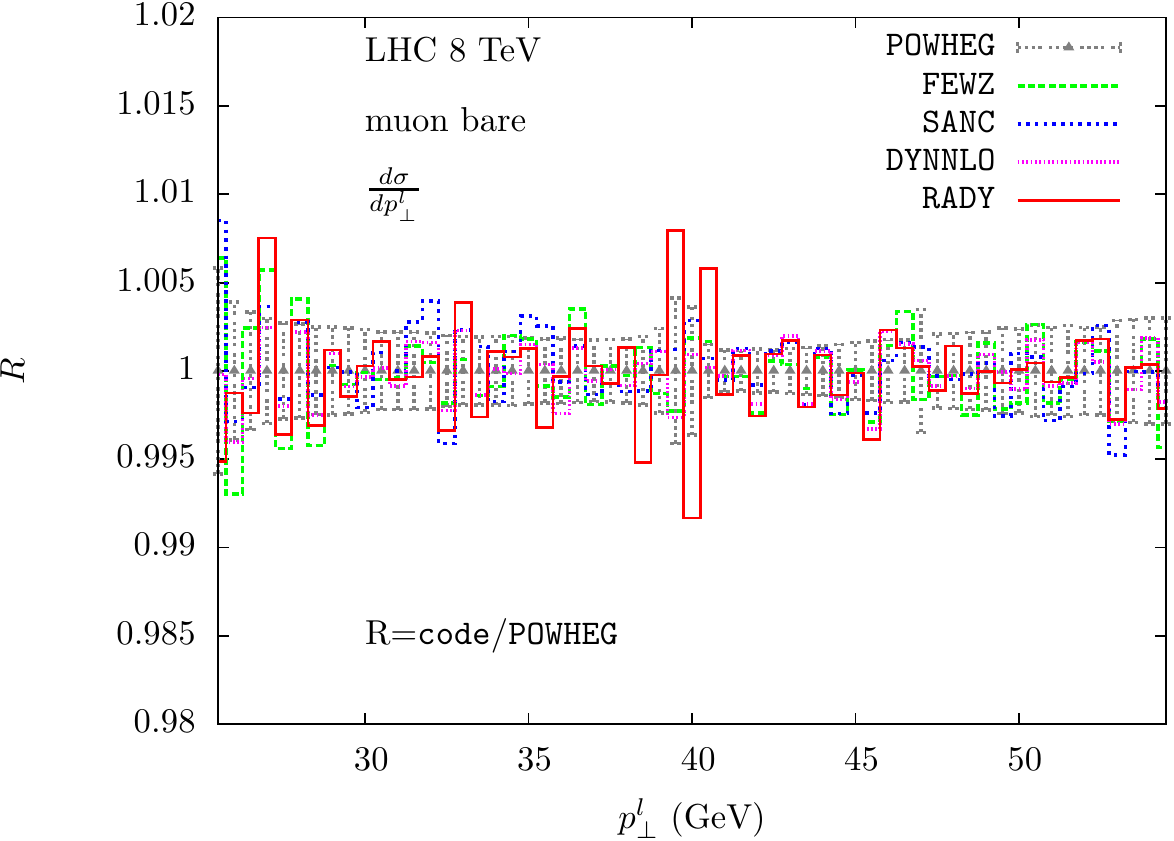}
\caption{Tuned comparison of the muon and muon neutrino transverse
  momentum distributions in $pp\to W^+ \to \mu^+\nu_\mu+X$ at the
  8~TeV LHC with ATLAS/CMS cuts in the {\em bare} setup, including NLO
  QCD corrections.}\label{fig:six}
\end{figure}

\begin{figure}[h]
\centering
\includegraphics[width=85mm,angle=0]{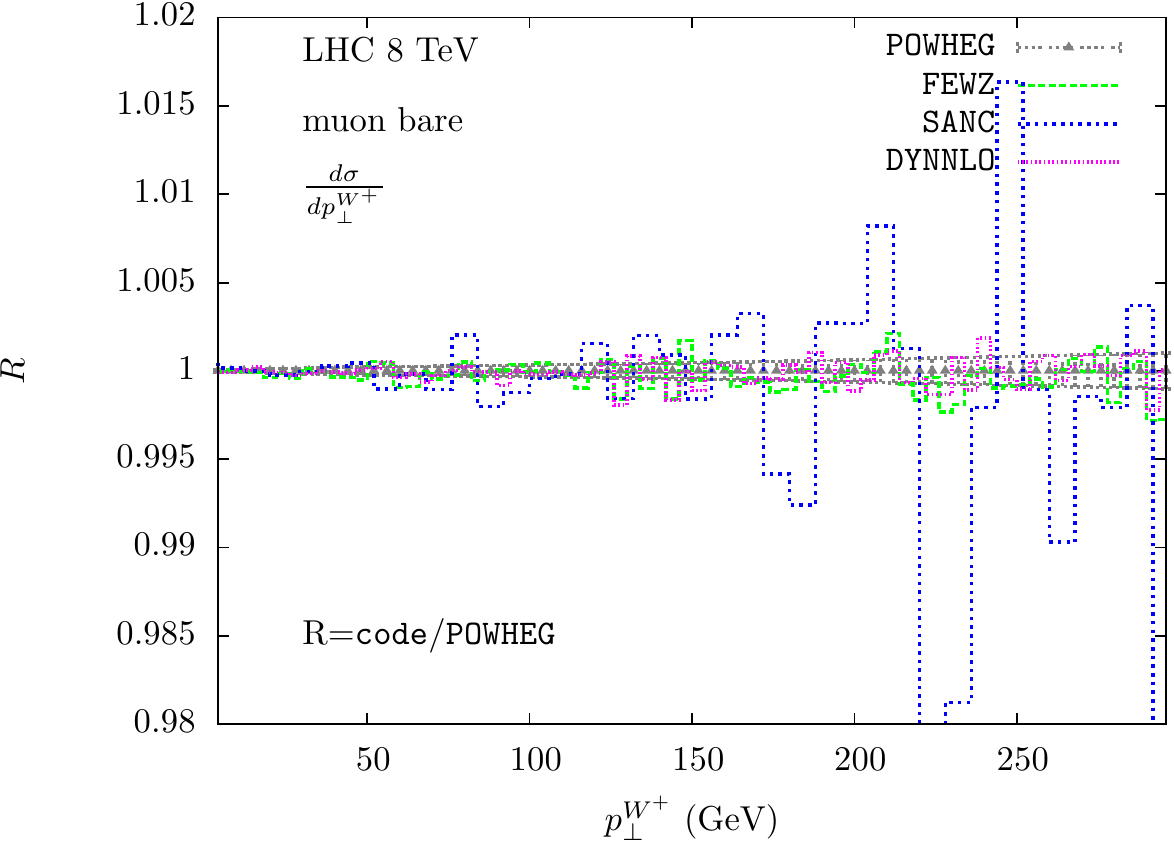}
\caption{Tuned comparison of the lepton-pair transverse momentum
  distribution in $pp\to W^+\to \mu^+\nu_\mu+X$ at the 8~TeV LHC with
  ATLAS/CMS cuts in the {\em bare} setup at high $p_\perp^W$,
  including NLO QCD corrections.}\label{fig:seven}
\end{figure}

\begin{figure}[h]
\includegraphics[width=75mm,angle=0]{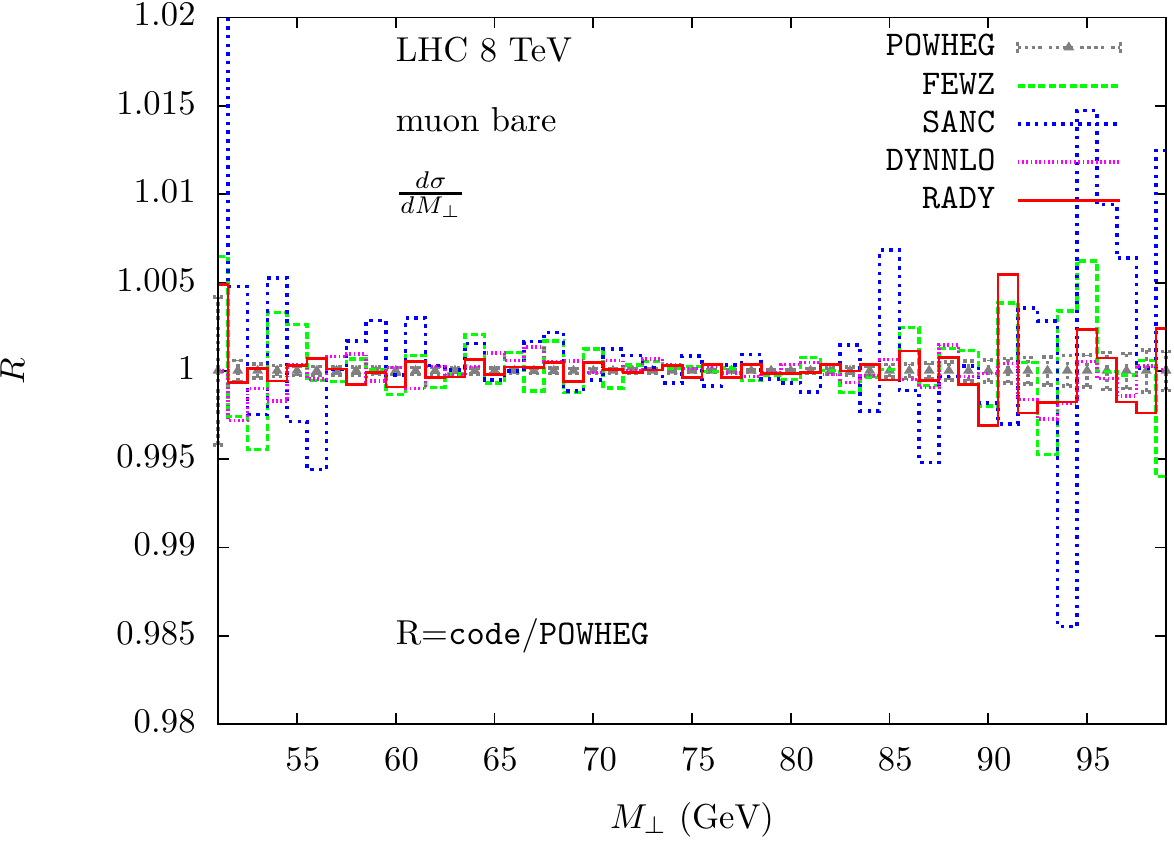}
\includegraphics[width=75mm,angle=0]{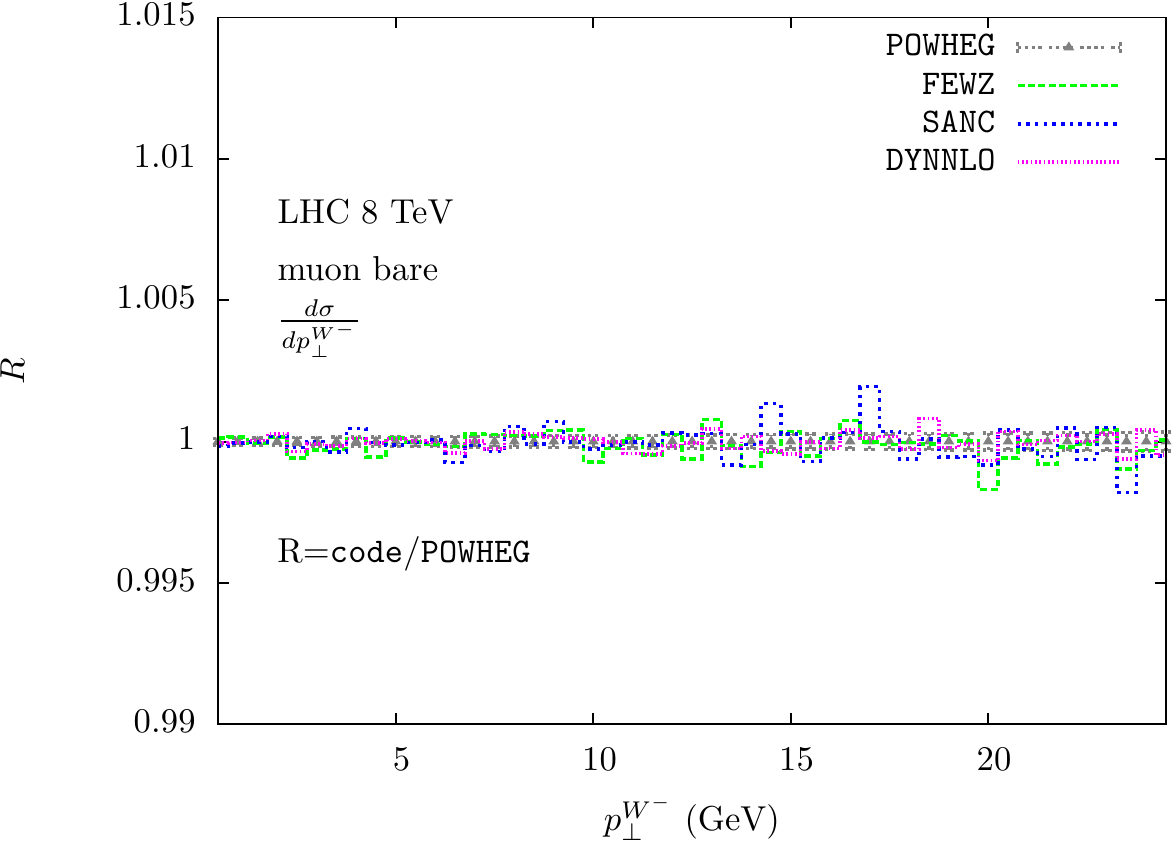}
\caption{Tuned comparison of the lepton-pair transverse mass and
  transverse momentum distribution in $pp\to W^- \to
  \mu^-\bar\nu_\mu+X$ at the 8~TeV LHC with ATLAS/CMS cuts in the {\em
    bare} setup, including NLO QCD
  corrections.}\label{fig:nlo-QCD-comp-one}
\end{figure}

\begin{figure}[h]
\includegraphics[width=75mm,angle=0]{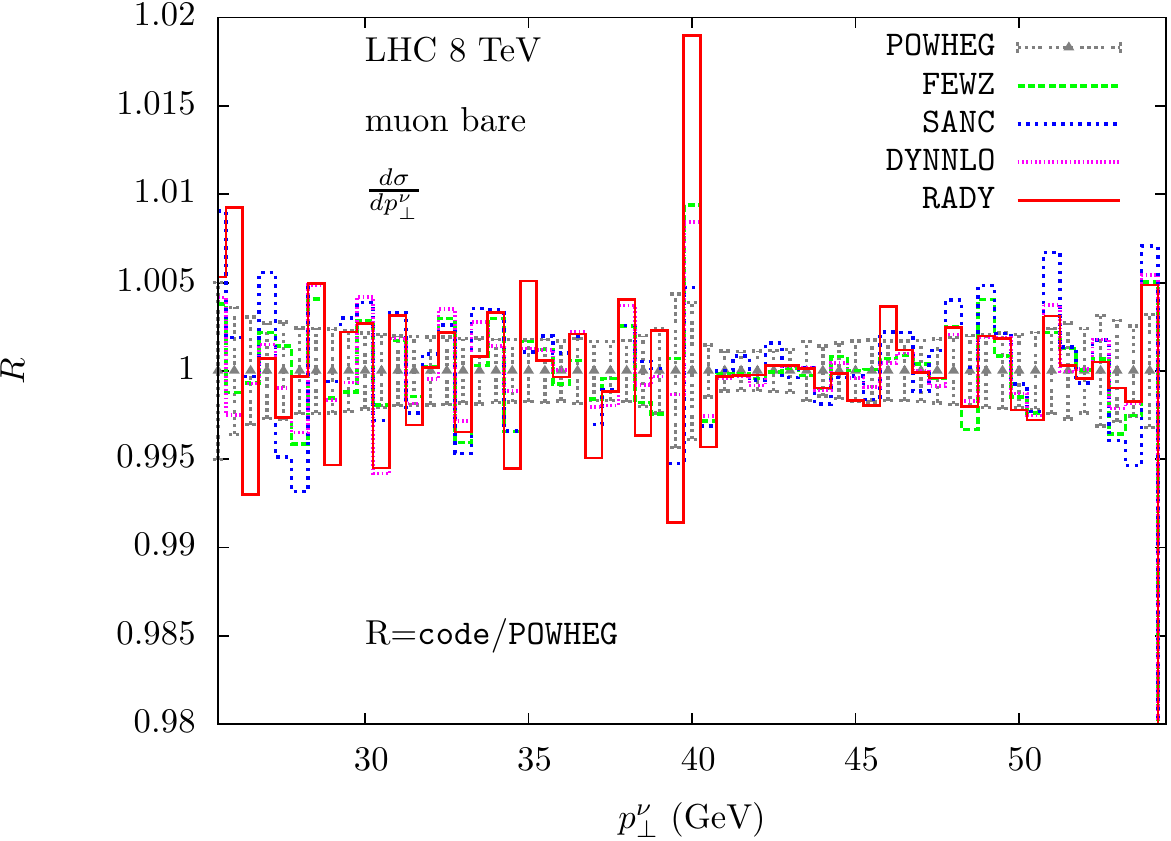}
\includegraphics[width=75mm,angle=0]{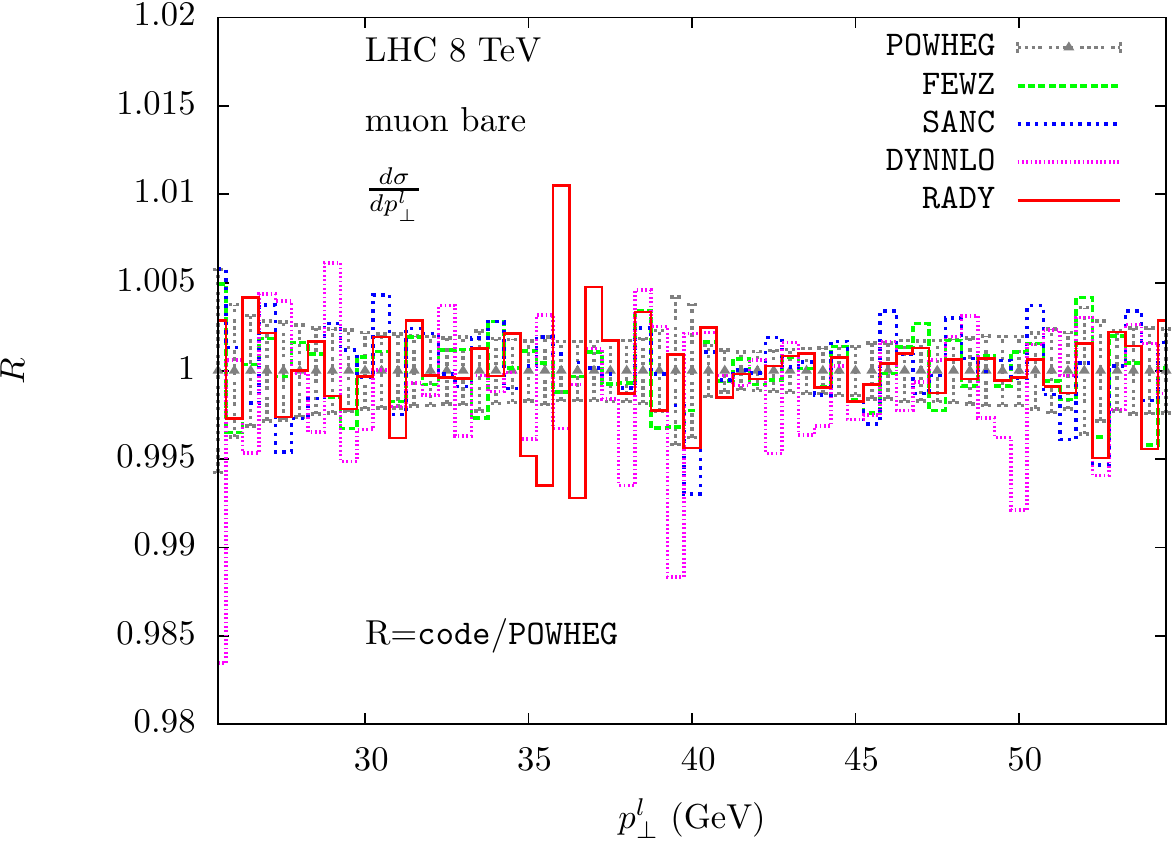}
\caption{Tuned comparison of the muon and muon neutrino transverse momentum
  distributions in $pp\to W^- \to \mu^-\bar\nu_\mu+X$ at
  the 8~TeV LHC with ATLAS/CMS cuts in the {\em bare} setup, including
  NLO QCD corrections.}\label{fig:nlo-QCD-comp-two}
\end{figure}

\begin{figure}[h]
\centering
\includegraphics[width=85mm,angle=0]{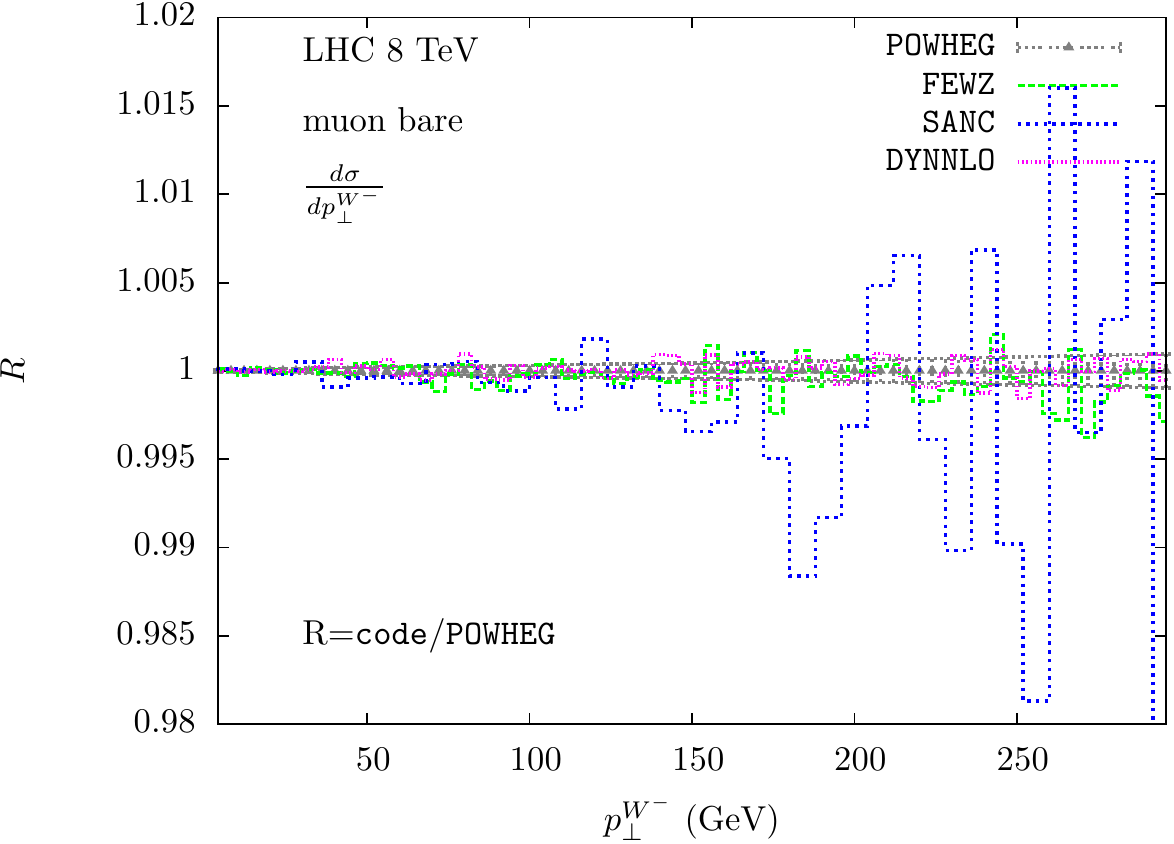}
\caption{Tuned comparison of the lepton-pair transverse momentum
  distributions in $pp\to W^-\to \mu^- \bar\nu_\mu+X$ at the 8~TeV LHC
  with ATLAS/CMS cuts in the {\em bare} setup at high $p_\perp^W$,
  including NLO QCD corrections.}\label{fig:nlo-QCD-comp-three}
\end{figure}

In the following we present a tuned comparison of results for the
$M_\perp, p_\perp^W$ and $p_\perp^l,p_\perp^\nu$ distributions for $W^\pm$
production in $pp\to\mu^\pm \nu_\mu+X$ at the 8~TeV LHC with ATLAS/CMS
cuts in the {\em bare} setup.  To compare the results of different
codes at NLO EW we show in Figs.~\ref{fig:one}-\ref{fig:four} the
ratios R=code/{\tt HORACE}, where code={\tt HORACE, POWHEG\_BMNNP,
POWHEG\_BW, RADY, SANC, WZGRAD}, and at NLO QCD we show in
Figs.~\ref{fig:five}-\ref{fig:nlo-QCD-comp-three} the ratios
R=code/{\tt POWHEG}, where code={\tt DYNNLO, FEWZ, POWHEG, RADY,
SANC}.

We observe that the agreement between different codes that include NLO
EW corrections is at the five per mill level or better in the
transverse mass of the lepton pair, $M_\perp$, and in the lepton
transverse momentum, $p_\perp^l$, in the relevant kinematic range under 
study. Some codes exhibit larger statistical fluctuations at larger values of the lepton transverse momenta, for instance, which can be improved by performing dedicated higher-statistics runs.  
For very small values of the transverse momentum of the
lepton pair, $p_\perp^W$, the agreement is only at the one percent
level and there are large statistical uncertainties at larger values
of $p_\perp^W$. We consider this level of agreement to be sufficient,
since there is only a very small $p_\perp^W$ kick due to photon
radiation, and it is not worthwhile to perform dedicated higher
statistics runs for higher values of $p_\perp^W$ to improve the
statistical uncertainty. Only the {\tt POWHEG\_BW} result for the $p_\perp^W$ distribution 
in the $W^-$ case shows a 
systematic difference, and its origin is presently under study. 
In any case, these results should be
considered just for technical checks, since $p_\perp^W$ receives large
contributions from QCD radiation. The combined effects of EW and QCD
corrections in $p_\perp^W$ can be studied for instance by using a
calculation of NLO EW corrections to $W+j$
production~\cite{Denner:2009gj} and the implementation of NLO EW
corrections in {\tt POWHEG}~\cite{Barze:2012tt,Bernaciak:2012hj} as
discussed in Section~\ref{sec:interplay}.

\subsubsection{Tuned comparison of $Z$ boson observables}
\label{sec:zcomp}

\begin{figure}[h]
\centering
\includegraphics[width=75mm,angle=0]{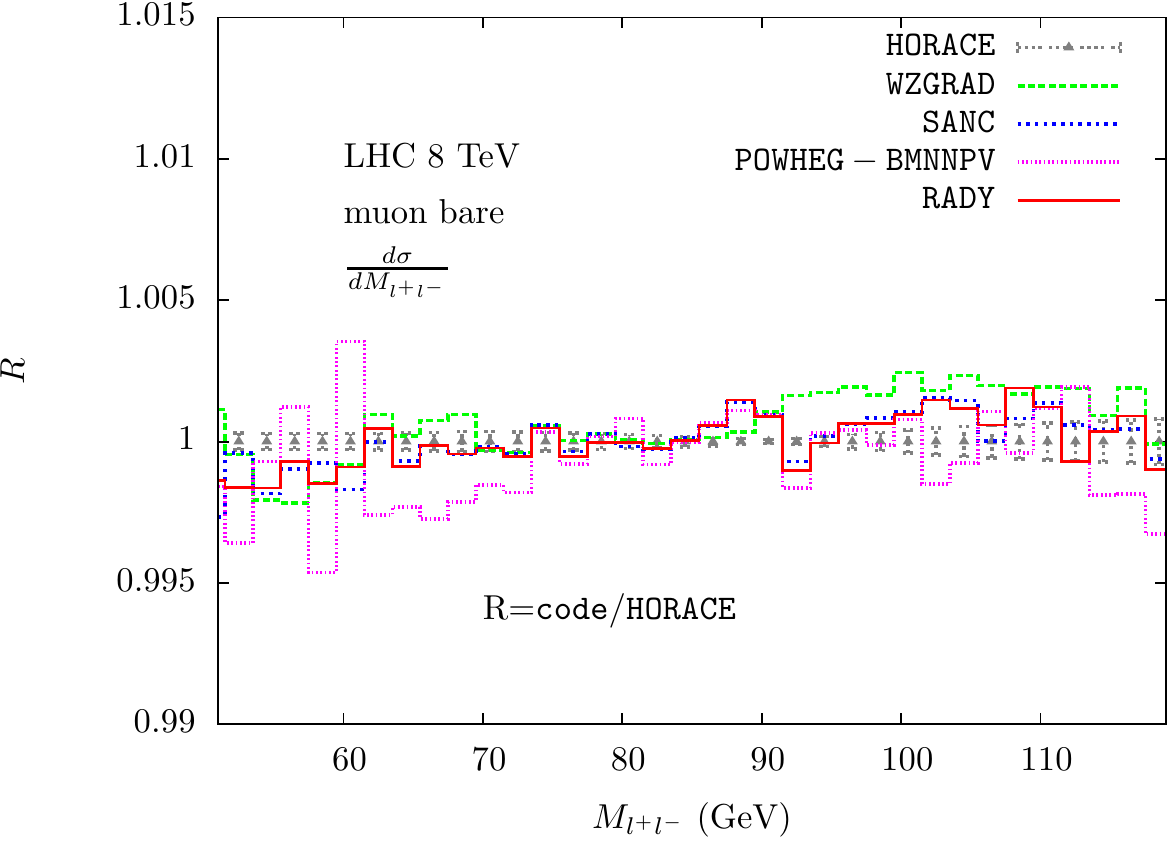}
\includegraphics[width=75mm,angle=0]{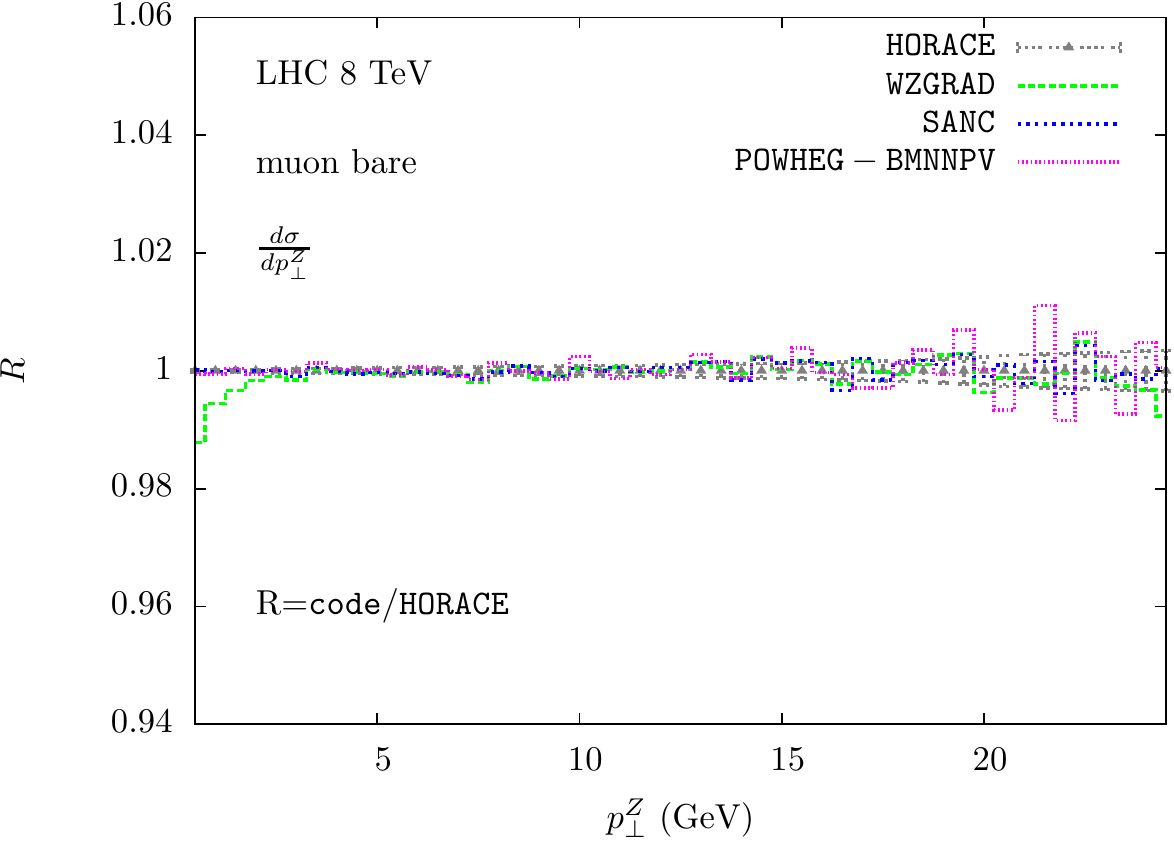}
\caption{Tuned comparison of the lepton-pair invariant mass and
  transverse momentum distributions in $pp\to \gamma,Z \to \mu^+\mu^-
  +X$ at the 8~TeV LHC with ATLAS/CMS cuts in the {\em bare} setup,
  including NLO EW corrections.}\label{fig:eight}
\end{figure}

\begin{figure}[h]
\includegraphics[width=75mm,angle=0]{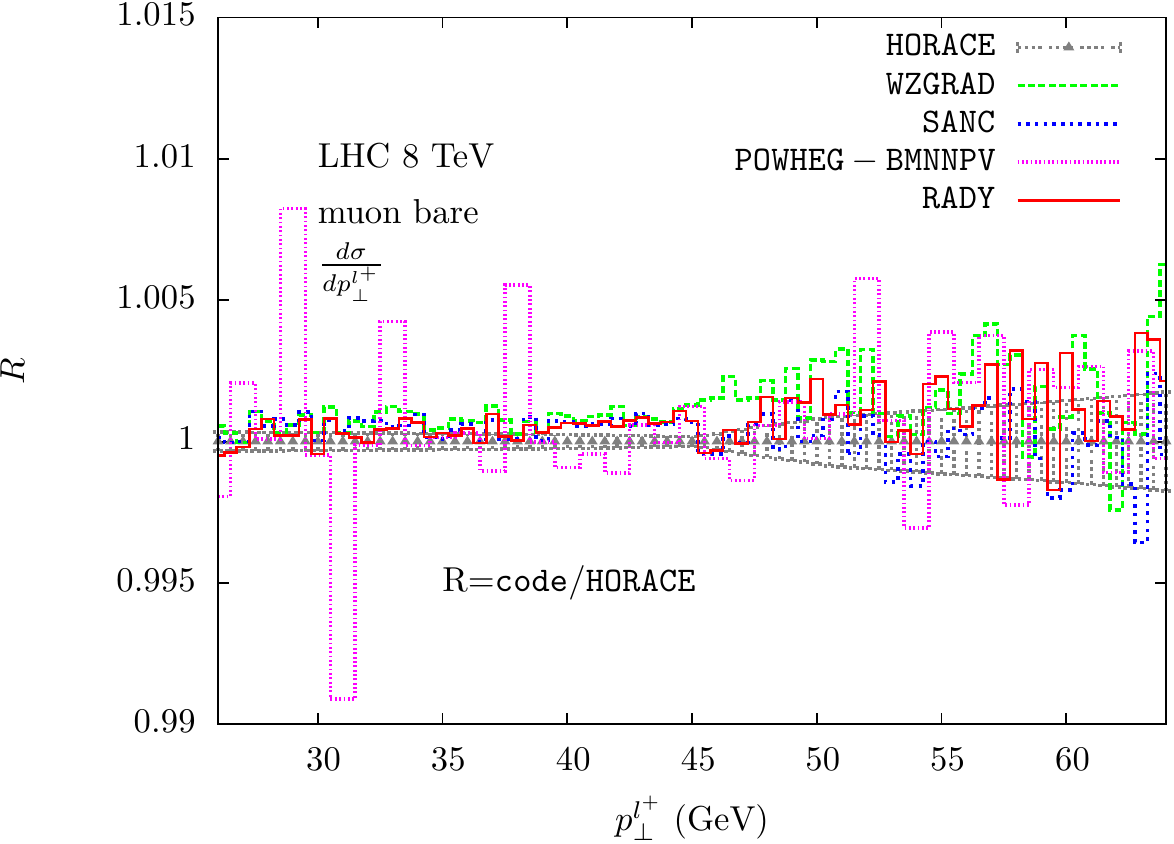}
\includegraphics[width=75mm,angle=0]{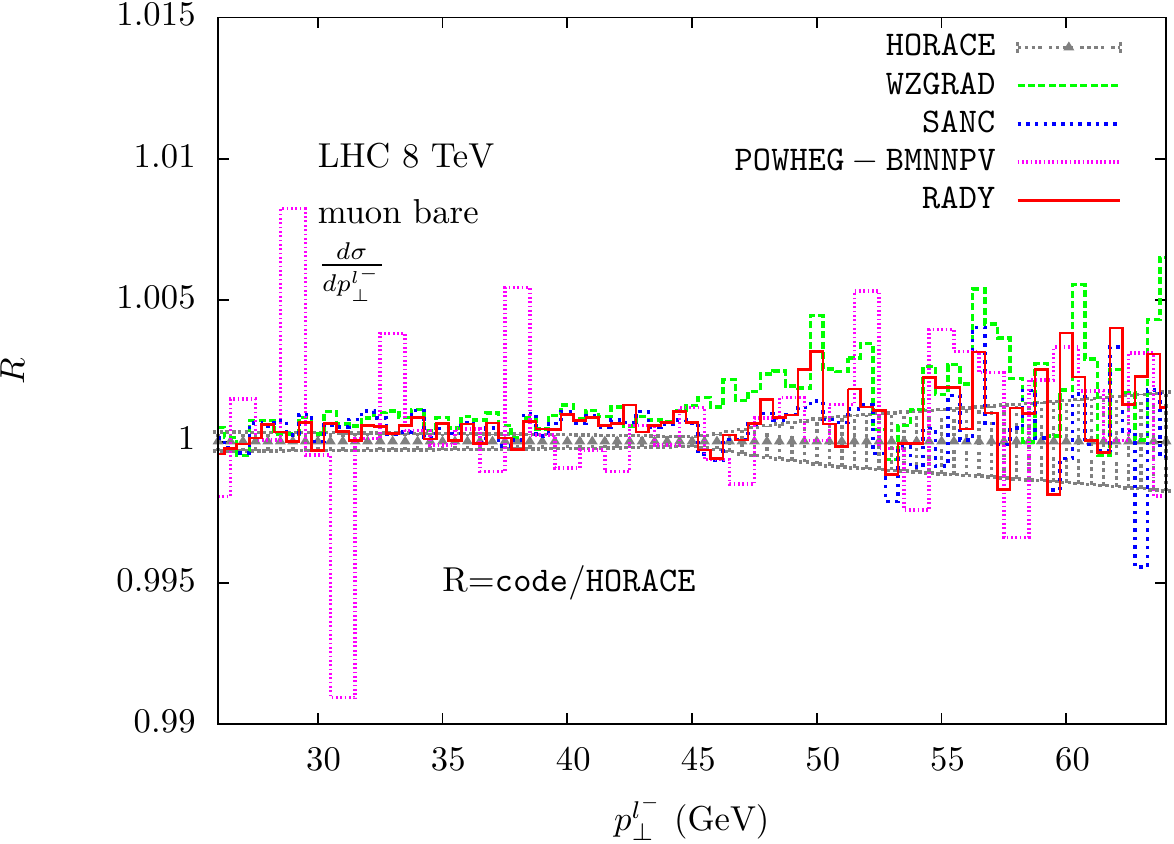}
\caption{Tuned comparison of the $\mu^+$ and $\mu^-$ transverse
  momentum distributions in $pp\to \gamma, Z\to \mu^+ \mu^- +X$ at the
  8~TeV LHC with ATLAS/CMS cuts in the {\em bare} setup, including NLO
  EW corrections.}\label{fig:nine}
\end{figure}

\begin{figure}[h]
\centering
\includegraphics[width=75mm,angle=0]{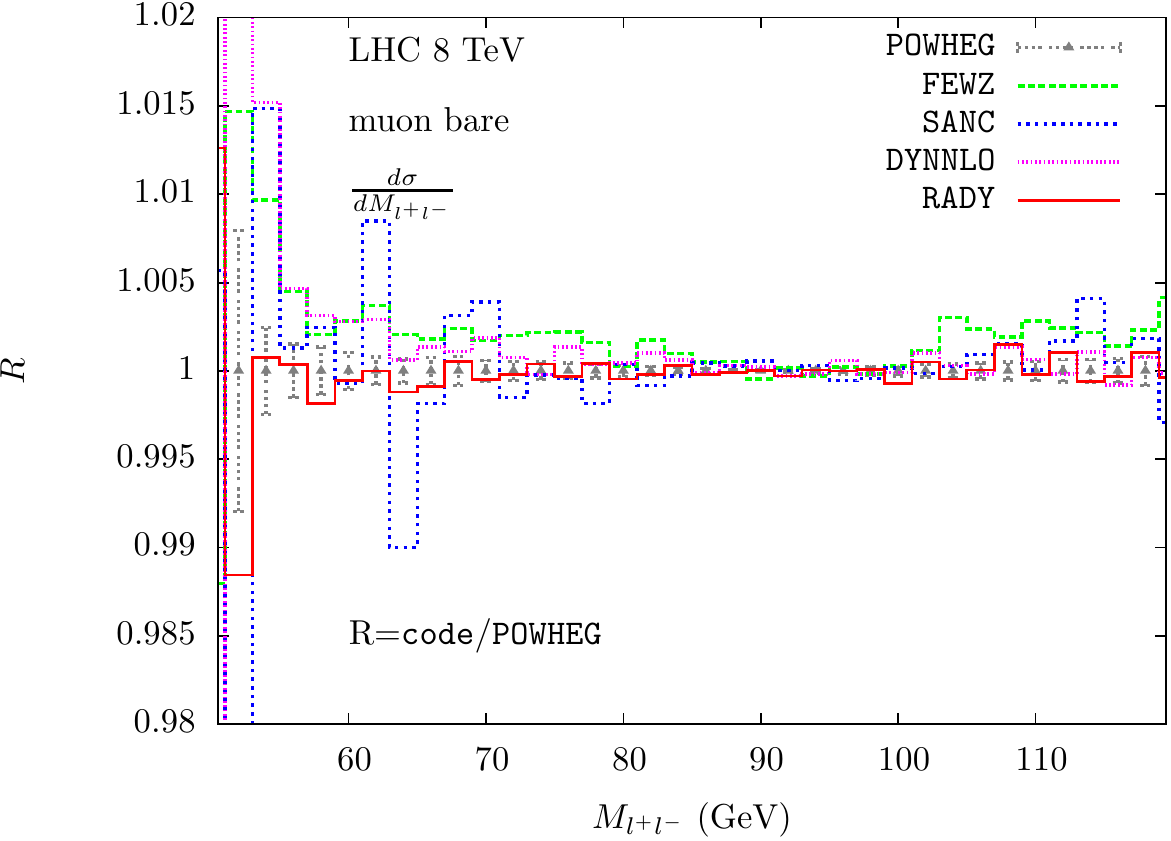}
\includegraphics[width=75mm,angle=0]{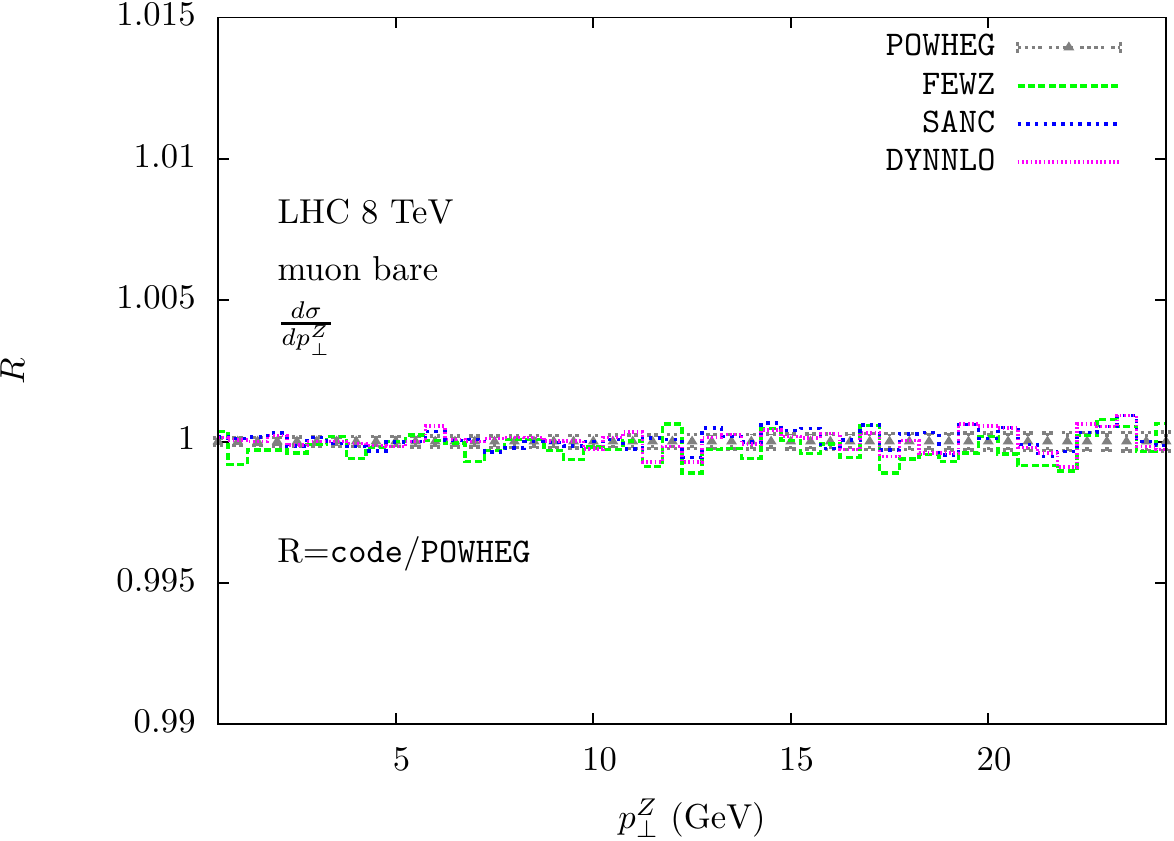}
\caption{Tuned comparison of the lepton-pair invariant mass and
  transverse momentum distributions in $pp\to \gamma,Z \to \mu^+\mu^-
  +X$ at the 8~TeV LHC with ATLAS/CMS cuts in the {\em bare} setup,
  including NLO QCD corrections.}\label{fig:ten}
\end{figure}

\begin{figure}[h]
\includegraphics[width=75mm,angle=0]{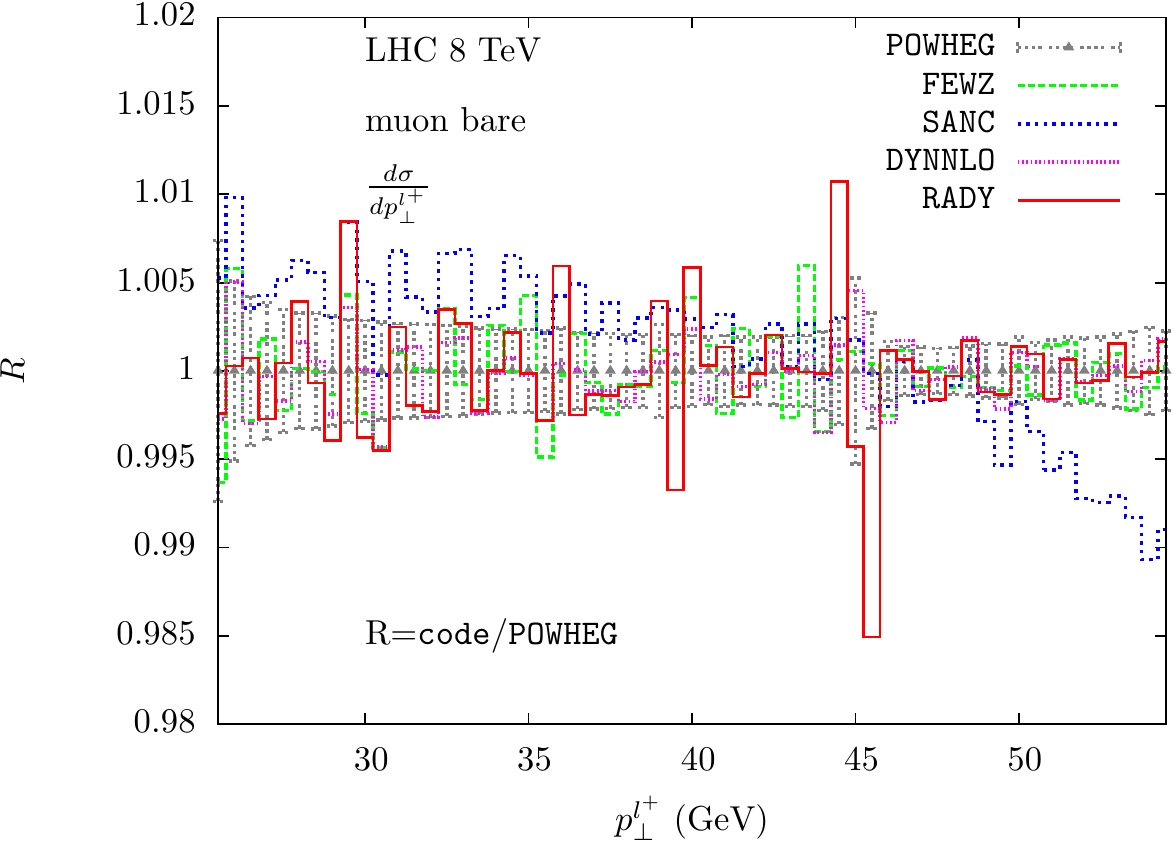}
\caption{Tuned comparison of the muon transverse momentum
  distributions in $pp\to \gamma, Z\to \mu^+ \mu^- +X$ at the 8~TeV
  LHC with ATLAS/CMS cuts in the {\em bare} setup, including NLO QCD
  corrections.}\label{fig:eleven}
\end{figure}

In Figs.~\ref{fig:eight}, \ref{fig:nine} and in
Figs.~\ref{fig:ten}, \ref{fig:eleven} we present a tuned comparison of
results for NLO EW and QCD predictions, respectively, for the
$M_{l^+l^-}, p_\perp^Z$ and $p_\perp^l$ distributions in $pp\to
\gamma,Z\to\mu^+\mu^- +X$ at the 8 TeV LHC with ATLAS/CMS cuts in the
{\em bare} setup of Section~\ref{sec:setup}.  The agreement of
different codes providing NLO EW predictions for these distributions
in the kinematic regions under study are at the five per mill level or
better, apart from a difference at the one per cent level in the
transverse momentum distribution of the lepton pair for small values
of $p_\perp^Z$.  As it is the case for CC DY, these results should be
considered just for technical checks, since $p_\perp^Z$ receives large
contributions from QCD radiation. The combined effects of EW and QCD
corrections in $p_\perp^Z$ can be studied for instance by using a
calculation of NLO EW corrections to $Z+j$
production~\cite{Denner:2011vu} and the implementation of NLO EW
corrections in {\tt POWHEG}~\cite{Barze':2013yca} as discussed in
Section~\ref{sec:interplay}.

\clearpage
\section{Impact of higher-order radiative corrections}
\label{sec:impact}
The setup described in Section \ref{sec:setup}, and used to perform the
tuned comparison of the codes participating in this study, has been
chosen with two main practical motivations: 1) the simplicity to
implement the renormalization of the NLO EW calculation and 2) the
possibility to rely and easily reproduce the results of previous
similar studies \cite{Buttar:2006zd,Gerber:2007xk}, where technical
agreement between different codes had already been demonstrated.

On the other hand, the setup of Section \ref{sec:setup} suffers for
two reasons, relevant from the phenomenological but also from the
theoretical point of view: 1) the choice of the fine-structure
constant as input parameter in the EW Lagrangian introduces an
explicit dependence on the value of the light-quark masses via the
electric charge renormalization; these masses are not well defined
quantities and introduce a non-negligible parametric dependence of all
the results; 2) the strength of the coupling of the weak currents is
best expressed in terms of the Fermi constant, whose definition
reabsorbs to all orders various classes of large radiative
corrections; when using the Fermi constant, the impact of the
remaining, process dependent corrections is thus reduced in size with
respect to other input schemes, like, e.g., the one of
Section \ref{sec:setup}.

We propose here to use a different input scheme, 
which absorbs known higher-order corrections already in the (N)LO
  predictions, thus minimalizing the impact of neglected orders in
  perturbation theory. This scheme will be called {\em benchmark} and
the corresponding numbers at NLO EW will be considered as our
benchmark results, relevant in particular for the discussion of the impact of
higher-order corrections.

\subsection{Setup for {\em benchmark} predictions}
\label{sec:setupbest}
We provide {\em benchmark} predictions for the 8~TeV LHC for muons in the
{\em bare} setup, i.e. when only applying acceptance cuts, and for
electrons in the {\em calo} setup as defined in the setup for the
tuned comparison in Section~\ref{sec:setup}. For the {\em benchmark}
results we made the following changes to the setup described in
Section~\ref{sec:setup}:
\begin{itemize}
\item[1)] In the case of $W$ boson production, in addition to the acceptance
  cuts we apply $M_\perp(l\nu) >40$~GeV.
\item[2)] To account for the fact that we are using the constant width
  approach, we have to adjust the $W,Z$ mass and width input
  parameters that have been measured in the $s$-dependent width
  approach accordingly, as
  follows~\cite{Wackeroth:1996hz,Dittmaier:2009cr}
  ($\gamma_V=\Gamma_V/M_V$):
\[M_V \to \frac{M_V}{\sqrt{1+\gamma_V^2}} \; \; ; \; \; \Gamma_V \to \frac{\Gamma_V}{\sqrt{1+\gamma_V^2}} \; .\]
Consequently, the input values for the $W,Z$ masses and widths change
to
\begin{eqnarray}
M_Z = 91.1535 \; {\rm GeV}, & \quad & \Gamma_Z = 2.4943  \; {\rm GeV}
\nonumber  \\ 
M_W = 80.358 \; {\rm GeV}, & \quad & \Gamma_W =  2.084 \; {\rm GeV} \; .
\end{eqnarray}
\item[3)] We use the following EW input scheme:\\ In the calculation
  of the tree-level couplings we replace $\alpha(0)$ by the effective
  coupling $\alpha_{G_\mu}=\sqrt{2} G_\mu M_W^2 (1-M_W^2/M_Z^2)/\pi$.
  The relative ${\cal O}(\alpha)$ corrections are calculated with the
  fine structure constant $\alpha(0)$.  At NLO EW this replacement
  implies an additional contribution of $\Delta r$ to the relative
  ${\cal O}(\alpha)$ corrections. The one-loop result for $\Delta r$
  has been calculated in Refs.~\cite{Sirlin:1980nh,Marciano:1980pb}
  and can be decomposed as follows:
\[\Delta r({\rm 1-loop})=\Delta \alpha-\frac{c_w^2}{s_w^2} \Delta \rho+\Delta r_{rem}(M_H) \; .\]
When using the input values of Eq.~\ref{eq:pars} and the values for
$M_W$ and $M_Z$ given in item 2) $\Delta r({\rm 1-loop})=0.0295633444$
($\Delta r=0.0296123554$ for the unshifted $W/Z$ masses of
Eq.~\ref{eq:pars}).
\end{itemize}
To be able to discuss the impact of higher order correction beyond NLO
in this setup, we successively included higher-order corrections,
i.e. we start with the NLO result using the changed setup as described
above, successively add different sources of higher order corrections,
such as multiple photon radiation and two-loop corrections to
$\Delta \rho$, and compare the resulting observables to the NLO
results.

\subsubsection{Setup for the evaluation of photon-induced contributions}

For the comparison of predictions for the photon-induced processes
$\gamma\gamma\to l^+l^-$ and $\gamma \smartqaq \to l^+l^-\smartqaq $
in NC DY, and $\gamma \smartqaq \to l \nu_l \smartqaq$ in CC DY we
make the following additional changes to the {\em benchmark} setup:
\begin{itemize}
\item
In order to have a modern parametrization of the photon density,
we used the central {\tt NNPDF2.3\_lo\_as\_130\_qed} PDF set~\cite{Ball:2013hta}.
\item
We use as input parameters $(\alpha(0), \mw, \mz)$ for all photon-induced processes.
\end{itemize}
In the NC DY case, we compute separately the contribution of the LO
$\gamma\gamma\to l^+l^-$ process and those of the
$\gamma \smartqaq \to l^+l^-\smartqaq $ processes.  To express the
percentage effect of these subprocesses, we present their ratio to the
LO $q\bar q$-initiated cross sections.

\clearpage

\subsection{Total cross sections in the {\em benchmark} setup at NLO EW and NLO QCD with ATLAS/CMS cuts}
\label{sec:total-xsec-best}

In Table~\ref{tab:xsec-wp-lhc8-atlcms-best}, \ref{tab:xsec-wm-lhc8-atlcms-best}, and \ref{tab:xsec-z-lhc8-atlcms-best} we provide a tuned
comparison of the total cross sections for $W^+$, $W^-$ and $Z$ boson production, respectively, 
computed at fixed order, namely
LO, NLO EW and NLO QCD, using the {\em benchmark} setup of Section~\ref{sec:setupbest} for
the choice of input parameters and ATLAS/CMS acceptance cuts.
We use the symbol
$\times$ in the tables to indicate that a particular correction is not
available in the specified code, and ($\times$) in cases where the result can be 
produced with the specified code but has not been provided for this report.

\subsubsection{Results for $W^{\pm}$ boson production}

\begin{table}[h]
\begin{center}
\begin{tabular}{|l|l|l|l|l|}
\hline
       & LO         & NLO         & NLO                  &  NLO                \\ 
code   &            & QCD         & EW $\mu$ {\em bare}  & EW $e$ {\em calo}      \\ 
\hline
HORACE & 3109.65(8) & $\times$    & 3022.8(1)            &  3039.5(2)   \\
\hline
WZGRAD & 3109.62(2) & $\times$    & 3022.68(4)           & 3039.13(5)   \\
\hline
SANC   & 3109.66(2) & ($\times$)  & 3022.53(4)           & 3038.94(4)  \\
\hline 
DYNNLO & 3109.5(2)  & 3092.3(9)   & $\times$             & $\times$            \\
\hline
FEWZ   & 3109.20(8) & 3089.1(3)   & $\times$             & $\times$     \\
\hline
POWHEG-w& ($\times$)          & 3090.4(2)   & $\times$             & $\times$  \\
\hline
POWHEG\_BMNNP   
       & 3109.68(7) & 3089.6(2)   & 3022.8(2)            &       ($\times$)        \\
\hline
\end{tabular}
\caption{\label{tab:xsec-wp-lhc8-atlcms-best}  $p p \to W^+\to l^+ \nu_l$
  total cross sections (in pb) at LO, NLO EW and NLO QCD at the 8 TeV LHC with ATLAS/CMS cuts in the {\em benchmark} setup.}
\end{center}
\end{table}

\begin{table}[h]
\begin{center}
\begin{tabular}{|l|l|l|l|l|}
\hline
         & LO          & NLO        & NLO        & NLO           \\ 
code     &             & QCD        &  EW $\mu$ {\em bare} & EW $e$ {\em calo}    \\ 
\hline
HORACE   & 2156.36(6)  & $\times$   & 2101.17(8) & 2111.1(2) \\
\hline
WZGRAD   & 2156.46(2)  & $\times$   & 2101.23(2) &  2110.65(4)\\
\hline
SANC     & 2156.46(2)  & ($\times$) & 2101.31(4) &  2110.69(4)  \\
\hline 
DYNNLO   & 2156.38(2)  & 2189.3(7)  &  $\times$  &   $\times$                     \\
\hline
FEWZ     & 2156.09(4)  & 2187.1(1)  &  $\times$  &       $\times$     \\
\hline
POWHEG-w &    ($\times$)         & 2187.72(6) & $\times$   & $\times$ \\
\hline
POWHEG\_BMNNP & 
           2156.44(4)  & 2187.5(1)  & 2101.5(1) &    ($\times$)         \\
\hline
\end{tabular}
\caption{\label{tab:xsec-wm-lhc8-atlcms-best}  $p p \to W^-\to l^- \bar\nu_l$
total cross sections (in pb) at LO, NLO EW and NLO QCD at the 8 TeV LHC with ATLAS/CMS cuts in the {\em benchmark} setup.}
\end{center}
\end{table}

\subsubsection{Results for $Z$ boson production}

\begin{table}[h]
\begin{center}
\begin{tabular}{|c|l|l|l|c|c|}
\hline
        & LO         & NLO       & NLO                  & NLO              \\ 
code    &            & QCD       & EW $\mu$ {\em bare}  & EW $e$ {\em calo}       \\ 
\hline
HORACE  & 462.663    &  $\times$ & 443.638              &            \\
\hline
WZGRAD  & 462.677(4) &  $\times$ & 443.950(6)           & 445.178(7)  \\
\hline
SANC    & 462.675(2) & ($\times$)& 443.794(4)           & 444.963(4)  \\
\hline  
DYNNLO  &  ($\times$)          & 491.94(5) &  $\times$            & $\times$    \\
\hline
FEWZ    & 462.631(9) & 491.62(4) &  443.84(2)           & 444.67(2)          \\
\hline
POWHEG-z&     ($\times$)       & 491.744(4)& $\times$             & $\times$  \\
\hline
POWHEG\_BMNNPV   
        & 462.67(1) & 491.3(8)   & 443.4(1)             & ($\times$)  \\
\hline
\end{tabular}
\caption{\label{tab:xsec-z-lhc8-atlcms-best}  $p p \to \gamma,Z \to l^-
  l^+$  total cross sections (in pb) at LO, NLO EW and NLO QCD at the 8 TeV LHC with ATLAS/CMS cuts in the {\em benchmark} setup. }
\end{center}
\end{table}

\clearpage

\clearpage

\subsection{Impact of QCD corrections on $W$ and $Z$ boson observables in the {\em benchmark} setup}
\label{sec:impact-qcd}
\subsubsection{NLO QCD corrections}
\label{sec:NLO-QCD}

At LO the DY processes are described in terms of quark-antiquark
annihilation subprocesses\footnote{We discuss separately the NC DY LO
  contribution given by $\gamma\gamma\to l^+l^-$ scattering, which
  receives a non-trivial QCD correction only starting from third
  perturbative order.}.  The NLO QCD corrections are due to real and
virtual corrections to the incoming quark-antiquark line, but they
receive a contribution also from the (anti)quark-gluon scattering
subprocesses.

Some observables, such as the lepton-pair transverse momentum, the
$\phi^*$ variable or the single-lepton transverse momentum, are
strongly sensitive to the details of real QCD radiation.  The
lepton-pair transverse momentum or the $\phi^*$ distributions are
indeed absent at LO ($p_\perp^V=0$ and $\phi^*=\pi$), so that for
these quantities NLO QCD is the first perturbative non-vanishing
order.  In the single-lepton transverse momentum case, the
distribution receives, on top of the LO value, a large contribution
from the recoil of the intermediate gauge boson against initial-state
QCD radiation, enhanced by its collinearly divergent behaviour.  Even
if this is not formally the case, NLO QCD is numerically the lowest
perturbative order which can be used to assess the impact of higher
order corrections.  On the contrary the (pseudo-)rapidity
distributions and the invariant/transverse mass distributions receive
a milder, slowly varying NLO QCD correction, close in size to the
value of the total NLO K-factor.

\subsubsection{NNLO QCD corrections: total cross section}
\label{sec:NNLO-QCD}

We study the predictions for DY processes with the inclusion of QCD
next-to-next-to-leading order (NNLO) corrections in the strong
coupling constant using~\footnote{Recently, an implementation of NNLO QCD corrections to $pp \to Z$ and $pp \to W$ including the decays of the unstable gauge bosons became also available in {\tt MCFM}~\cite{Boughezal:2016wmq}.} the following three MC codes, 
\dynnlo \cite{Catani:2009sm},
\fewz \cite{Anastasiou:2003yy,Gavin:2012sy}, and 
\sherpafo \cite{Hoeche:2014aia}. 

These three codes have the same perturbative accuracy, in the sense
that they include the same set of radiative corrections, but differ in
the explicit implementation of the combination of real and virtual
corrections, in particular for what concerns the cancellation of soft
and collinear divergences.  In principle the differences between these
codes are at the technical level and should not affect physical
predictions.  The comparison of their results should thus be
understood as a tuned comparison at NNLO QCD level.  The results for
the evaluation of the total cross section in the {\em benchmark} setup
described in Section \ref{sec:setupbest} are reported in
Table \ref{tab:xsec-nnloqcd}.  The agreement between the three codes
is at the 0.5\%  level, for the three processes (NC and CC)
under consideration.  

\begin{table}[h]
\begin{center}
\begin{tabular}{|c|l|l|l|l|}
\hline
process                & \dynnlo         & \fewz       & \sherpafo        \\ 
\hline
$pp\to l^+\nu_l+X$     &  3191(7)               & 3207(2)           &  3204(4)            \\
\hline
$pp\to l^-\bar\nu_l+X$ &  2243(6)               & 2238(1)           &   2252(3)           \\
\hline
$pp\to l^+l^-+X$       & 502.4(4)                &  504.6(1)          &    502.0(6)          \\
\hline
\end{tabular}
\caption{\label{tab:xsec-nnloqcd} Tuned comparison of NNLO QCD total cross sections (in pb) at the 8 TeV LHC in the {\em benchmark} setup with ATLAS/CMS cuts.}
\end{center}
\end{table}

\begin{figure}[!ht]
\includegraphics[width=75mm,angle=0]{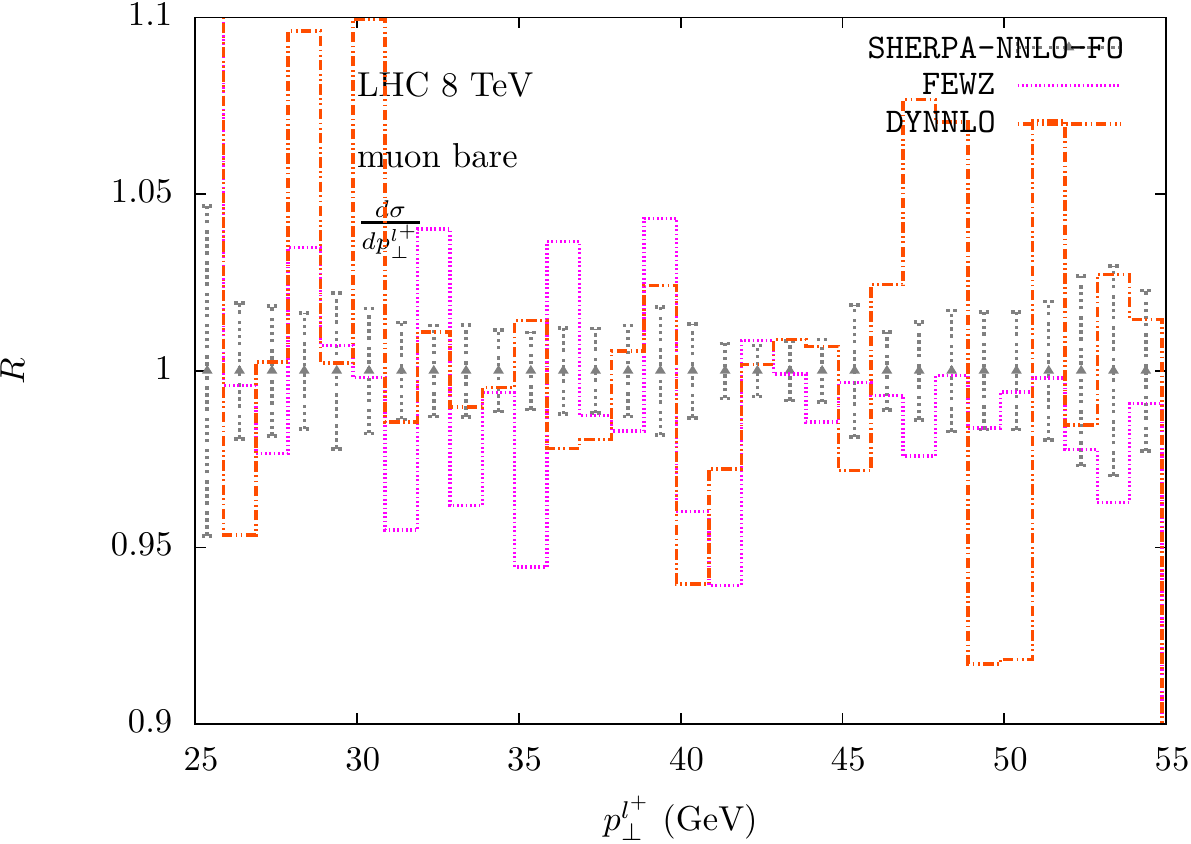}
\includegraphics[width=75mm,angle=0]{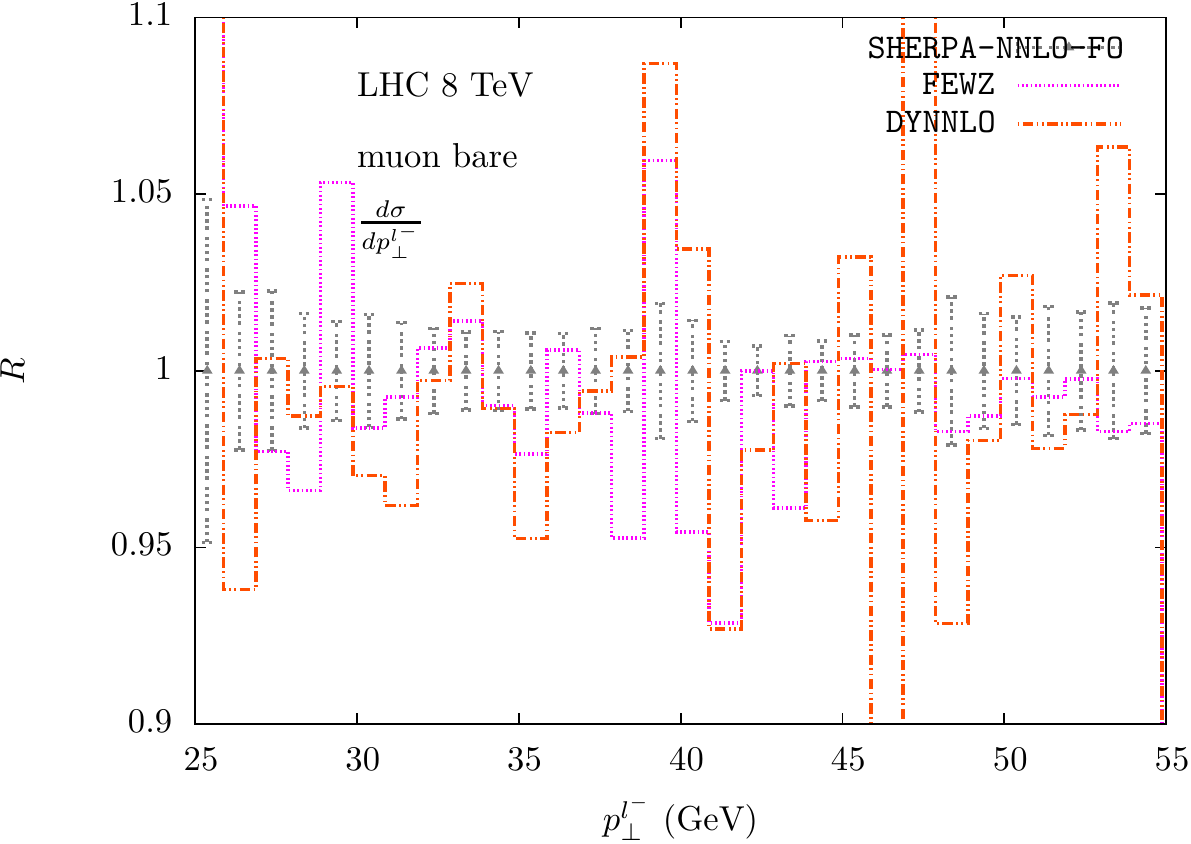}\\
\includegraphics[width=75mm,angle=0]{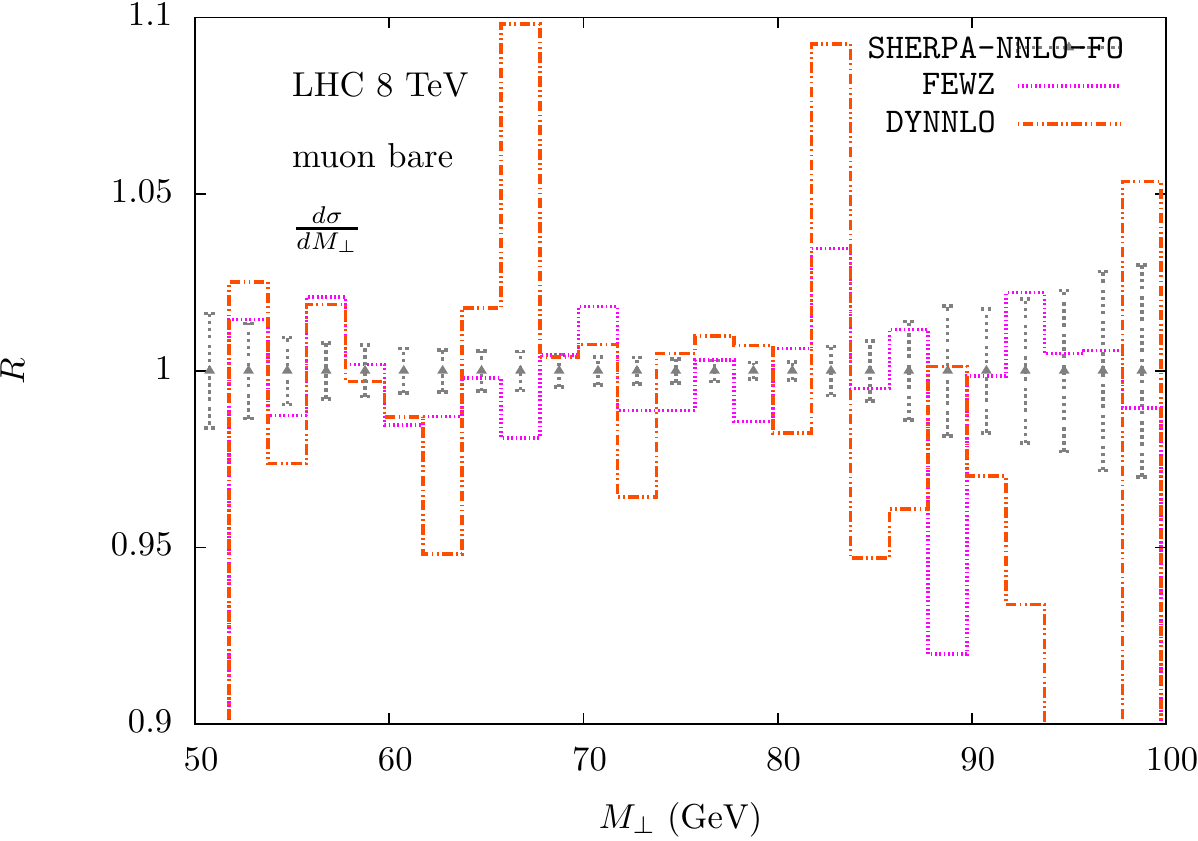}
\includegraphics[width=75mm,angle=0]{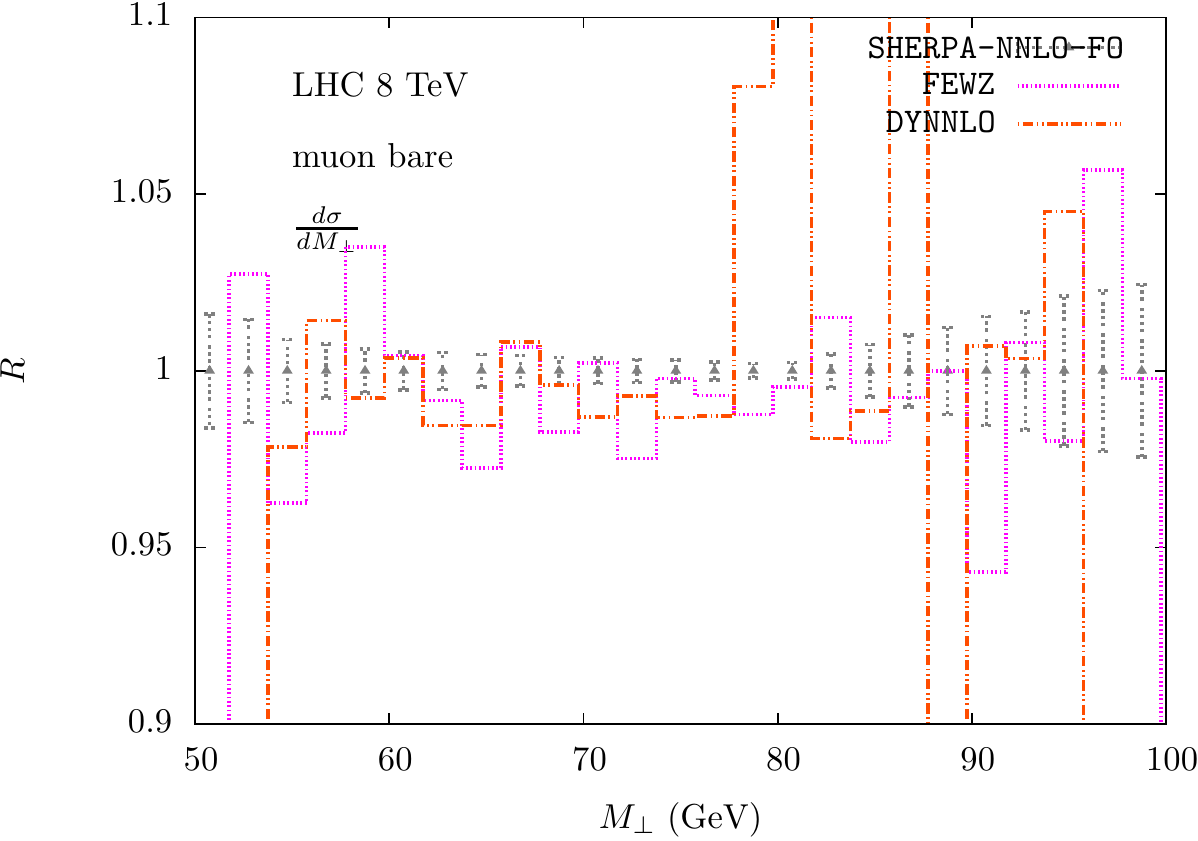}\\
\includegraphics[width=75mm,angle=0]{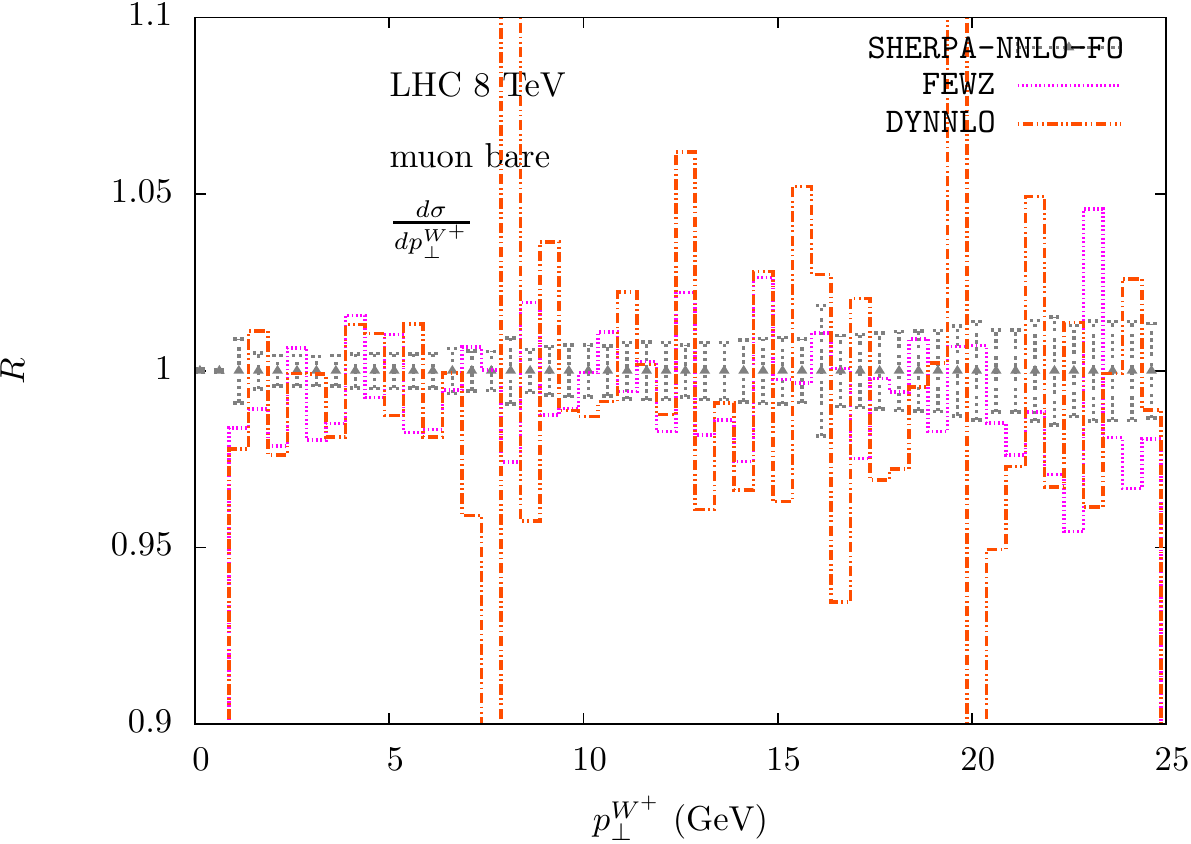}
\includegraphics[width=75mm,angle=0]{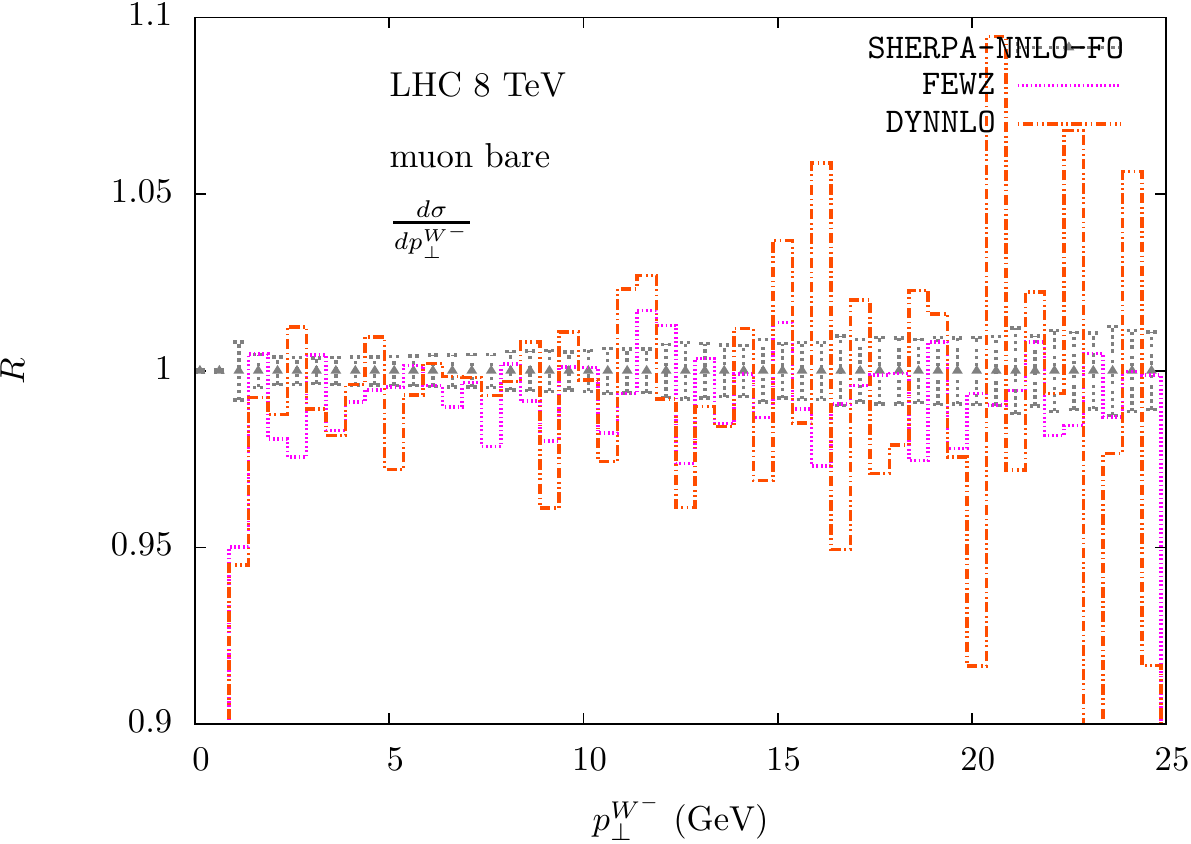}\\
\includegraphics[width=75mm,angle=0]{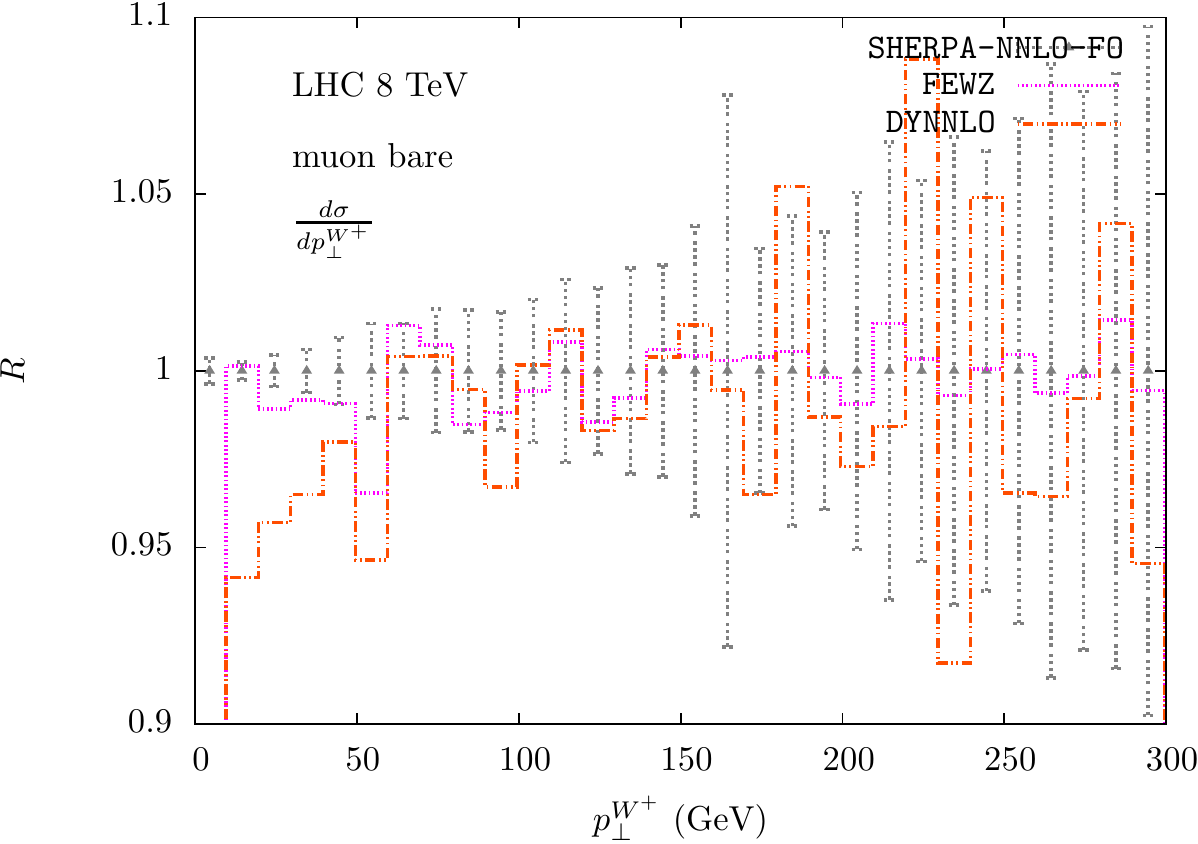}
\includegraphics[width=75mm,angle=0]{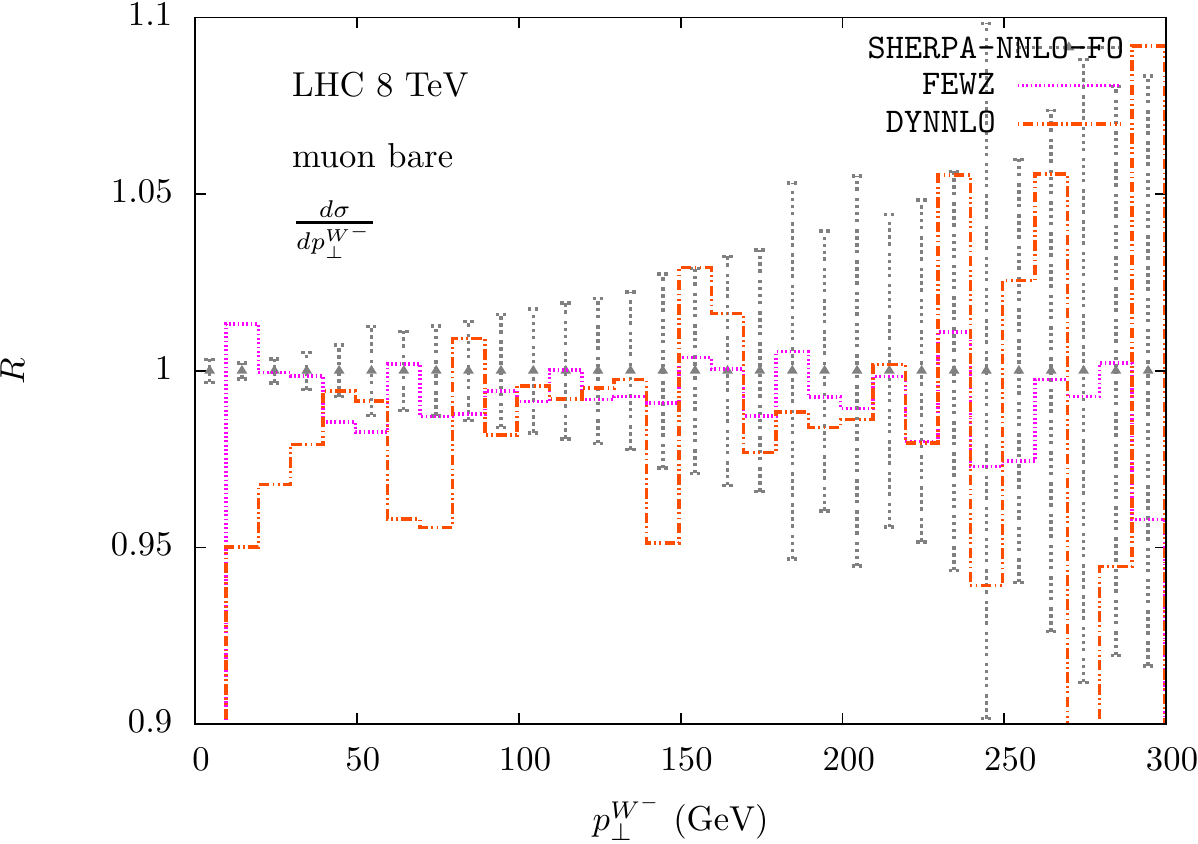}
 \caption{Comparison of NNLO QCD predictions by \dynnlo, \fewz and
\sherpafo for $pp\to W^\pm \to \mu^{\pm}\nu_\mu+X$ for $\mu^+\nu_{\mu}$
(left plots) and $\mu^-\bar\nu_{\mu}$ (right plots) final states.
Comparison of the lepton transverse momentum (upper plots), transverse
mass (middle plots) and lepton-pair transverse momentum (lower plots)
distributions, obtained in the {\em benchmark} setup with ATLAS/CMS cuts at the 8 TeV LHC.
\label{fig:Wpm-nnlo-qcd-tuned}
}
 \end{figure}
\begin{figure}[!ht]
\includegraphics[width=75mm,angle=0]{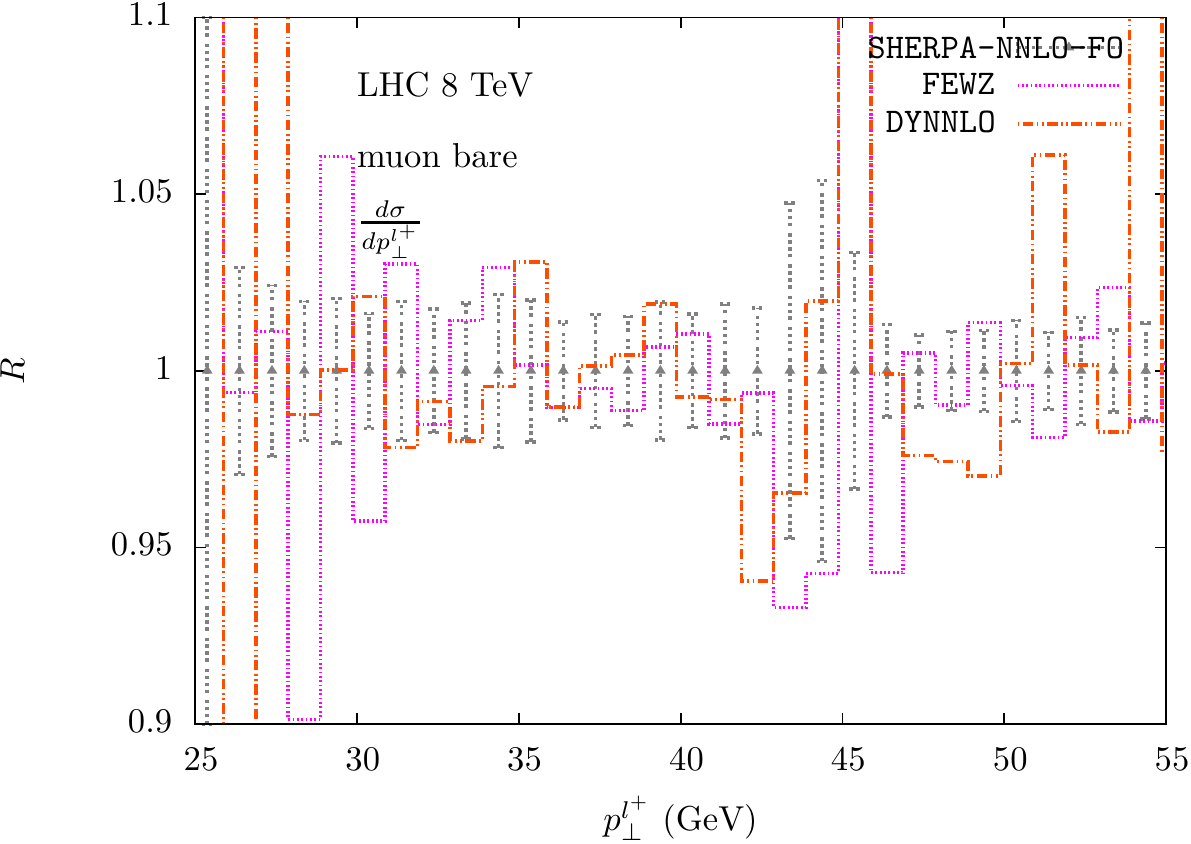}
\includegraphics[width=75mm,angle=0]{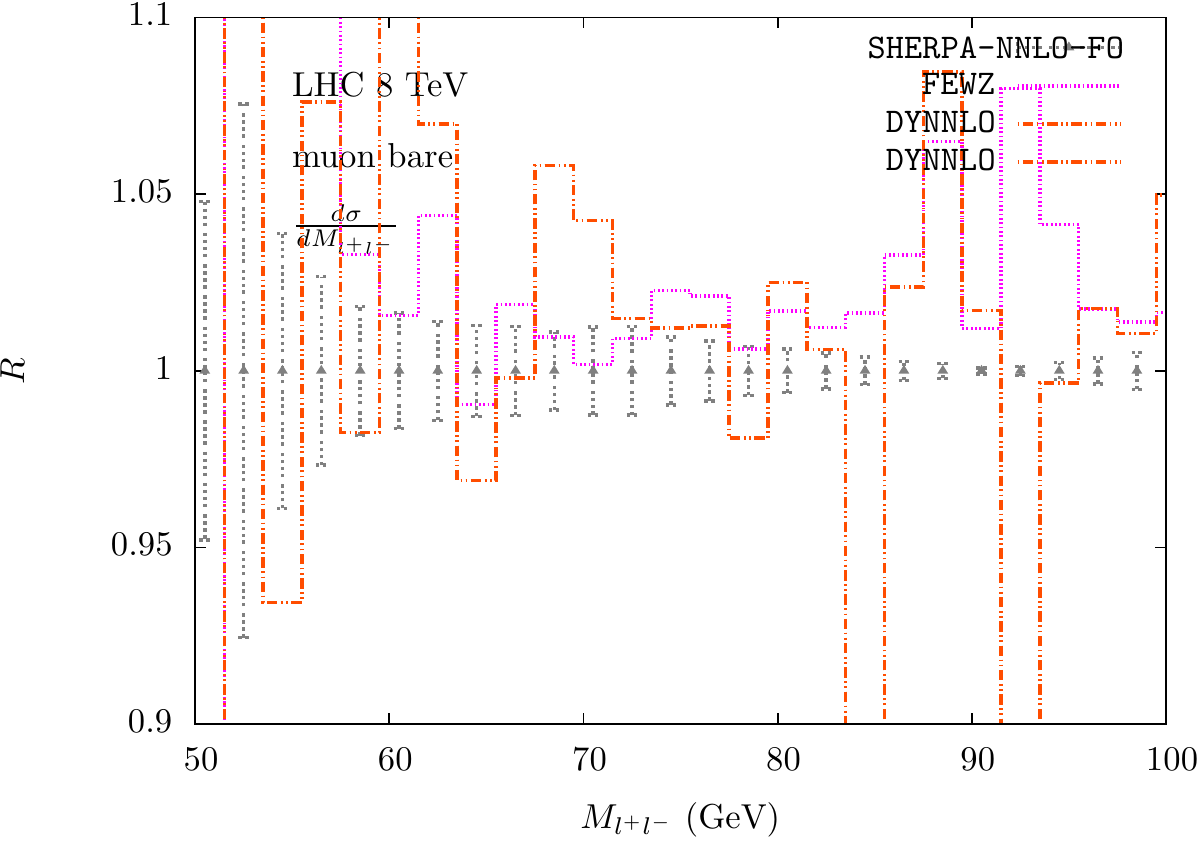}\\
\includegraphics[width=75mm,angle=0]{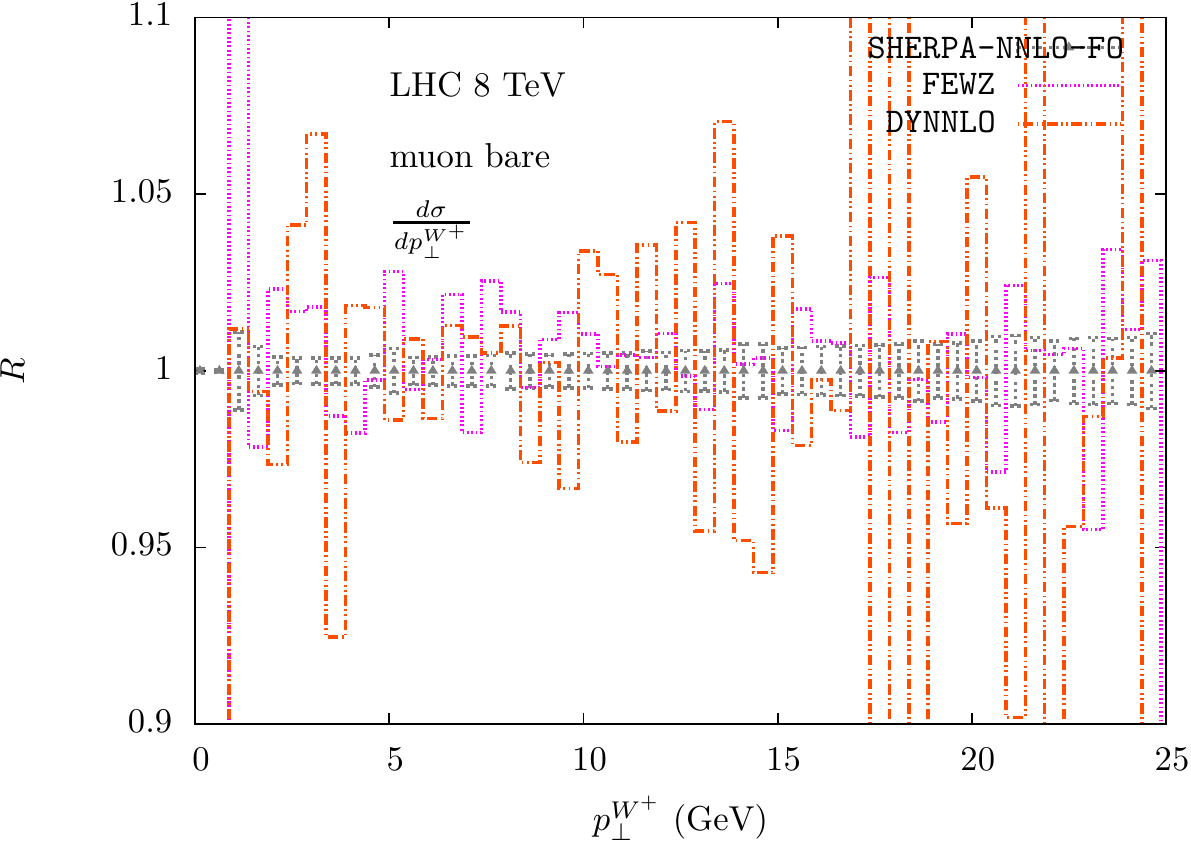}
\includegraphics[width=75mm,angle=0]{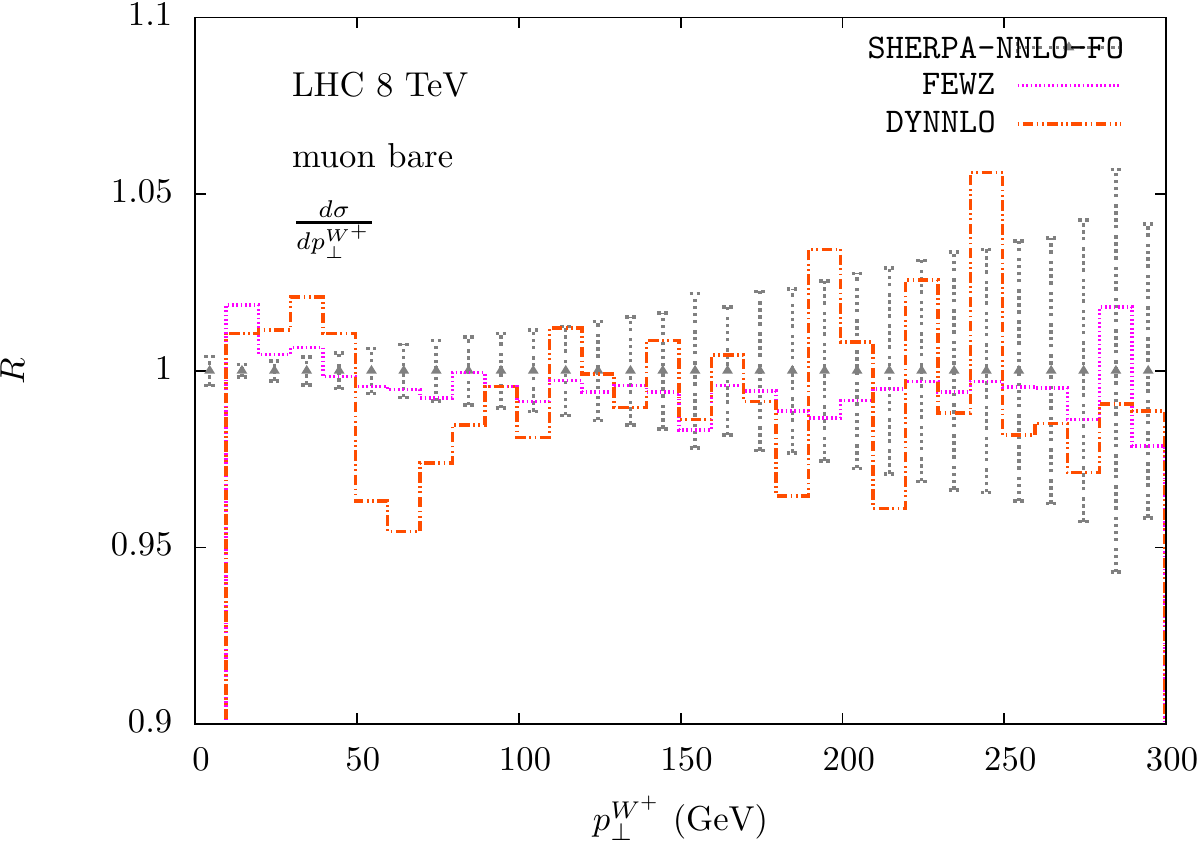}
 \caption{Comparison of NNLO QCD predictions by \dynnlo, \fewz and \sherpafo for $pp\to \gamma, Z \to \mu^+\mu^-+X$.
Comparison of the lepton transverse momentum (upper left),
lepton-pair invariant mass (upper right)
and lepton-pair transverse momentum (lower plots) distributions, obtained in the {\em benchmark} setup with ATLAS/CMS cuts at the 8 TeV LHC.
\label{fig:Z-nnlo-qcd-tuned}
}
 \end{figure}

The impact of NNLO QCD corrections on the total cross section of the
DY processes depends on the corrections to the lower-order processes 
but also on a small contribution from new partonic channels. 
The second order corrections reduce the
renormalization/factorization scale dependence of the final result,
with respect to NLO QCD, and bring it down to the 1\%
level \cite{Anastasiou:2003yy,Catani:2009sm}.  

The small differences between the results of Table
\ref{tab:xsec-nnloqcd} can be partially understood by an analysis of
the behavior of the subtraction methods implemented in the three codes
in the setup of the report.  The integrated cross section in presence
of symmetric cuts on the transverse momentum of lepton and missing
energy suffers from the pathological behavior first described
in~\cite{Frixione:1997ks}.  Let us assume staggered cuts, where
$p_{T,l}\ge E_T^{\rm cut}$ and $E_{T,miss}\ge E_T^{\rm cut}+\Delta$,
i.e.\ the difference in the minimum transverse momentum is
parametrized as $\Delta$.  The real-emission contribution to the
integrated NLO cross section then behaves as~\cite{Frixione:1997ks}
\begin{equation}\label{eq:frixione_ridolfi_realxs}
  \sigma^{(r)} = A(\Delta,\delta)+B\log\delta-C(\Delta+\delta)\log(\Delta+\delta)\;.
\end{equation}
Here, $\delta$ denotes the regulator in a phase-space slicing method.
In subtraction methods, $\delta$ is zero. $A(\Delta,\delta)$ and its
first derivative with respect to $\Delta$ are regular in $\Delta = 0$
for any $\delta$, including $\delta = 0$~\cite{Frixione:1997ks}.  $B$
and $C$ are coefficients, with $B$ identifying the collinear
singularity, which is canceled by the corresponding singular terms in
the two-body contribution to the total cross section.  The term of
interest is therefore $-C(\Delta+\delta)\log(\Delta+\delta)$.  It is
possible to verify numerically that it describes the behavior of the
NLO cross section in the Drell-Yan process as a function of $\Delta$.
The maximal deviation of the cross section from the expected behavior
based on phase-space considerations is $O(1\%)$.  The important point
to notice, however is the dependence on the slicing parameter
$\delta$.  Its value must be chosen small enough to suppress any
residual effect on the total cross section as $\Delta\to 0$, i.e.\ in
the presence of symmetric cuts. The relevance to the present
comparison arises from the fact that both \sherpa and DYNNLO use a
phase-space slicing technique at NNLO, while FEWZ employs a subtraction
method.  The NNLO calculation shows a feature similar to
Eq.~(\ref{eq:frixione_ridolfi_realxs}), although the magnitude and
functional dependence on $\Delta$ and $\delta$ cannot be predicted due
to the intricate interplay between real-virtual and double-real
corrections.  A variation of the $q_T$ slicing parameter in \sherpa in
the range $0.15\ldots 1~{\rm GeV}$, yields a residual effect on the
total cross section of $O(0.2\%)$, which is of the same order as the
numerical accuracy in our NNLO calculations. The {\tt SHERPA-NNLO-FO} result shown in 
Table~\ref{tab:xsec-nnloqcd} is obtained with a $q_T$ slicing parameter of 0.01~GeV.

\subsubsection{NNLO QCD corrections: kinematic distributions}
\label{sec:NNLO-QCD-diff}

The NNLO QCD predictions for kinematic distributions are compared
for a subset of observables in Figures \ref{fig:Wpm-nnlo-qcd-tuned}
and \ref{fig:Z-nnlo-qcd-tuned}, where the ratio to the \sherpafo
prediction is shown. As it can be seen, the predictions agree within
the statistical uncertainties of the MC integration.

The impact of NNLO QCD corrections on the kinematic distributions
of the DY processes depends on the observable under study.  Since some
observables such as the lepton-pair transverse momentum, the
single-lepton transverse momentum or the $\phi^*$ variable are
strongly sensitive to the details of real QCD radiation at NLO, they
are significantly modified by the second order QCD corrections.  On
the contrary the (pseudo-)rapidity distributions and the
invariant/transverse mass distributions receive a milder corrections,
closer in size to the value of the total NNLO K-factor.

\begin{figure}[!ht]
 \includegraphics[width=75mm,angle=0]{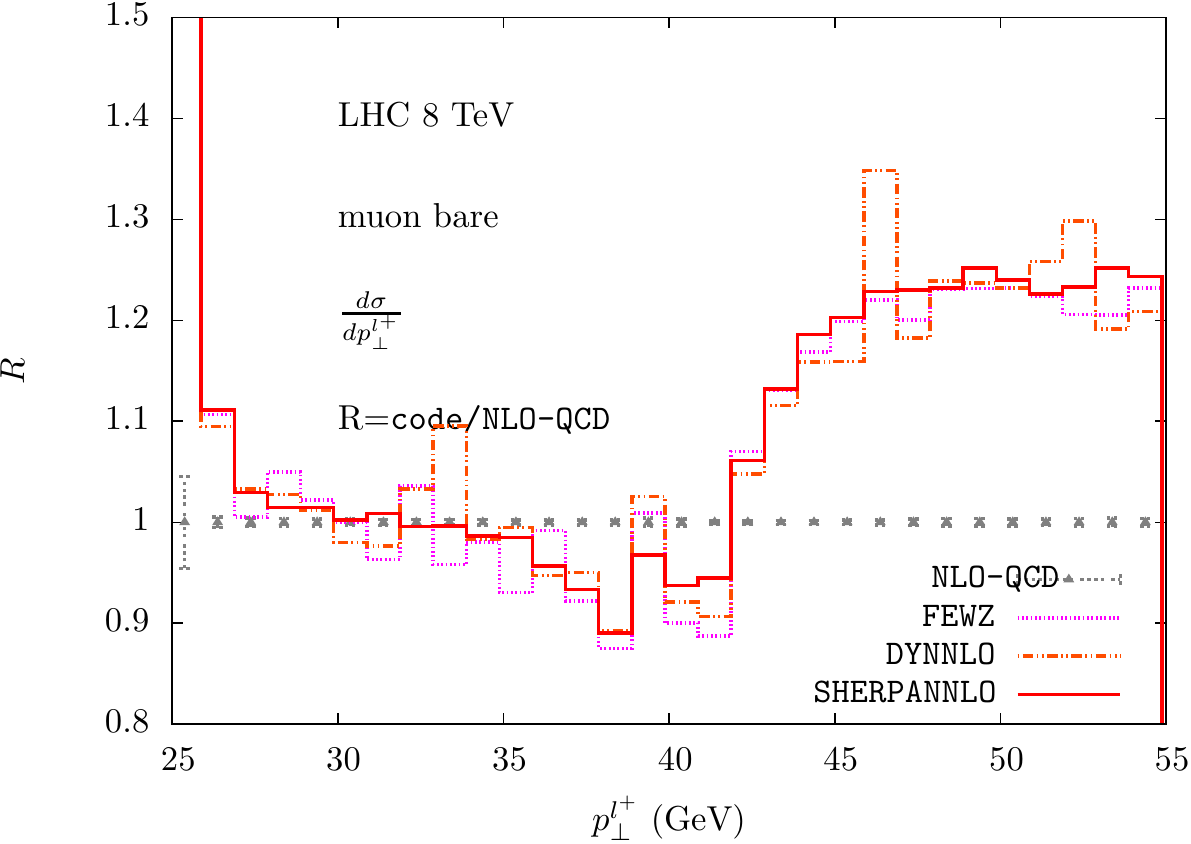}
 \includegraphics[width=75mm,angle=0]{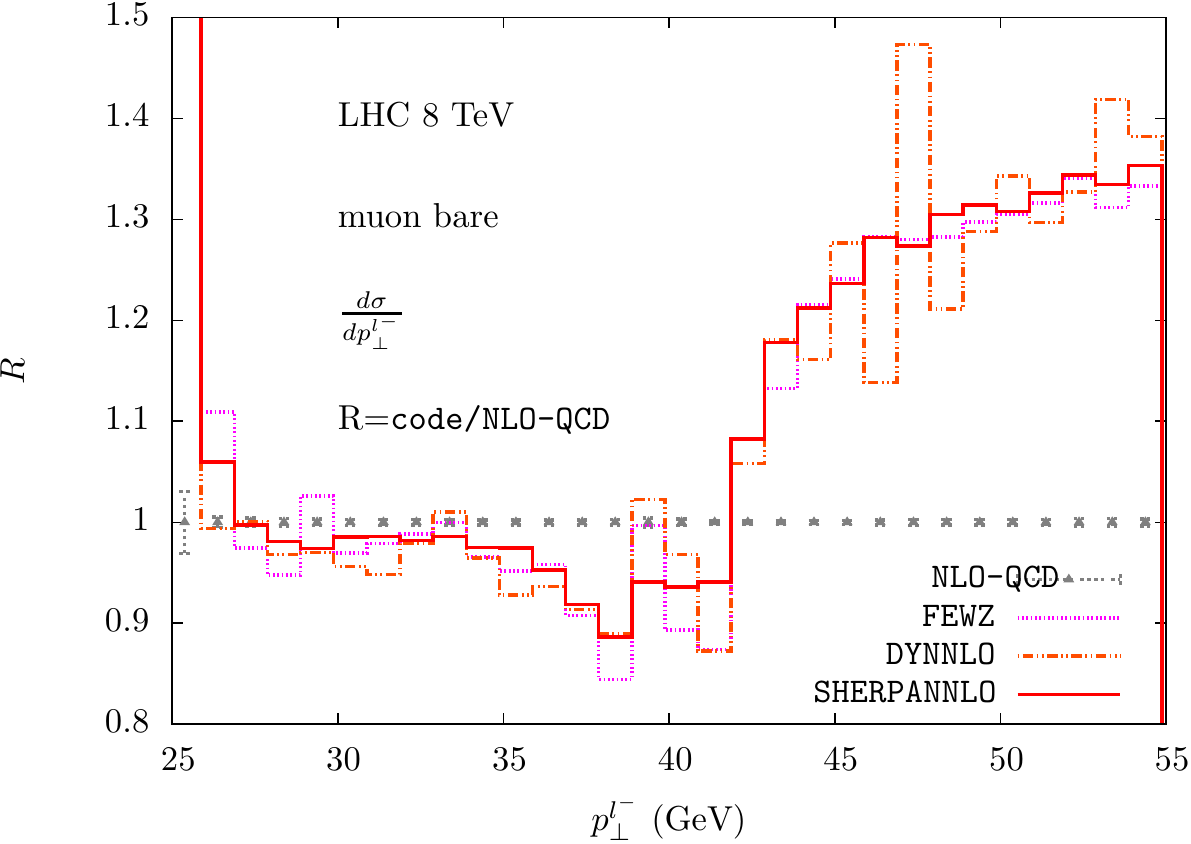}\\
 \includegraphics[width=75mm,angle=0]{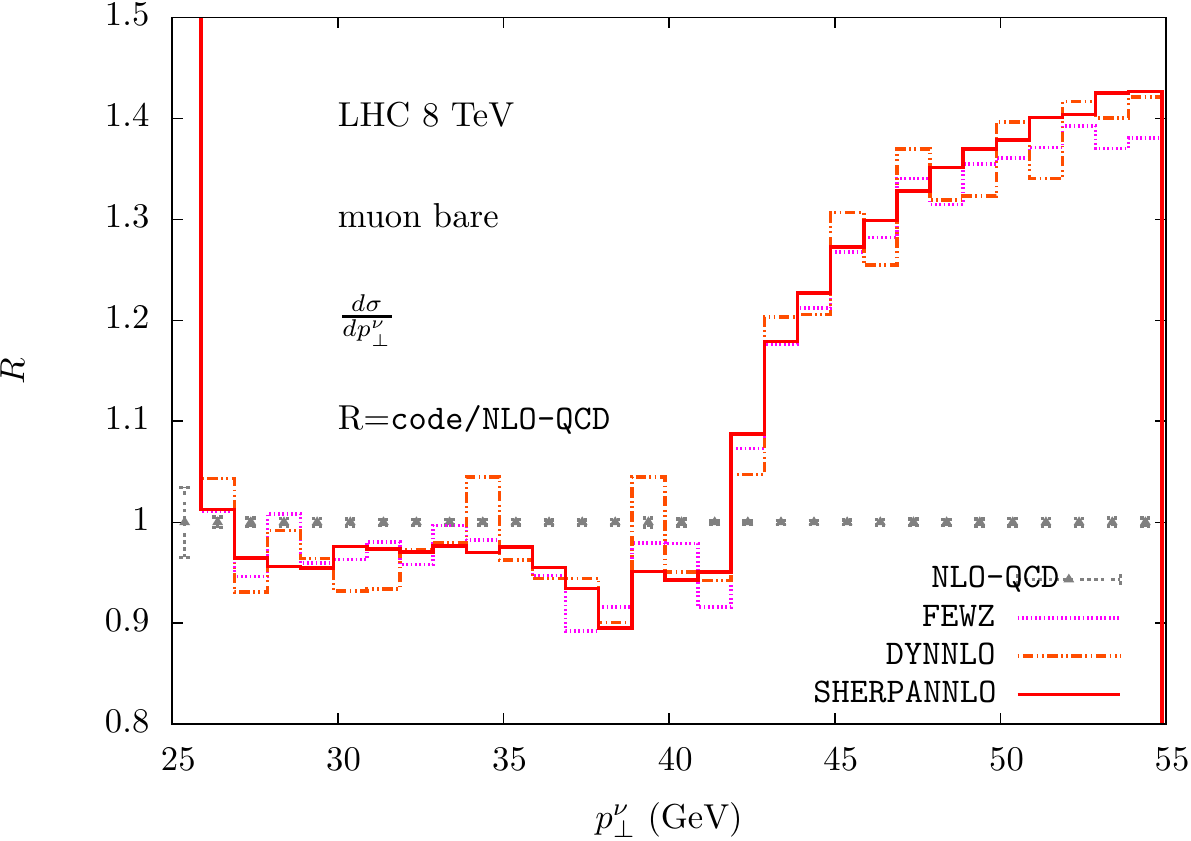}
 \includegraphics[width=75mm,angle=0]{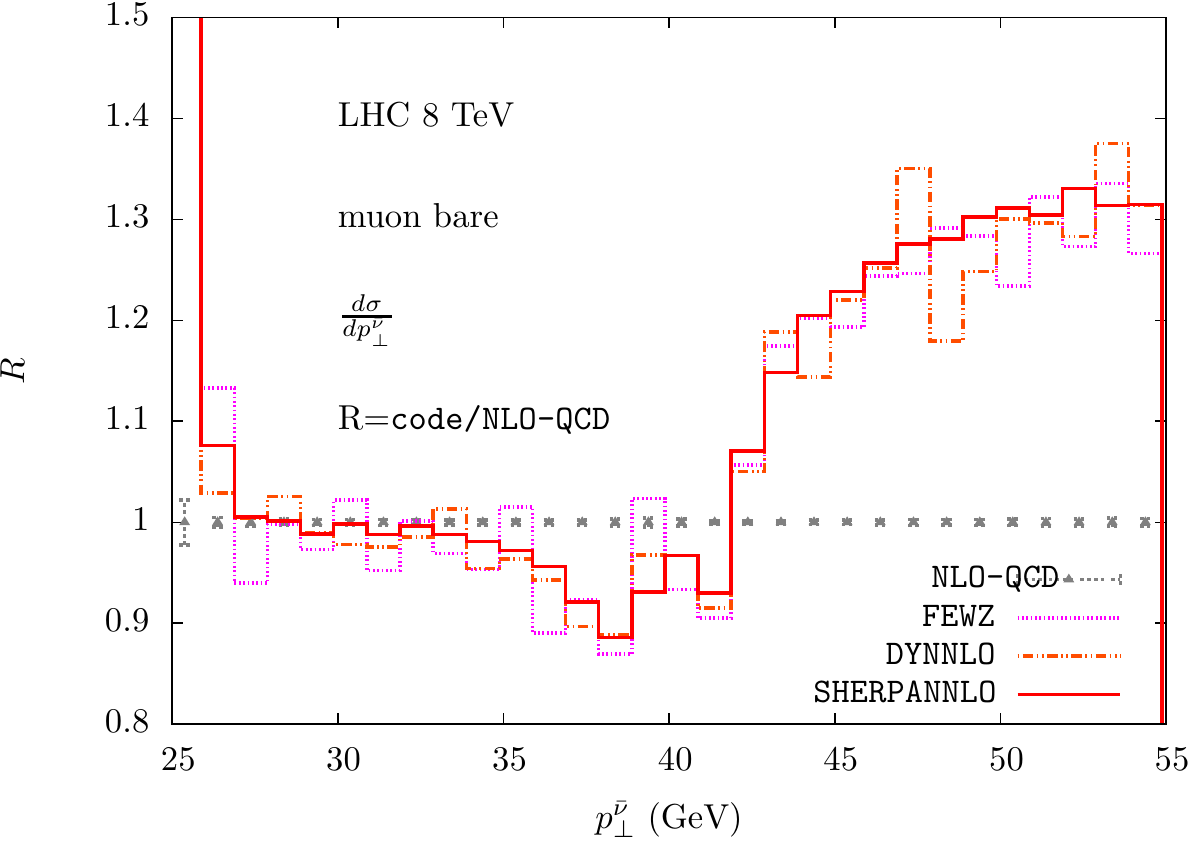}\\
 \includegraphics[width=75mm,angle=0]{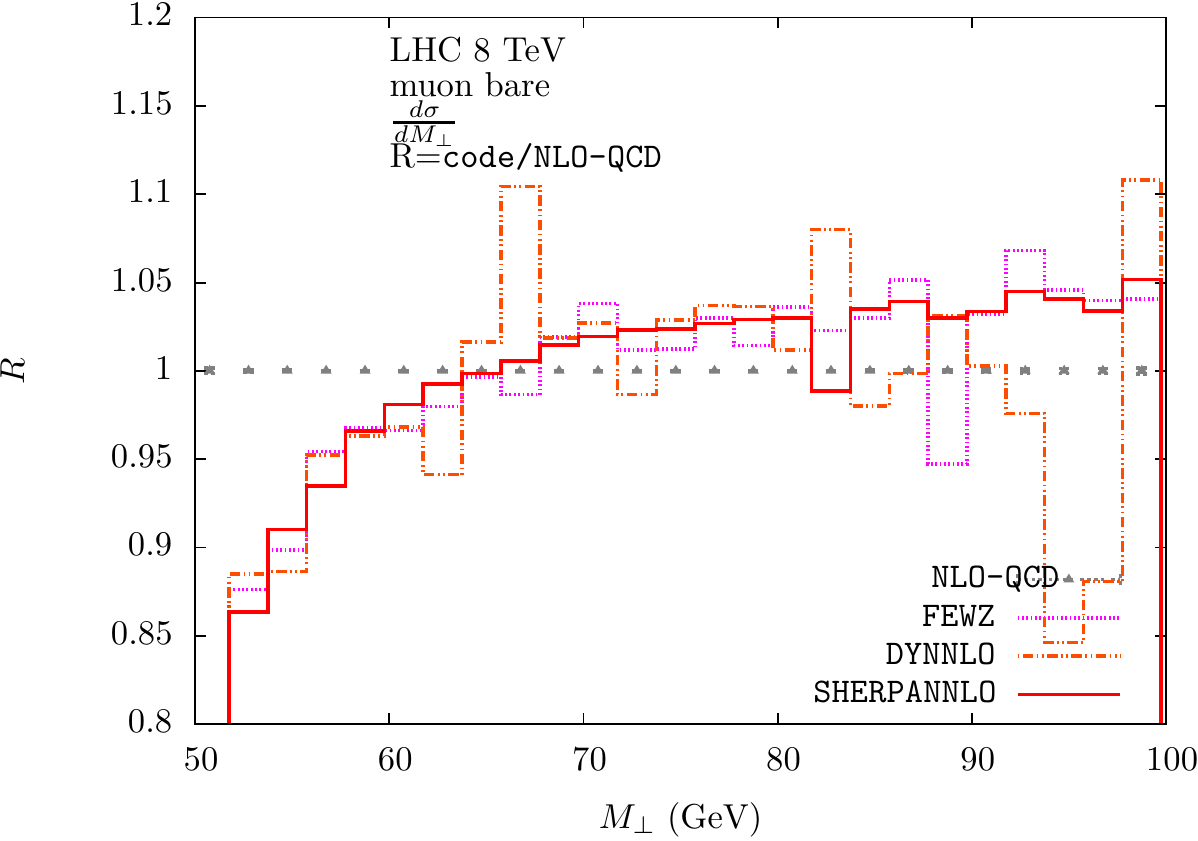}
 \includegraphics[width=75mm,angle=0]{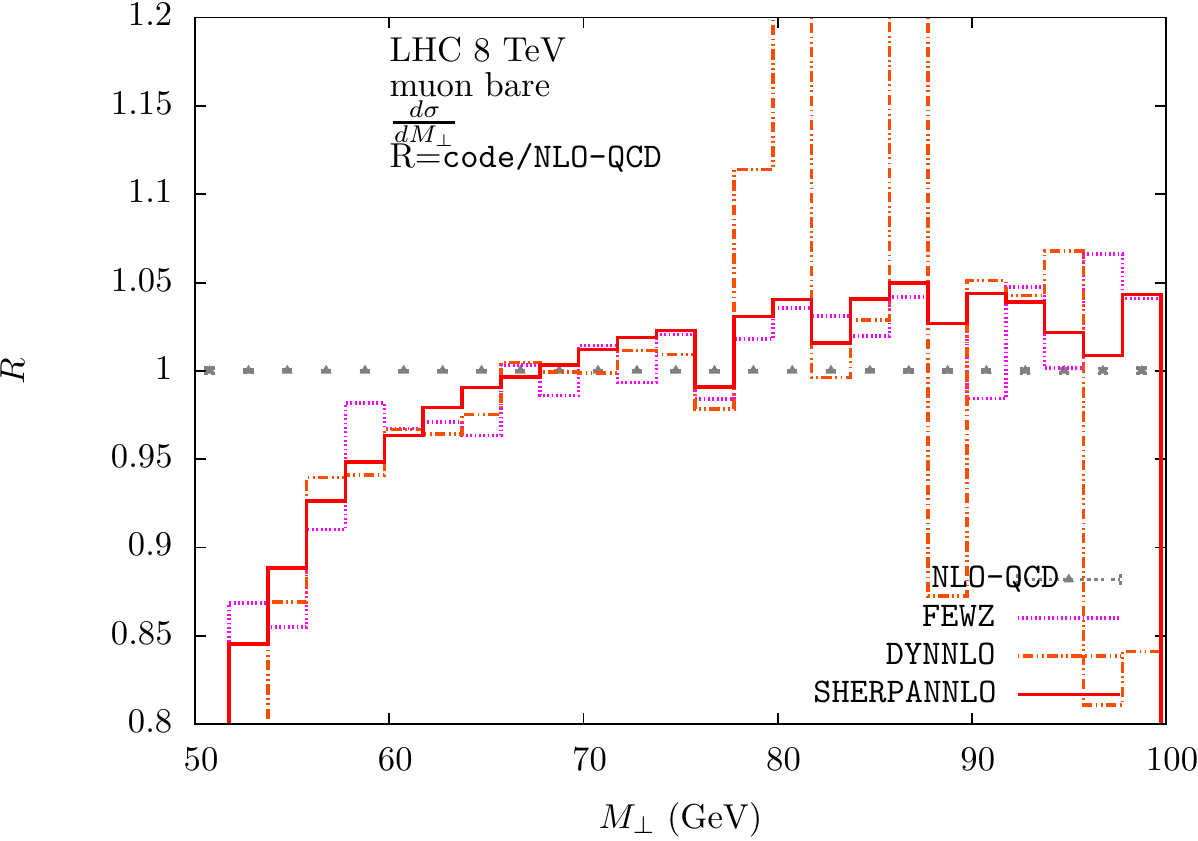}
 \caption{NNLO QCD effects, expressed in units of NLO QCD, in $pp\to\mu^{\pm}\nu_\mu+X$ in the {\em benchmark} setup with ATLAS/CMS cuts at the 8 TeV LHC,
for $\mu^+\nu_{\mu}$ (left plots) 
and  $\mu^-\bar\nu_{\mu}$ (right plots) final states. 
Comparison of the lepton transverse momentum (upper plots),
neutrino transverse momentum (middle plots) and transverse mass (lower plots) distributions, as predicted by
\sherpafo (red), \fewz (pink) and \dynnlo (orange dashed).
\label{fig:Wpm-ho-qcd-ptlptnumt}
}
 \end{figure}

\begin{figure}[!ht]
 \includegraphics[width=75mm,angle=0]{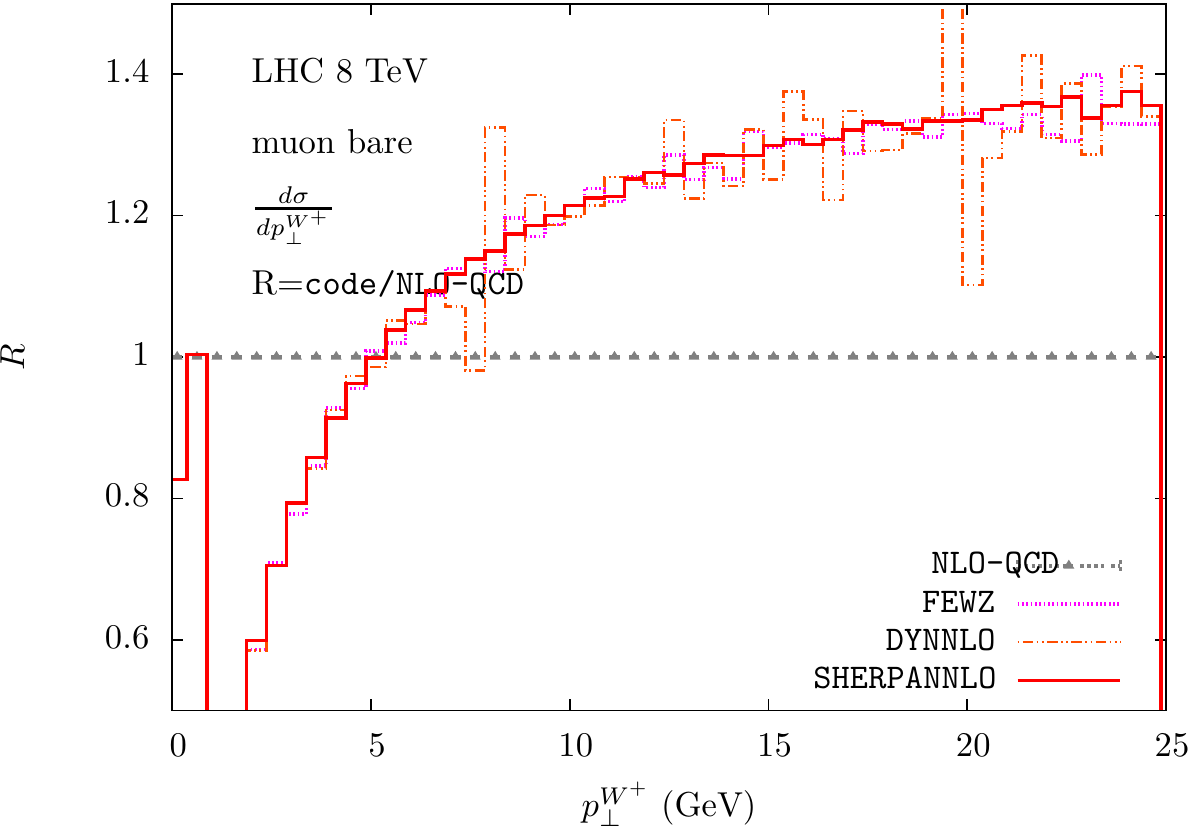}
 \includegraphics[width=75mm,angle=0]{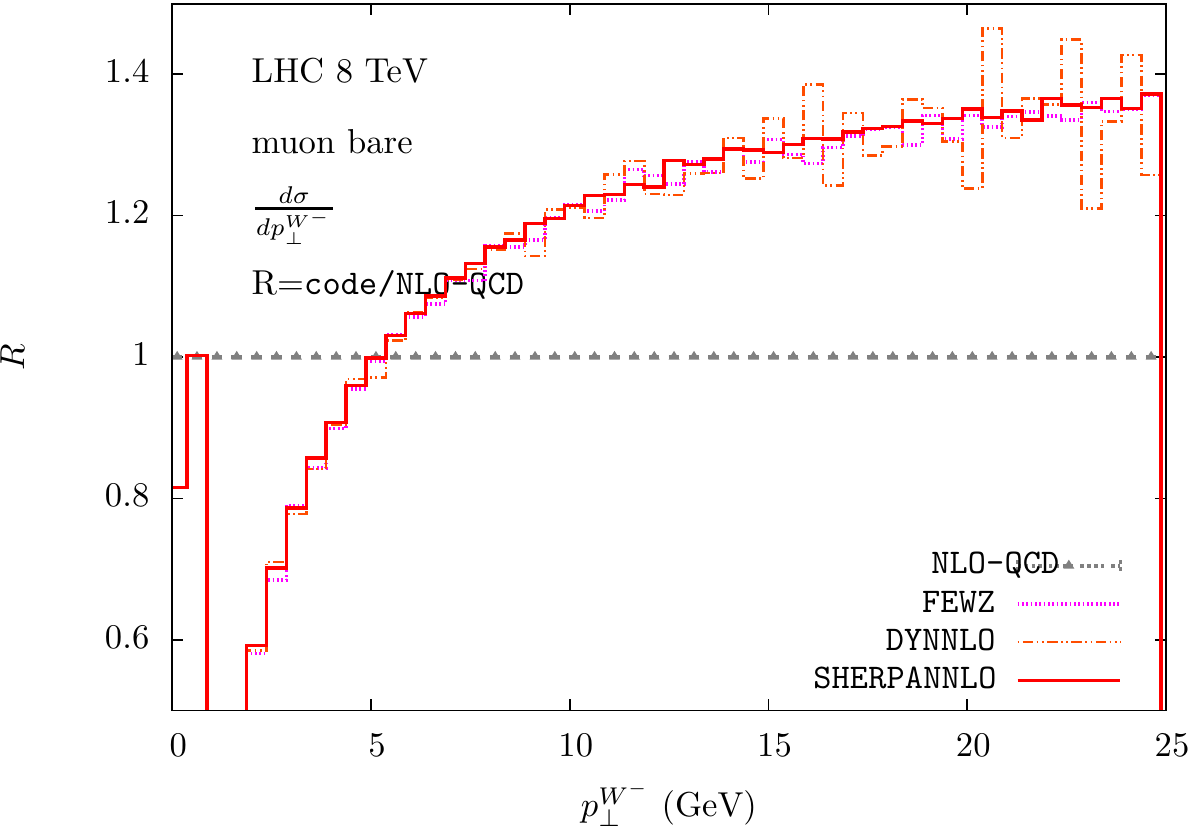}
 \includegraphics[width=75mm,angle=0]{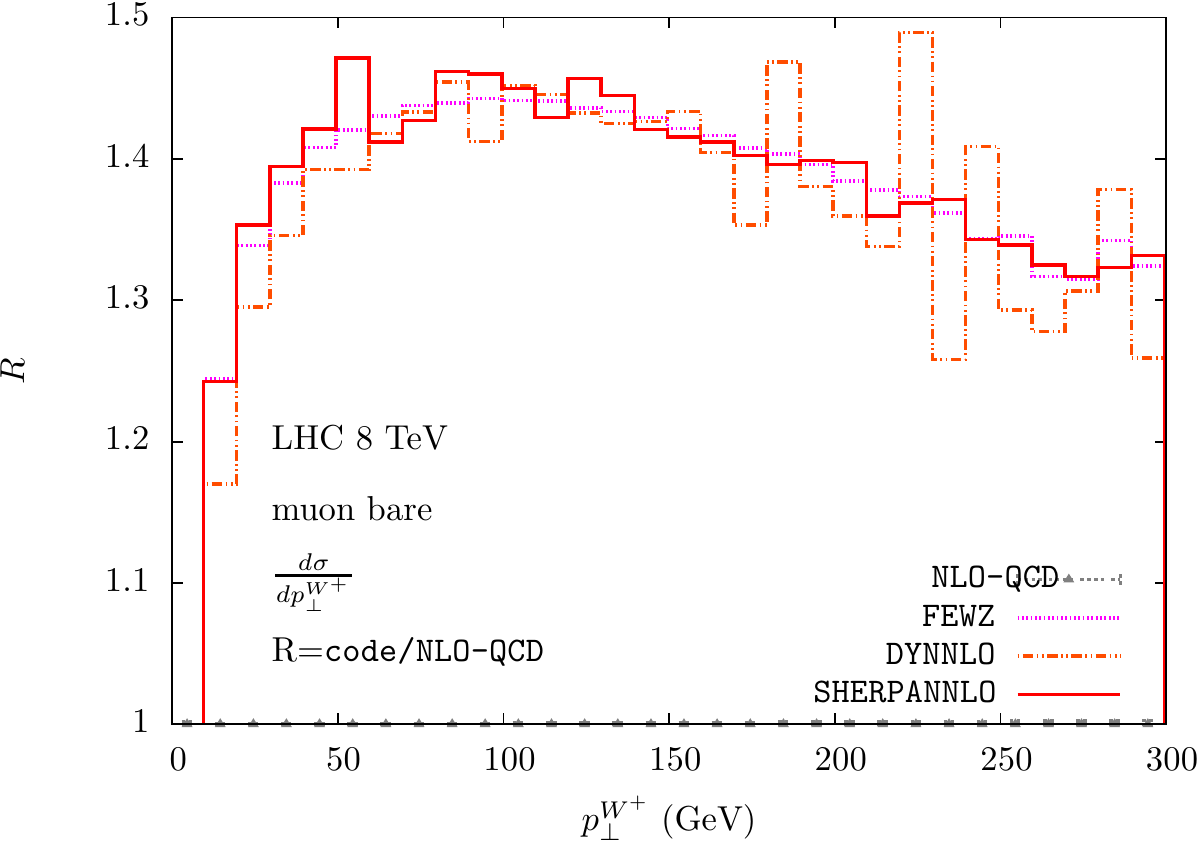}
 \includegraphics[width=75mm,angle=0]{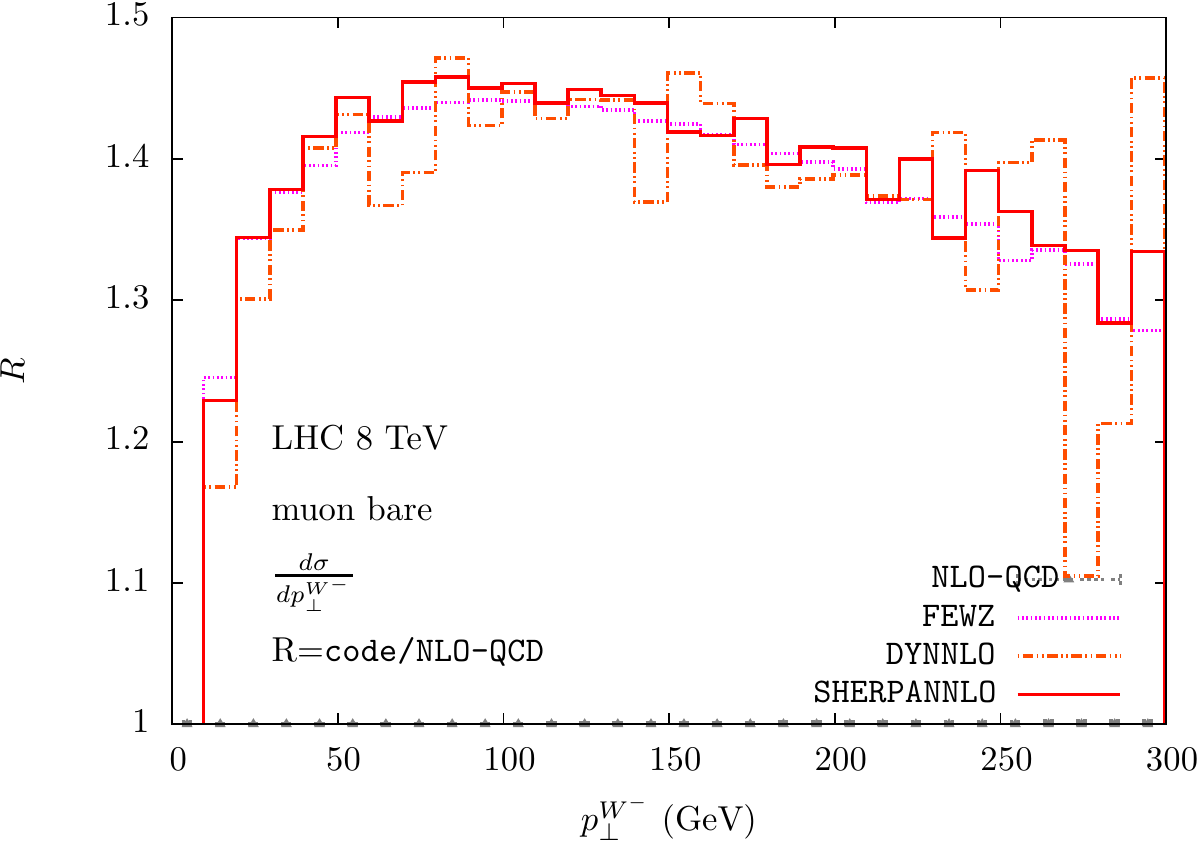}
 \caption{NNLO QCD effects, expressed in units of NLO QCD, in $pp\to\mu^{\pm}\nu_\mu+X$, in the {\em benchmark} setup with ATLAS/CMS cuts at the 8 TeV LHC,
for $\mu^+\nu_{\mu}$ (left plots) 
and  $\mu^-\bar\nu_{\mu}$ (right plots) final states. 
Comparison of the lepton-pair transverse momentum distributions
in the range $[0,25]$ GeV (upper plots) and
in the range $[0,300]$ GeV (lower plots), as predicted by
\sherpafo (red), \fewz (pink) and \dynnlo (orange dashed).
\label{fig:Wpm-ho-qcd-ptV}
}
 \end{figure}

\begin{figure}[!ht]
\includegraphics[width=75mm,angle=0]{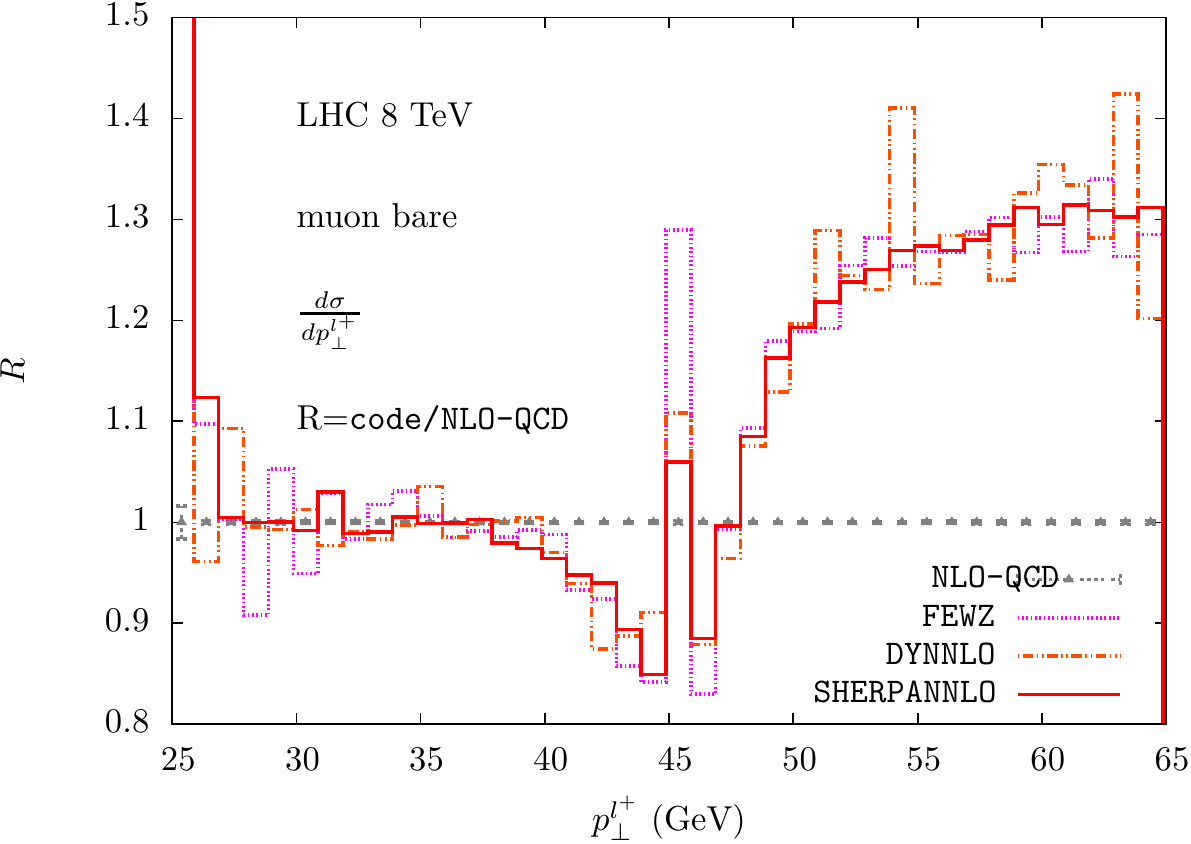}
\includegraphics[width=75mm,angle=0]{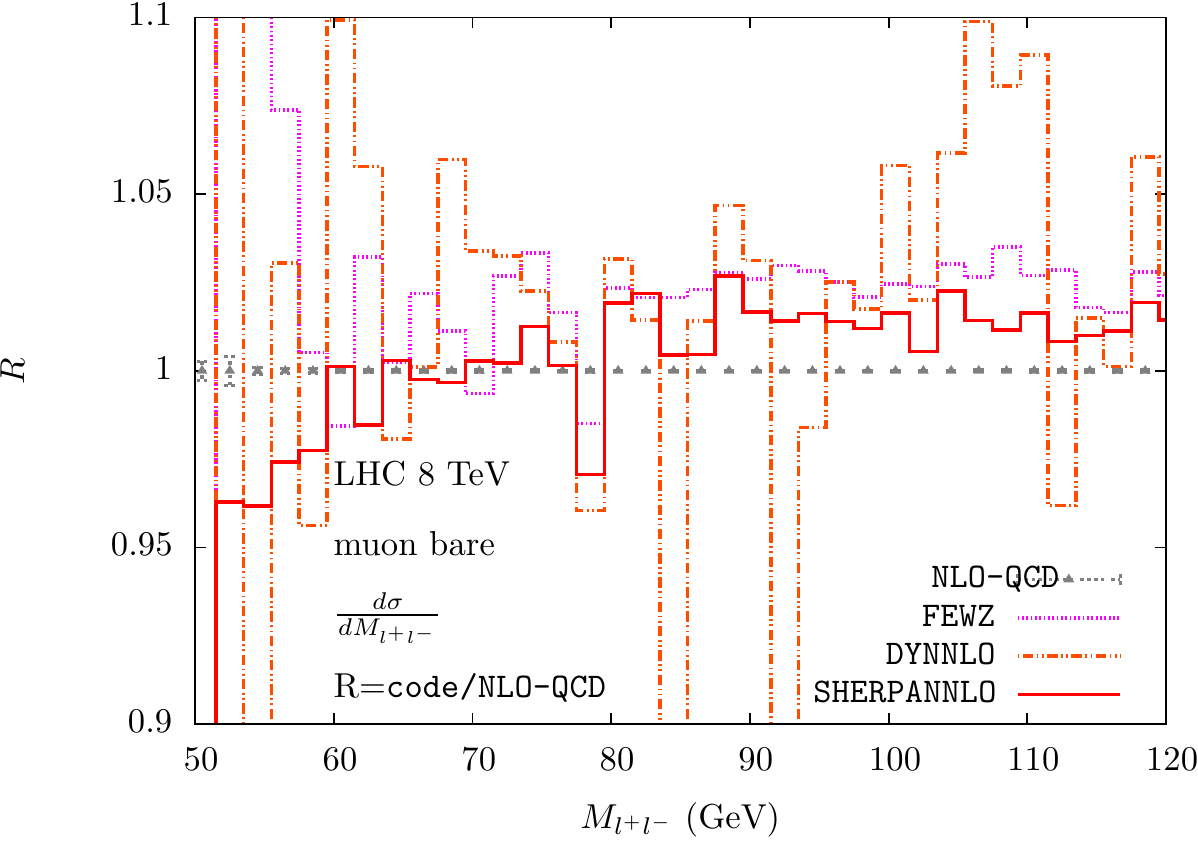}\\
\includegraphics[width=75mm,angle=0]{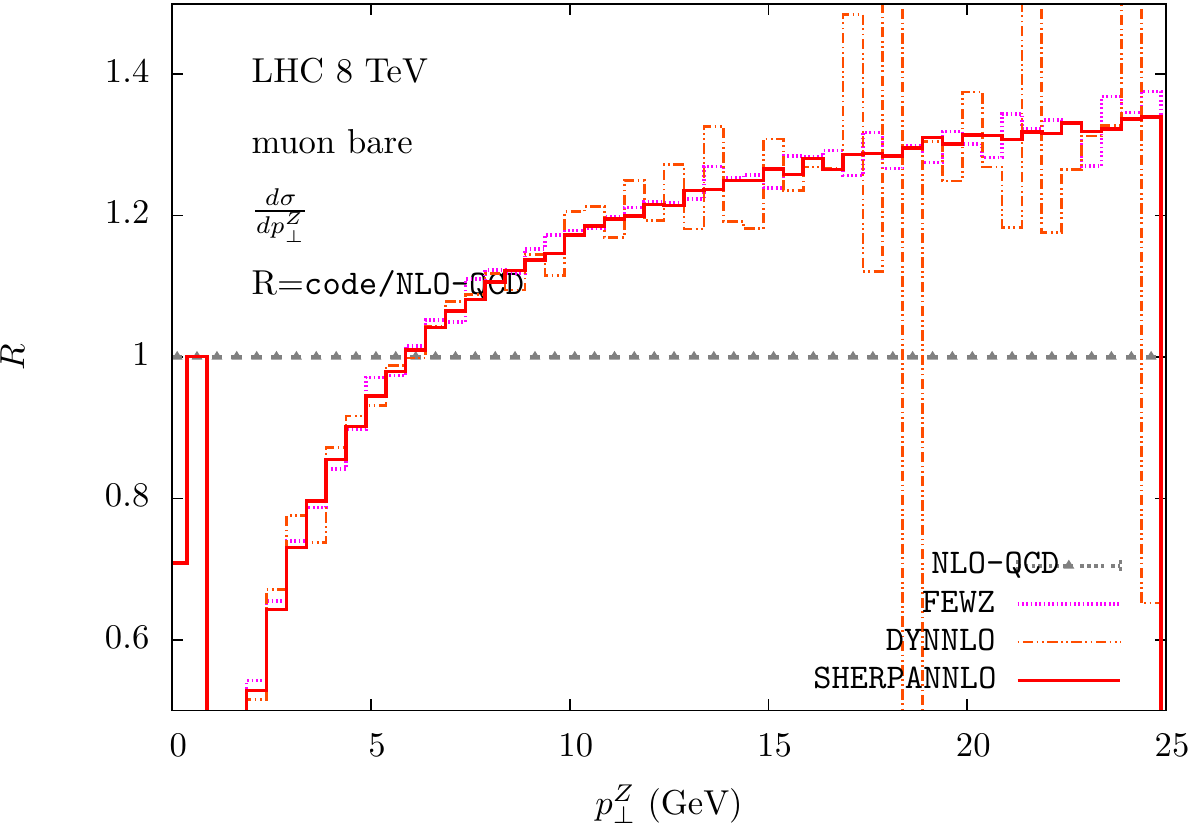}
\includegraphics[width=75mm,angle=0]{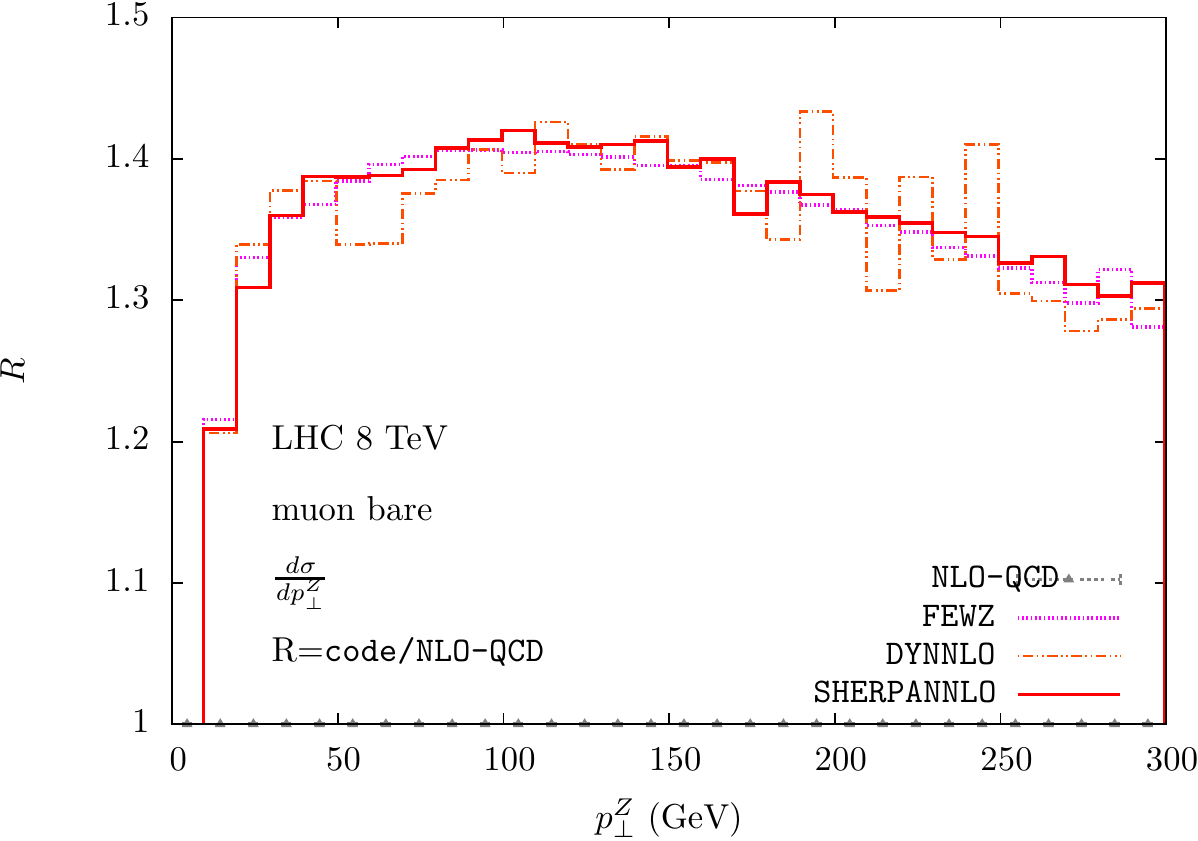}\\
\includegraphics[width=75mm,angle=0]{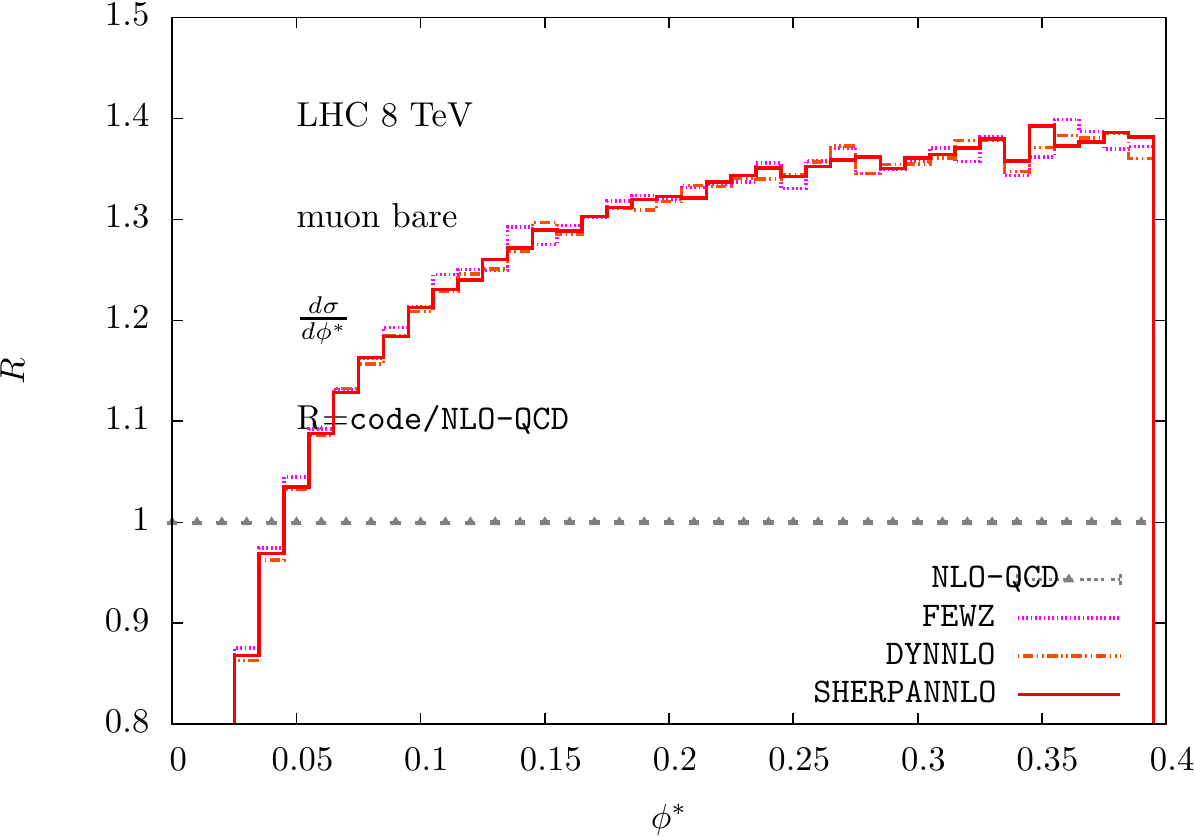}
\caption{NNLO QCD effects, expressed in units of NLO QCD, in $pp\to\mu^+\mu^-+X$, in the {\em benchmark} setup with ATLAS/CMS cuts at the 8 TeV LHC.
Comparison of the lepton transverse momentum (upper left), 
lepton-pair invariant mass (upper right),
lepton-pair transverse momentum (middle plots),
$\phi^*$ (lower plot) distributions, as predicted by
\sherpafo (red), \fewz (pink) and \dynnlo (orange dashed).
\label{fig:Z-ho-qcd}
}
\end{figure}

To illustrate the impact of the NNLO QCD corrections we compute for a
given observable ${\cal O}$ the ratio $R_{{\cal
    O}}=\left(\frac{d\sigma^{NNLO}}{d{\cal O}}\right) /
\left(\frac{d\sigma^{NLO}}{d{\cal O}} \right)$ with the same
distribution evaluated respectively with NNLO QCD and NLO QCD
accuracy.  We consider the distributions at NLO QCD as perfectly tuned
and neglect here the differences introduced by the choice in the
denominator of one NLO QCD code with respect to another one.  We
present the results in Figures
\ref{fig:Wpm-ho-qcd-ptlptnumt}-\ref{fig:Z-ho-qcd}.

We observe in Figures \ref{fig:Wpm-ho-qcd-ptlptnumt} and
\ref{fig:Z-ho-qcd} that the NNLO corrections have a mild impact on the
invariant-mass (NC DY) or transverse-mass (CC DY) distributions; the
correction is almost flat over the entire mass range considered.  The
more pronounced corrections that appear at the lower end of the
distributions can be understood as an effect of the acceptance cuts.

Figures~\ref{fig:Wpm-ho-qcd-ptlptnumt} and \ref{fig:Z-ho-qcd} show the
relative correction to the lepton and to the neutrino transverse
momentum distributions.  The NNLO QCD corrections, expressed in terms
of the NLO QCD result, are quite flat and moderate (smaller than 10\%)
below the Jacobian peak, they have a sharply peaked behaviour about
the Jacobian peak, where fixed order perturbation theory breaks down,
while they are of ${\cal O}(20\%)$ and are growing for increasing
transverse momentum above the Jacobian peak.  Again, the pronounced
corrections that appear at the lower end of the distributions can be
understood as an effect of the acceptance cuts.

In Figures \ref{fig:Wpm-ho-qcd-ptV} and \ref{fig:Z-ho-qcd} we show the
relative corrections to the lepton-pair transverse momentum
distributions, for the three processes (NC and CC) under
consideration, in two ranges of transverse momentum ($\ptv \in [0,25]$
GeV and $\ptv \in [0,250]$ GeV).  In fixed-order perturbation theory
the distribution is divergent in the limit of vanishing transverse
momentum; the sign of the first bin and the slope of the distributions
in this limit depend on the perturbative order, so that a comparison
between NLO QCD and NNLO QCD predictions is merely of technical
interest.  At large lepton-pair transverse momentum, where the
perturbative regime of QCD allows to study the convergence of the
perturbative expansion, the NNLO QCD corrections are large, of ${\cal
  O}(40\%)$, and quite flat in the range $50\leq \ptv \leq 300$ GeV.

The relative correction to the lepton-pair $\phi^*$ distribution in
the NC DY process is shown in Figure \ref{fig:Z-ho-qcd}.  Since in the
limit $\phi^* \to 0$ we probe the same phase-space region where the
lepton-pair has small transverse momentum, the distribution suffers of
the break-down of perturbation theory, so that the comparison between
the NNLO QCD and the NLO QCD predictions is again merely of technical
interest in this region.

\clearpage

\clearpage

\subsubsection{Higher-order QCD corrections to all orders: generalities}
\label{sec:QCDallorders}

As already mentioned in Section \ref{sec:NLO-QCD}, there are
observables whose description in fixed-order QCD is not adequate, so
that the resummation to all orders of logarithmically enhanced
contributions is necessary to obtain a physically sensible prediction.
The solution of this problem requires a certain number of choices,
which can be understood as potential sources of uncertainty.

\begin{itemize}
\item Matching a resummed and a (N)NLO fixed-order expressions
  requires a procedure that avoids double countings and possibly
  allows for the MC simulation of events with a probabilistic
  interpretation.  The solution of this problem at NLO was developed
  in \cite{Frixione:2002ik,Nason:2004rx} and more recently in
  \cite{Hamilton:2012np,Hamilton:2012rf,Hoeche:2014aia} also for the
  inclusion of NNLO partonic results.  Each approach solves the
  matching problem in a different way, yielding predictions that
  respect the nominal perturbative accuracy for observable that are
  stable under the inclusive evaluation of radiative effects, but
  differ in the treatment of higher-order terms.  The matching
  ambiguity, parametrized in different ways, should be considered as
  an additional source of theoretical uncertainty, together with the
  one usually expressed by the choice of the
  renormalization/factorization scales.

\item In the MC codes the resummation to all orders of some classes of
  contributions is done by means of a Parton Shower (PS) approach,
  with leading logarithmic (LL) accuracy in the log of the gauge boson
  transverse momentum.  There are differences of subleading
  logarithmic order in the available PS algorithms, which yield a
  difference in the final predictions.

\item The PS codes are usually interfaced with models that describe
  non-perturbative effects of the strong interaction at low energy
  scales; the parameters of these models are usually tuned to
  reproduce some relevant distribution, but their choice (and the
  corresponding quality of the description of the data) represents an
  additional source of ambiguity in the predictions.

\end{itemize}

In the study of the codes which match resummed and fixed-order results~\footnote{We note that 
the public codes {\tt ResBos}~\cite{Ladinsky:1993zn,Balazs:1995nz,Balazs:1997xd} 
and {\tt MC@NLO}~\cite{Frixione:2002ik} also 
provide predictions for matched NLO QCD+resummed initial-state contributions for
DY processes but are not used in this study.},
the presence of the entangled sources of differences listed above does
not allow a tuned comparison of 'central' values, as done with fixed
order results, and requires a careful interpretation of observed
differences.

In Figures~\ref{fig:Wpm-shower-qcd-NLO}-\ref{fig:Z-shower-qcd-NNLO} we
expose the impact of higher-order corrections, ${\cal O}(\alpha_s^2)$
and higher, in units of the NLO QCD results. In this way we appreciate
where the higher orders play a crucial role, how well the NNLO QCD
results are approximated by a NLO+PS formulation (Figures
\ref{fig:Wpm-shower-qcd-NLO},\ref{fig:Z-shower-qcd-NLO}), and the
impact of matching the NNLO QCD fixed-order calculation and a QCD-PS
(Figures \ref{fig:Wpm-shower-qcd-NNLO}-\ref{fig:Z-shower-qcd-NNLO}).
The disadvantage of this choice of presenting the results is that for
some observables the NLO QCD is not a sensible lowest order
approximation.

\clearpage

\subsubsection{Comparison of (NLO+PS)-QCD vs NNLO QCD results}
\label{sec:matching-NLOPS-QCD}
The \powheg+\pythia and the \sherpanlo NLO+PS predictions are based on the
same exact matrix elements present in all the codes that have NLO QCD
accuracy for the total cross section, but they add the
higher-order effects due to multiple parton emissions to all orders
via a QCD-PS, with two different matching procedures.  At ${\cal
O}(\alpha_S^2)$ they both have a partial overlap with those by the
fixed-order NNLO results, because of the inclusion of the LL terms. It
should be stressed that the \powheg+\pythia and the \sherpanlo NLO+PS codes
do not have NNLO QCD accuracy for the total cross section nor do they have an
accurate description of the large lepton-pair transverse momentum region,
where exact matrix element effects for the second emission are
important.  On the other hand, they include the resummation to all
orders of multiple parton emissions, which is important to yield a
sensible description of the small lepton-pair transverse momentum
region, of the low-$\phi^*$ region of the $\phi^*$ distribution or of
the Jacobian peak of the single lepton transverse momentum
distribution.

\begin{figure}[!h]
\includegraphics[width=75mm,angle=0]{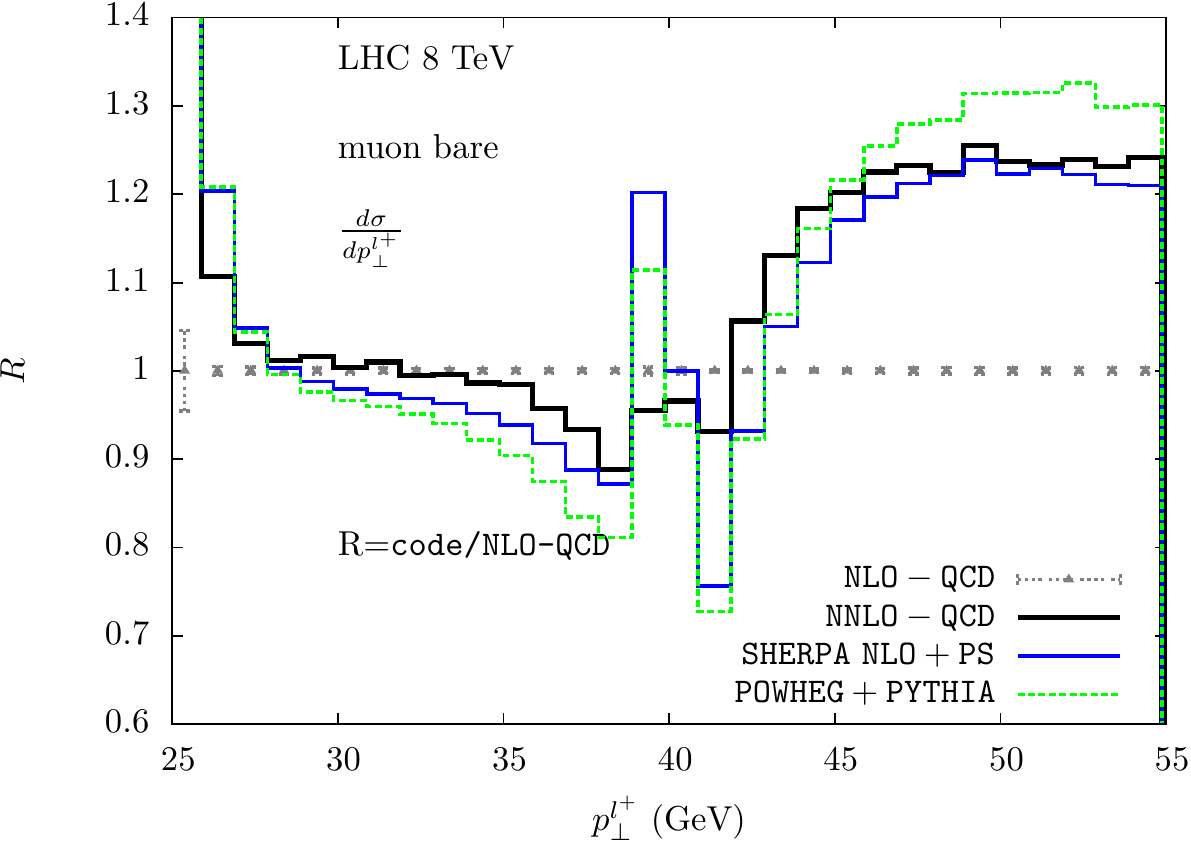}
\includegraphics[width=75mm,angle=0]{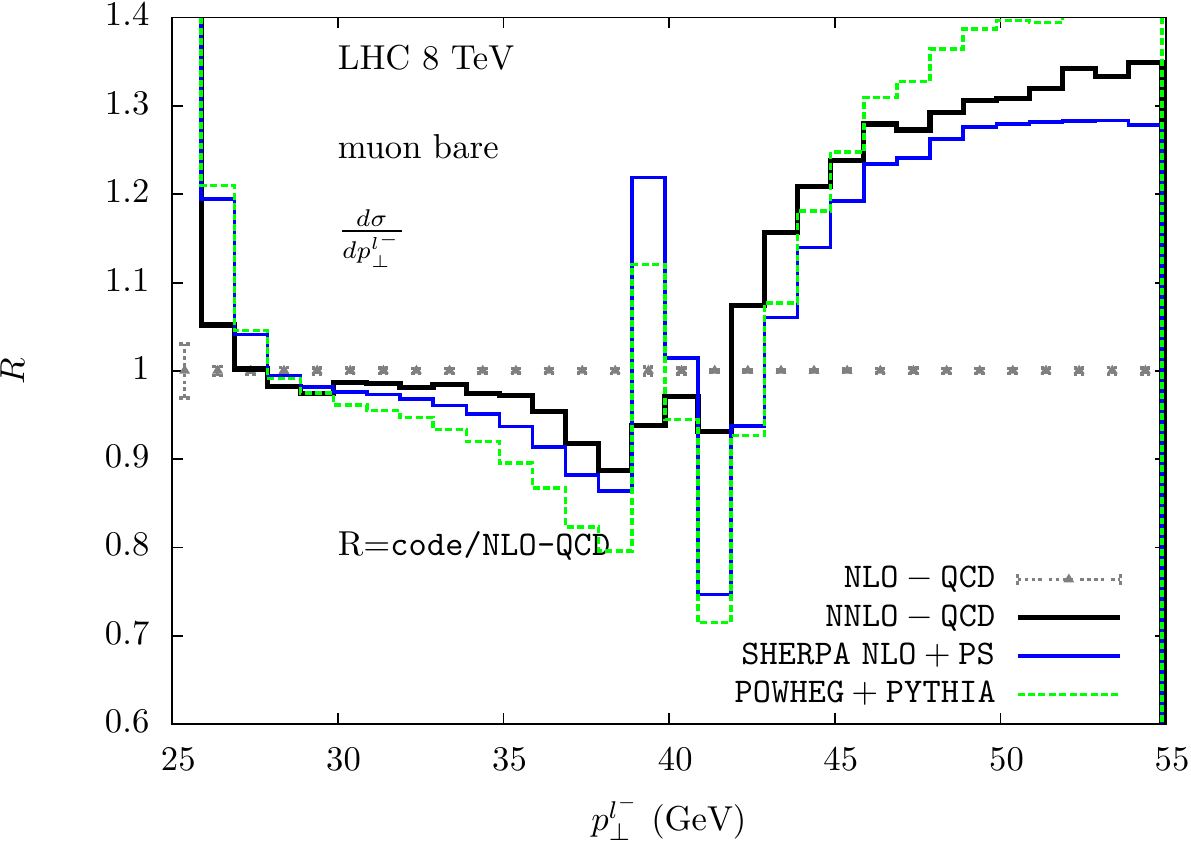}\\
\includegraphics[width=75mm,angle=0]{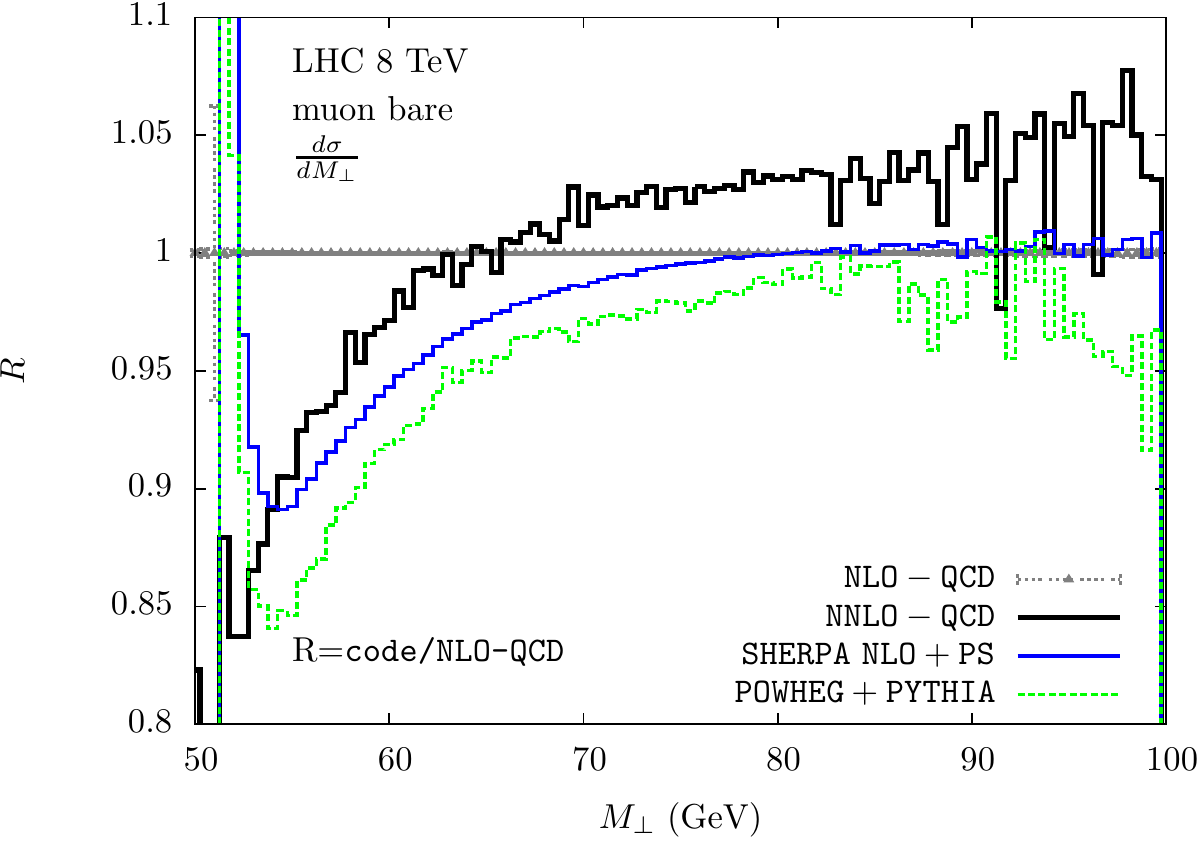}
\includegraphics[width=75mm,angle=0]{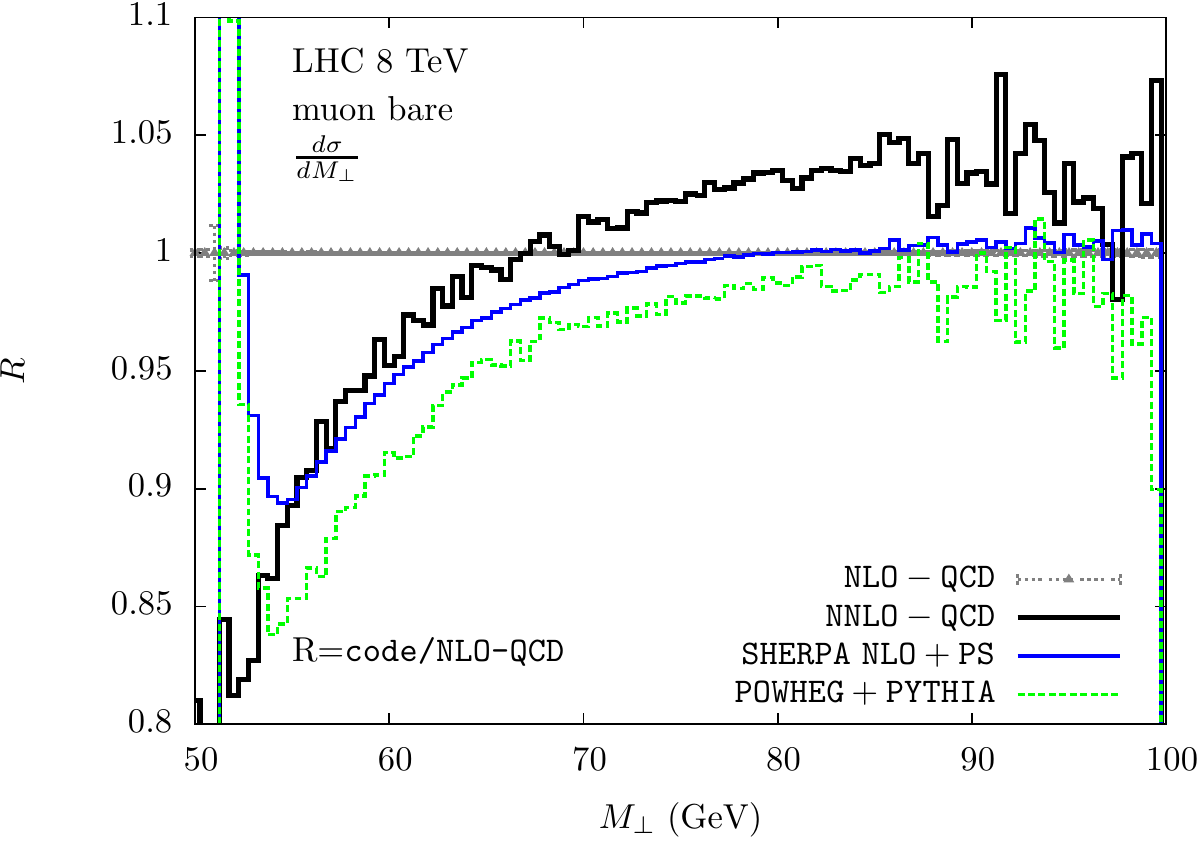}\\
\includegraphics[width=75mm,angle=0]{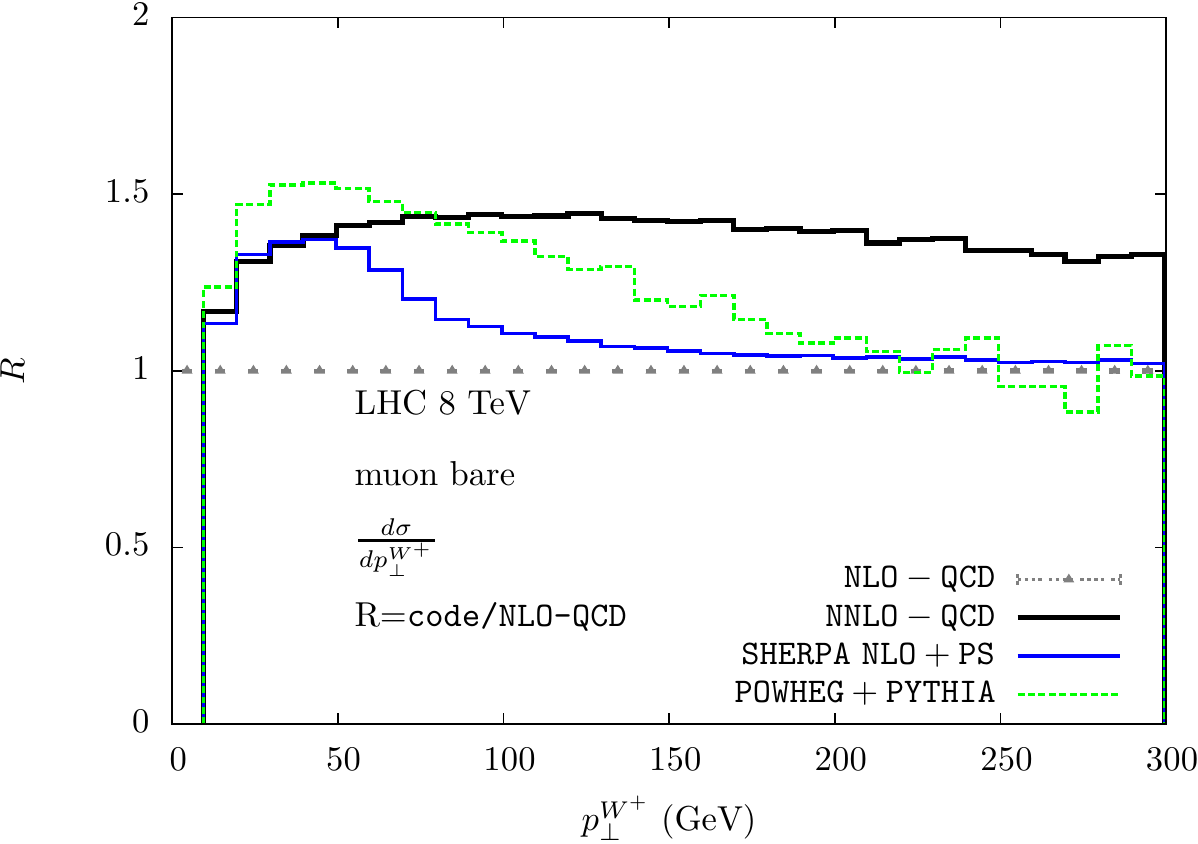}
\includegraphics[width=75mm,angle=0]{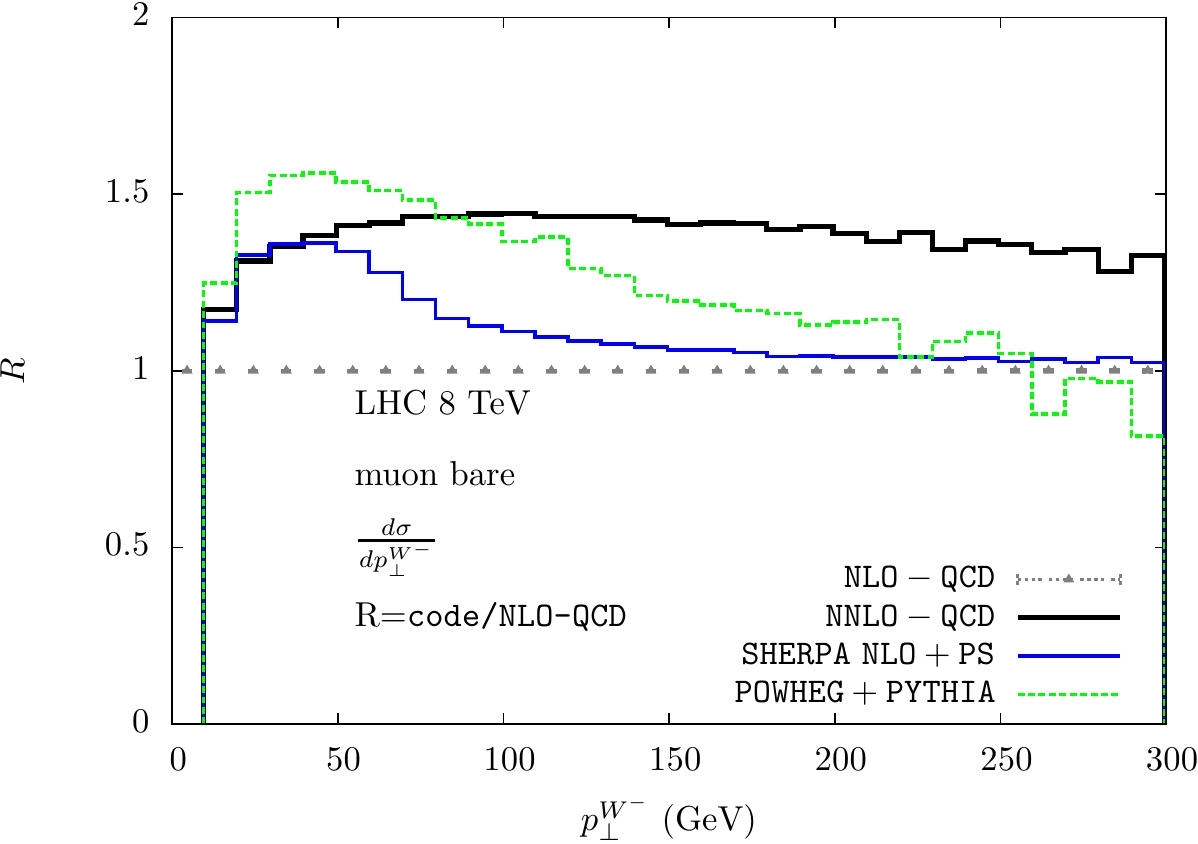} 
\caption{Higher-order QCD effects, expressed in units of NLO QCD, in $pp\to \mu^{\pm}\nu_\mu+X$, due to the matching of resummed and fixed order results, in codes with NLO accuracy:
\powheg+\pythia (green) and \sherpanlo (blue). The fixed-order NNLO QCD results are shown in black.
Comparison of results for the lepton transverse momentum (upper plots),
lepton-pair transverse mass (middle plots) and transverse momentum (lower plots) distributions,
for $\mu^+\nu_{\mu}$ (left plots) and  $\mu^-\bar\nu_{\mu}$ (right plots) final states, 
obtained with ATLAS/CMS cuts at the 8 TeV LHC.
\label{fig:Wpm-shower-qcd-NLO}
}
 \end{figure}

\begin{figure}[!h]
\includegraphics[width=75mm,angle=0]{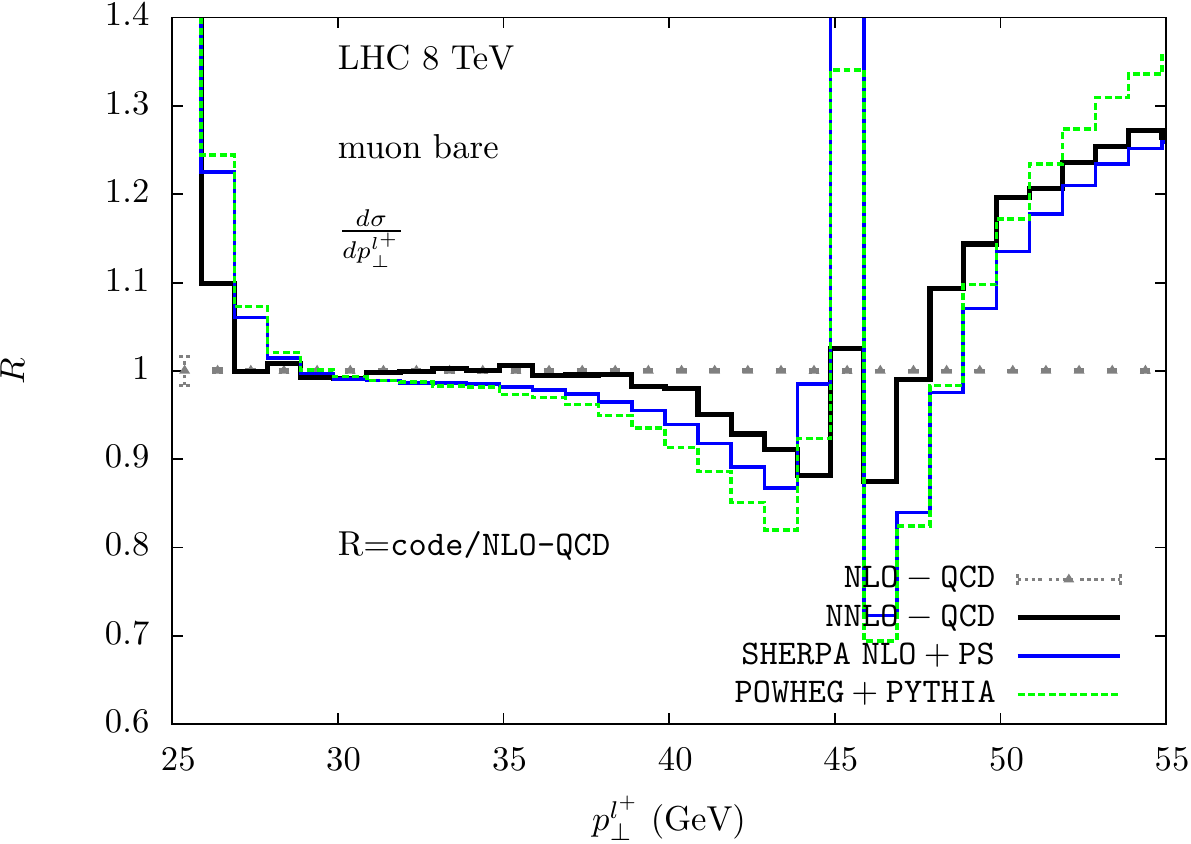}
\includegraphics[width=75mm,angle=0]{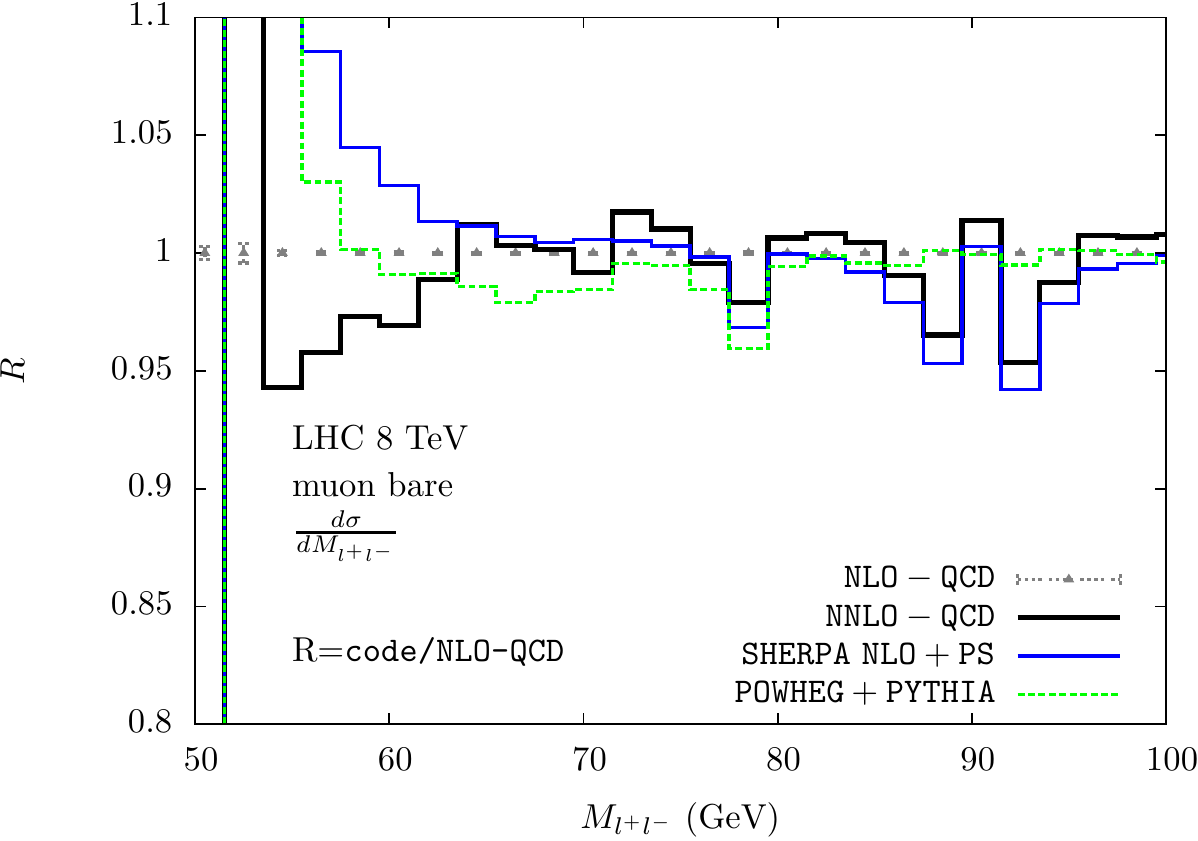}\\
\includegraphics[width=75mm,angle=0]{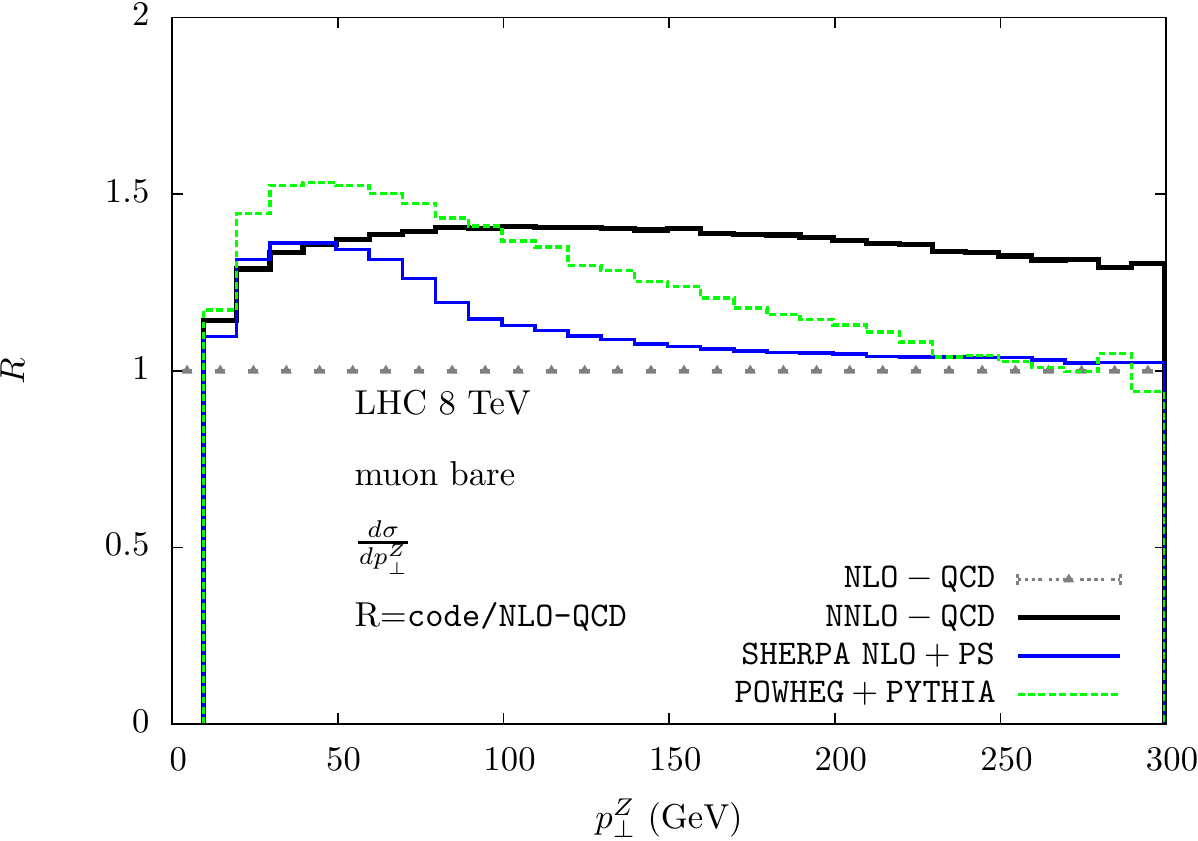} 
\includegraphics[width=75mm,angle=0]{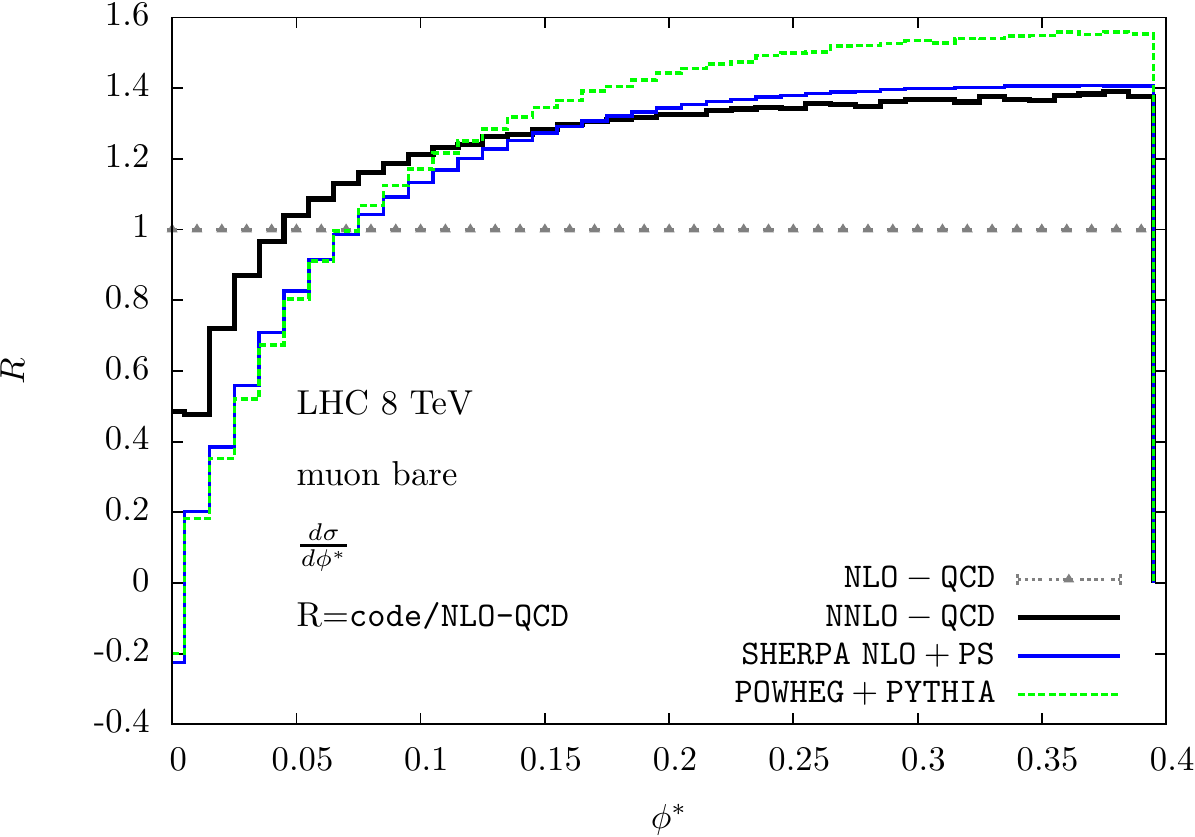} 
\caption{Higher-order QCD effects, expressed in units of NLO QCD, in $pp\to \mu^+\mu^-+X$, due to the matching of resummed and fixed order results, in codes with NLO accuracy:
\powheg+\pythia (green) and \sherpanlo (blue). 
The fixed-order NNLO QCD results are shown in black.
Comparison of results for the 
lepton transverse momentum (upper left),
lepton-pair invariant mass (upper right), 
transverse momentum (lower left) and
$\phi^*$ (lower right)
distributions, obtained with ATLAS/CMS cuts at the 8 TeV LHC.
\label{fig:Z-shower-qcd-NLO}
}
 \end{figure}

We observe in Figures
\ref{fig:Wpm-shower-qcd-NLO}-\ref{fig:Z-shower-qcd-NLO}
that the QCD-PS corrections in \powheg+\pythia have a small impact on
the invariant-mass (NC DY) or transverse-mass (CC DY) distributions
(middle plots); the correction is slowly varying over the entire mass
range, with the exception of the lower end of the
distribution, where the acceptance cuts yield a distinction between
one-emission and multiple-emissions final states.

In the same figures, we show the corrections to the lepton transverse
momentum distribution (upper plots).  We observe at the jacobian peak
the distortion due to the fact that in this region a fixed order
description is not sufficient to describe this observable.  Below the
jacobian peak the corrections of ${\cal O}(\alpha_S^2)$ and higher
become smaller for decreasing values of the transverse momentum, 
before reaching the acceptance cut.  
Above the jacobian peak, the QCD-PS
effects follow those obtained at NNLO QCD.  This result can be
interpreted by observing that the lepton transverse momentum has two
components, one from the gauge boson decay at LO and one due to the
gauge-boson recoil against QCD radiation; immediately above the
jacobian peak, the recoil component is characterized by a small value
of the lepton-pair transverse momentum; in this region the collinear
approximation on which the PS is based is quite accurate,
and thus the second real emission in the PS approximation is
close to the exact result.  For larger values of the lepton-pair
transverse momentum the QCD-PS becomes inadequate to describe the
spectrum; the role of the first and second order exact matrix element
corrections is shown in the lower plots of
Figures \ref{fig:Wpm-shower-qcd-NLO}-\ref{fig:Z-shower-qcd-NLO}.  The
difference between the two approximations vary between zero and 40\% in
the interval $\ptv\in [70,300]$ GeV.

The resummation of multiple parton emissions to all orders via the 
PS makes the distribution vanish in the limit of vanishing
lepton-pair transverse momentum, as it is physically expected (Sudakov
suppression).  The size of the QCD-PS correction in units NLO QCD is
infinitely negative when $\ptv\to 0$; this peculiar result is a
consequence of the choice of the NLO QCD prediction as unit to express
the higher-order effects, which is inappropriate in this specific corner of the
phase-space.  This comment is at variance with respect to the one for
the NNLO QCD corrections: also in that case the size of the correction
is infinitely large, but only because at each fixed order the
distribution diverges, each time with a different coefficient.

\clearpage

\clearpage
\subsubsection{Comparison of different (NNLO+PS)-QCD matching schemes}
\label{sec:nnlops}

The matching of NNLO QCD results with a QCD-PS has been achieved first
in the \minlo approach \cite{Hamilton:2012np,
  Hamilton:2012rf,Hamilton:2013fea}.
In the DY case the calculation has been implemented in a code based on
\powheg +\minlo combined with \dynnlo, and henceforth denoted
\dynnlops \cite{Karlberg:2014qua}.  This method is based on the NLO+PS
formulation of the original hard process plus one-jet, and supplements
it with Sudakov form factors that lead to finite predictions as the
additional jet becomes unresolved.  The NNLO accuracy is achieved by
reweighing via a pre-tabulated phase-space dependent K-factors.
 
Another NNLO+PS matching approach is called UN2LOPS
\cite{Hoeche:2014aia,Hoche:2014dla} and it is a variant of the UNLOPS
\cite{Lonnblad:2012ix} method.  UNLOPS is one of the unitary merging
techniques recently developed to merge multi-jet NLO calculations
while preserving the inclusive cross section of the process with the
lowest jet multiplicity.  In UN2LOPS, by only keeping events with
resolvable QCD emissions, which are available as part of the NNLO
calculation, the description of the DY processes at large transverse
momentum becomes equivalent to the study of $W$($Z$) plus one
additional jet at NLO.  The remainder of the phase space is filled by
a calculation at NNLO, with a corresponding veto on any QCD activity,
forming the zero jet bin.  This is essentially the phase space slicing
method, and the goal of the UN2LOPS approach is to merge the two parts
after the PS is added.  Only the part of $W$($Z$) plus one jet at NLO
is matched with PS, where any standard methods could be used.  Events
in the zero jet bin should not be showered to avoid double counting
because QCD radiation has already been described by the PS matched
$W$($Z$) plus one jet process at NLO\footnote{Except for the pure
  two-loop virtual contribution, which contributes to $W$($Z$) plus
  one jet at NNLO if showered.}.  The merging is done by suppressing
the divergence in $W$($Z$) plus one jet via the shower veto algorithm
in which the vetoed events are added back to the zero jet bin to
preserve the inclusive cross section.  In order to generate physically
meaningful results, the separation cut scale $q_\perp$ must be smaller
than the terminating scale of the parton shower.  In contrast to the
\minlo method, real-emission configurations do not receive a
contribution from the NNLO calculation because two-loop virtual
contributions in the 0-jet bin are not showered.  The resulting
difference is beyond NNLO accuracy for the original hard process.
Formally the resummation of UN2LOPS is limited by the accuracy of the
parton shower, while in the \minlo method, a higher logarithmic
accuracy of the first emission can be achieved with analytic Sudakov
form factor for the corresponding observable\footnote{The analytic
  Sudakov form factor is generally observable-dependent (not fully
  differential); in the application to DY here, the relevant
  observable used by \minlo is the $W$($Z$) transverse momentum
  \ptv).}.  Nevertheless, for other observables or subsequent
emissions, resummation in \minlo is only as accurate as the parton
shower can provide.  The calculation of the DY processes in the
UN2LOPS approach has been implemented in the code \sherpa.

Both these two matching approaches should not be considered as a final
answer to the problem of matching NNLO fixed order with PS results,
but rather as a first step towards more general methods.

We note that results for Drell-Yan production at NNLL'+NNLO matched to a PS in 
the {\tt GENEVA} Monte-Carlo framework are presented in Ref.~\cite{Alioli:2015toa}, 
but not included in this study.

\begin{figure}[!h]
\includegraphics[width=75mm,angle=0]{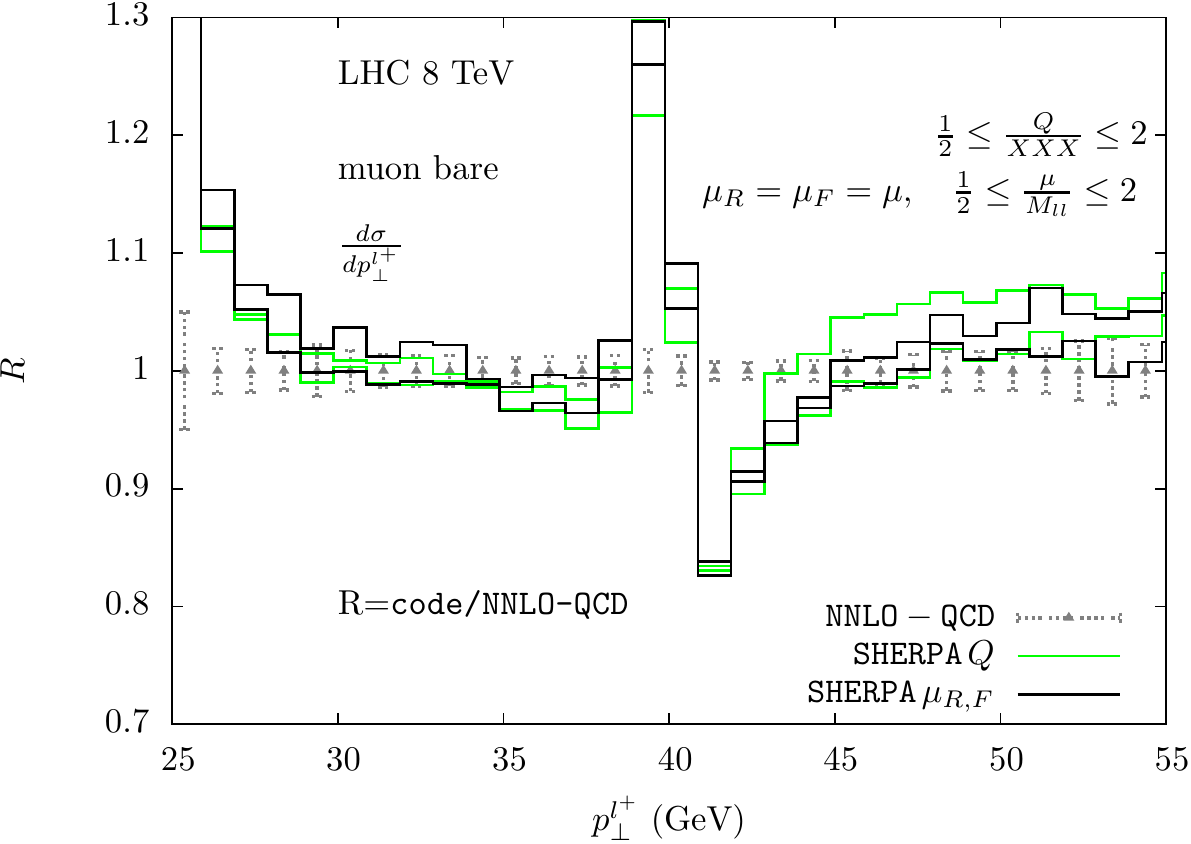}
\includegraphics[width=75mm,angle=0]{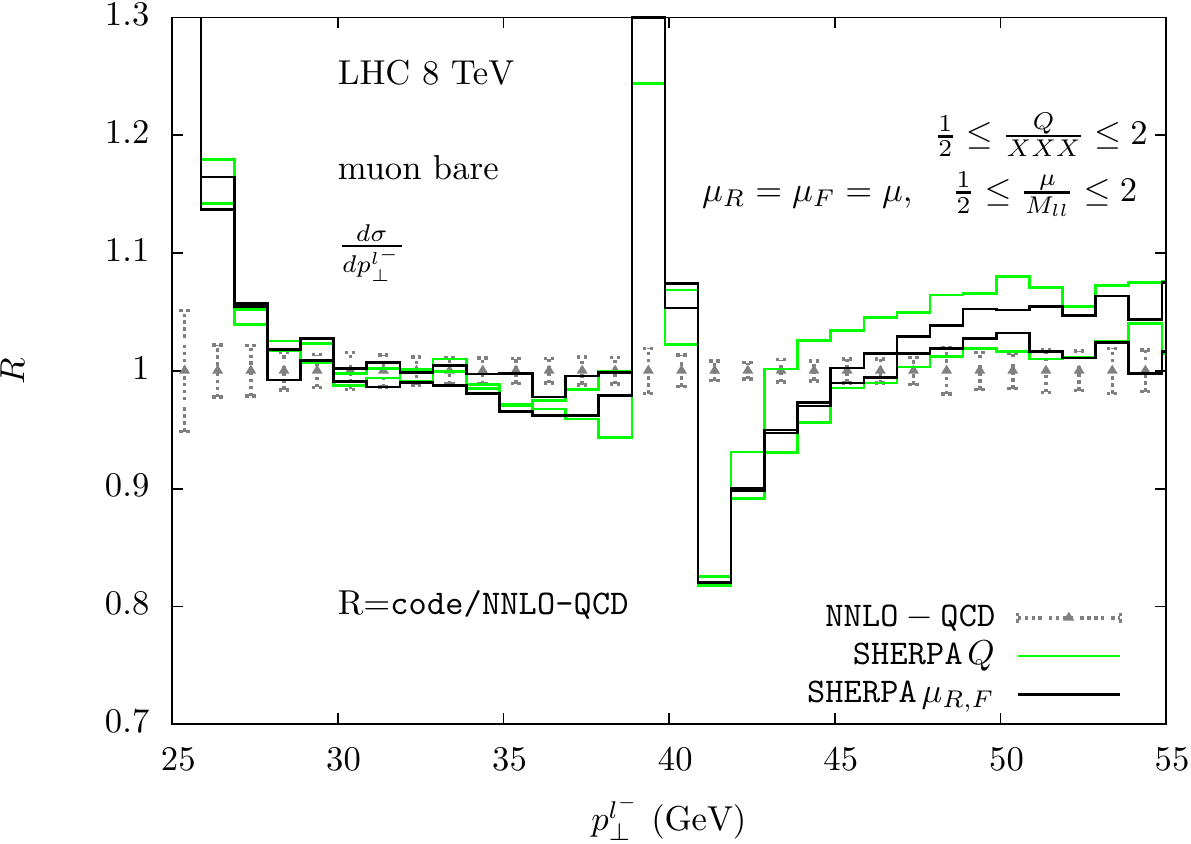}\\
\includegraphics[width=75mm,angle=0]{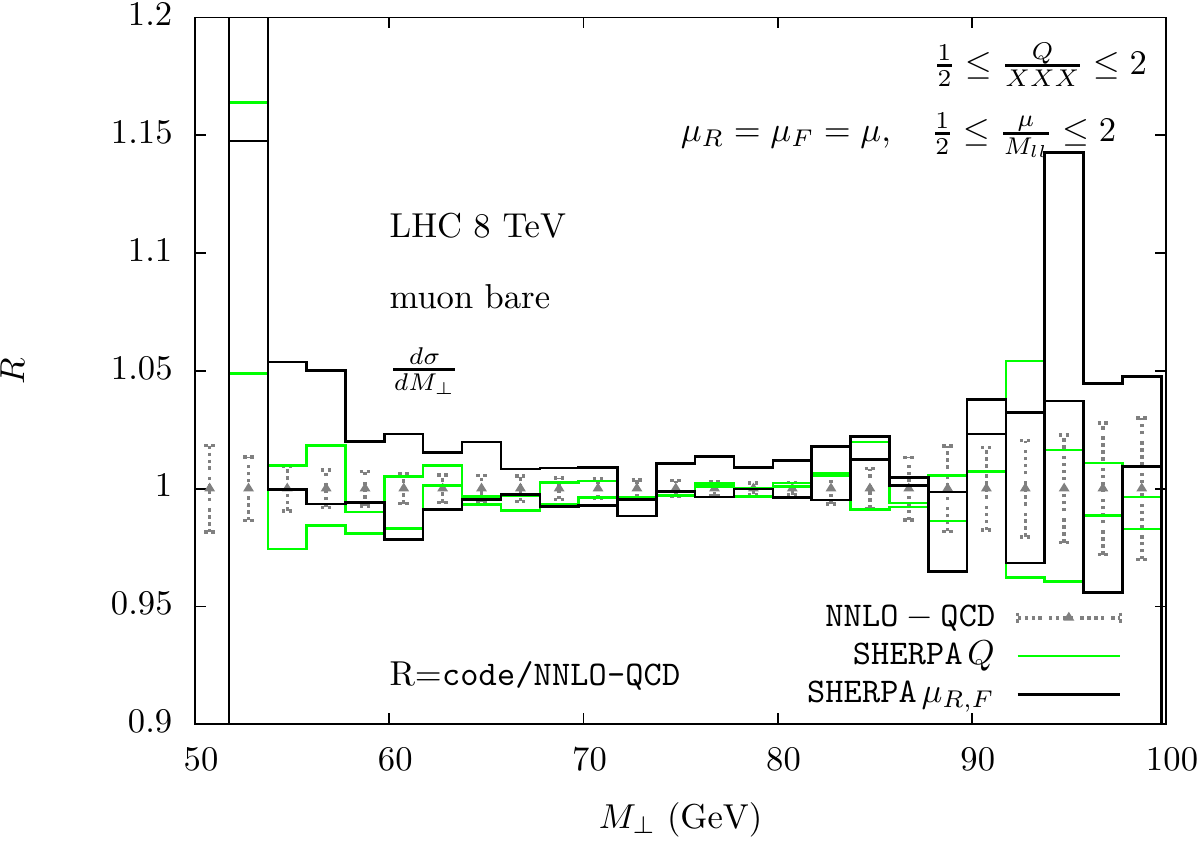}
\includegraphics[width=75mm,angle=0]{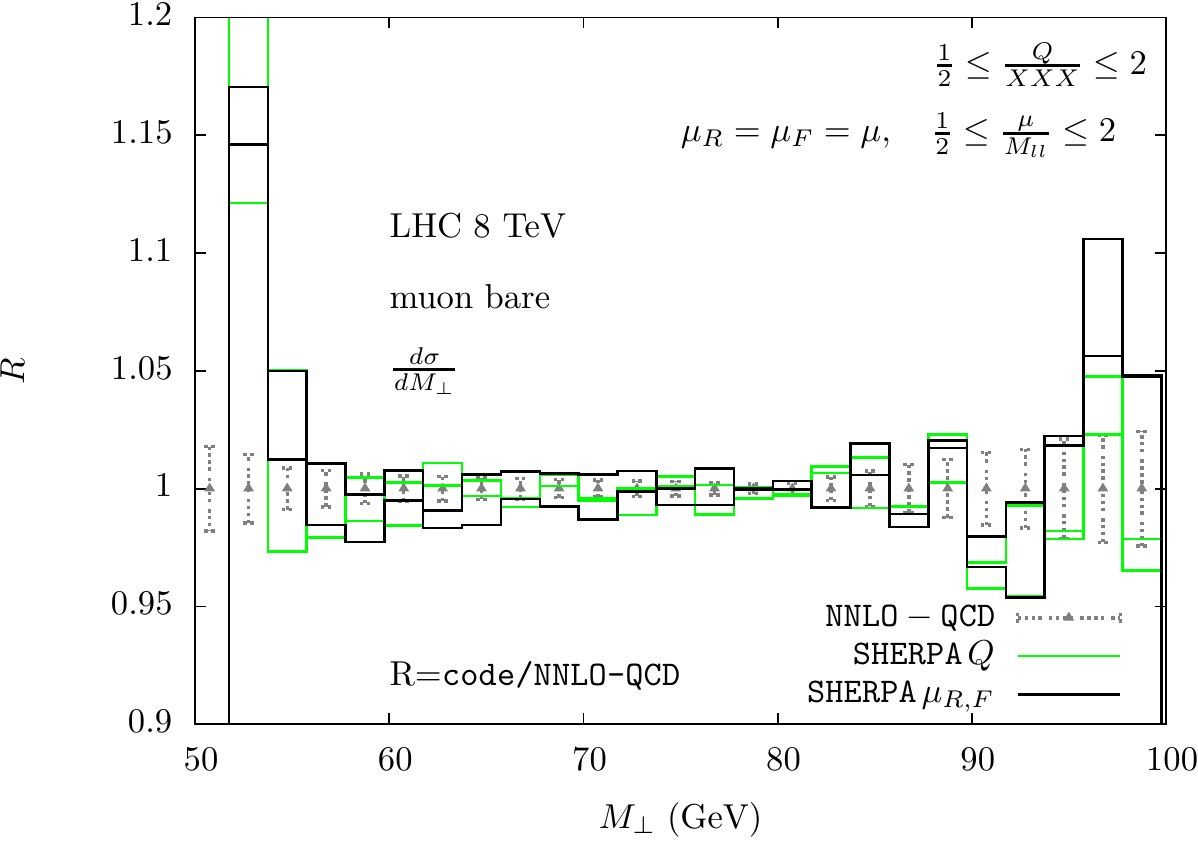}\\
\includegraphics[width=75mm,angle=0]{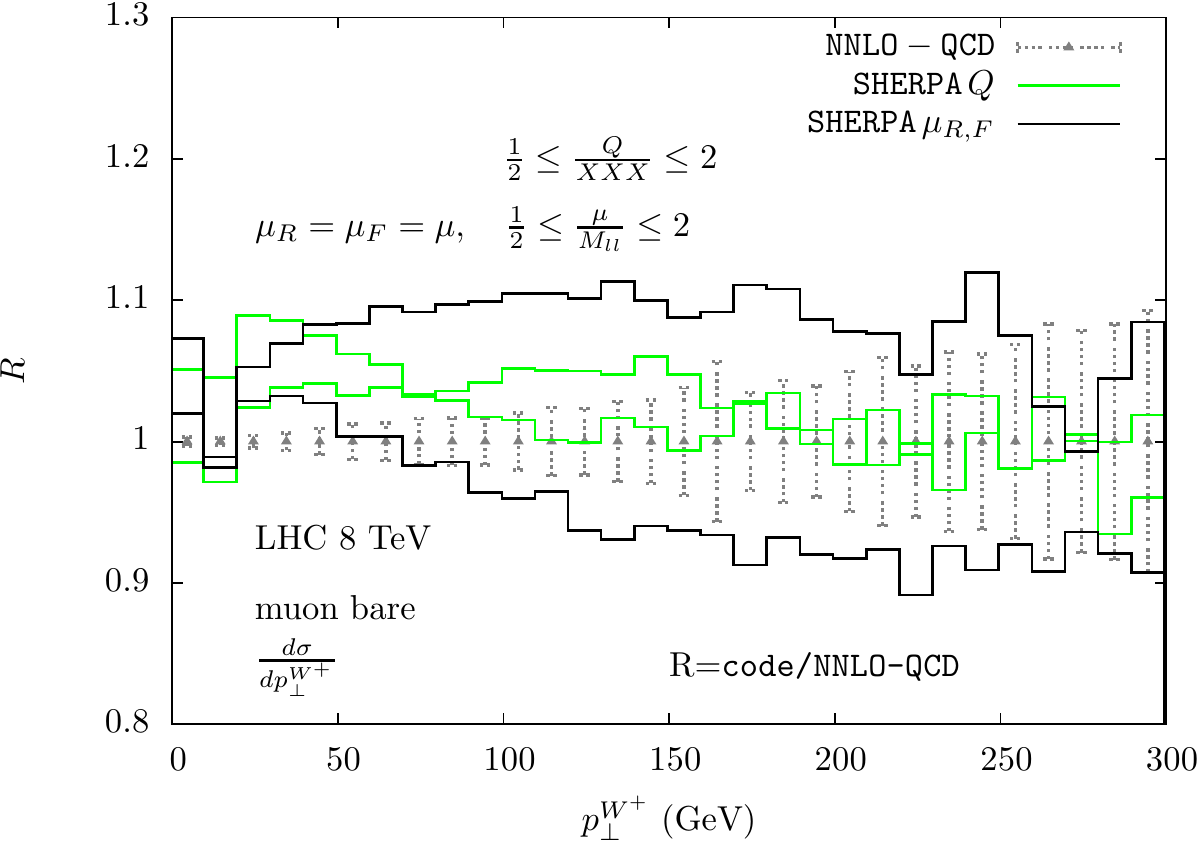}
\includegraphics[width=75mm,angle=0]{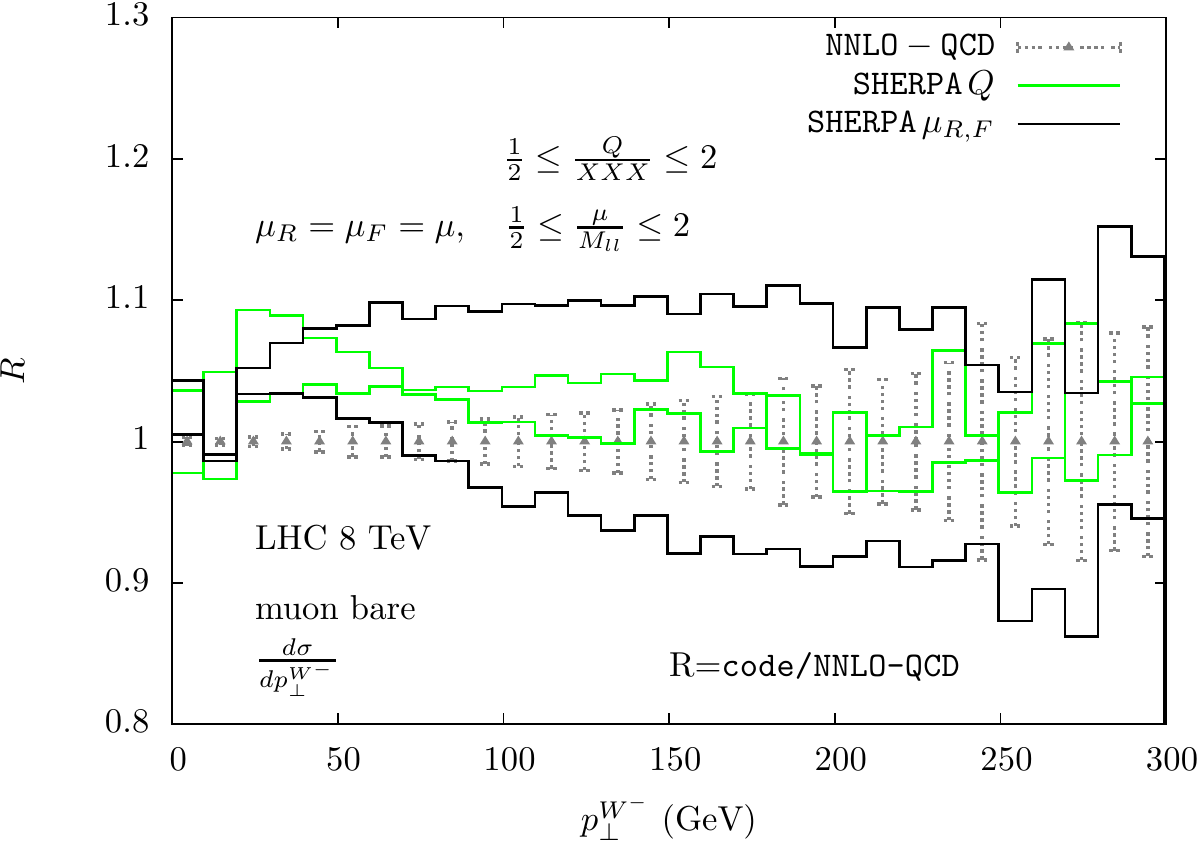}
\caption{
Higher-order QCD effects, expressed in units of NNLO QCD,  due to the matching of resummed and fixed order results, in codes with NNLO accuracy, for the processes $pp\to \mu^+\nu_\mu+X$ (left plots) and $pp\to \mu^-\bar\nu_\mu+X$ (right plots), obtained with ATLAS/CMS cuts at the 8 TeV LHC.
The {\tt SHERPA NNLO+PS} uncertainty bands due to renormalization/factorization scales (black) and
shower scale (green) variations are shown for
the lepton transverse momentum (upper plots),
neutrino transverse momentum (middle plots) and 
transverse mass (lower plots) distributions.
\label{fig:Wpm-shower-qcd-NNLO}
}
 \end{figure}

In Figure \ref{fig:Wpm-shower-qcd-NNLO} we show the results obtained
with the \sherpa code, in the case of CC DY, and compare them to the
corresponding NNLO fixed-order predictions.  We present two different
uncertainty bands: the first one, in black in the plots, is obtained
by varying the renormalization $\mu_R$ and factorization $\mu_F$
scales of the underlying fixed order calculation, with $\mu_R=\mu_F$
and $1/2\le \mu_R/M_{ll} \le 2$; the second one, in green in the
plots, is obtained by varying the shower scale $Q$ of the QCD-PS in
the interval $1/2 \le Q/M_{ll} \le 2$.

\begin{figure}[!h]
 \includegraphics[width=75mm,angle=0]{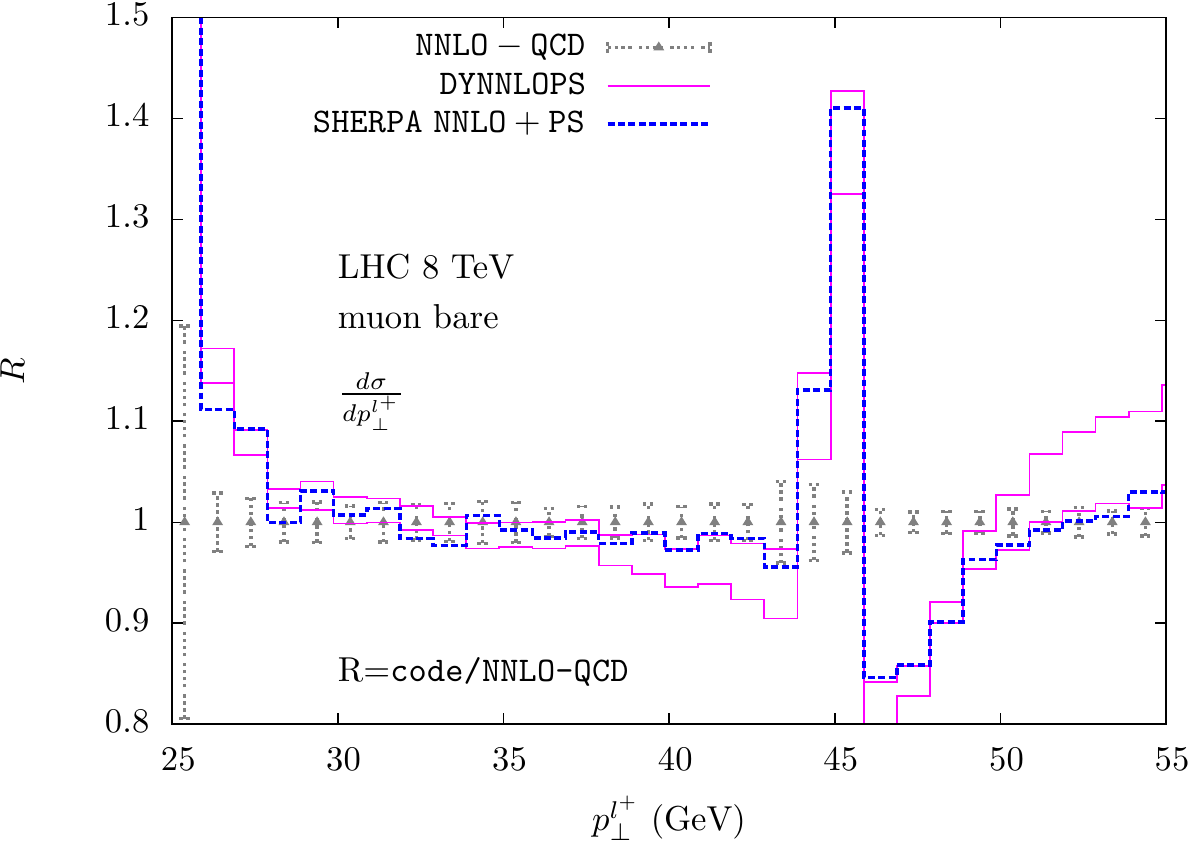}
 \includegraphics[width=75mm,angle=0]{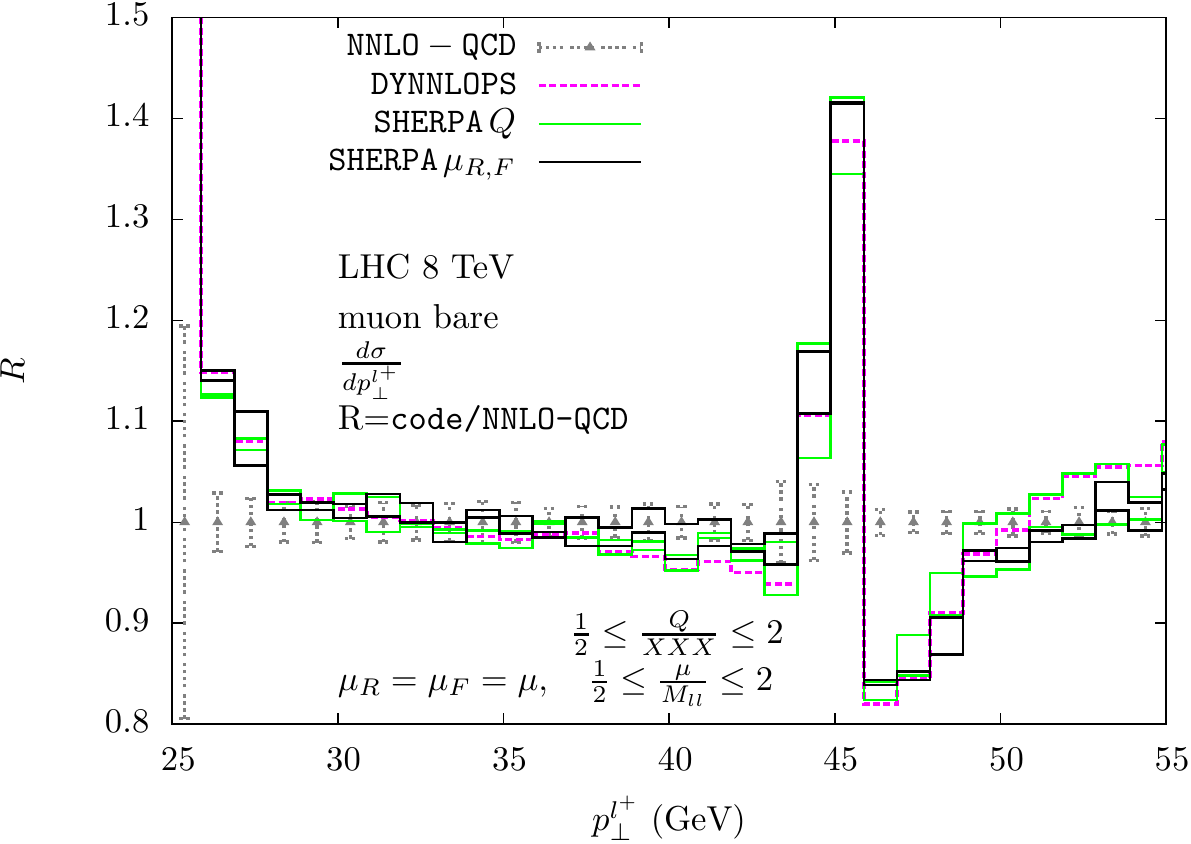}\\
\includegraphics[width=75mm,angle=0]{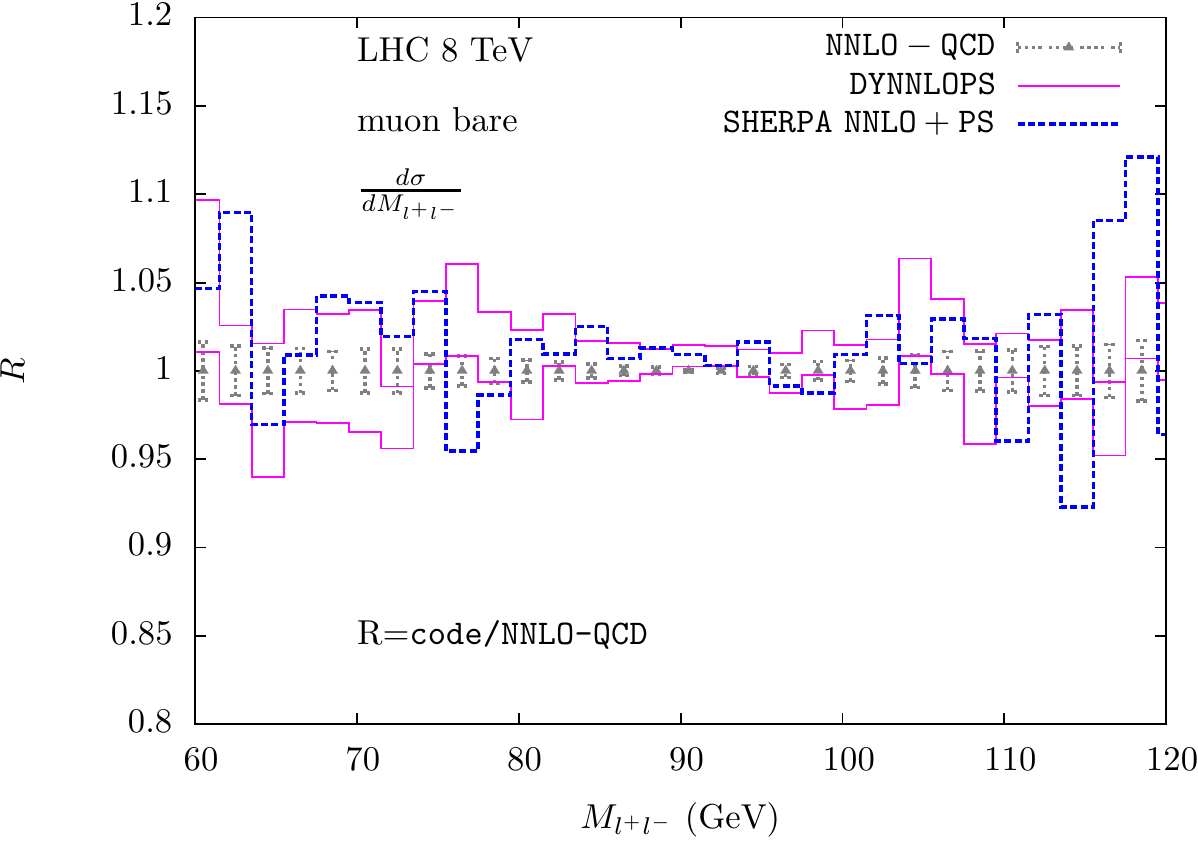}
\includegraphics[width=75mm,angle=0]{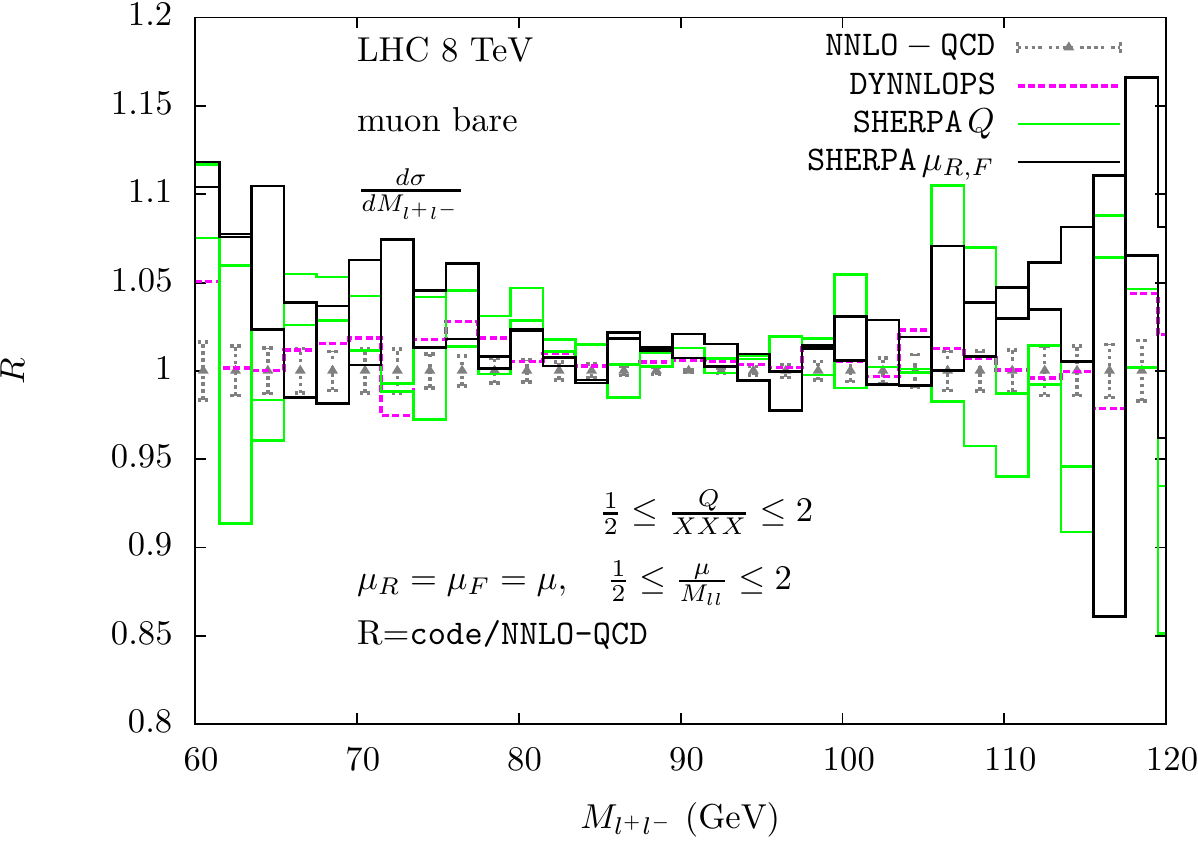}\\
\includegraphics[width=75mm,angle=0]{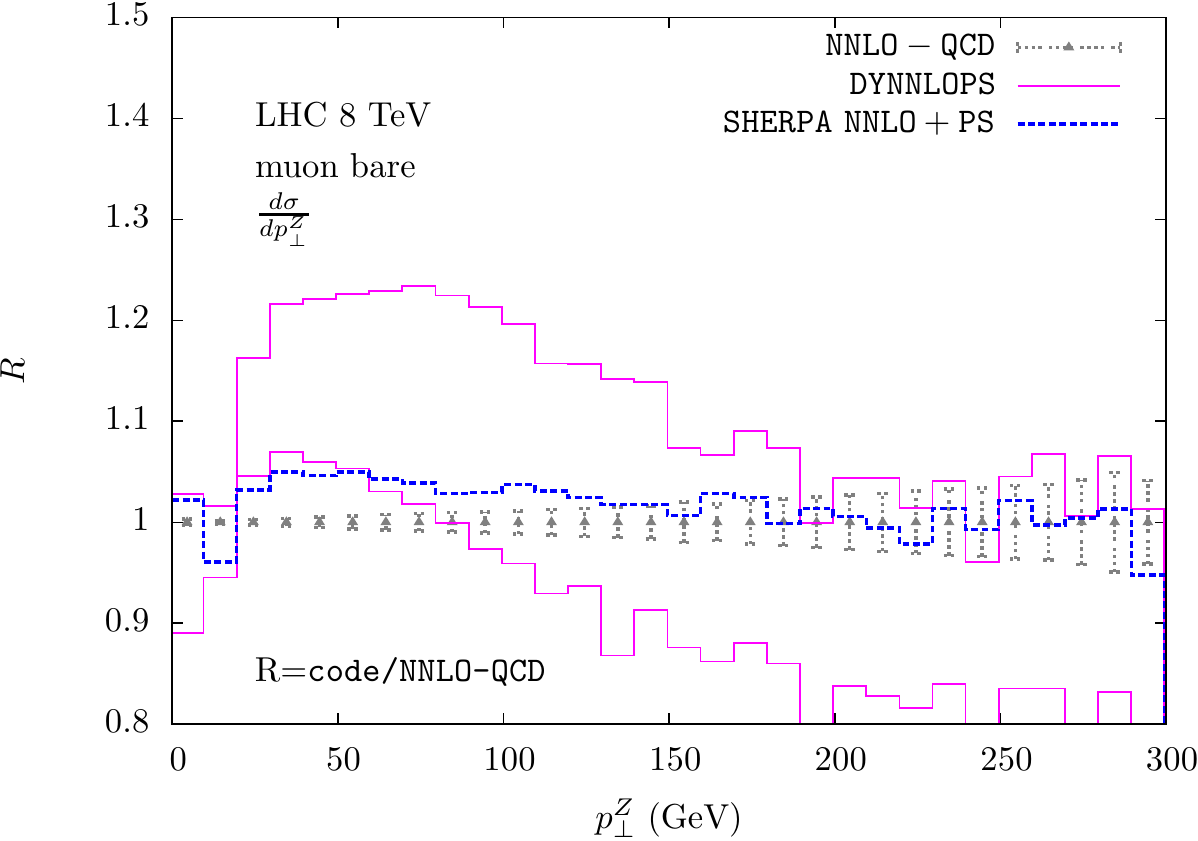} 
\includegraphics[width=75mm,angle=0]{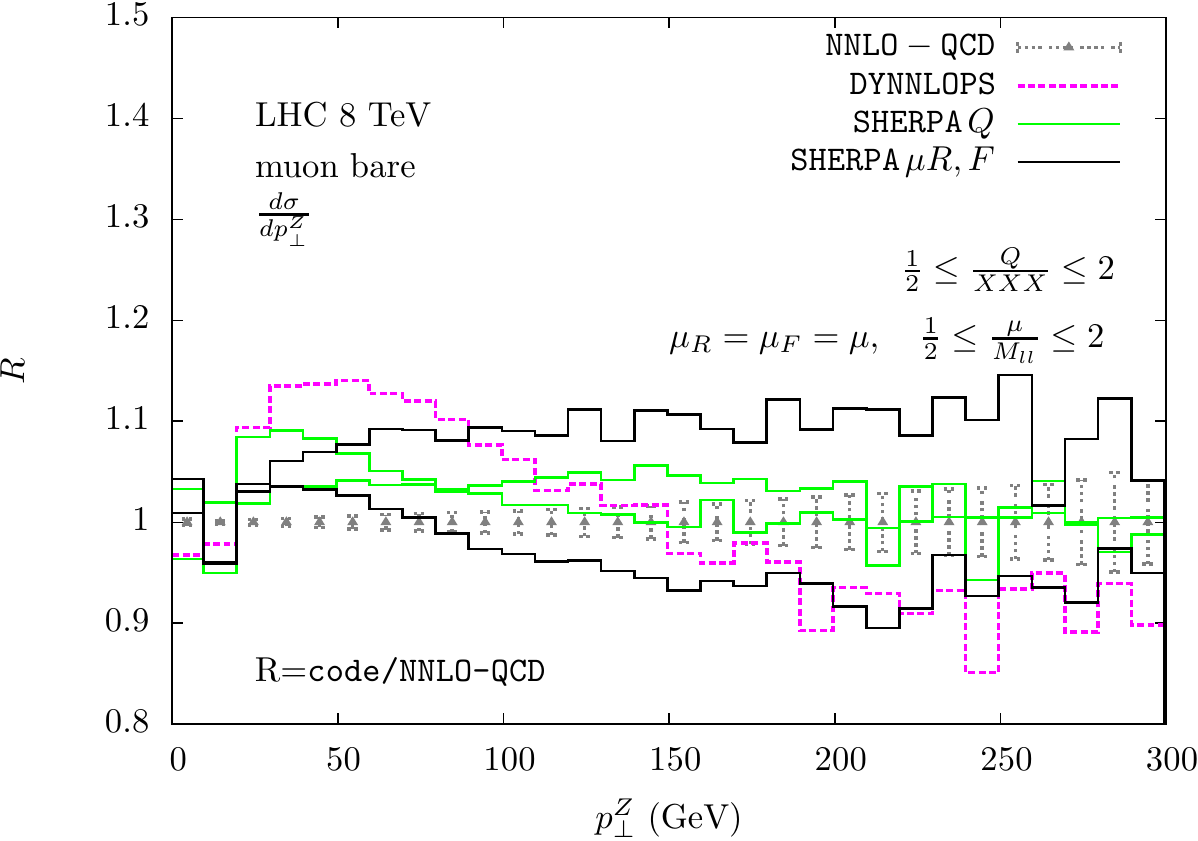} 
\caption{Higher-order QCD effects, expressed in units of NNLO QCD,  due to the matching of resummed and fixed order results, in codes with NNLO accuracy, for the process $pp\to \mu^+\mu^-+X$,
obtained with ATLAS/CMS cuts at the 8 TeV LHC.
The \sherpa uncertainty bands for renormalization/factorization scales (black) and shower scale (green) variations are shown in the right plots. The \dynnlops (pink) uncertainty bands are shown in the left plots. 
Cfr. the text for details about the definition of the bands.
The central scales results are presented with dashed lines for \sherpa (blue) and \dynnlops (pink).
Results are shown for the lepton transverse momentum (upper plot), lepton-pair invariant
mass (middle plots) and transverse momentum (lower plot)
distributions.
\label{fig:Z-shower-qcd-NNLO}
}
 \end{figure}

In Figure \ref{fig:Z-shower-qcd-NNLO} we show the results obtained
with the two codes \sherpa and \dynnlops, in the case of NC DY, and
compare them with each other and with the corresponding NNLO
fixed-order predictions.  The \sherpa uncertainty bands have been
computed as described above, while in the \dynnlops case the band is
obtained by varying by a factor 2 up and down independently all
renormalization and factorization scales appearing in the underlying
\minlo procedure (at variance with the report setup, in the \minlo
approach both renormalization and factorization scales are set equal
to the gauge boson transverse momentum), keeping their ratio between
$1/2$ and $2$. This leads to 7 different scale choices. Independently
of this we vary by a factor 2 up and down the renormalization and
factorization scale in the underlying \dynnlo calculation keeping the
two equal. This leads to 3 different scale choices. As these scale
choices are taken to be independent, this leads to $3\cdot 7=21$ scale
choices of which the envelope is taken as the uncertainty band. The
procedure is described in more detail in \cite{Karlberg:2014qua}.
Since the procedures used to evaluate the uncertainty bands are
different for the two codes, we present separately in the two columns:
the \dynnlops band and the central scales \sherpa prediction (left
plots) and the two \sherpa bands and the central scales \dynnlops
prediction (right plots).

As expected, for the invariant mass distribution of the lepton pair,
in Figure \ref{fig:Z-shower-qcd-NNLO}, all predictions agree very
well. In particular in the central region, closer to the peak, the
large statistics allow us to appreciate that also uncertainty bands
are very similar among the two NNLO+PS results, and that the central
line of one lies well within the (very narrow) uncertainty band of the
other tool. For smaller and larger invariant masses, the conclusions
are similar, although the limited statistics do not allow such a
precise comparison.

Turning to the lepton transverse momentum, \ptl, spectrum, in Figure
\ref{fig:Z-shower-qcd-NNLO} one observes that in the range where this
distribution is NNLO accurate (i.e. where \ptl is less than half the
mass of the Z boson), the results of the two NNLO+PS codes are again
in good agreement with each other and with the NNLO QCD reference
line. The uncertainty band is very thin, as expected, until one
approaches the Jacobian peak region. As explained in the previous
section, in this region resummation effects are important. Although
the two NNLO+PS results are obtained with very different approaches,
the mutual agreement is very good. One should notice however, that to
the left of the Jacobian peak, the NNLO+PS result from \dynnlops seems
to depart from the pure fixed-order results a few bins earlier than
the one from \sherpa. These differences are likely to be due to the
differences in how events are generated close to the Sudakov peak in
\ptz, which is a phase-space region where resummation is crucial, and
the two NNLO+PS calculations perform it using very different
approaches. Therefore differences at the few percent level are not
unexpected.  The differences between the NNLO+PS and the fixed-order
results at the lower end of the \ptl spectrum have already been
noticed and commented on earlier in this chapter.  For transverse
momenta larger than $\mz/2$, the two NNLO+PS results rapidly start to
re-approach the fixed-order line, which in this region is NLO QCD
accurate.  However, towards the end of the plotted range, some
differences among the results can be observed: firstly, the \dynnlops
result exhibits a moderately harder spectrum, which would probably be
more evident at higher \ptl values.  Secondly, the uncertainty band of
the two NNLO+PS results (the one due to the $\mu_R,\mu_F$ scale
variation only) is larger in the \dynnlops result than in the \sherpa
one.  Both these differences can be understood by looking at the
differences amongst the results for the vector-boson transverse
momentum in the medium to low range ($[0,50]$ GeV), which is the phase
space region where the bulk of the events with \ptl approximately
equal to [55,60] GeV are generated.

The transverse momentum spectrum \ptz of the lepton pair is the
observable that exposes most clearly the differences between the two
results.  For the purpose of this comparison, the more relevant
difference to explain is the difference in shape (and absolute value)
for $\ptz \in [20,100]$ GeV, that we will address in the next
paragraph. At very high \ptz, differences are also fairly large, but
in that region they can be mostly attributed to the \minlo scale
choice: when \ptz is large (above $M_Z$), the \minlo Sudakov form
factor switches off, but the strong coupling is evaluated at \ptz,
whereas in \sherpa and in the fixed-order calculation it is evaluated
at the dilepton invariant mass $m_{ll}$.

The range $\ptz\in [20,50]$ GeV is a ``transition'' region, since it
is the region where higher-order corrections (of fixed-order origin as
well as from resummation) play a role, but none of them is dominant.
Due to Sudakov suppression, in \dynnlops the first two bins of the
\ptz distribution are suppressed compared to the fixed-order results;
in turn, the unitarity fulfilled by the matching procedure, in order
to respect the total cross section normalization, spreads part of the
cross section close to the singular region across several bins in
\ptz, including those to the right of the Sudakov peak.

The \sherpa results instead are closer to the fixed-order prediction
in the first bins, which is may be a consequence of the PS not being
applied to the events of the 0-jet bin.

Since the first bins are the region where most of the cross-section is
sitting, a relatively small difference among the two NNLO+PS results
in the peak region will show up, greatly amplified, in the transition
region (to preserve the total cross section).  At, say, 50 GeV, both
the NNLO+PS results have a cross section larger than the pure
fixed-order, with \dynnlops larger than \sherpa.
Moreover, although at large \ptz the cross section is small, 
the \dynnlops result is, by construction, below the others, as explained previously. 
This difference must also be compensated, 
and this takes place in the transition region too.

For the \dynnlops results, the scale choice in the transition region
is inherited from the underlying \minlo simulation.  This means that
the conventional factor 1/2 or 2 is applied to a dynamical scale
choice ($\mu = \ptz$), and this fact helps in explaining why not only
the result is larger than the fixed order and the \sherpa
distributions, but it also exhibits a different shape and uncertainty
band.  In the \sherpa approach, effects similar to the latter in the
transition region are mainly taken into account by the variation of
the resummation scale, as the corresponding plot supports.  In fact,
this is the dominant uncertainty of the \sherpa result in the
transition region.

In spite of all the aforementioned details, one should also notice
that for \ptz, the two NNLO+PS results are mutually compatible
over almost all the entire spectrum, once the uncertainty bands are
considered.

\clearpage

\subsection{Impact of EW corrections on $W$ and $Z$ boson observables in the {\em benchmark} setup}
\label{sec:impact-ew}
In Section \ref{sec:impact-qcd} 
we presented the impact of higher-order QCD corrections,
using the fixed-order NLO QCD results 
(which have been demonstrated to be fully under control) 
as unit to express the relative effect of different subsets..
We follow the same approach now to discuss the EW corrections.

We discuss in Section \ref{sec:NLO-EW} the main features of the NLO EW corrections, with special emphasis on the observables that are relevant to EW precision measurements.
In Sections \ref{sec:gammainduced}-\ref{sec:fermionpair},
we present the impact of different subsets and combinations of higher-order corrections and if not stated otherwise express their effect using as a unit the results computed
at NLO EW.

\subsubsection{NLO EW corrections}
\label{sec:NLO-EW}

At LO the DY CC and NC processes are purely of EW nature (the cross section is
of ${\cal O}(G_{\mu}^2)$\,).  The typical size of the impact of NLO EW
corrections on the total cross section is of ${\cal O}(\alpha)$,
i.e. at the per cent level. However, it is important to
stress that the real radiation may have a much larger impact on the
differential distributions, in particular in the presence of
acceptance cuts.  At NLO EW all the electrically charged particles may
radiate a real photon.  The distinction between initial state, final
state and interference effects has been discussed not only in the NC,
but also in the CC case \cite{Wackeroth:1996hz}.  
It is important to stress that the
potentially large effects due to initial state collinear emissions are
re-absorbed in the definition of the physical proton PDFs, leaving a
numerically small remnant.  On the other hand the final state radiation effects are
phenomenologically very important, because they modify the momenta of
the final state leptons, affecting all the relevant distributions.  We
distinguish between observables whose line shape is relevant for the
determination of the gauge boson masses and widths and other
quantities whose normalization is important to constrain the proton
PDFs or to correctly describe the background to new physics searches.

\begin{figure}[h]
\includegraphics[width=75mm,angle=0]{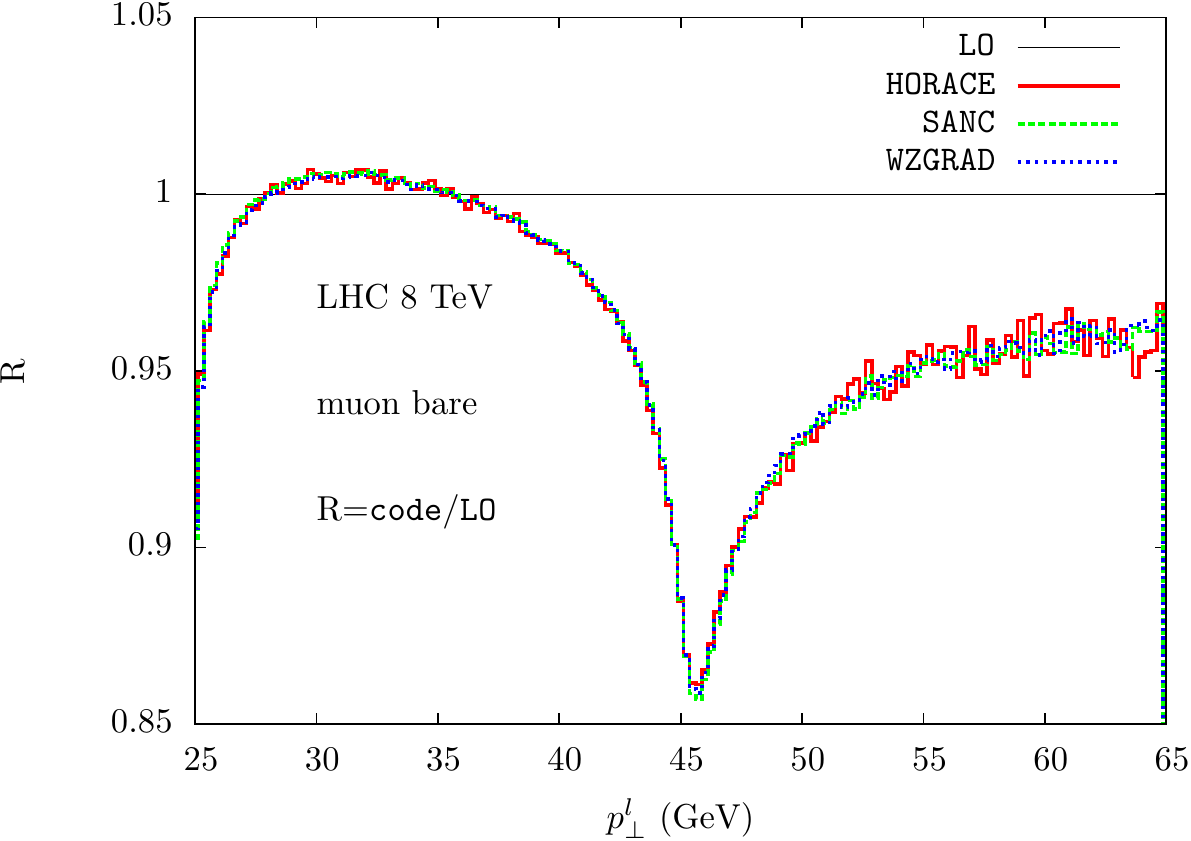}
\includegraphics[width=75mm,angle=0]{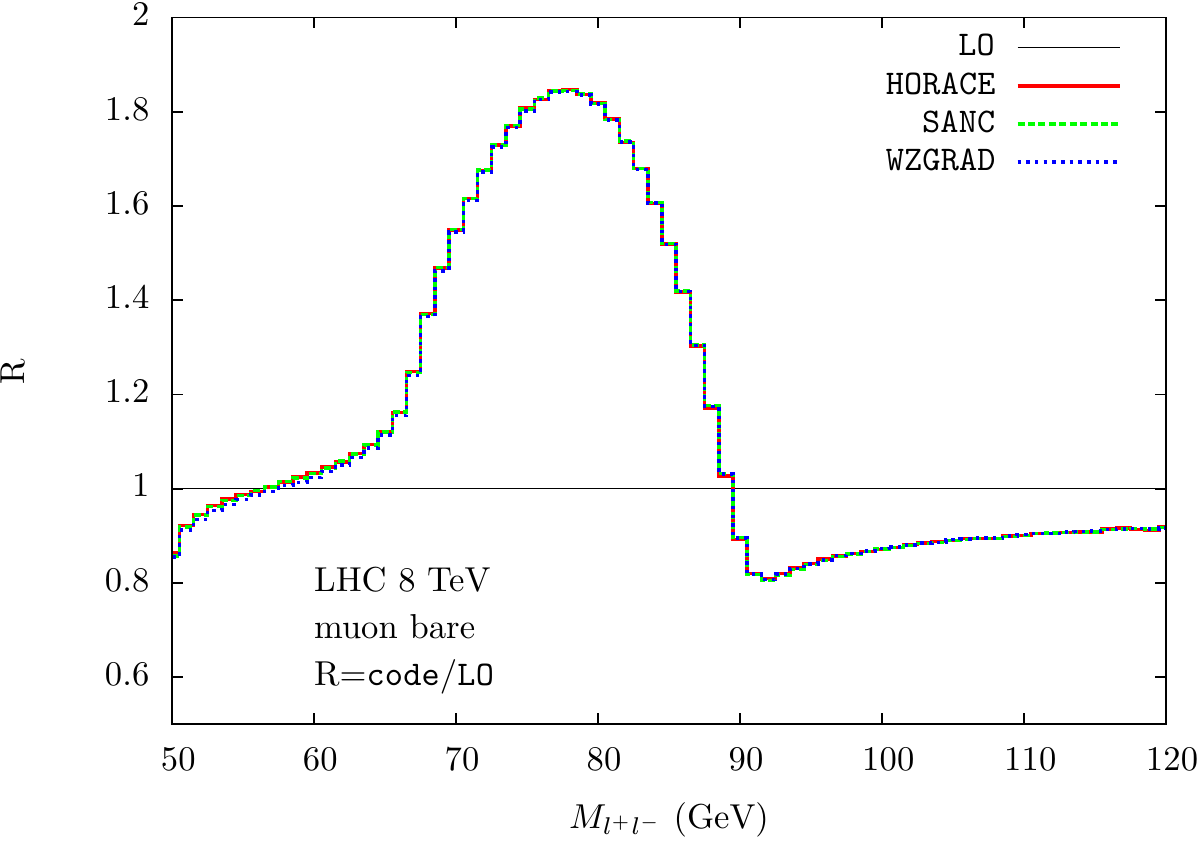}\\
\includegraphics[width=75mm,angle=0]{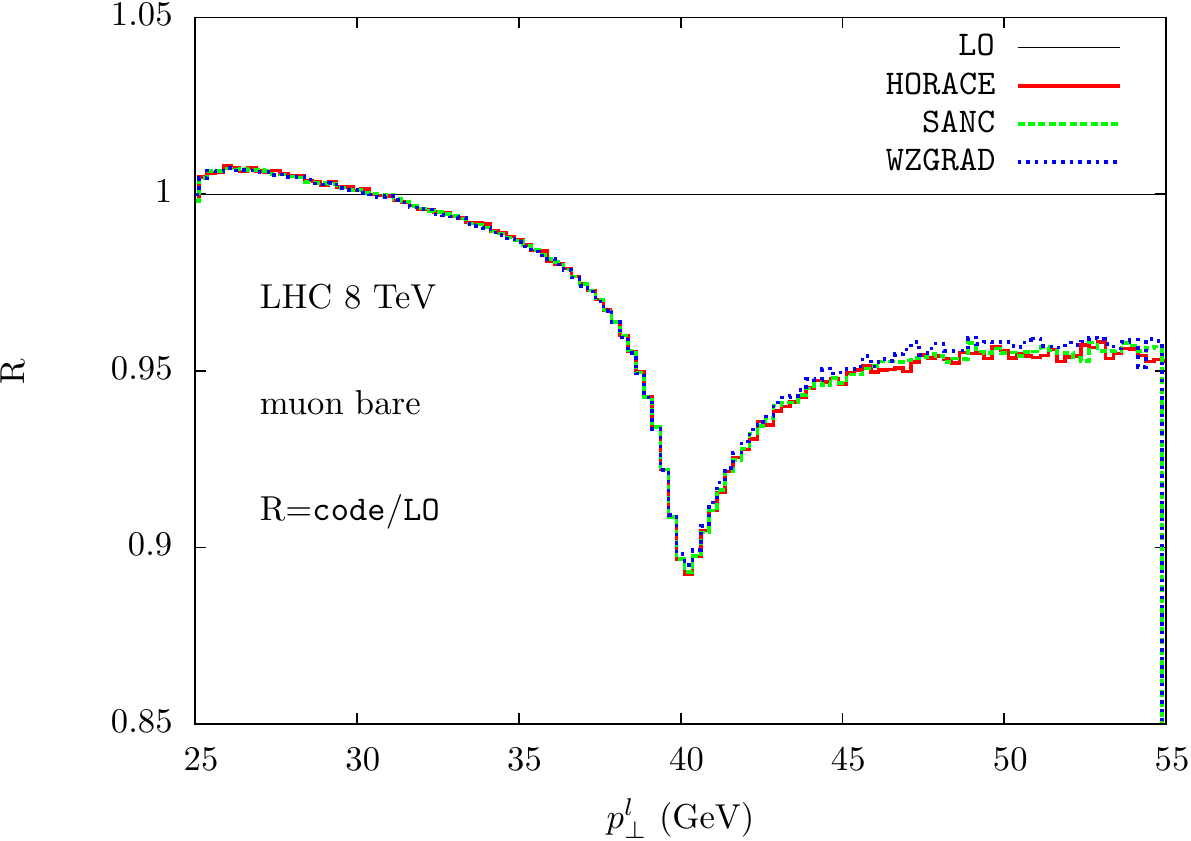}
\includegraphics[width=75mm,angle=0]{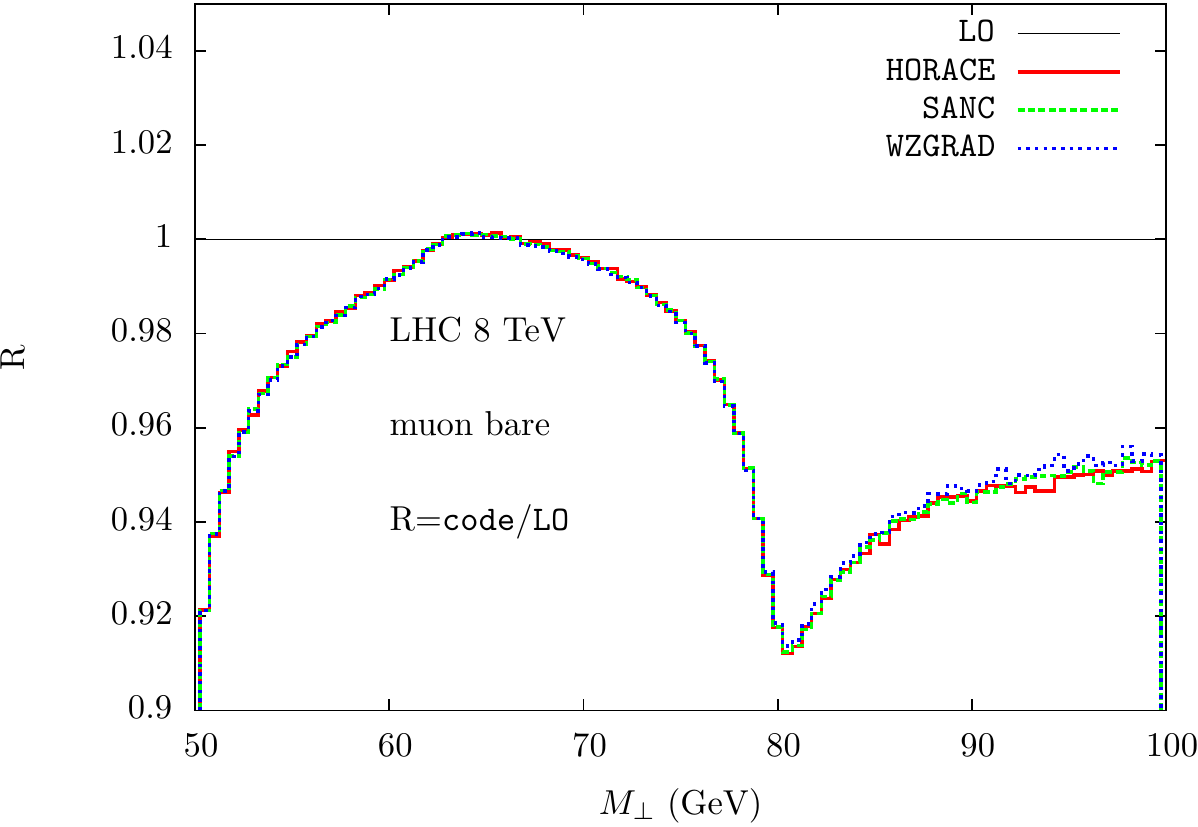}
\caption{Impact of NLO EW corrections in NC and CC DY processes with bare muon(s) at the 8 TeV LHC with ATLAS/CMS cuts, expressed in units of the corresponding LO, 
in the {\em benchmark} setup, evaluated with different codes. 
In the upper panels, for the $pp\to\mu^+\mu^-+X$ process, the lepton transverse momentum (left) and the lepton-pair invariant mass distributions are shown;
in the lower panels, for the $pp\to\mu^+\nu_\mu+X$ process, the lepton transverse momentum (left) and the lepton-pair transverse mass distributions are shown.
\label{fig:nlo-ew}
}
\end{figure}

To the first group belongs the single lepton transverse momentum
distributions and the lepton-pair transverse mass distributions around
the $W$ ($Z$) Jacobian peak, and, in the NC channel, at the $Z$
resonance, the lepton-pair invariant mass distribution.  
In Figure~\ref{fig:nlo-ew}, we show the impact of NLO EW corrections relative to LO on these distributions.
The largest,
negative, corrections arise at the (Jacobian) peak of each
distribution.  The effect can be understood as a combination of the
properties of the gauge boson production mechanism, which is peaked at
the ($W$) $Z$ boson mass, with the energy/momentum loss due to final
state radiation; the latter reduces the actual value of the measured
observables, depleting the peak and enhancing the left tail of the
resonant shape.  Since after QED mass factorization there are no large
logarithms due to ISR, the impact of initial state radiation on the lepton-pair
and on the single lepton transverse momentum distributions is
suppressed by the smaller coupling constant with respect to the QCD
case; in the QED case the largest fraction of the corrections to these
observables is due to final state radiation.

Among the observables which are sensitive to the absolute
normalization of the process, we have the single lepton pseudo-rapidity
and the lepton-pair rapidity distributions, and also the large-mass
tail of the lepton-pair invariant mass distribution.  The former
receive a correction which is very close in size to the one of the
total cross section, and which is quite flat along the whole
(pseudo-)rapidity range (the FSR corrections and the redefinition of
the couplings via renormalization do not modify the LO kinematics,
yielding, in first approximation, a global rescaling of the
distributions).

The NLO EW virtual corrections become large and negative in the 
tails of the single-lepton transverse momentum, lepton-pair invariant
and transverse-mass distributions, when at least one kinematical
invariant becomes large, because of the contribution of the purely
weak vertex and box corrections.  This effect of the so-called EW Sudakov 
logarithms can not be re-absorbed in
a redefinition of the couplings and is process dependent. 
A recent discussion of the DY processes in the Sudakov regime can be found, e.g., in Ref.~\cite{Chiesa:2013yma,Mishra:2013una}.

The size of the effects due to the emission of real photons depends on
the experimental definition of the lepton, i.e. on the recombination
procedure of the momenta of the lepton with those of the surrounding
photons.  The radiation of photons collinear to the emitting lepton
has a logarithmic enhancement, with a natural cut-off provided by the
lepton mass. These mass logarithms cancel completely in the total
inclusive cross section (Kinoshita-Lee-Nauenberg theorem), but leave an effect on the
differential distributions.  The recombination of the photons and
lepton momenta effectively acts like the integration over the
collinear corner of the photon phase space, yielding a cancellation of
the singular contribution from that region; as a consequence, the
logarithmic enhancement of the corrections is reduced, as if the
lepton had acquired a heavier effective mass.


\subsubsection{Photon-induced processes}
\label{sec:gammainduced}

The ${\cal O}(\alpha)$ corrections develop initial-state QED
collinear singularity, which have to be subtracted from the partonic
cross section and can be re-absorbed in the definition and
evolution of the proton PDFs, in close analogy to what is done in QCD.  
In turn, the QED terms present in the
evolution kernel of these PDFs imply the existence of a photon
density inside the proton, which allows the contribution of partonic
subprocesses initiated by photons.  The latter are present already at
LO in the case of the NC DY process, $\gamma\gamma\to l^+l^-$, or they
appear at NLO in both the NC and CC DY processes, $\gamma q(\bar q) \to l^+ l^-
q(\bar q)$ and $\gamma q(\bar q) \to l\nu
q'(\bar q')$.

\begin{figure}[h]
\includegraphics[width=75mm,angle=0]{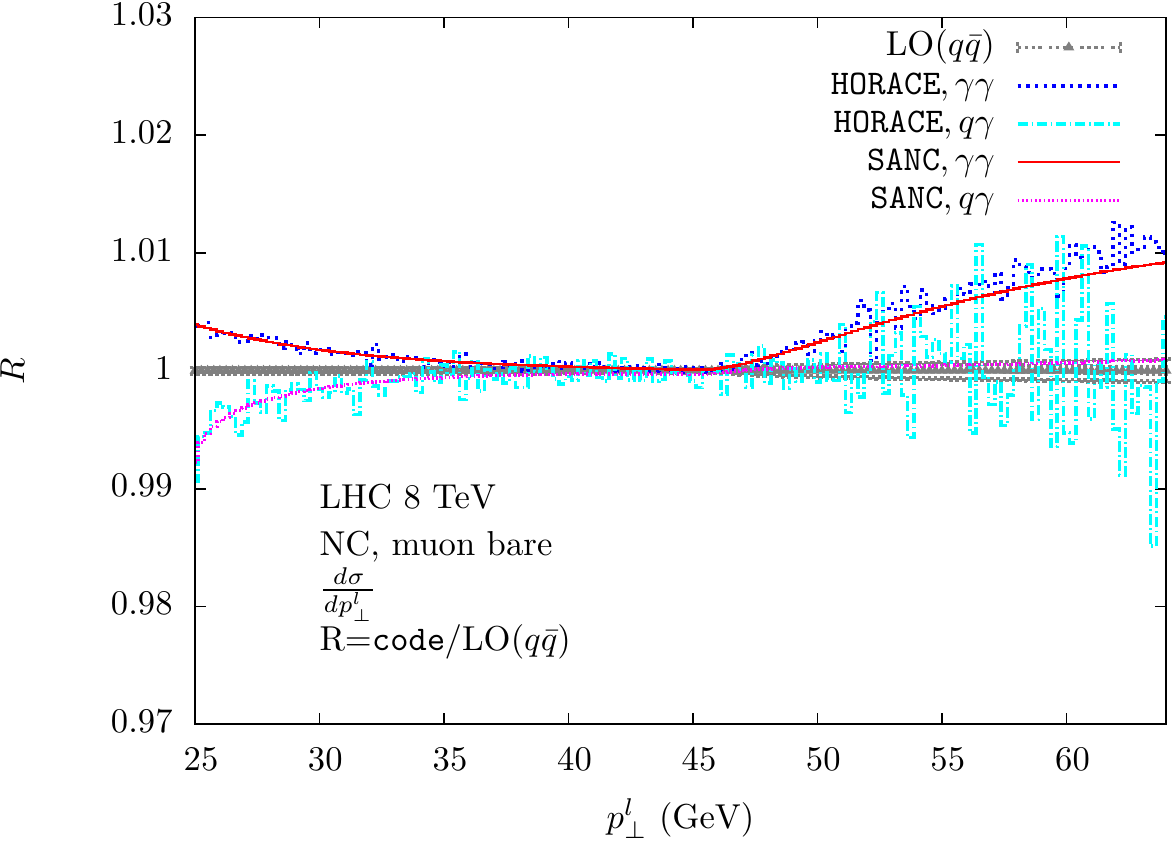}
\includegraphics[width=75mm,angle=0]{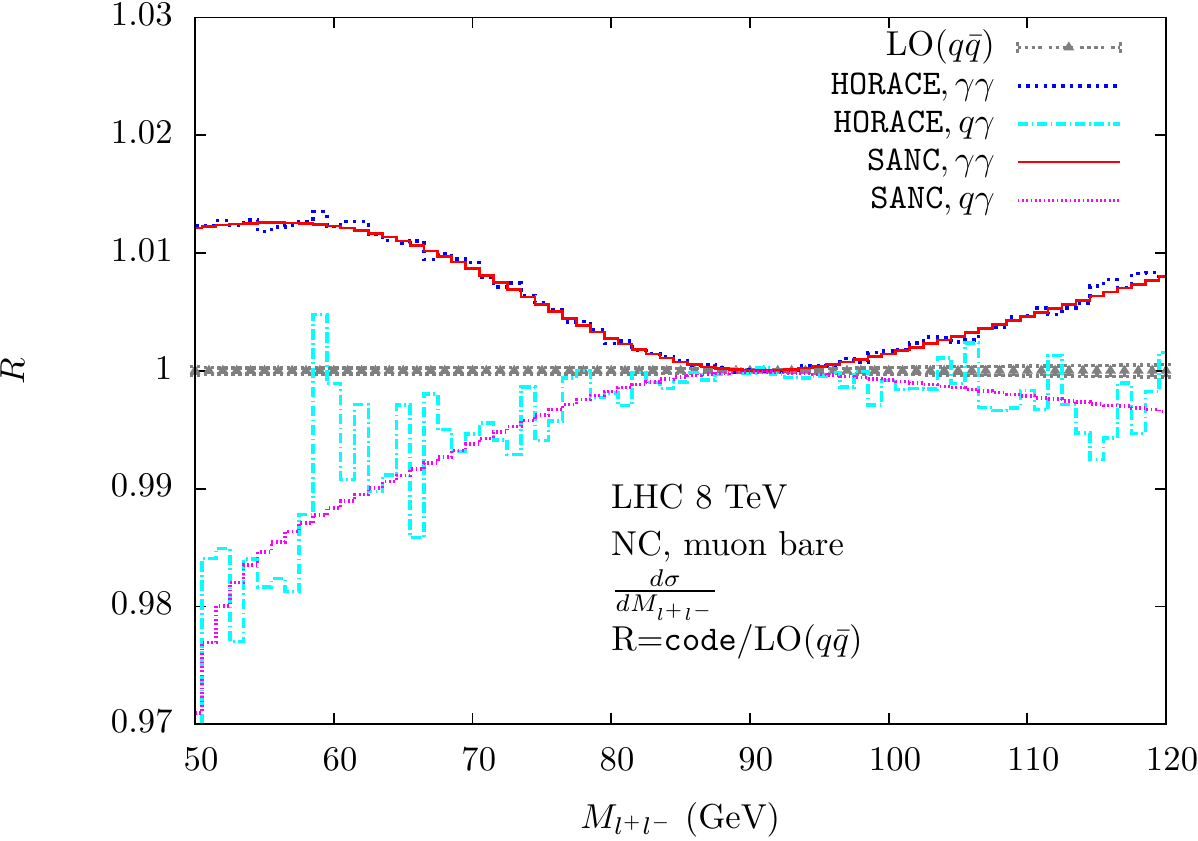}
\caption{
Relative effect of photon-induced subprocesses in $pp\to\mu^+\mu^-+X$, compared to the LO $q\bar q$ results in the $G_\mu$ scheme, both evaluated with the
{\tt NNPDF23\_lo\_as\_0130\_qed} PDF set at the 8 TeV LHC with ATLAS/CMS cuts. Results are shown for the lepton transverse momentum (left plot) and the lepton-pair invariant mass (right plot) and are obtained with \horace and \sanc. Separately shown are the contributions of the LO $\gamma\gamma\to\mu^+\mu^-$ subprocess and the ${\cal O}(\alpha)$ $\gamma\smartqaq\to\mu^+\mu^-\smartqaq$ subprocesses.
\label{fig:ho-photoninduced}
}
\end{figure}

In Figure \ref{fig:ho-photoninduced} we present the evaluation at hadron level of these contributions in the case of the NC DY process,
done with the proton PDF set {\tt NNPDF2.3\_lo\_as\_0130\_qed},
using the codes \horace and \sanc. We show the ratios $R=1+d\sigma(\gamma \gamma, \gamma q)/d\sigma(q\bar q)$ to illustrate the relative effect of including the photon-induced processes in the LO prediction. 
The reason for the contribution of the $\gamma\smartqaq\to\mu^+\mu^-\smartqaq$ subprocess to be negative, i.e. values smaller than 1 in the plots,
can be understood as being due to the presence of subtraction terms for the collinear divergences, which are necessary in a NLO calculation.

\subsubsection{EW input scheme choices}
\label{sec:renormalization}

The calculation of the NLO EW set of corrections to the DY processes,
requires the renormalization of EW couplings and masses, which is
typically done by imposing on-shell conditions on the relevant Green's
functions.  The choice of the set of physical observables
necessary to evaluate the parameters $(g,g',v)$ of the gauge sector of
the Lagrangian is done following two main criteria: 1) the quantities
which are best determined from the experimental point of view minimize
the parametric uncertainties affecting all the predictions; 2) some
observables automatically include in their definition important
classes of radiative corrections, so that their use reduces the impact
of the radiative corrections to the scattering process under study.

A convenient set of parameters that describes EW processes at hadron
colliders is $(G_\mu,\mw,\mz)$, the so called $G_\mu$ scheme. The
Fermi constant $G_\mu$ measured from muon decay naturally parameterize
the CC interaction, while the $W$ and $Z$ masses fix the scale of EW
phenomena and the mixing with the hyper-charge field.  A drawback of
this choice is the fact that the coupling of real photons to charged
particles is computed from the inputs and in lowest order is equal to
$\alpha_{G_\mu}=G_\mu \sqrt{2} \mw^2 (1-\mw^2/\mz^2)/\pi \sim 1/132$
much larger than the fine structure constant $\alpha(0)\sim 1/137$,
which would be the natural value for an on-shell photon.

The alternative choice $(\alpha(0),\mw,\mz)$, the so-called
$\alpha(0)$ scheme, does not suffer of the problem with real photon
radiation, but introduces: $i)$ a dependence on the unphysical
quantities, light-quark masses, via the electric charge
renormalization, and
$ii)$ it leaves large radiative corrections at NLO and in higher orders.

These drawbacks of the two above mentioned schemes can be circumvented
by a use of modified $G_\mu$ scheme when only LO couplings are
re-expressed in terms of $\alpha_{G_\mu}$
\begin{equation}
\alpha\equiv\alpha(0)\to \alpha_{G_\mu}(1-\Delta r)
\end{equation} 
and Sirlin's parameter $\Delta r$ \cite{Sirlin:1980nh}, representing
the complete NLO EW radiative corrections of $\cal{O}(\alpha)$ to the
muon decay amplitude.  Both real and virtual relative
$\cal{O}(\alpha)$ corrections are calculated at the scale $\alpha(0)$,
therefore such an approach may be referred as NLO at ${\cal O}(\alpha
G_\mu^2)$. This choice is adopted in the {\em benchmark} setup of
Section~\ref{sec:setupbest} both for NC and CC DY processes. In this
scheme leading universal corrections due to the runnning of $\alpha$
and connected to the $\rho$ parameter are absorbed in the LO
couplings.

Further modifications may be considered. For NC DY the gauge
invariant separation of complete EW radiative corrections into
pure weak (PW) and QED corrections (involving virtual or real photons) is
possible.  Therefore, these two contributions may be considered at
different scales, PW at ${\cal O}(G_\mu^3)$, and QED still at ${\cal
O}(\alpha G_\mu^2)$.  These different scales seem to be most natural
for PW and QED contributions correspondingly.
For CC DY PW and QED corrections are not separately gauge invariant,
so that usually the complete NLO EW contribution (PW+QED) is
considered using the same overall scale, either ${\cal O}(G_\mu^3)$ or
${\cal O}(\alpha G_\mu^2)$.  More refined modifications may be
considered, for instance based on defining gauge invariant subsets by
using the Yennie-Frautschi-Suura approach~\cite{yfs:1961}.  The spread of
predictions with different modifications of the $G_\mu$ scheme may be
considered as an estimate for the uncertainty due to missing
higher-order EW effects.


\subsubsection{Impact of different gauge boson mass definitions}
\label{sec:massdef}

\begin{figure}[h]
\centering
\includegraphics[width=75mm,angle=0]{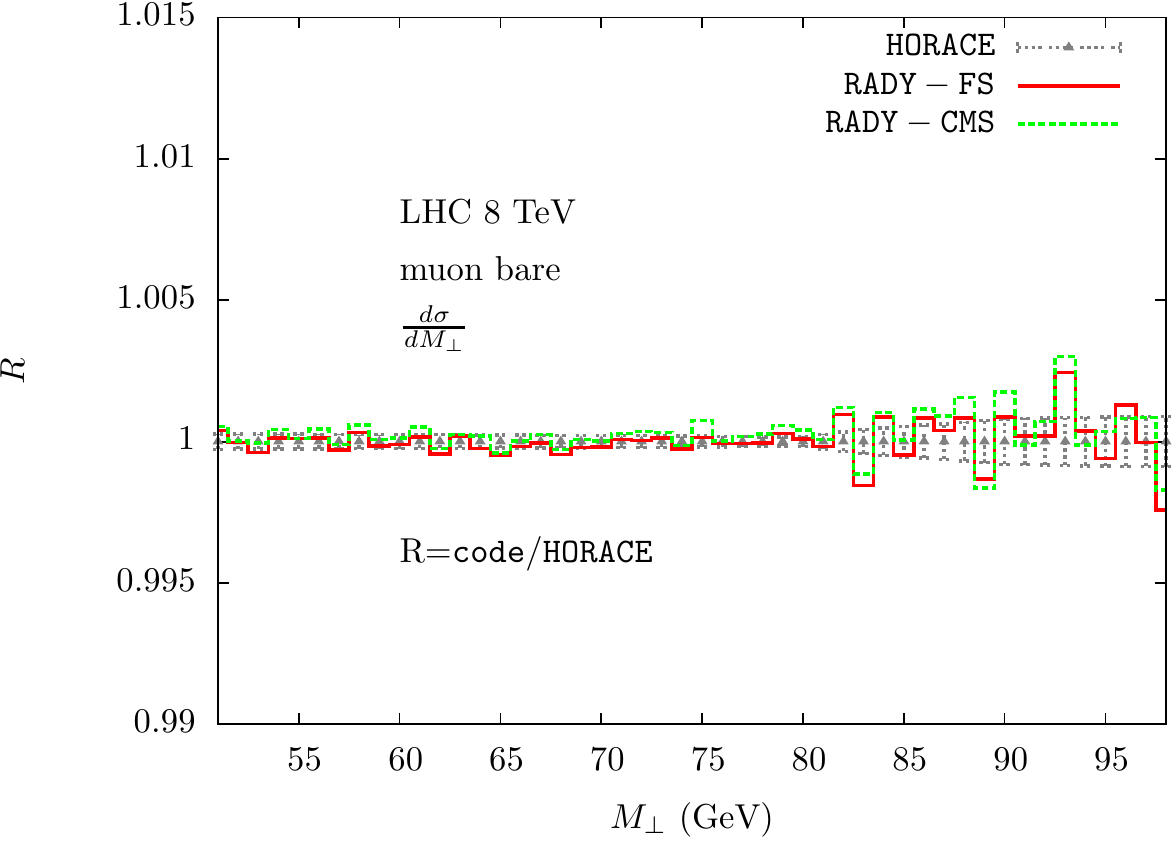}
\includegraphics[width=75mm,angle=0]{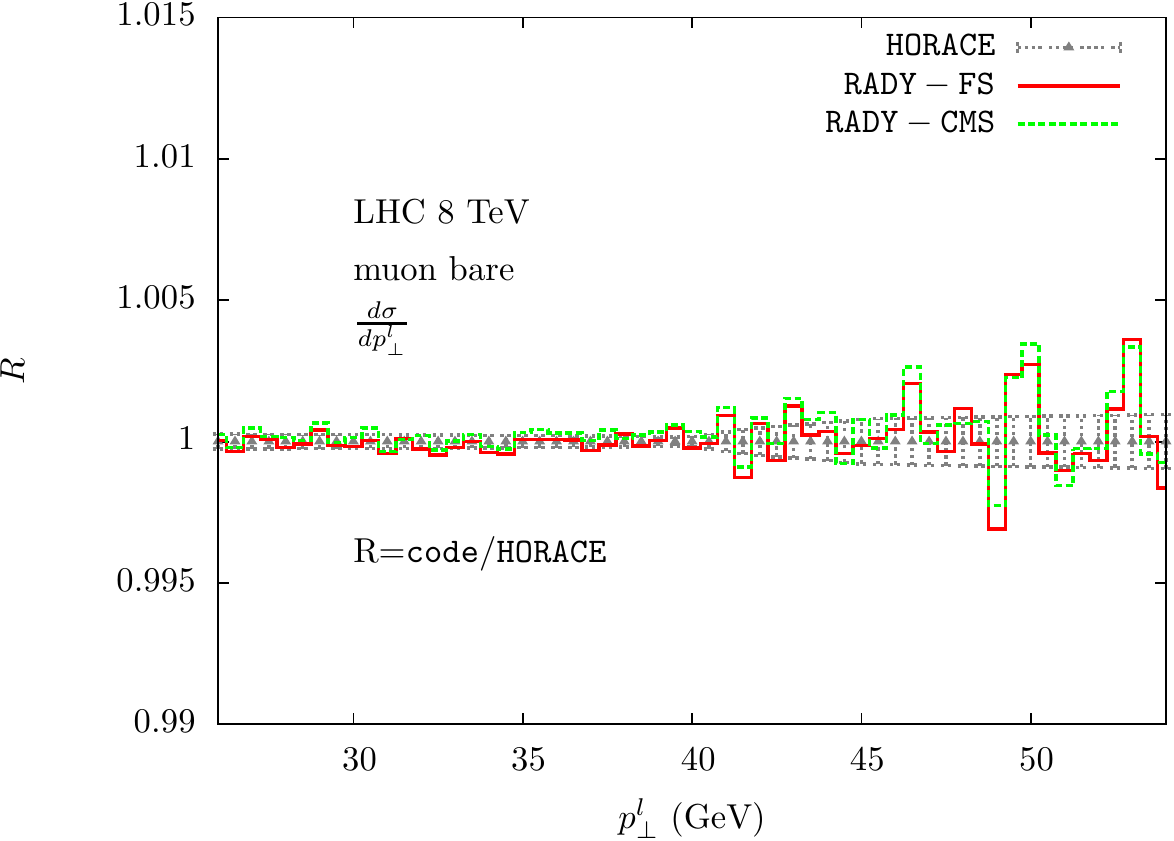}
\caption{Comparison of RADY NLO EW predictions when using different schemes for treating the $W$ resonance. The plots show the transverse mass and momentum distribution of the final-state
  charged lepton in $pp\to W^+ \to \mu^+\nu_\mu+X$ at the 8~TeV LHC with ATLAS/CMS cuts in the {\em bare} setup. The definitions of the FS, CMS and PS schemes can be found in Ref.~\cite{Dittmaier:2009cr}.}\label{fig:rady_w}
\end{figure}

\begin{figure}[h]
\centering
\includegraphics[width=75mm,angle=0]{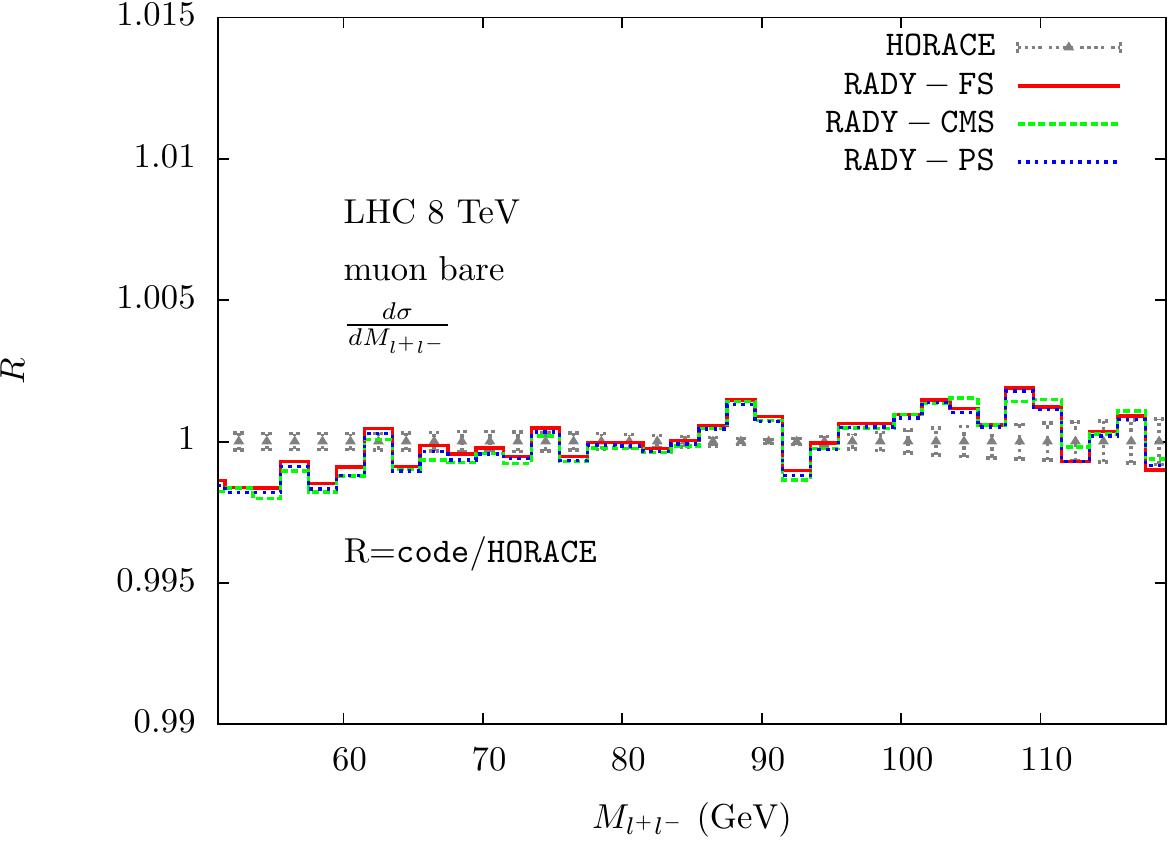}
\includegraphics[width=75mm,angle=0]{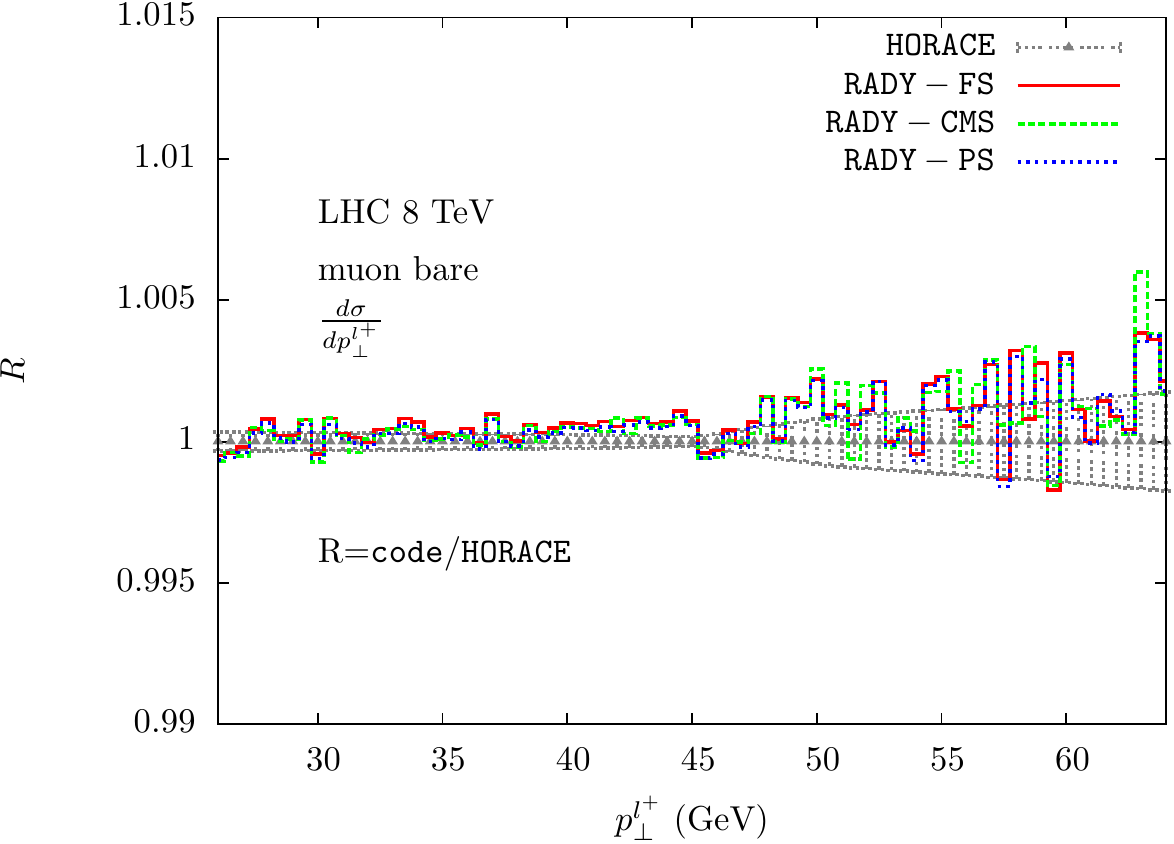}
\caption{Comparison of RADY NLO EW predictions when using different schemes for treating the $Z$ resonance. The plots show the inverse mass and momentum distribution of the final-state
  lepton in $pp\to \gamma,Z \to \mu^+\mu^- +X$ at the 8~TeV LHC with ATLAS/CMS cuts in the {\em bare} setup. The definitions of the FS, CMS and PS schemes can be found in Ref.~\cite{Dittmaier:2009cr}.}\label{fig:rady_z}
\end{figure}

In Ref.~\cite{Dittmaier:2009cr} the evaluation of the LO and NLO EW
cross sections for the NC DY process has been performed in different
schemes for treating the $Z$-boson resonance, denoted as the factorization
scheme (FS), complex-mass scheme (CMS) and pole scheme (PS).  We refer
to Ref.~\cite{Dittmaier:2009cr} for a detailed description of these
various procedures. Here we provide in Figs.~\ref{fig:rady_w} and \ref{fig:rady_z} a comparison of
predictions for CC and NC Drell-Yan processes, respectively, obtained in these
different schemes in the {\em tuned} comparison setup of
Section~\ref{sec:setup}. As also concluded in
Ref.~\cite{Dittmaier:2009cr}, the numerical differences between the
CMS and FS/PS schemes are small. We observe that the predictions for
the observables under study in this report obtained by using the FS, CMS and PS schemes agree within the
statistical uncertainties of the MC integration.

\subsubsection{Universal higher-order corrections in NC DY}
\label{sec:renormalization-ho-universal}

\newcommand{\sw}{s^2_\smallw}
\newcommand{\cw}{c^2_\smallw}
\newcommand{\swb}{\bar{s}^2_\smallw}
\newcommand{\cwb}{\bar{c}^2_\smallw}
\newcommand{\drho}{\Delta\rho}
\newcommand{\xt}{x_t}
\newcommand{\mt}{m_t}
\newcommand{\smallh}{{\scriptscriptstyle H}}
\newcommand{\mh}{m_\smallh}
\newcommand{\rhoone}{\rho^{(1)}}
\newcommand{\rhotwo}{\rho^{(2)}}

In the following the starting point is the modified $G_\mu$ scheme
(the {\em benchmark} scheme in this report) and we discuss two
possible ways to include leading universal higher-order corrections,
i.e. corrections beyond ${\cal O}(\alpha)$. In both cases the LO prediction is at
${\cal O}(G_\mu^2)$ and higher orders start at ${\cal
O}(G_\mu^3)+{\cal O}(G_\mu^2\alpha_s)$.
\begin{itemize}
\item
Following Ref.~\cite{Dittmaier:2009cr}, the leading $G_\mu\mt^2$
universal higher order corrections are taken into account via the
replacements:
\begin{equation}
\sw\to\swb\equiv\sw+\drho\,\cw\,,\qquad\cw\to\cwb\equiv 1-\swb=(1-\drho)\,\cwb
\label{eq:replacements}
\end{equation}
in the LO expression for the NC DY cross section.  As was argued in
Refs.~\cite{Consoli:1989pc,Consoli:1989fg}, this approach correctly
reproduces terms up to ${\cal O}(\drho^2)$.

The quantity $\drho$ 
\begin{equation}
\label{eq:rhoho}
\drho=3\xt\,\left[1+\rhotwo\,\left(\mh^2/\mt^2\right)\,\xt\,\right]\,
            \left[1-\frac{2\alpha_s(\mt^2)}{9\pi}(\pi^2+3)\right]
\end{equation}
contains two contributions:\\
i) the two-loop EW part at ${\cal O}(G_\mu^2)$, second term in the first
square
brackets, \cite{Barbieri:1992nz,Barbieri:1992dq,Fleischer:1993ub,Fleischer:1994cb},
with $\rhotwo$ given in Eq.~(12) of
Ref.~\cite{Fleischer:1993ub,Fleischer:1994cb} (actually, after the
discovery of the Higgs boson and the determination of its mass it became sufficient
to use the low Higgs mass asymptotic, Eq.~(15) of
Ref.~\cite{Fleischer:1993ub,Fleischer:1994cb}); \\
ii) the mixed
EW$\otimes$QCD at ${\cal O}(G_\mu\alpha_s)$, second term in the second
square brackets, \cite{Djouadi:1987gn,Djouadi:1987di}.

The quantity $\Delta\rhoone$ 
\begin{equation}
\label{eq:rho1l}
\Delta\rhoone\Big|^{G_\mu}=3\xt=\frac{3\sqrt{2}G_\mu\mt^2}{16\pi^2}
\end{equation}
represents the leading NLO EW correction to $\drho$ at ${\cal O}(G_\mu)$ 
and should be subtracted from higher-order effects.
Therefore, the contribution of higher-order effects has the following generic form: 
\begin{equation}
\sum_i\,c_i\left[2\left(\drho-\Delta\rhoone\right)\,R_{1i}+\drho^2\,R_{2i}\,\right],
\label{eq:genericho}
\end{equation}
where $c_i$ and $R_{1i,2i}$ are combinations of $Z(\gamma)f\bar{f}$ couplings and the ratio $\cw/\sw$, and their explicit 
form depends on the parametrization of the LO cross section where the replacements $(\ref{eq:replacements})$ 
are performed (cf. Eq.~(3.49) of \cite{Dittmaier:2009cr}).

This approach is implemented in \rady and \sanc.
\item
As described in Ref.~\cite{Baur:2001ze}, the implementation of the NC DY in 
{\tt WZGRAD} closely follows Ref.~\cite{Bardin:1997xq,Hollik:1988ii}
for a careful treatment of higher-order corrections, which is important
for a precise description of the $Z$ resonance.
The NLO differential parton cross section 
including weak ${\cal O}(\alpha)$ and leading ${\cal O}(\alpha^2)$
has the following form
\begin{equation}\label{eq:xsecweak}
{\rm d} \hat \sigma^{(0+1)}={\rm dP_{2f}} \, \frac{1}{12} \, \sum 
|A_{\gamma}^{(0+1)}+ A_Z^{(0+1)}|^2(\hat s,\hat t,\hat u) + 
{\rm d} \hat \sigma_{{\rm box}}(\hat s,\hat t,\hat u) \; .
\end{equation} 
${\rm d} \hat \sigma_{{\rm box}}$ describes
the contribution of the box diagrams and
the matrix elements
$A_{\gamma,Z}^{(0+1)}$ comprise the Born matrix elements, $A_{\gamma,Z}^0$, 
the $\gamma,Z, \gamma Z$ self energy insertions, 
including a leading-log resummation of the terms involving the light
fermions, and the one-loop vertex corrections.
$A_{\gamma,Z}^{(0+1)}$ can be expressed in terms of effective
vector and axial-vector couplings $g_{V,A}^{(\gamma,Z),f}, f=l,q$, including vertex corrections and
self energy insertions. 
Moreover, the $\mz$ renormalization constant
$\delta\mz^2={\cal R}e\left(\Sigma^Z(\mz^2)\right)$
is replaced by
$\delta\mz^2={\cal R}e\left(
\Sigma^Z(\mz^2)-
\frac{(\hat\Sigma^{\gamma Z}(\mz^2))^2}{\mz^2+\hat\Sigma^\gamma(\mz^2)}
\right)$
where $\Sigma^V\, (\hat\Sigma^V)$ denotes the transverse part of the
unrenormalized (renormalized) gauge boson self energy corrections.
Higher-order (irreducible) corrections connected to the $\rho$ parameter 
are taken into account by performing the replacement
\begin{equation}
\frac{\delta \mz^2}{\mz^2}-
\frac{\delta \mw^2}{\mw^2}
\to
\frac{\delta \mz^2}{\mz^2}-
\frac{\delta \mw^2}{\mw^2}
-\Delta\rho^{h.o.}
\end{equation}
where $\Delta\rho^{h.o.}=\Delta \rho-\Delta\rhoone\Big|^{G_\mu}$ with $\Delta\rho$ of Eq.~\ref{eq:rhoho} and
$\Delta \rhoone\Big|^{G_\mu}$ of Eq.~\ref{eq:rho1l}.
\end{itemize}

\begin{figure}[h]
\includegraphics[width=75mm,angle=0]{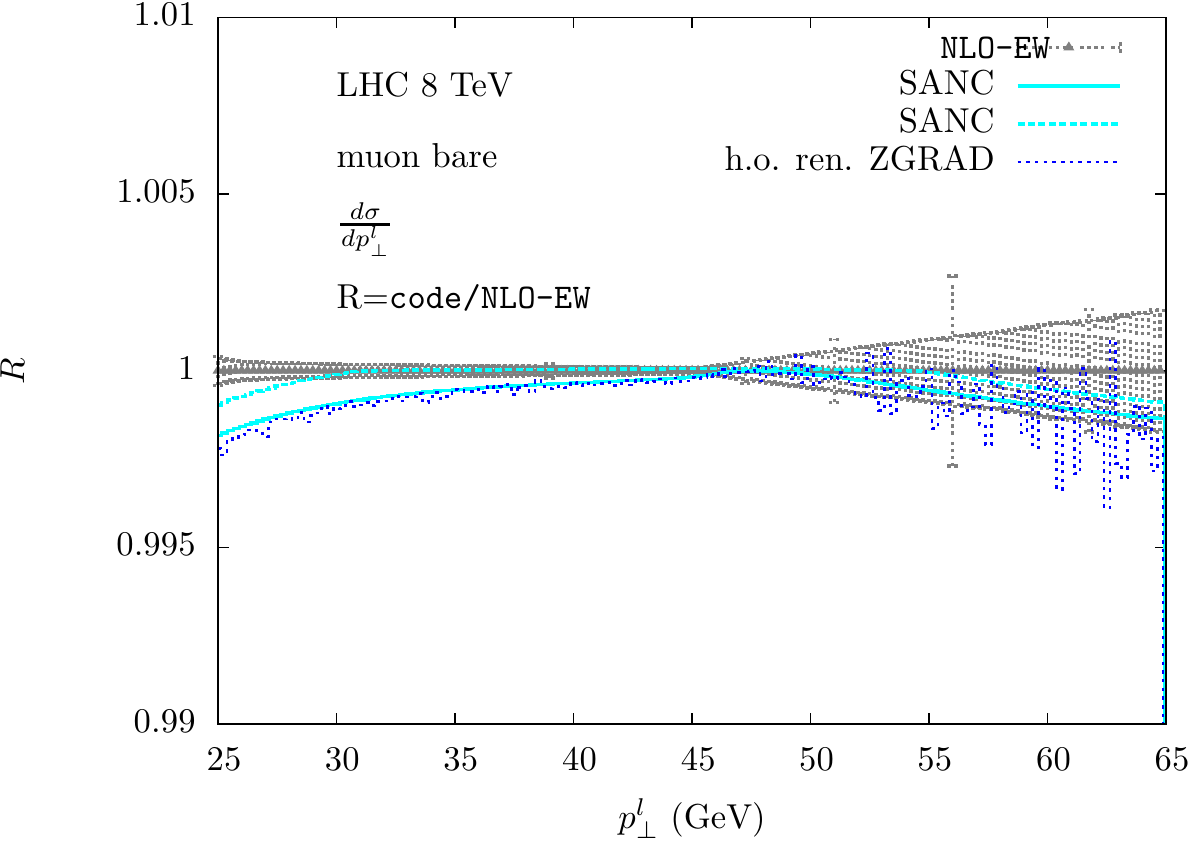}
\includegraphics[width=75mm,angle=0]{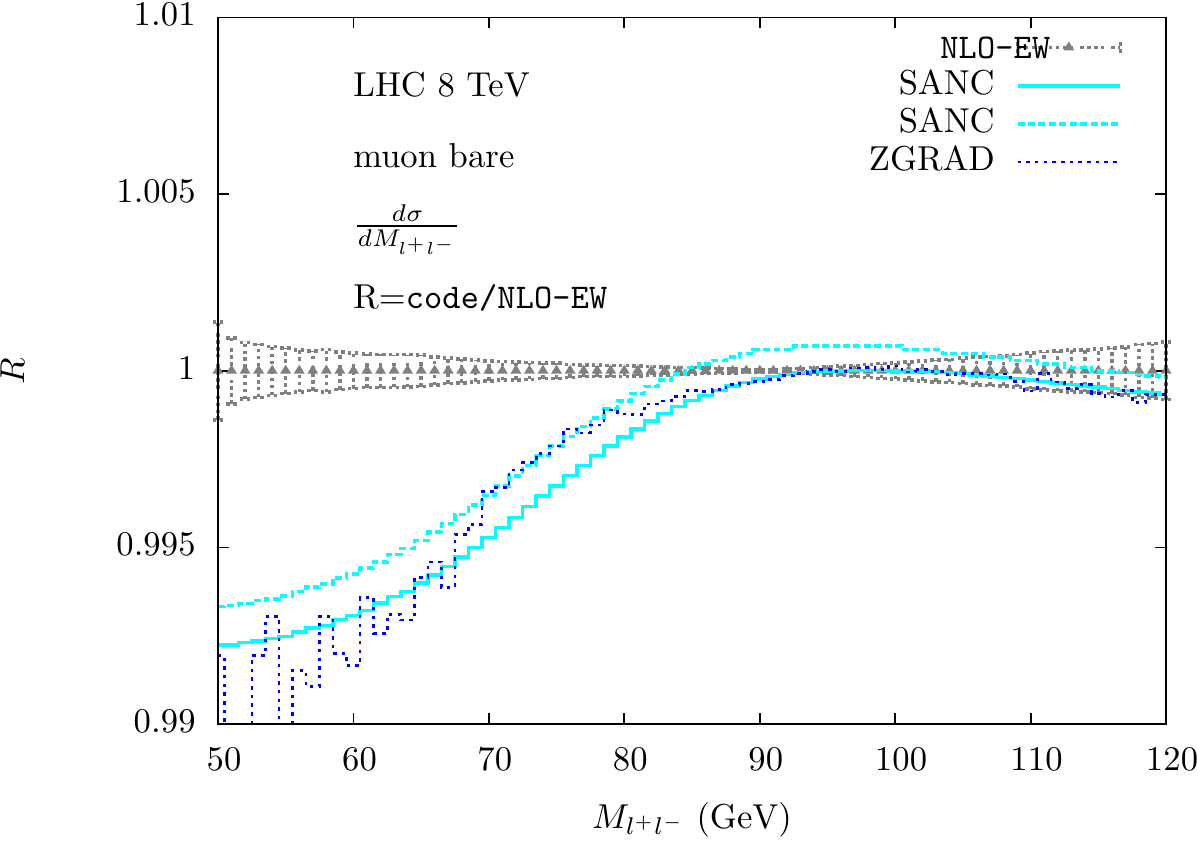}\\
\caption{
Relative effects of higher-order (${\cal O}(G_\mu^2)$ and higher) EW corrections 
in $pp \to \gamma, Z \to \mu^+ \mu^-$, 
due to the inclusion of universal corrections using the $\rho$ parameter as described in the text.
Shown are the lepton transverse momentum (left) and lepton-pair invariant mass (right) distributions, obtained for the 8 TeV LHC with ATLAS/CMS cuts.
In blue the \wzgrad results; 
in light blue the \sanc results obtained in a linear (solid)
[first term of Eq.~(\ref{eq:genericho})] and quadratic (dashed) 
[both terms of Eq.~(\ref{eq:genericho})] implementation.
\label{fig:ho-rho}
}
\end{figure}

The impact of these universal higher-order EW corrections as implemented in 
SANC and WZGRAD is shown in Fig.~\ref{fig:ho-rho}. 

\subsubsection{Higher-order effects to all orders via running couplings in NC DY}
\label{sec:renormalization-ho-running}

The purely EW fixed-order results, in the case of the NC DY process,
can be improved with the systematic inclusion of some classes of
universal higher-order corrections.  The strategy to achieve this
result is given by the matching of an Improved Born Approximation
(IBA) of the LO description of the process, together with the full
${\cal O}(\alpha)$ calculation, avoiding any double counting.

The IBA for reactions of the class $2f\to 2f$ has been extensively
discussed at LEP \cite{Altarelli:1989hv}; here we discuss a specific
implementation in the \horace event generator.  We can write the
LO scattering amplitude in a symbolic compact form as
\be
{\cal M}^{LO}={\cal M}_\gamma+{\cal M}_Z 
        = \alpha(0) \frac{J^\gamma_{q\bar q}\cdot J^\gamma_{l^+l^-}}{q^2+i\varepsilon} +
          \frac{g^2}{\cos\theta_W} \frac{J^Z_{q\bar q}\cdot J^Z_{l^+l^-}}{q^2-\mz^2+i\gz\mz} \; ,
\label{eq:improvedcouplings1}
\ee
where $J^{\gamma,Z}_{f\bar f}$ are the fermionic currents coupling to
photons and to $Z$ bosons and $\cos\theta_W$ is the cosinus of the
electroweak mixing angle.  An improved expression of the amplitude
${\cal M}_{IBA}^{LO}$ is obtained with the following replacement of
the coupling constants:
\bea
\alpha(0)&\to&\alpha(M_{ll}^2)
\quad\quad\quad\quad\quad\quad 
\rm{photon-exchange}\nonumber\\
\frac{g^2}{\cos\theta_W} &\to& 4 \sqrt{2} G_\mu \mz^2 
\frac{\rho_{fi}(M_{ll}^2)}{1-\delta\rho_{irr}}
\quad\quad
\rm{Z-exchange} \; ,
\label{eq:improvedcouplings2}
\eea
where $\alpha(M_{ll}^2)$ is the on-shell running electromagnetic
coupling constant, while $\delta\rho_{irr}$ represents universal
corrections to the neutral current coupling and $\rho_{fi}(M_{ll}^2)$
is a compact notation for all those process dependent corrections that
can be cast as an overall factor multiplying the $Z$-exchange
amplitude (more details can be found in
Refs. \cite{Degrassi:1990ec,CarloniCalame:2007cd}).  The factors
$\alpha(M_{ll}^2)$ and $\frac{1}{1-\delta\rho_{irr}}$ include
universal corrections to all orders while $\rho_{fi}(M_{ll}^2)$ is of
${\cal O}(\alpha)$\footnote{
For a discussion on the definition of an effective electromagnetic coupling at the 2-loop level see Ref. \cite{Degrassi:2003rw}.
}.

The use of the amplitudes in
Eqs.~(\ref{eq:improvedcouplings1}-\ref{eq:improvedcouplings2}) to
compute the cross section represents an approximation of the exact
NLO EW calculation for the non radiative part of the cross section;
since they contain terms beyond NLO EW, one can also read a partial
improvement over pure NLO.  Their matching with the exact NLO EW
expressions allows to recover this perturbative accuracy, but also to
have a systematic inclusion of universal higher-order terms.  Double
counting is avoided by subtracting the ${\cal O}(\alpha)$ part of the
effective couplings in Eq.(\ref{eq:improvedcouplings2}), in that part
of the virtual corrections where the UV counterterms are introduced.

The events are generated with the full NLO EW results computed with
$(\alpha(0),\mw,\mz)$ as input parameters, with a weight $d\sigma^{NLO-EW}$ for
each phase space point.  The latter is rescaled by the factor
$K_{IBA}(\Phi_B)\equiv |{\cal M}^{LO}_{IBA}(\Phi_B)|^2/|{\cal
M}^{LO}(\Phi_B)|^2$, that accounts for all the higher-order effects
and depends on the Born kinematical variables $\Phi_B$\footnote{ In
the case of a radiative event, an effective Born configuration is
computed to evaluate $K_{IBA}$.  }.
\be
d\sigma^{NLO-EW}_{IBA} = K_{IBA}(\Phi_B) d\sigma^{NLO-EW} \; .
\ee
We remark that this rescaling is motivated by the factorization of the
leading contributions due to soft and collinear QED radiation; in
these phase-space regions the exact matrix element is well
approximated by a factorized expression proportional to the underlying
Born.  The rescaling generates several factorizable terms of ${\cal
O}(\alpha^2)$: among them, those due to the emission of a real photon
enhanced by the effective couplings may have a sizeable impact on the
differential distributions.

\begin{figure}[h]
\includegraphics[width=75mm,angle=0]{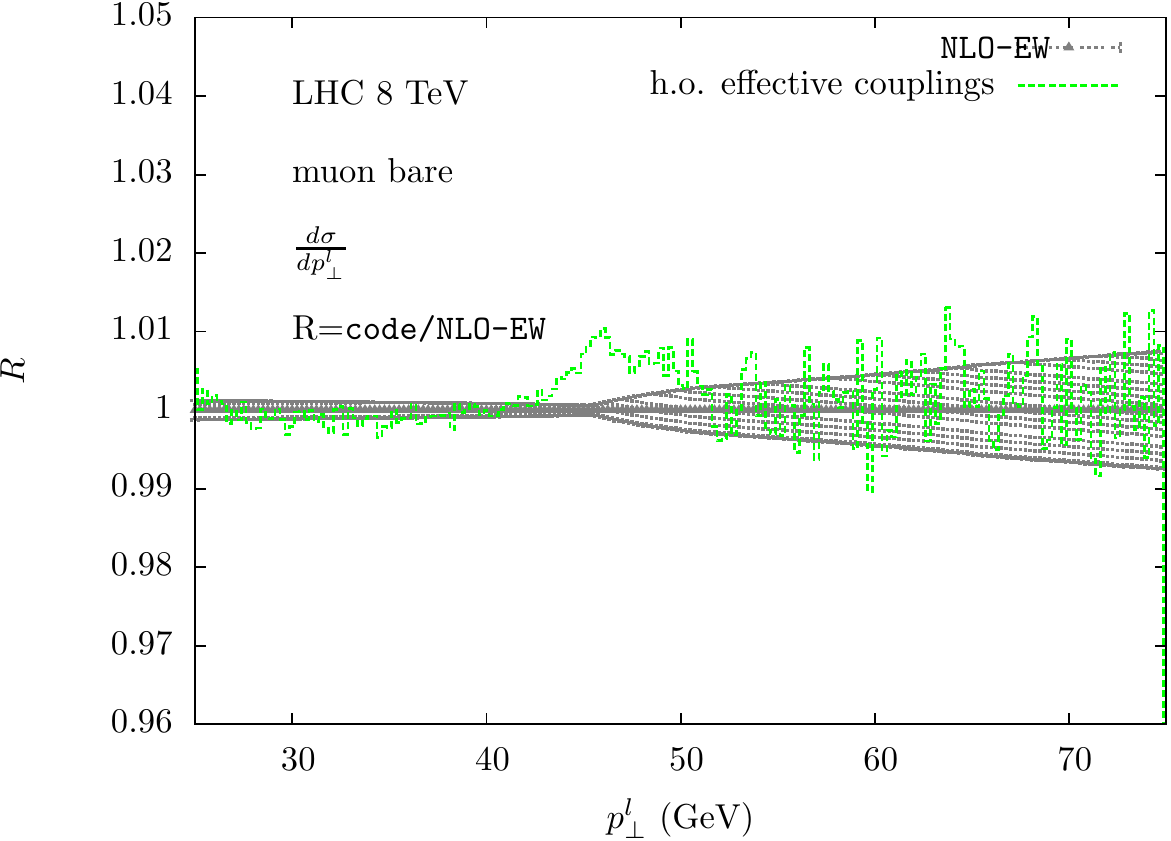}
\includegraphics[width=75mm,angle=0]{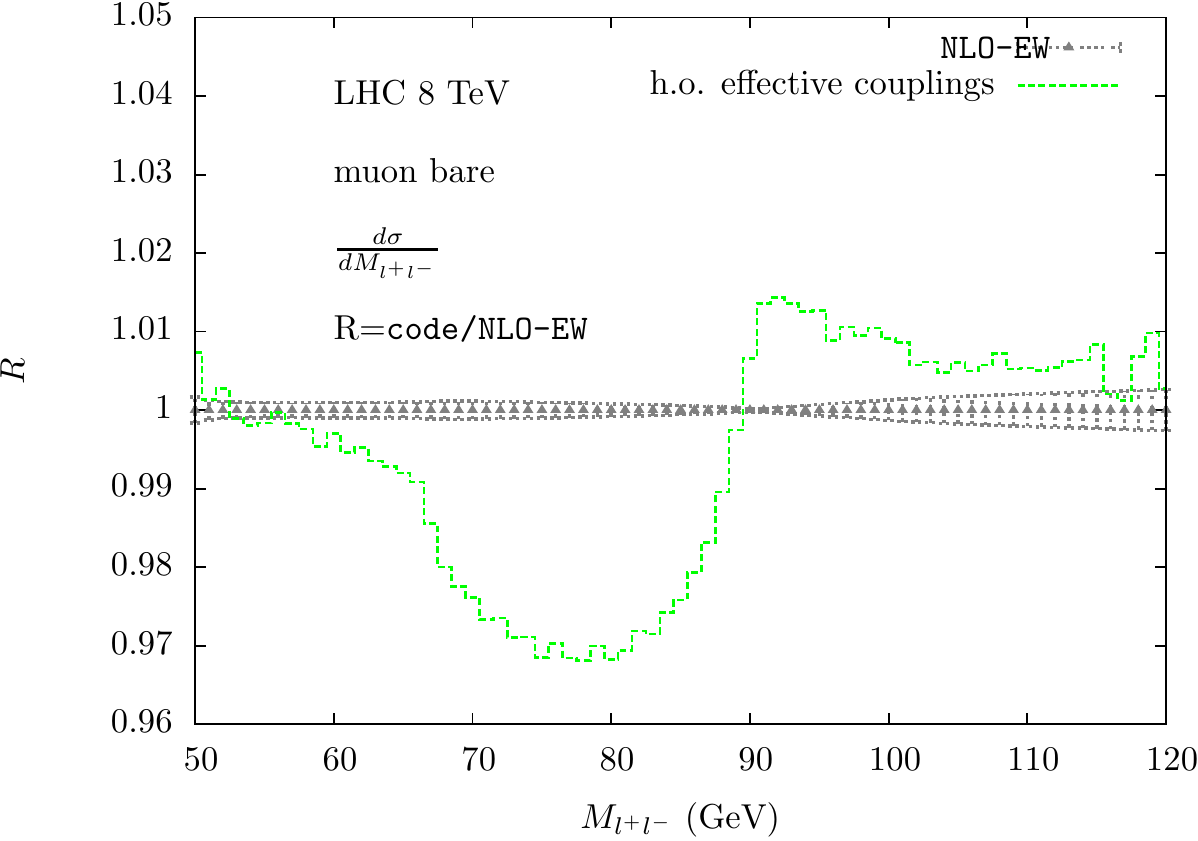}\\
\caption{
Relative effect of higher-order (${\cal O}(\alpha^2)$ and higher) EW
corrections in $pp\to\mu^+\mu^-+X$ due to the inclusion of universal
corrections via effective couplings.  Shown are the lepton transverse
momentum (left), lepton-pair invariant mass (right) distributions for
the 8 TeV LHC with ATLAS/CMS cuts. The results have been obtained with \horace.
\label{fig:ho-runningalpha}
}
\end{figure}

In the invariant mass region below the $Z$ resonance the QED
corrections increase the cross section by up to 100\% of the
fixed-coupling LO result.  The introduction of the effective couplings
yields a net effect at the few per cent level of the LO result.  The
impact of this redefinition of the LO couplings is demonstrated in
Figures \ref{fig:ho-runningalpha}, where we take the ratio of these
improved predictions with those computed at NLO EW in the best setup
of Section~\ref{sec:setupbest}; the deviation from 1 is entirely due
to terms of ${\cal O}(\alpha^2)$ or higher, present in the effective
couplings.

The corrections described in this section are a reducible, gauge
invariant subset, part of the full NNLO EW calculation of the NC DY
process.  They represent a sizeable contribution, due to the
combination of two effects which, separately, are numerically leading
on their own.

\subsubsection{QED shower matched to NLO EW matrix elements}
\label{sec:QEDshowers}

\begin{figure}[h]
\includegraphics[width=75mm,angle=0]{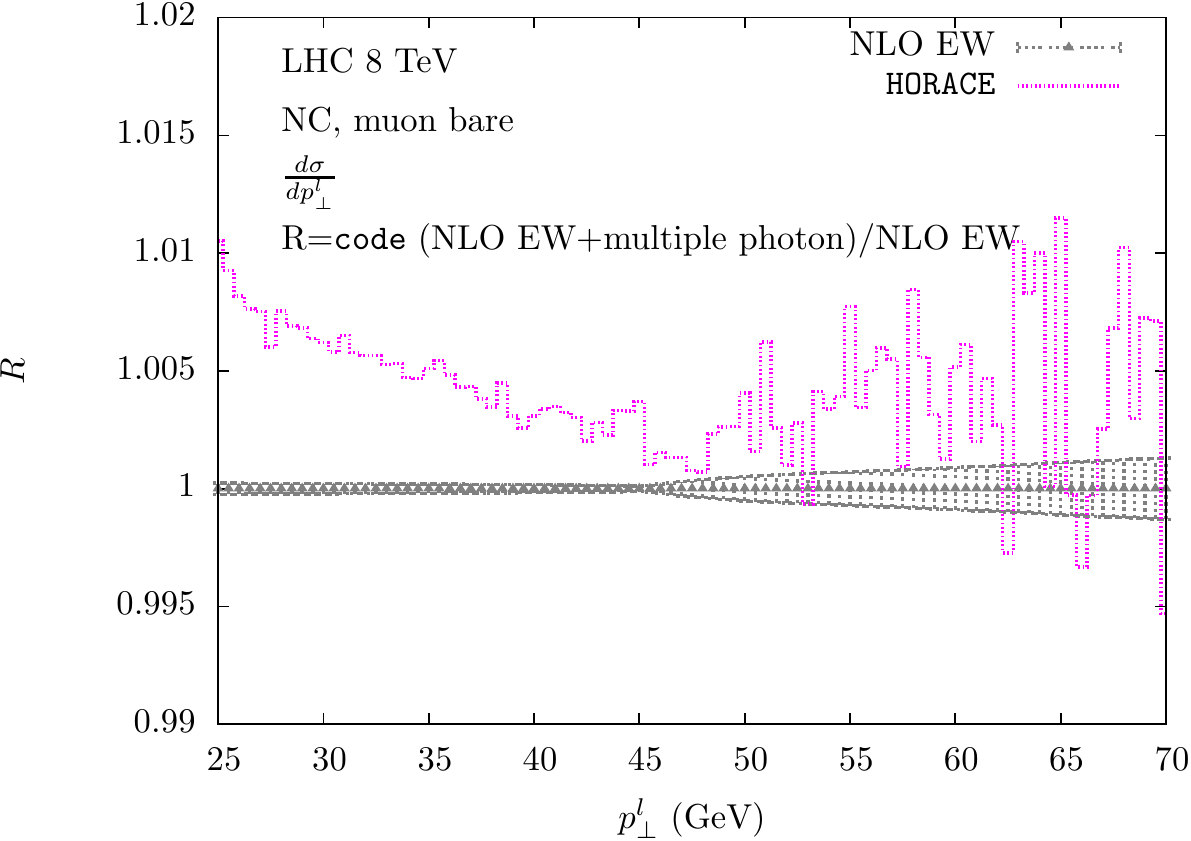}
\includegraphics[width=75mm,angle=0]{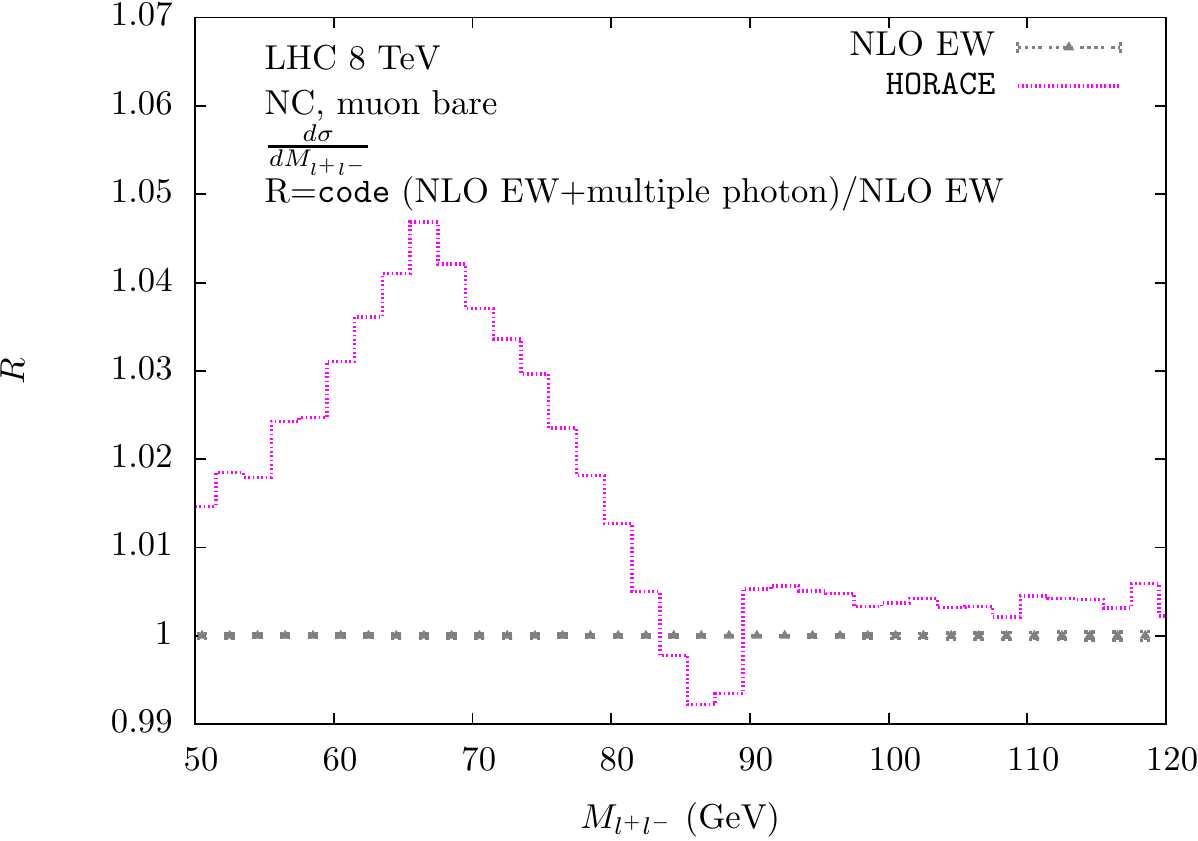}\\
\includegraphics[width=75mm,angle=0]{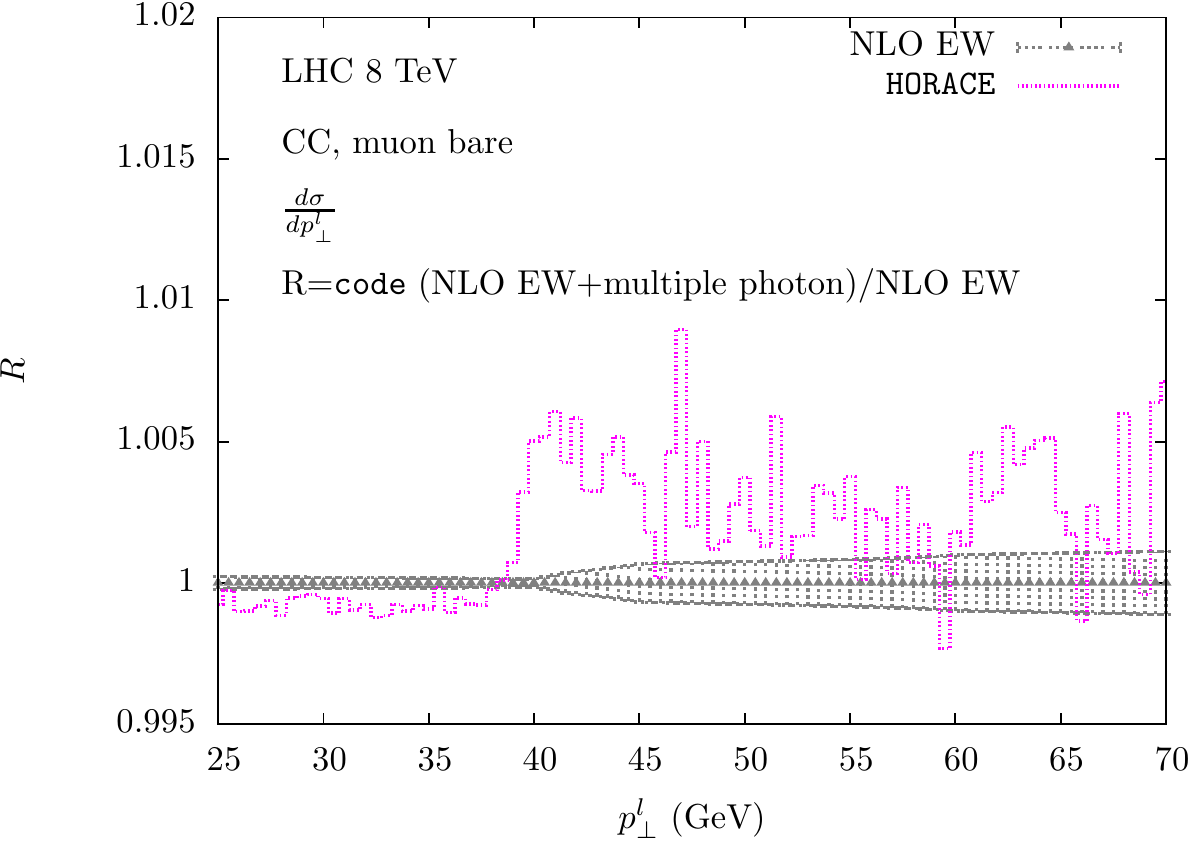}
\includegraphics[width=75mm,angle=0]{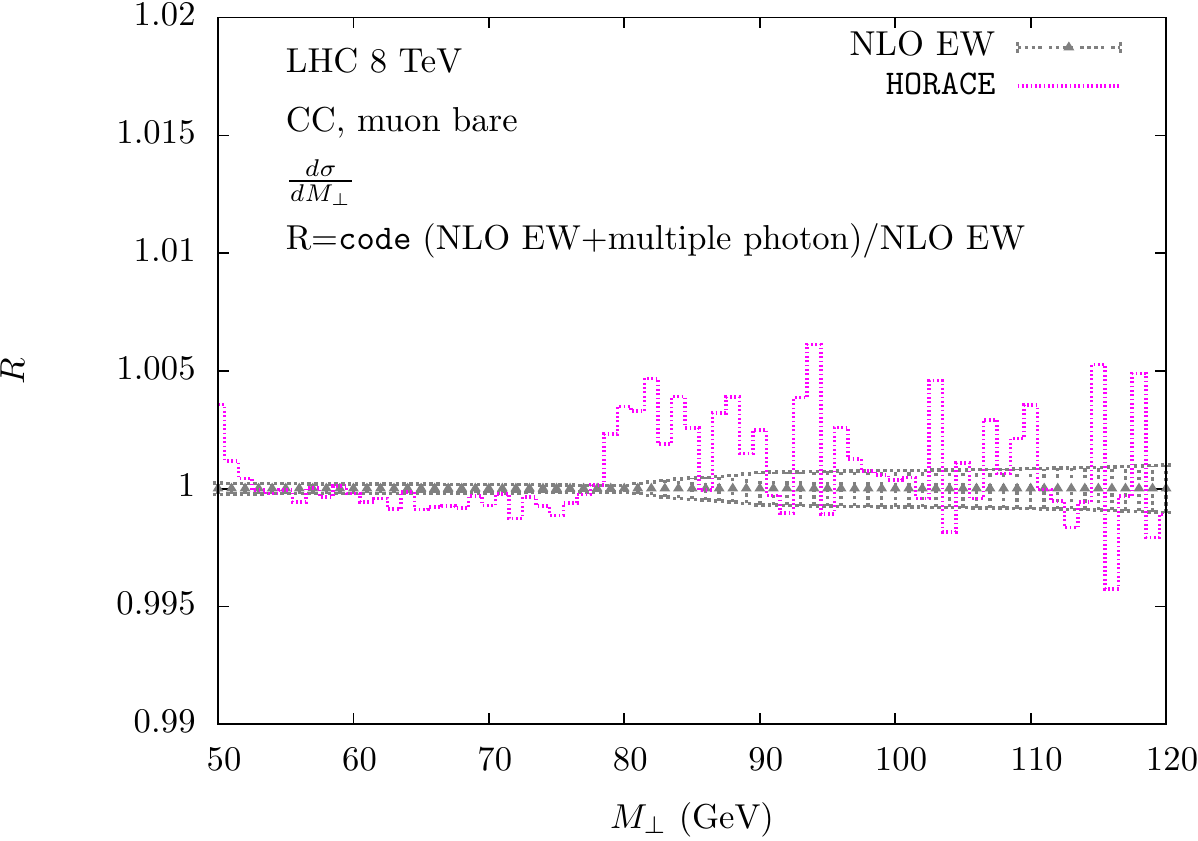}
\caption{
Relative effect of higher-order (${\cal O}(\alpha^2)$ and higher) EW
corrections in $pp\to\mu^+\mu^-+X$ (upper plots) and
$pp\to\mu^+\nu_\mu+X$ (lower plots), due to multiple-photon radiation
matched to the NLO EW results, expressed in units of the pure NLO EW
calculation evaluated in the {\em benchmark} setup for {\em bare} muons.  Shown are the
lepton transverse momentum in NC DY (upper left), lepton-pair
invariant mass in NC DY (upper right), 
lepton transverse momentum in CC DY (lower left), and
lepton-pair transverse mass in CC DY (lower right) for the 8 TeV LHC
with ATLAS/CMS cuts.  The results are obtained in the \horace formulation of
matching NLO EW corrections to multiple-photon emission.
\label{fig:ho-multiphoton}
}
\end{figure}

The inclusion of multiple photon radiation in the presence of NLO EW
matrix elements requires a matching procedure to avoid double
counting.  Several examples have been proposed in the literature
following different algorithms, which have been implemented in the
codes \horace, \powheg, and \winhac, for instance.  
In Fig.~\ref{fig:ho-multiphoton} we use \horace to illustrate 
the effect of all photon emissions beyond the first one 
in the NC (upper plots) and CC (lower plots) 
processes in the {\em benchmark} setup
of Section~\ref{sec:setupbest} for the case of {\em bare} muons. 
The ratio shows the impact of the improved NLO EW prediction, 
when the NLO EW correction is matched to multiple photon radiation,
over the NLO EW prediction;
thus a deviation from 1 is entirely due to terms of ${\cal O}(\alpha^2)$ or higher. 
The impact of ${\cal O}(\alpha)$ corrections on the LO distributions shown in Fig.~\ref{fig:nlo-ew} is largely due to 
photon radiation and thus we also observe a non-negligible effect on the shape from higher-order multiple photon radiation 
in Fig.~\ref{fig:ho-multiphoton};
the size of these effects, as expected, is in the 1\% per cent ballpark,
and depends on the shape of the observable.
For example, while the ${\cal O}(\alpha)$ corrections to the lepton-pair transverse mass distribution
can be as large as $-8\%$ of the LO prediction around the Jacobian peak, 
the ${\cal O}(\alpha^2)$ corrections of multiple photon radiation
are $<0.5\%$ of the NLO EW prediction. 
The lepton-pair invariant mass is the only observable that significantly
changes because of multiple photon radiation:
in fact the ${\cal O}(\alpha)$ radiative effect is of ${\cal O}(85\%)$ below the $Z$ resonance, 
while at ${\cal O}(\alpha^2)$ the effects are a fraction of the previous order correction and can be as large as $5\%$.

\begin{figure}[h]
\includegraphics[width=75mm,angle=0]{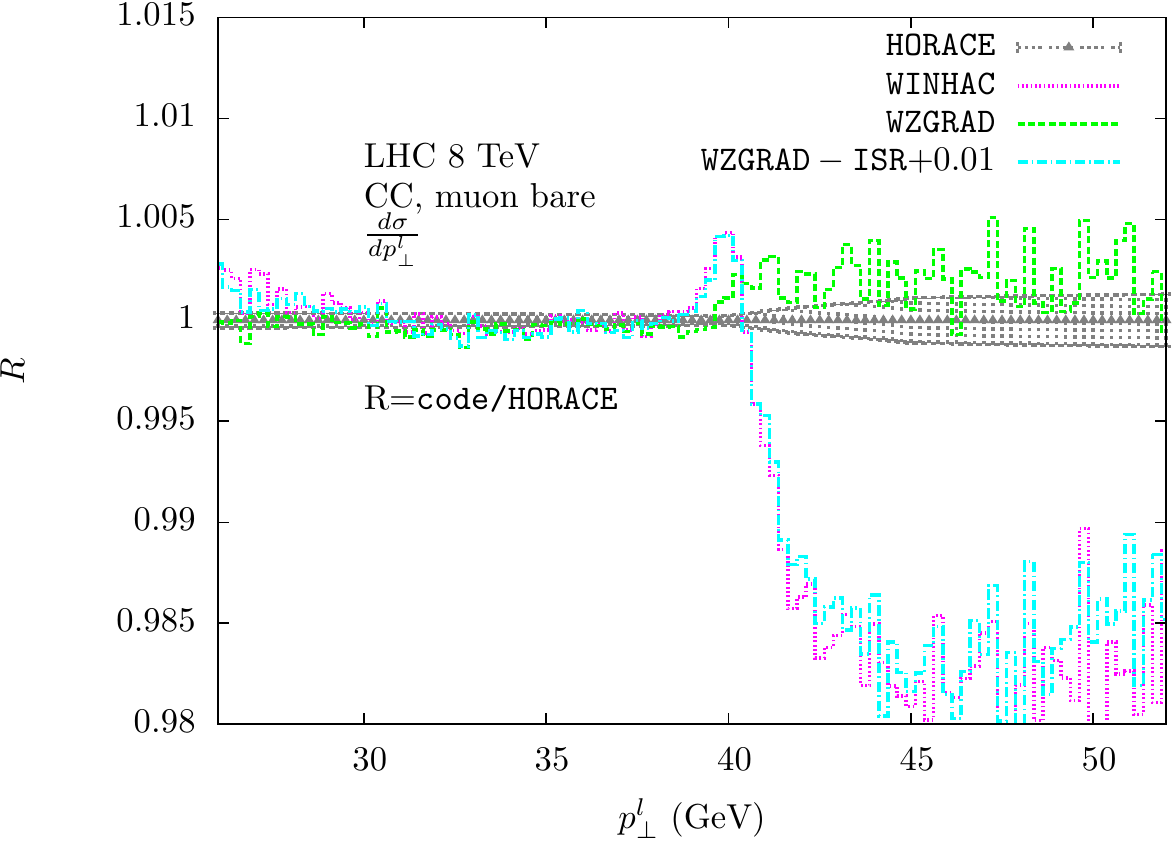}
\includegraphics[width=75mm,angle=0]{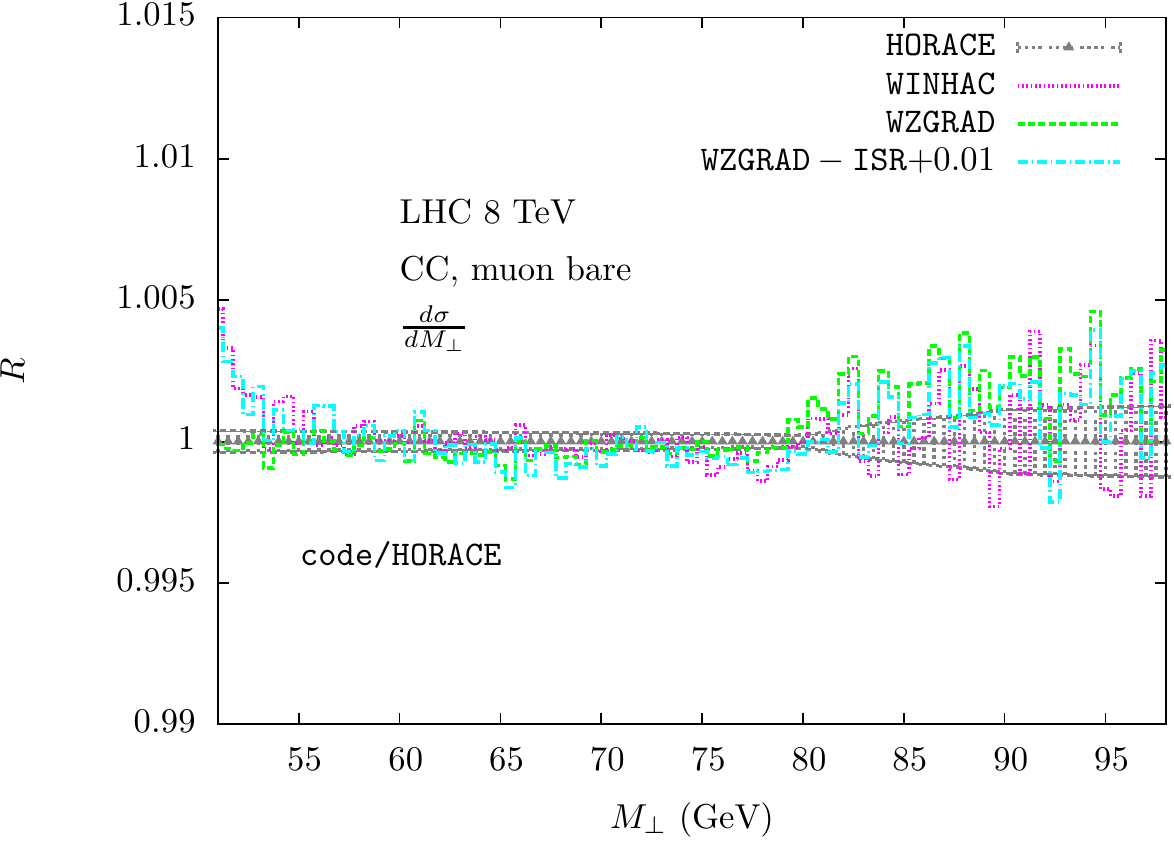}
\caption{
Tuned comparison of the 
the lepton transverse momentum (left) and 
lepton-pair transverse mass (right) distributions in $pp\to\mu^+\nu_\mu+X$ for the 8 TeV LHC with ATLAS/CMS
cuts in the {\em tuned comparison} setup for {\em bare} muons, including NLO EW corrections (\horace, \wzgrad (green curve)) and
ISR-QED subtracted NLO EW corrections by \winhac (pink curve) and \wzgrad (blue curve).
\label{fig:winhac}
}
\end{figure}

\begin{figure}[h]
\includegraphics[width=75mm,angle=0]{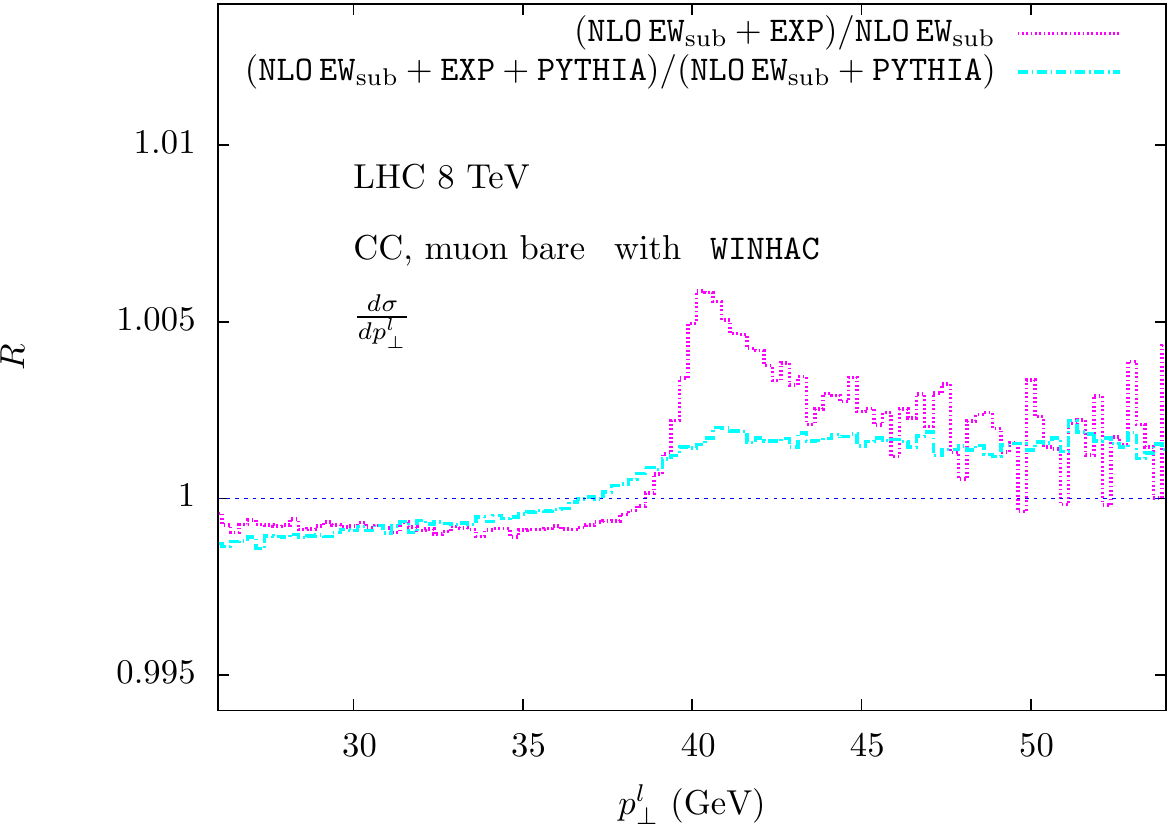}
\includegraphics[width=75mm,angle=0]{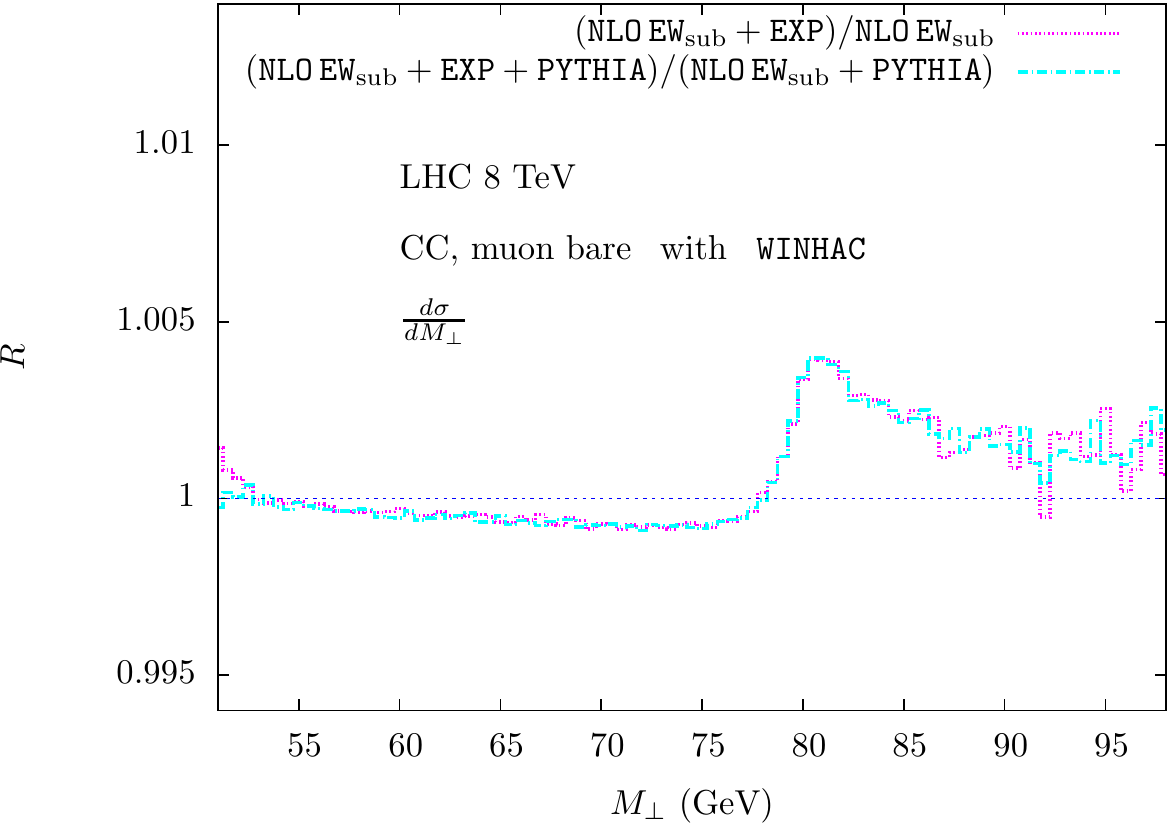}
\caption{
Relative effect of higher-order (${\cal O}(\alpha^2)$ and higher) EW
corrections in $pp\to\mu^+\nu_\mu+X$ due to multiple-photon radiation
in the YFS exponentiation scheme (denoted as {\tt EXP}) matched to the ${\tt NLO \, EW_{\rm
sub}}$ result, expressed in units of the pure ${\tt NLO \, EW_{\rm sub}}$
calculation evaluated in the {\em benchmark} setup for {\em bare}
muons, with and without taking into account the \pythia parton
shower for initial-state photon and parton radiation.  Shown are the
lepton transverse momentum (left), lepton-pair transverse mass (right)
for the 8 TeV LHC with ATLAS/CMS cuts.  The results are obtained in
the \winhac formulation of matching ISR-QED subtracted NLO EW corrections to
multiple-photon emission.
\label{fig:winhac2}
}
\end{figure}

In Fig.~\ref{fig:winhac2} we study the impact of multiple-photon
radiation in the CC DY process as described by \winhac, which is
based on the Yennie-Frautschi-Suura (YFS) exponentiation
scheme~\cite{yfs:1961} matched to a NLO EW contribution, which  
leaves the generation of initial-state photon radiation (ISR) to a parton shower MC.
This ISR-QED contribution is subtracted from the NLO EW prediction
in a gauge-invariant way according to the YFS prescription, and the resulting prediction 
is denoted here as ${\tt NLO \, EW_{\rm sub}}$.  As can be seen
in Fig.~\ref{fig:winhac}, the resulting modified relative NLO EW prediction of \winhac 
agrees with the corresponding
modified relative NLO EW prediction of \wzgrad, {\tt WZGRAD-ISR} in
Fig.~\ref{fig:winhac}, in shape but differs in the normalization by a constant value of 0.01.
This difference can be understood by comparing with the explicit expression for the  
ISR QED ${\cal O}(\alpha)$ correction of \wzgrad as defined in
 Ref.~\cite{Wackeroth:1996hz}, but is left to a future study. The 
results for this comparison have been obtained in the setup of the
{\em tuned comparison} of Section~\ref{sec:setup}.
 
The {\em best} results of \winhac for the CC DY process are obtained when
interfaced with a parton shower MC (here: \pythia), which also handles the initial-state
photon radiation, and when including multiple-photon radiation in the
YFS scheme.  The impact of the YFS exponentiation is shown in
Fig.~\ref{fig:winhac2} on the example of the $p_T$
distribution of the charged lepton and the transverse mass distribution of
the $l\nu$ pair with and without taking into account the \pythia
shower for initial-state photon and parton radiation. 

The impact of YFS 
exponentation observed in Fig.~\ref{fig:winhac2} is very similar to the multiple-photon radiation effects
obtained with \horace as shown in Fig.~\ref{fig:ho-multiphoton}, i.e.
also in the YFS exponentiation scheme of \winhac
the ${\cal O}(\alpha^2)$ corrections (and higher) amount to at most $0.5\%$
of the ${\tt NLO \, EW_{\rm sub}}$ prediction. As expected, 
in the presence of the QCD PS the multiple-photon radiation effects are less pronounced in 
the lepton $p_T$ distribution but are unchanged in the lepton-pair transverse mass distribution (see also Section~\ref{sec:interplay} for a discussion of the interplay of QCD and QED effects in these observables).
 

\subsubsection{Additional light-fermion-pair emission}
\label{sec:fermionpair}

\begin{figure}[h]
\includegraphics[width=75mm,angle=0]{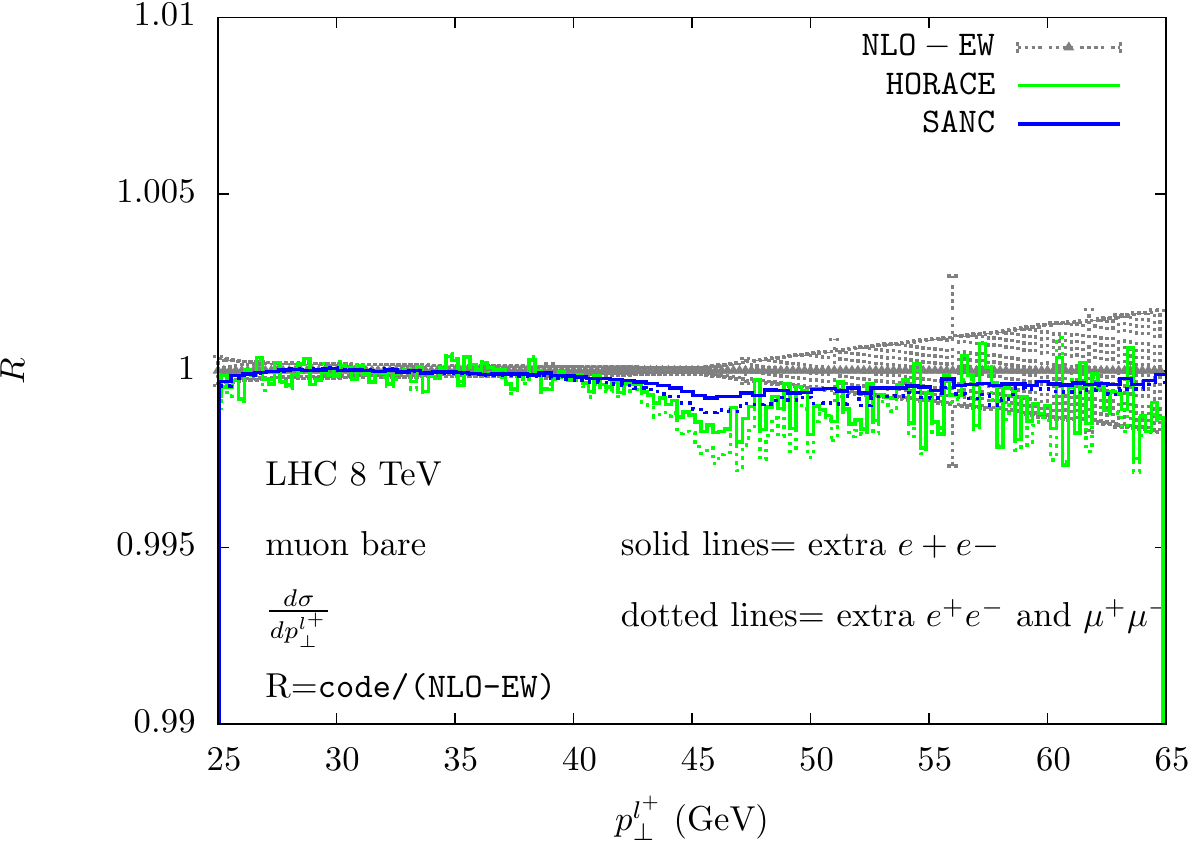}
\includegraphics[width=75mm,angle=0]{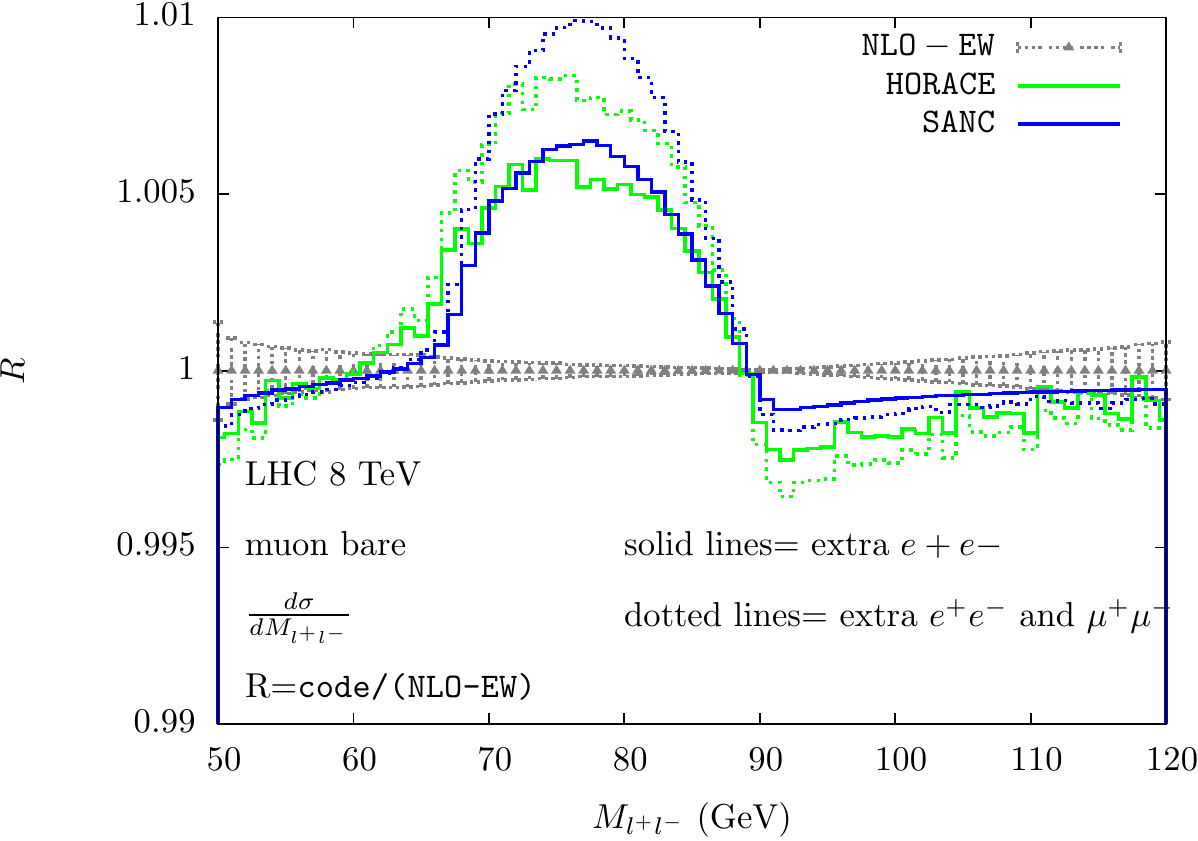}
\caption{
Relative effect of the emission of an additional light-fermion pair in
$pp\to\mu^+\mu^-+X$, compared to the NLO EW cross section in the {\em
benchmark} setup.  The solid lines represent the effect of emitting an
additional $e^+e^-$ pair, while the dashed lines account for the
possibility of emitting a $e^+e^-$ and a $\mu^+\mu^-$ pair. The \horace results are shown in green and the \sanc results in blue.  The lepton transverse momentum
(left) and the lepton-pair invariant mass (right)
distributions are shown for the 8 TeV LHC with ATLAS/CMS cuts.
\label{fig:ho-extrapairs}
}
\end{figure}

We used the MC codes \sanc and \horace to 
study the impact of the emission of an additional light-fermion pair
in the NC DY process. In Fig.~\ref{fig:ho-extrapairs} the relative effect
with respect to the NLO EW result is shown for the lepton transverse mass and 
lepton-pair invariant mass distributions. The effect of additional light-fermion pair emission in 
the CC DY process has also been studied with the \sanc code and was found 
to be less numerically important compared to the NC DY case.

\clearpage

\section{Interplay of QCD and EW corrections}
\label{sec:interplay}
A precise description of DY observables requires the simultaneous inclusion of QCD and EW corrections and control over mixed QCD and EW effects, which is the topic of this section. 
To set the stage, we formally write a fixed-order double perturbative expansion for the fully differential DY cross section\footnote{
We understand that the phase-space factors are properly included in the definition of the various $d\sigma$ coefficients.
},
in the strong and in the weak coupling constants, $\alpha_s$ and $\alpha$, as follows:
\be
d\sigma=d\sigma_{LO}+
\alpha d\sigma_{\alpha} + \alpha^2 d\sigma_{\alpha^2} +\dots
+\alpha_s d\sigma_{\alpha_s} + \alpha_s^2 d\sigma_{\alpha_s^2} +\dots
+\alpha \alpha_s d\sigma_{\alpha \alpha_s} + \alpha \alpha_s^2 d\sigma_{\alpha \alpha_s^2} +\dots
\label{eq:xsectot}
\ee
We identify purely EW ($d\sigma_{\alpha,\alpha^2}$), purely QCD ($d\sigma_{\alpha_s,\alpha_s^2}$) and mixed QCDxEW corrections ($d\sigma_{\alpha \alpha_s, \alpha \alpha_s^2}$).
The exact ${\cal O}(\alpha^2)$ and  ${\cal O}(\alpha\alpha_s)$ results are not yet available, only some subsets are known (see Section~\ref{sec:approximations} for a detailed discussion).
In an effort to provide the most precise prediction including mixed EW and QCD effects, we 
identify two distinct problems that, to some extent, overlap:
\begin{enumerate}
\item
As already discussed in the previous sections,
many observables relevant for precision EW measurements
require a formulation 
that goes beyond fixed-order perturbation theory
and includes the resummation to all orders of some logarithmically enhanced terms,
preserving with a matching procedure the (N)NLO accuracy on the total cross section.
This problem, which was discussed separately for QCD and for EW corrections, is present also once we consider the effect of mixed QCDxEW terms: in other words we need a matching procedure that preserves the NLO-(QCD+EW) accuracy on the total cross section
and that describes the emission of the hardest parton (gluon/quark/photon) with exact matrix elements, leaving the remaining emissions to a Parton Shower algorithm.

\item
As long as the exact ${\cal O}(\alpha\alpha_s)$ corrections to the four-fermion process are not fully known, 
we need to assess the accuracy of the recipes that combine QCD and EW effects available from independent calculations,
e.g., the validity of an ansatz which factorizes QCD and EW terms.
\end{enumerate}
In the Sections \ref{sec:combi} and \ref{sec:approximations} 
we will address both the above issues, in presence of a matching between fixed NLO and all-orders results.

In Section \ref{sec:yfs-ps-nloew-sherpa-pythia}
we additionally show a comparison of different ways to simultaneously include
QCD and QED/EW corrections to all orders on top of a LO description of the observables (with LO accuracy for the total cross section) and compare these results with the fixed order NLO predictions, in the case of calorimetric electrons in the final state.

\subsection{Combination of QED/EW with QCD results in the {\tt POWHEG} framework}
\label{sec:combi}

\begin{figure}[!ht]
\includegraphics[width=75mm,angle=0]{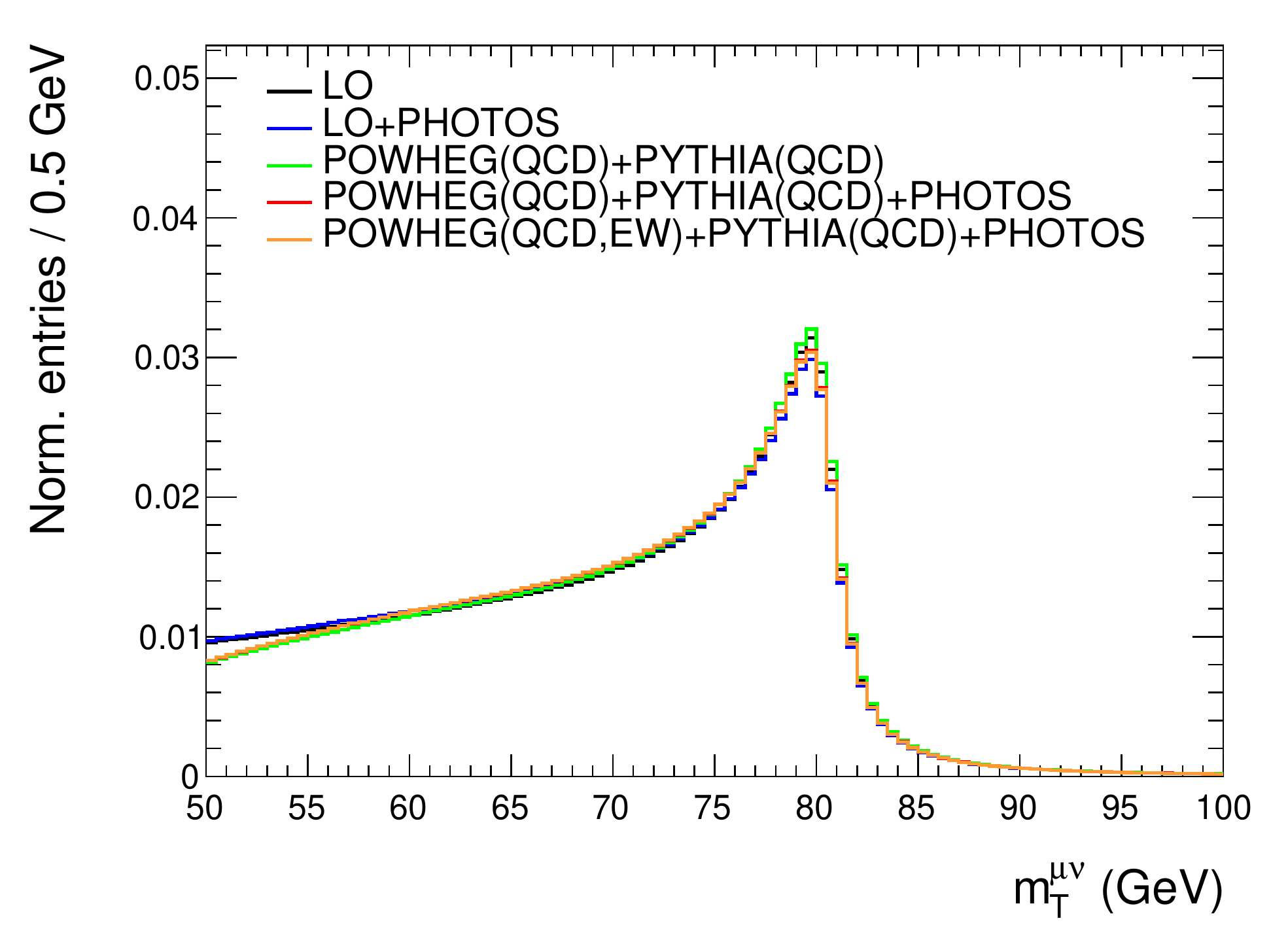}
\includegraphics[width=75mm,angle=0]{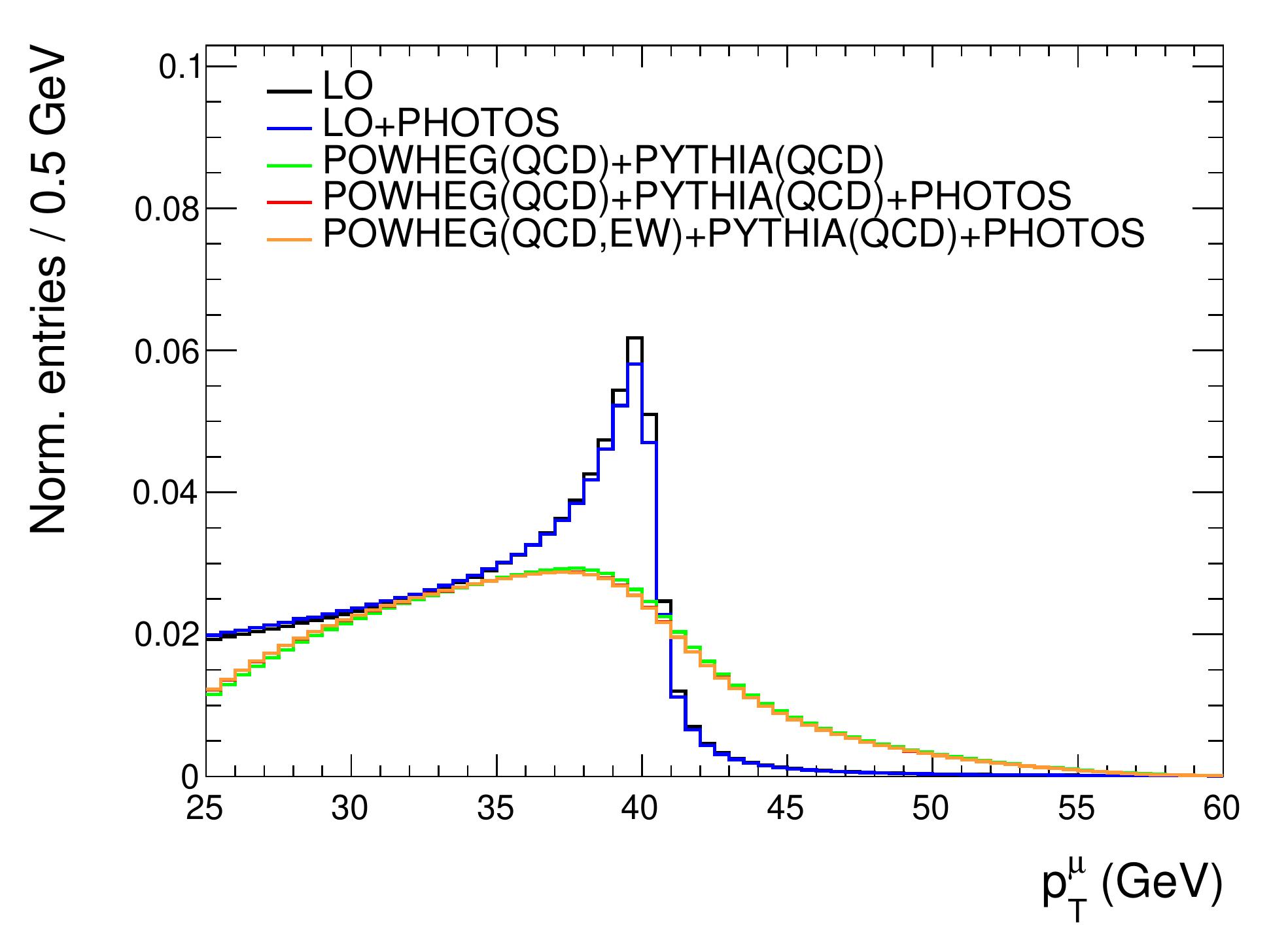}\\
\includegraphics[width=75mm,angle=0]{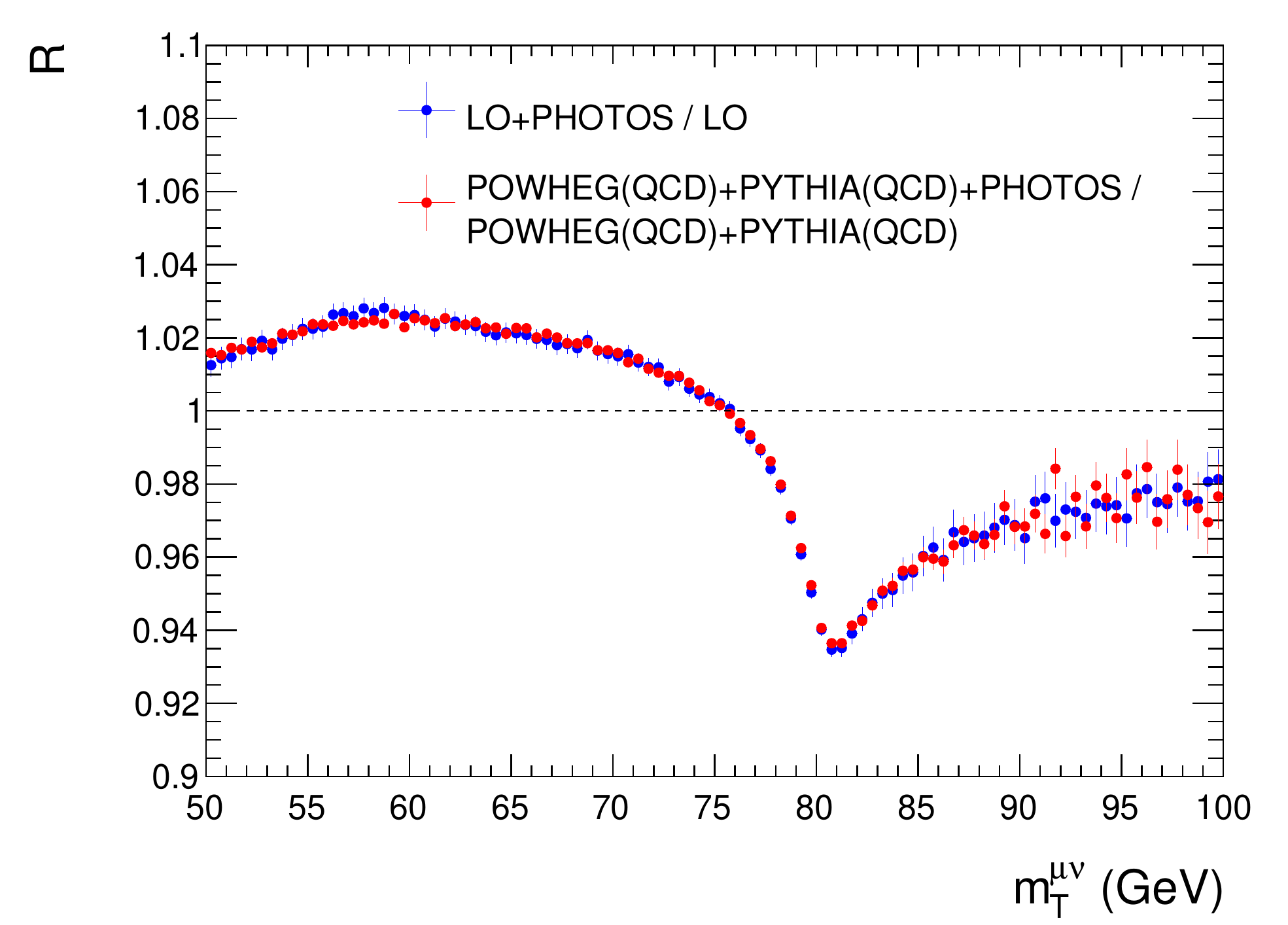}
\includegraphics[width=75mm,angle=0]{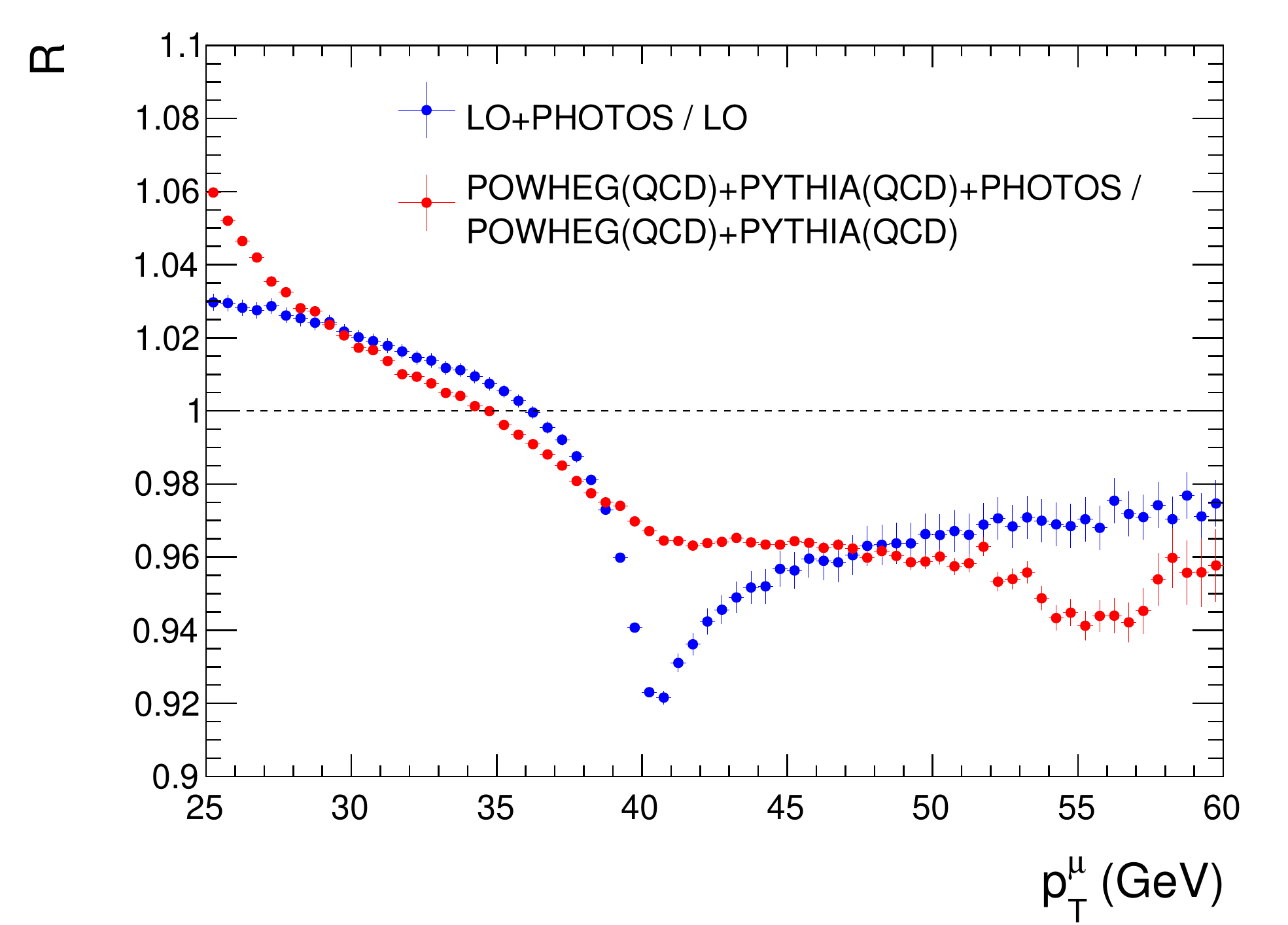}\\
\includegraphics[width=75mm,angle=0]{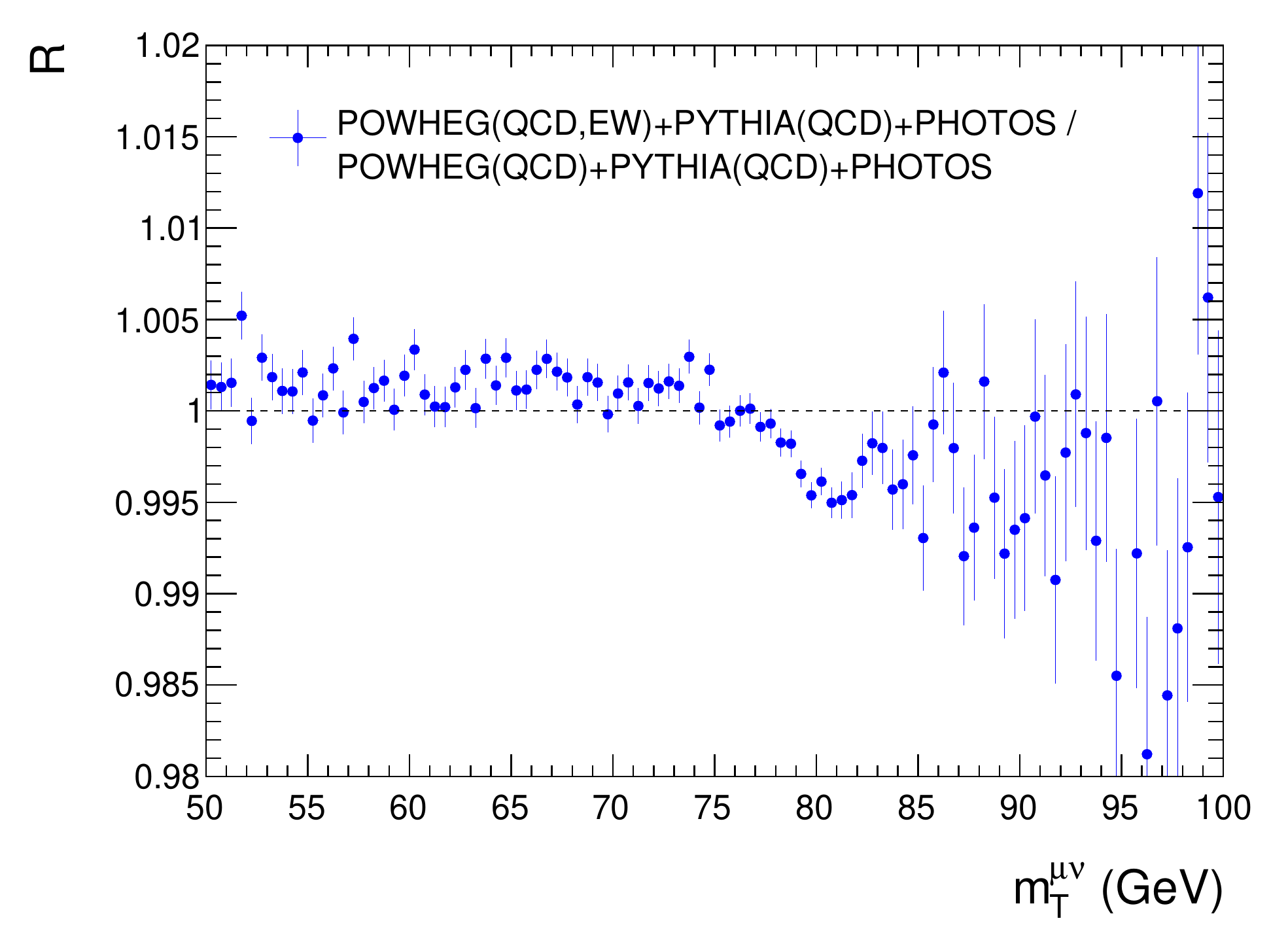}
\includegraphics[width=75mm,angle=0]{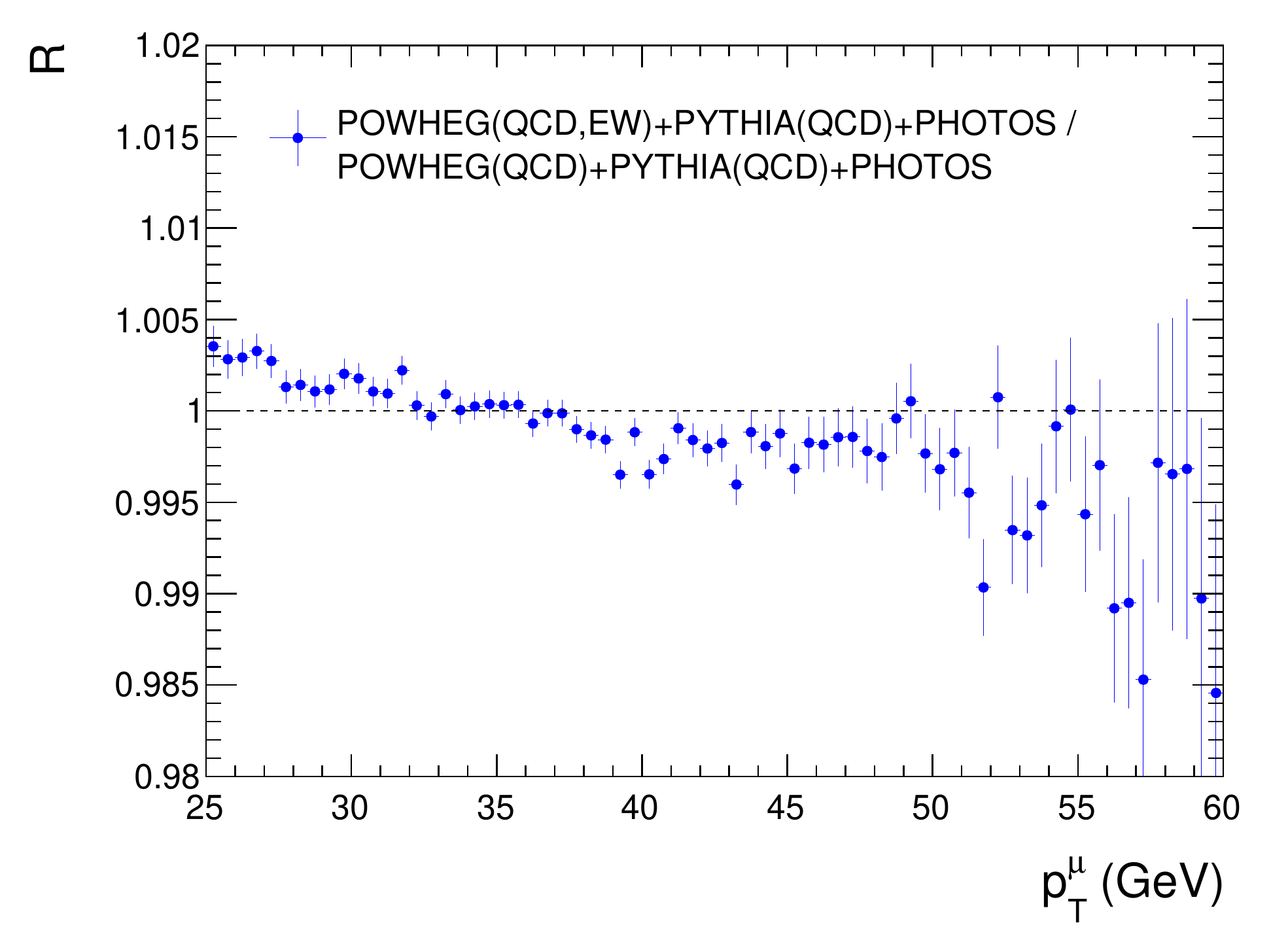}
 \caption{Combination of QCD and EW corrections in the process
$pp\to\mu^+\nu_\mu+X$ at the 14 TeV LHC with standard ATLAS/CMS cuts 
for the lepton-pair transverse mass (left plots) 
and the charged lepton transverse momentum (right plots) distributions.
The normalized distributions in different perturbative approximations are shown in the upper plots.
The ratio of these distributions including a QED FSR PS and the corresponding quantities where the QED shower has been switched off is shown in the middle plots.
The ratio of distributions including full NLO-(QCD+EW) corrections matched with (QCD+QED)-PS
and the corresponding quantities where only NLO-QCD + (QCD+QED)-PS has been retained
is shown in the lower plots.
\label{fig:qcdxew-powheg}
}
 \end{figure}

The study of the DY observables that are relevant for high-precision measurements
requires the inclusion of QED-FSR effects to all orders and of QCD-ISR effects to all orders,
in order to obtain a description stable upon inclusion of further higher-order corrections.

The impact of multiple parton radiation has been discussed 
in Sections \ref{sec:impact-qcd} and \ref{sec:impact-ew},
separately in the QCD and QED cases,
in codes that match the PS algorithm with NLO fixed-order results.

PS codes are often used as stand-alone tools, 
since they provide a good approximation of the shape of the differential distributions.
When QCD-PS and QED-PS are combined together,
the resulting description has an exact treatment of the kinematics of each individual QCD/QED parton emission,
but lacks the exact matrix element corrections and the normalization which are instead available in a fixed-order NLO-accurate calculation.

In the following we discuss in two steps the impact of the inclusion of different higher-order corrections,
taking as representative examples the lepton-pair transverse mass (cfr. Fig. \ref{fig:qcdxew-powheg} left plots)
and the lepton transverse momentum distributions (cfr. Figure \ref{fig:qcdxew-powheg} right plots),
in the process $pp\to\mu^+\nu_\mu+X$ at the 14 TeV LHC with standard ATLAS/CMS cuts and bare muons.
In Figure \ref{fig:qcdxew-powheg} we show the normalized distributions, $d\sigma/dX / \sigma_{tot}$ ($X=m_T^{\mu \nu_mu},p_T^\mu$), in different perturbative approximations (upper plots),
we expose the impact of QED-FSR corrections applied to different underlying hard processes (middle plots)
and the impact of mixed QCD-EW effects in a simulation with full NLO-(QCD+EW) accuracy (lower plots). 

We first start from the LO distributions of these two quantities,
which show the sharply peaked behavior due to the jacobian factor.
The QED-FSR emissions are simulated with the \photos code 
and yield effects which are similar for the two observables,
with a negative correction of ${\cal O}(-8\%)$ at the jacobian peak, as shown in the middle plots by the blue points.

We then consider the role of NLO-QCD corrections and of a QCD-PS in the \powheg+\pythia code and remark 
(cfr. the upper plots) that, 
while the shape of the transverse mass distribution is preserved, to a large extend, by QCD corrections, 
the lepton transverse momentum distribution is instead strongly smeared, with a much broader shape around the jacobian peak.
The inclusion of the \photos corrections on top of the \powheg+\pythia simulation has now a different fate, compared to the LO case (cfr. middle plots, red points): 
the shape and the size of the QED corrections are similar to the LO case for the transverse mass;
in the lepton transverse momentum case instead the QED correction is reduced in size and flatter in shape, with respect to the LO case.
The comparison of the percentage corrections due to QED-FSR in the two examples discussed above
(blue and red points in the middle plots)
shows a difference which is due to mixed QCDxQED corrections,
since the set of pure QED corrections is common to the two simulations.

The code \powhegew has been validated, separately in its QCD and EW components, in Section \ref{sec:tuned}.
Its use allows to reach the NLO-(QCD+EW) accuracy for the total cross section
but it also has an impact on the differential distributions.
In Figure \ref{fig:qcdxew-powheg} (lower plots) we show the ratio of the distributions obtained with \powhegew+\pythia+\photos
and with \powheg+\,\pythia+\,\photos. These ratios expose the size of mixed QCD-EW corrections
present in the \powhegew+\, \pythia+\, \photos\, prediction but absent in \powheg+\,\pythia+\,\photos.

\subsection{Towards exact ${\cal O}(\alpha\alpha_s)$: assessment of the accuracy of current approximations}
\label{sec:approximations}

As mentioned earlier, the question how to properly combine QCD and EW corrections in
predictions will only be settled by a full NNLO calculation of the
${\cal O}(\alpha\alpha_s)$ corrections that is not yet
available, although first steps in this direction have been taken by
calculating two-loop
contributions~\cite{Kotikov:2007vr,Bonciani:2011zz,Kilgore:2011pa,Bonciani:2016ypc},
the full ${\cal O}(\alpha\alpha_s)$ correction to the
W/Z-decay widths~\cite{Czarnecki:1996ei,Kara:2013dua}, and the full
${\cal O}(\alpha)$ EW corrections to W/Z+jet production including the
W/Z decays~\cite{Denner:2009gj,Denner:2011vu,Denner:2012ts}. 

\begin{figure}
\epsfig{file=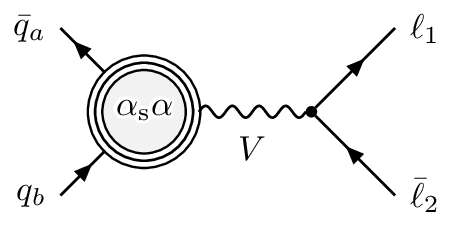,scale=0.75}
\hfill
\epsfig{file=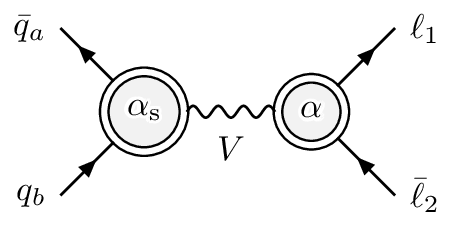,scale=0.75}
\hfill
\epsfig{file=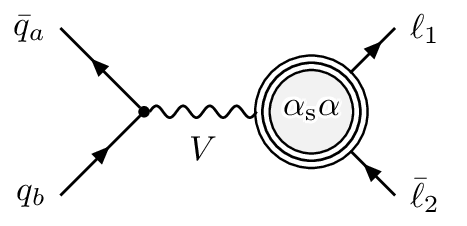,scale=0.75}
\hfill
\epsfig{file=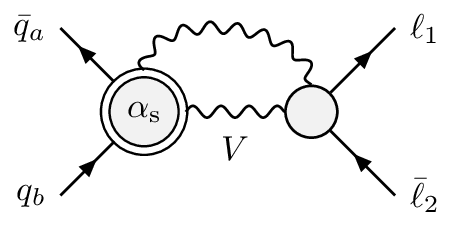,scale=0.75}
\\[-.5em]
\hspace*{3.5em} (a) \hfill (b) \hfill (c) \hfill (d) \hspace*{4em}
\caption{Generic diagrams for the various contributions to the virtual 
factorizable (a--c) and non-factorizable (d)
corrections of ${\cal O}(\alpha\alpha_s)$ in PA, 
with $\alpha_s$, $\alpha$, and $\alpha\alpha_s$ in the blobs
indicating the order of the included loop corrections.}
  \label{fig:NNLOcontrib}
\end{figure}

Results for mixed EW-QCD ${\cal O}(\alpha\alpha_s)$ corrections to the
charged- and neutral-current DY processes have been recently obtained in the
so-called {\it pole
  approximation}~(PA)~\cite{Dittmaier:2014qza,Dittmaier:2014koa,Dittmaier:2015rxo}.
This allows to assess the validity of simple prescriptions for the combination
of EW and QCD corrections.  The PA provides a systematic approximation of
radiative corrections near the W- or Z-boson resonances, which is important
for precision physics such as the $M_{\mathrm{W}}$ measurement.  Applications
of the PA to NLO EW
corrections~\cite{Wackeroth:1996hz,Baur:1998kt,Dittmaier:2001ay,Dittmaier:2014qza}
have been validated by a comparison to the complete EW NLO calculations and
show excellent agreement at the order of some $0.1\%$ in kinematic
distributions dominated by the resonance region.
Therefore the PA is expected to be a reliable tool for the calculation
of the ${\cal O}(\alpha\alpha_s)$ corrections for resonant
W/Z production. In the framework of the PA, radiative corrections
are classified into factorizable corrections to W/Z production and
decay sub-processes, and non-factorizable corrections that link
production and decay by soft-photon exchange. The application to the
${\cal O}(\alpha\alpha_s)$ corrections results in four
types of contributions illustrated in Fig.~\ref{fig:NNLOcontrib} for the
case of the double-virtual corrections.  The initial--initial
factorizable corrections (a) are given by two-loop ${\cal
  O}(\alpha\alpha_s)$ corrections to on-shell W/Z
production. The factorizable initial--final corrections (b)
consist of one-loop QCD corrections to W/Z production multiplied by
one-loop EW corrections to the decay.  Factorizable final--final
corrections (c) only arise from the vertex counterterm involving QCD
corrections to the vector-boson self-energies, 
but are phenomenologically negligible~\cite{Dittmaier:2015rxo}.  In the
non-factorizable two-loop corrections (d), the soft-photon corrections
connecting the initial state, the intermediate vector boson, and the
final-state leptons are dressed with gluon loop corrections to the
initial quark--antiquark pair.  For each class of contributions with
the exception of the final--final corrections (c), also the
associated real--virtual and double-real corrections have to be
computed, obtained by replacing one or both of the labels $\alpha$ and $\alpha_s$ in
the blobs in Fig.~\ref{fig:NNLOcontrib} by a real
photon or gluon, including crossed partonic channels, e.g.\ with
quark--gluon initial states.
In Ref.~\cite{Dittmaier:2014qza}
the non-factorizable ${\cal O}(\alpha\alpha_s)$ corrections
to W/Z production have been computed in terms of soft-photon
correction factors to squared tree-level or one-loop QCD matrix
elements by using gauge-invariance arguments.
The numerical impact of these corrections was found to be below the
$0.1\%$ level and is therefore phenomenologically negligible.

The ${\cal O}(\alpha\alpha_s)$
initial--final state corrections have been computed
in~Ref.~\cite{Dittmaier:2015rxo}.  Because of the large effect of
real-photon emission off the final-state leptons at NLO, this class is
expected to capture the dominant part of the full ${\cal
  O}(\alpha\alpha_s)$ corrections on kinematic
distributions in the resonance region.  Therefore the sum
of the NLO QCD cross section $\sigma_{\mathrm{NLO}_s}$ and the NLO EW corrections 
can be improved by adding the
initial--final-state corrections in the PA,
$\sigma_{\alpha\alpha_s}^{\mathrm{prod}\times\mathrm{dec}}$:
\begin{equation}
\sigma_{\mathrm{NNLO}_{\mathrm{s\otimes ew}}}
= \sigma_{\mathrm{NLO}_{\mathrm{s}}}+ \alpha\,\sigma_\alpha
+ \alpha\alpha_s\,\sigma_{\alpha\alpha_s}^{\mathrm{prod}\times\mathrm{dec}}.
\label{eq:sig-a-as-if}
\end{equation}
The last term in Eq.~(\ref{eq:sig-a-as-if}), in particular, includes the
double-real contribution that is given in terms of the exact matrix elements
for gluon or photon emission in vector-boson production and decay,
respectively, treated without kinematic approximation on the photon or gluon
momenta. In the {\tt POWHEG} implementation discussed in
Section~\ref{sec:combi}, these effects are approximated by treating the first
emission exactly and generating the second emission by a QCDxQED shower in the
collinear approximation. On the other hand, this approach includes multiple
collinear photon and gluon emissions which are not included in the fixed-order
prediction~(\ref{eq:sig-a-as-if}).

In the numerical results shown below, all terms of
Eq.~(\ref{eq:sig-a-as-if}) are consistently evaluated using the NNPDF2.3QED
NLO set~\cite{Ball:2013hta}, which includes $\mathcal{O}(\alpha)$
corrections.
We consider  the case of ``bare muons'' without any photon recombination.
Results obtained assuming a recombination of leptons with collinear photons
can be found in Ref.~\cite{Dittmaier:2015rxo} and show the same overall
features, with corrections that typically reduced by a factor of two.

\begin{figure}
\hspace*{1.5em}

%
  \includegraphics[width=.5\textwidth]{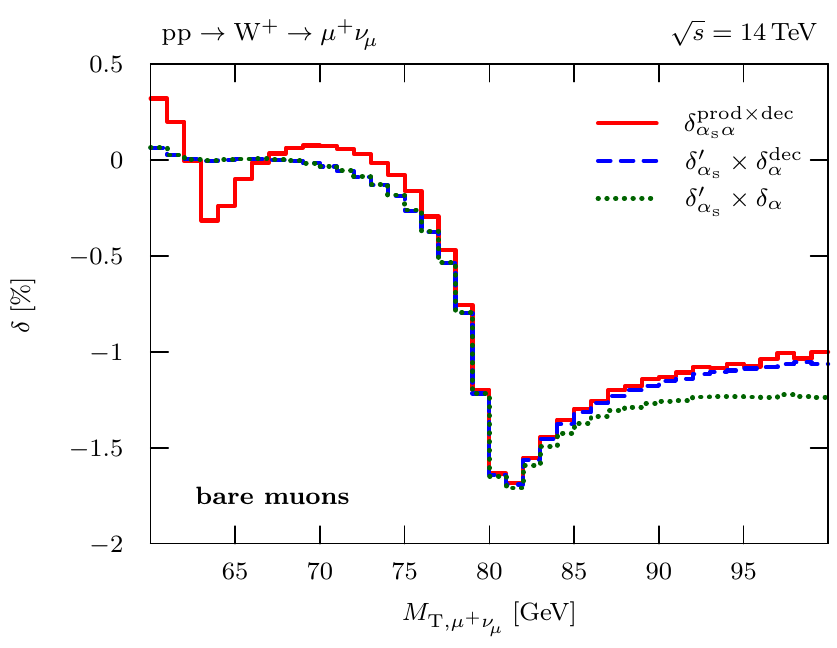} 
  \includegraphics[width=.5\textwidth]{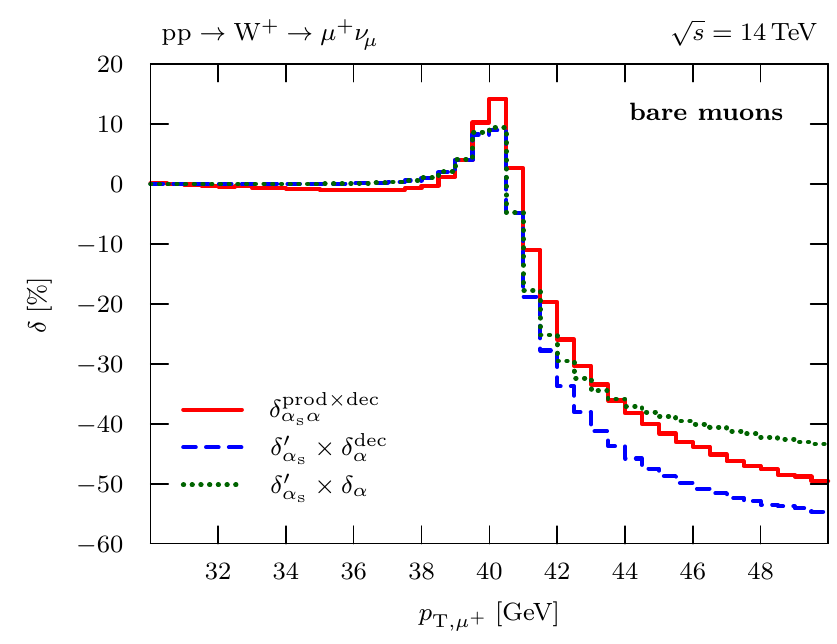}
  \caption{Relative factorizable corrections (in red) of ${\cal
      O}(\alpha\alpha_s)$ induced by initial-state QCD and final-state EW
    contributions to the transverse-mass distribution (left) and the
    transverse-lepton-momentum distribution (right) for $\mathrm{W^+}$
    production at the LHC. The naive products of the NLO correction factors
    $\delta_{\alpha}$ and $\delta_{\alpha_s}^\prime$ are shown for comparison
    (taken from Ref.~\cite{Dittmaier:2015rxo}).}
\label{fig:Oaas-IFfact-Wp}
%
\vspace{2em}
  \includegraphics[width=.5\textwidth]{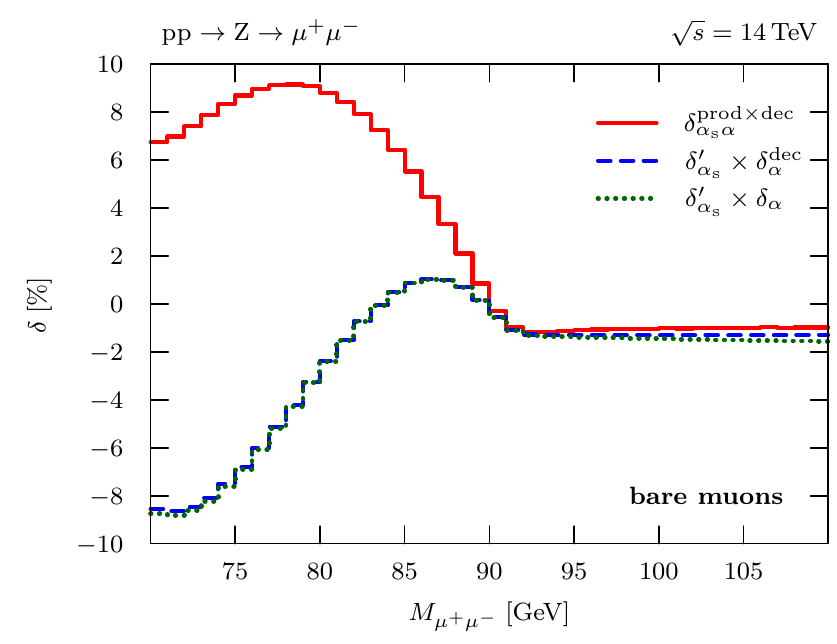} 
  \includegraphics[width=.5\textwidth]{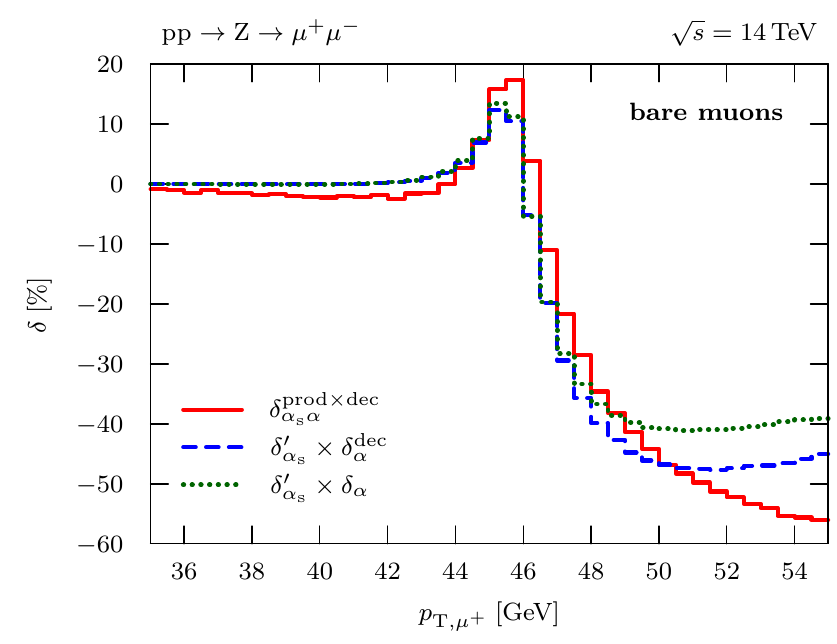} 
  \caption{Relative factorizable corrections (in red) of ${\cal
      O}(\alpha\alpha_s)$ induced by initial-state QCD and final-state EW
    contributions to the lepton-invariant-mass
    distribution~(left) and a transverse-lepton-momentum distribution~(right)
    for Z~production at the LHC.  The naive products of the NLO correction
    factors $\delta_{\alpha_s}'$ and $\delta_\alpha$ are shown for
    comparison (taken from Ref.~\cite{Dittmaier:2015rxo}.)}
  \label{fig:Oaas-IFfact-Z}
\end{figure}

Predictions for the transverse-mass and transverse-lepton-momentum
distributions for $\mathrm{W^+}$ production at the LHC with $\sqrt
s=14\,\mathrm{TeV}$ are shown in Fig.~\ref{fig:Oaas-IFfact-Wp}.  For
Z~production, Fig.~\ref{fig:Oaas-IFfact-Z} displays the results for the
lepton-invariant-mass distribution and a transverse-lepton-momentum
distribution.  The red curves are given by the factorizable initial--final
${\cal O}(\alpha\alpha_s)$ corrections, normalized to the LO cross-section
prediction,
\begin{equation}
\delta^{\mathrm{prod}\times\mathrm{dec}}_{\alpha\alpha_s}
=\frac{\alpha\alpha_s\, \sigma^{\mathrm{prod}\times\mathrm{dec}}_{\alpha\alpha_s}}{\sigma_{\mathrm{LO}}}, 
\label{eq:delta-a-as-if}
\end{equation}
where $\sigma_{\mathrm{LO}}$ is computed using the NNPDF2.3QED LO PDFs.  One
observes corrections beyond NLO of approximately $-1.7\%$ in the
$M_{\mathrm{T},\nu l}$ distribution (left plot in
Fig.~\ref{fig:Oaas-IFfact-Wp}).  As can be anticipated from the size of the
NLO QCD corrections, corrections to the transverse-lepton-momentum spectrum
(right plots in Figs.~\ref{fig:Oaas-IFfact-Wp} and~\ref{fig:Oaas-IFfact-Z})
can be much larger, rising to about $15\%$ ($20\%$) above the Jacobian peak
for the case of the W$^+$ boson (Z boson) and dropping to almost $-50\%$
above.  In fact, a realistic description of the $p_{\mathrm{T},l}$ spectrum
near resonance requires the inclusion of higher-order gluon-emission effects.
In case of the $M_{l^+l^-}$ distribution for Z production (left plot in
Fig.~\ref{fig:Oaas-IFfact-Z}), corrections up to $10\%$ are observed below the
resonance, consistent with the large EW NLO corrections from FSR in this
region.

The result of the PA~(\ref{eq:delta-a-as-if}) allows to assess the validity 
of a naive product ansatz
of the  ${\cal O}(\alpha\alpha_s)$  correction,
\begin{equation}
\sigma^{\mathrm{naive\,fact}}_{\mathrm{NNLO}_{\mathrm{s\otimes ew}}}
= \sigma_{\mathrm{NLO}_{\mathrm{s}}}(1+\delta_\alpha).
\label{eq:sigma-naive-fact}
\end{equation}
Here the relative EW correction factor
$\delta_\alpha=\alpha\,\sigma_{\alpha}/\sigma_0$  is introduced 
as the ratio of the NLO EW correction and the
LO contribution $\sigma_0$ to the NLO cross section, both evaluated with NLO PDFs, 
so that PDF effects cancel in this factor.
The difference of the prediction~(\ref{eq:sig-a-as-if}) to the
product ansatz~(\ref{eq:sigma-naive-fact}), normalized to the LO cross section, reads
\begin{equation}
\frac{\sigma_{\mathrm{NNLO}_{\mathrm{s\otimes ew}}} -
\sigma^{\mathrm{naive\,fact}}_{\mathrm{NNLO}_{\mathrm{s\otimes ew}}}
}{\sigma_{\mathrm{LO}}}
=\delta^{\mathrm{prod}\times\mathrm{dec}}_{\alpha\alpha_s} -\delta_\alpha \delta_{\alpha_s}^\prime,
\end{equation}
with the relative QCD correction factor
$\delta_{\alpha_s}^\prime=(\sigma_{\mathrm{NLO}_{\mathrm{s}}}-\sigma_0)/\sigma_{\mathrm{LO}}$.%
\footnote{ Note that this correction factor differs from that in the standard
  QCD $K$~factor $ K_{\mathrm{NLO}_{\mathrm{s}}}
  =\sigma_{\mathrm{NLO}_{\mathrm{s}}}/\sigma_{\mathrm{LO}}\equiv 1
  +\delta_{\alpha_s}$ due to the use of different PDF sets in the Born
  contributions. See Ref.~\cite{Dittmaier:2014koa} for further discussion.}
The agreement of the correction factor~(\ref{eq:delta-a-as-if}) with the
product $\delta_\alpha \delta_{\alpha_s}'$ therefore provides an estimate for
the accuracy of the naive product ansatz.  In Figs.~\ref{fig:Oaas-IFfact-Wp}
and~\ref{fig:Oaas-IFfact-Z} two different versions of the EW correction factor
are used for the product approximation, first based on the full NLO correction
($\delta_\alpha$, black curves), and second based on the dominant EW
final-state correction of the PA ($\delta^{\mathrm{dec}}_\alpha$, blue
curves). The difference of these curves provides an estimate for the size of
the remaining as yet uncalculated ${\cal O}(\alpha\alpha_s)$ corrections
beyond the initial--final corrections considered in the calculation 
of Refs.~\cite{Dittmaier:2014qza,Dittmaier:2014koa,Dittmaier:2015rxo}
and therefore also provides an error estimate of the PA, and in particular of the
omission of the corrections of initial--initial type.

In the case of the $M_{\mathrm{T},\nu l}$ distribution (left plot in
Fig.~\ref{fig:Oaas-IFfact-Wp}), which is rather insensitive to W-boson recoil
due to jet emission, both versions of the naive product ansatz approximate the
PA prediction quite well near the Jacobian peak and below. Above the peak, the
product $\delta_{\alpha_s}' \delta_\alpha$ based on the full NLO EW correction
factor deviates from the other curves, which signals the growing importance of
effects beyond the PA.  In contrast, the product ansatz fails to provide a
good description for the lepton $p_{\mathrm{T},l}$ distributions (right plots
in Figs.~\ref{fig:Oaas-IFfact-Wp} and~\ref{fig:Oaas-IFfact-Z}), which are
sensitive to the interplay of QCD and photonic real-emission effects.  In this
case one also observes a larger discrepancy of the two different
implementations of the naive product, which indicates a larger impact of the
missing ${\cal O}(\alpha\alpha_s)$ initial-initial corrections of
Fig.~\ref{fig:NNLOcontrib}~(a), and in particular the real-emission
counterparts.  For the $M_{l^+l^-}$ distribution for Z production (left plot
in Fig.~\ref{fig:Oaas-IFfact-Z}), the naive products approximate the full
initial--final corrections reasonably well for $M_{l^+l^-}\geq M_{\mathrm{Z}}$, but
completely fail already a little below the resonance where they do not even
reproduce the sign of the full correction
$\delta^{\mathrm{prod}\times\mathrm{dec}}_{\alpha_s\alpha}$.  This failure can
be understood from the fact that the naive product ansatz multiplies the
corrections locally on a bin-by-bin basis, while a more appropriate treatment
would apply the QCD correction factor at the resonance,
$\delta_{\alpha_s}'(M_{l^+l^-}=M_{\mathrm{Z}})\approx 6.5\%$, for the events that are
shifted below the resonance by photonic FSR. The observed mismatch is further
enhanced by a sign change in the QCD correction $\delta_{\alpha_s}'$ 
at $M_{l^+l^-}\approx 83\,\mathrm{GeV}$.

These examples show that a naive product approximation has to be used with
care and does not hold for all distributions. The results are also sensitive
to the precise definition of the correction factors $\delta_\alpha$ and
$\delta_{\alpha_s}$~\cite{Dittmaier:2014koa}.  As shown in
Ref.~\cite{Dittmaier:2015rxo}, a more suitable factorized approximation of the
dominant $\mathcal{O}(\alpha\alpha_s)$ effects can be obtained by combining
the full NLO QCD corrections to vector-boson production with the
leading-logarithmic approximation for FSR through a structure-function or a
parton shower approach such as used in PHOTOS~\cite{Golonka:2005pn}.  In this way the interplay of the recoil effects from
jet and photon emission is properly taken into account, while certain
non-universal, subleading, effects are neglected.

\newcommand{\sherpaprog}{{\tt SHERPA\,}}
\newcommand{\pythiaprog}{{\tt PYTHIA8\,}}
\newcommand{\openloopsprog}{{\tt OPENLOOPS\,}}

\subsection{Comparing different ansatzes of higher-order QED/EW corrections combined with QCD parton showers}
\label{sec:yfs-ps-nloew-sherpa-pythia}

In this section we compare the higher-order QED
corrections predicted by \sherpaprog's Yennie-Frautschi-Suura (YFS) 
soft-photon resummation \cite{yfs:1961,Schonherr:2008av}, 
the standard DGLAP collinear higher-order QED corrections as implemented in \pythiaprog \cite{Sjostrand:2014zea}, 
and the exact NLO EW calculation performed by \sherpaprog using one-loop 
matrix elements from \openloopsprog \cite{Kallweit:2015dum,Kallweit:2014xda,
  Cascioli:2011va}. 
In Ref.~\cite{Badger:2016bpw}, for the case of the NC DY process, 
the quality of the YFS implementation of \sherpaprog 
has been checked against the exact NLO EW $\mathcal{O}(\alpha)$ calculation 
and the NNLO QCD-EW mixed $\mathcal{O}(\alpha_s\alpha)$ calculation in the pole approximation of \cite{Dittmaier:2014qza,Dittmaier:2015rxo}; we point to this reference for the quantitative results. 
In the following, the calculations including YFS exponentiation, standard DGLAP QED and fixed-order NLO-EW corrections
have been performed also for the CC DY process and shall be compared 
among each other in a realistic scenario. 
We consider electrons dressed with 
the surrounding $\Delta R=0.1$, which are required to have $p_{\rm T}>25\,{\rm GeV}$ 
and $|y|<2.4$, and a missing transverse momentum of at least $25\,{\rm GeV}$.

\begin{figure}[t!]
  \centering
  \includegraphics[width=0.48\textwidth]{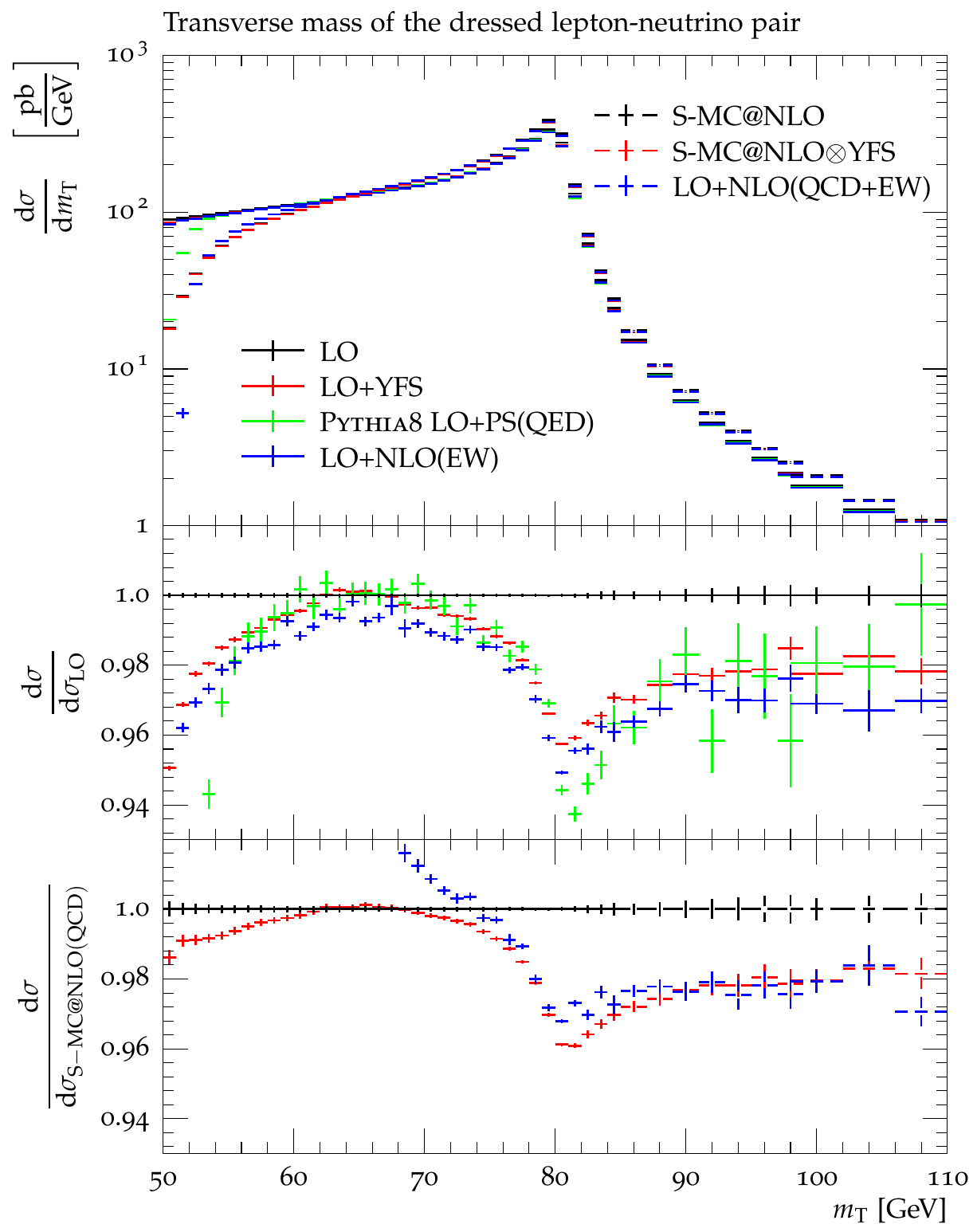}
  \hfill
  \includegraphics[width=0.48\textwidth]{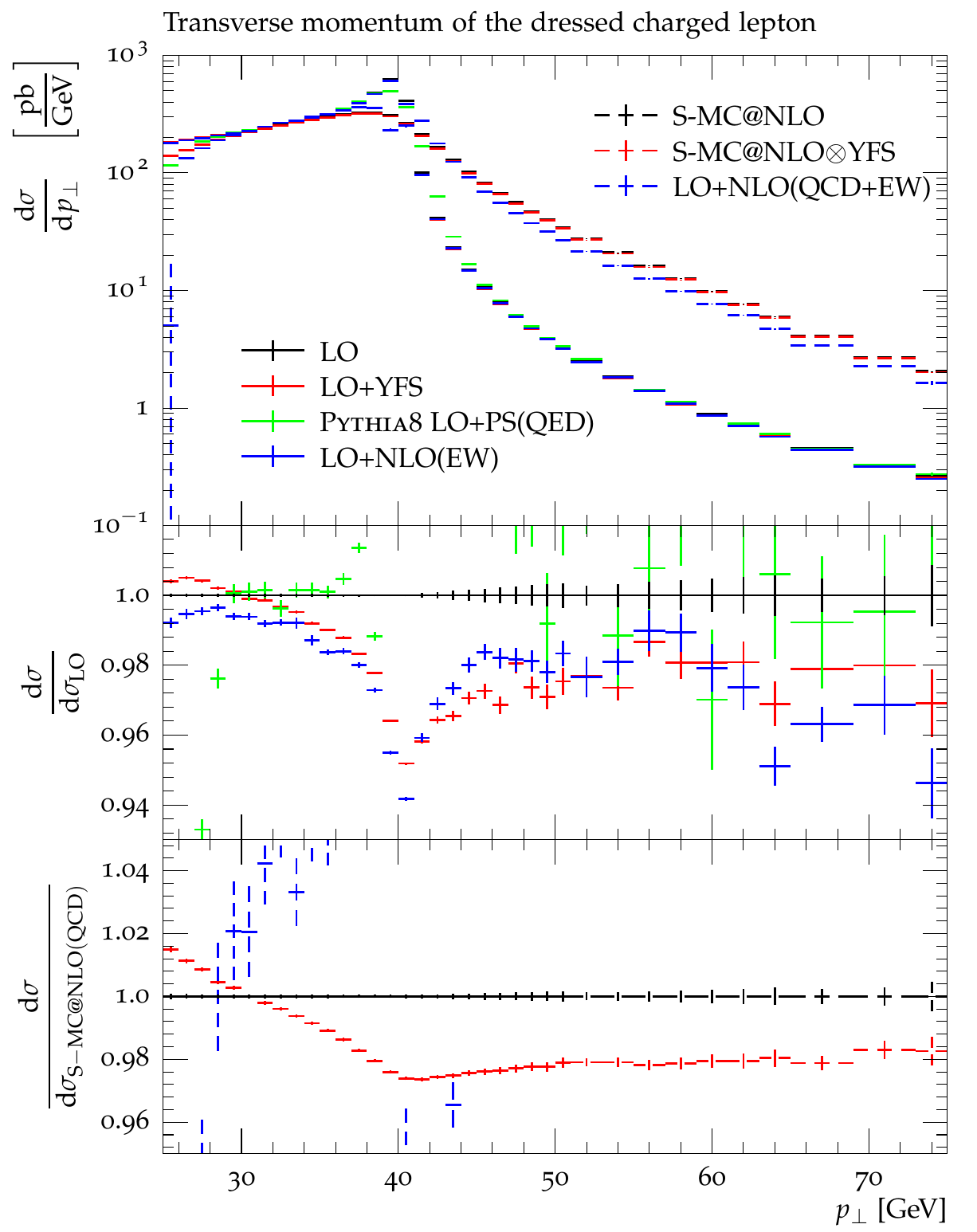}
  \caption{
            Comparison of the description of the transverse mass of 
            the dressed electron-neutrino pair (left) and the dressed 
            electron transverse momentum (right) in electron-neutrino-pair 
            production in the CC DY process with fiducial cuts (see text for more details).
          }
  \label{fig:yfs-ps-nloew-sherpa-pythia:e}
\end{figure}

Figure \ref{fig:yfs-ps-nloew-sherpa-pythia:e} (left) shows the comparison of the 
different calculations for the reconstructed transverse mass of the $W$ boson. 
Besides the leading QCD higher-order corrections, the higher-order EW
corrections between either the YFS resummation or the parton-shower approach 
agree well with the fixed-order result (see the central inset), 
only \pythiaprog's QED parton shower 
predicts a stronger correction around the peak and near the threshold. 
The differences with respect to the NLO EW 
correction can be traced to multi-photon emissions present in the all-order results and to genuine weak effects only present in the NLO EW calculation. 
The same findings were reported for the case 
of lepton pair production in Ref. \cite{Badger:2016bpw}. 
Applying the YFS resummation in 
addition to higher-order QCD corrections, the implementation corresponds to 
a multiplicative combination of both effects and preserves these findings for the lepton-pair transverse mass distribution (lower inset), 
as already observed in Section \ref{sec:combi}. 
Again, subpercent level agreement is found with the fixed-order calculation in the peak region. 
At low transverse masses the resummation of QCD corrections is 
important and drives the difference to the fixed-order result.

Figure \ref{fig:yfs-ps-nloew-sherpa-pythia:e} (right) details the comparison of 
the different calculations for the transverse momentum of the dressed 
electron. Again, the exact $\mathcal{O}(\alpha)$ calculation is in subpercent 
level agreement with the YFS resummation, and again, the general offset can 
be attributed to both multiple photon emission corrections and genuine weak 
corrections (central inset). 
The \pythiaprog QED parton shower shows a different behavior in 
the peak region. 
Once NLO QCD effects are also taken into 
account (lower inset), 
the importance of their resummation with respect to their simple 
fixed-order treatment, as already observed in Section \ref{sec:QCDallorders}, overwhelms the comparison between the YFS soft photon
resummation and the fixed-order NLO EW calculation 
for this observable.

The investigation of the observed difference in the behavior of the QED parton shower in \pythiaprog and the YFS soft-photon resummation is left to a future study. 



\clearpage

\section{Conclusions}
\label{sec:conclusions}

What we did:
\begin{itemize}
\item In this report we compared several public codes which simulate
  the Drell-Yan processes in different perturbative approximations.
  All these codes are at least NLO accurate in the description of
  inclusive observables in either the EW or strong interaction, or
  possibly with respect to both.
\item This common level of accuracy allowed to consistently compare
  the codes, testing their respective numerical implementations and
  the resulting level of agreement (see Section \ref{sec:tuned}).
\item Relying on this NLO-accurate framework, it has been possible to
  define a way to quantify the impact of higher-order corrections, i.e. beyond NLO,  
  which may differ from code to code (see Section \ref{sec:impact}).
  The study of the impact of different sets of corrections has been
  performed separately for the EW and strong interactions.
\item Some codes provide, in the same implementation, QCD and EW
  corrections, which have been separately tested in Sections
  \ref{sec:tuned} and \ref{sec:impact}.  The interplay of both sets of
  corrections is discussed in Section \ref{sec:interplay}.
\end{itemize}

\noindent
What we computed and observed:
\begin{itemize}

\item The impact of all the higher-order corrections, which are
  available in some but not in all codes, is expressed as a percentage
  effect, using a common unit, namely the distribution obtained in the
  calculation which has NLO accuracy for the total cross section and
  uses the inputs of the {\it benchmark} setup.

\item The distribution used as common unit may not be the most suitable
  choice for all the observables: in fact in some phase-space corners
  perturbation theory breaks down and the fixed-order distribution
  provides only a technical reference rather than a sensible estimate
  of the physical observable.

\item The problem of a consistent matching of fixed- and all-orders
  results emerges in several cases discussed in Section
  \ref{sec:impact}, both in the EW and in the QCD sectors.  Different
  matching procedures may agree on the accuracy on the observables
  inclusive over radiation (NLO or NNLO) but differ by the inclusion
  of higher-order subleading terms; the latter, despite their
  subleading classification, might nevertheless have a sizable impact
  on some differential distribution, sensitive to radiation effects.

\item The analytical expression of the terms by which two matching
  procedures differ is not always available, leaving open only the
  possibility of a numerical comparison.

\end{itemize}

\noindent
Comments on the numerical comparisons:
\begin{itemize}
\item In a tuned comparison at NLO, where all the input parameters and
  the simulation setup are identical and the matrix elements have the
  same accuracy for all the codes, we observe that the total cross
  sections agree at the 0.03\% level both in the NLO EW and in the
  NLO QCD calculations; the differential distributions differ at most
  at the 0.5\% level.

\item The spread of the predictions at differential level reflects the
  impact of different choices in the numerical implementation of
  exactly the same calculation, in particular the handling of the
  subtraction of infrared and collinear divergences.

\item In a tuned comparison of codes that share NNLO QCD accuracy for
  the observables inclusive over radiation (cfr. Section
  \ref{sec:NNLO-QCD}), the level of agreement for the total cross
  sections is at the 0.4\% level and for the differential
  distributions is at the ${\cal O}(1\%)$ level, depending on the observable and on the range considered, but always with compatibility within the statistical error bands.

\end{itemize}

\noindent
Comments on the hierarchy of the different higher-order effects:
\begin{itemize}
\item All the EW higher-order effects are of ${\cal O}(\alpha^2)$ or
  higher.  Their size is in general at the few per mill level, with
  some exceptions like the lepton-pair invariant mass distribution,
  which receives corrections up to 5\%.  This particularly large size
  is due to the combination of two elements: on the one side to the
  steeply falling shape of the $Z$ boson resonance; on the other side,
  to the fact that most of the events are produced at the $Z$ peak,
  but final state radiation reduces the eventual invariant mass of the
  lepton pair, so that the lower-mass bins are populated.  At \oal the
  effect is of ${\cal O}(100\%)$ and multiple photon radiation still
  yields an additional corrections of several per cent.

\item In the absence of a full NNLO EW calculation, all the
  higher-order EW effects are necessarily subsets of the full result.
  They thus may not be representative of the full result, and care
  should be taken in using these partial results to estimate the
  effects of missing higher-order corrections.

\item The size of the QCD radiative corrections strongly depends on
  the observable: the differential distributions which require a
  resummation to all orders in some phase-space corners should be
  discussed separately from those that are stable upon inclusion of
  radiative effects.  Given our reference results obtained with codes
  that have NLO QCD accuracy for the total cross section, we studied
  higher-order effects due to NNLO QCD corrections, NLO QCD
  corrections matched with a QCD PS, and NNLO QCD corrections matched
  with a QCD PS. In case of the matched calculations we compared two
  different matching formulations.

\item The NNLO QCD corrections to the invariant (transverse) mass
  distribution of the lepton pair are small in size, at the few per
  cent level over the whole spectrum. The same codes predict a large
  positive correction of ${\cal O}(40-50\%)$ of the lower-order result
  for the lepton-pair transverse momentum distribution\footnote{ We
    remind the reader that the codes that have NNLO QCD accuracy for the total
    cross section are only NLO QCD accurate in the prediction of the
    large momentum tail of the lepton-pair transverse momentum
    distribution. For the same reason our reference results, which are
    NLO QCD accurate for the total cross section, are only LO accurate
    for this observable.  }, as the effect of having the exact
  description of two hard real parton emissions.  The latter show to
  play an important role also in the description of the hard tail,
  above the Jacobian peak, of the single-lepton transverse momentum
  distribution, with effects again at the ${\cal O}(30-40\%)$ level.

\item Matching fixed- and all-order results is necessary to obtain a
  sensible description of the Jacobian peak in the single lepton
  transverse momentum distribution or the low-momentum tail of the
  lepton-pair transverse momentum distribution.  Even if this goal is
  achieved, nevertheless two codes that share the same accuracy for
  the total cross section (in the absence of acceptance cuts),
  i.e. NLO QCD or NNLO QCD, still exhibit sizable differences in the
  prediction of these same observables, in the intermediate ranges of
  the spectra.  It should be stressed that these differences can be,
  in the NLO+PS matching, as large as few percent at the Jacobian peak
  or even several tens of percent for the lepton-pair transverse
  momentum distribution.  The size of these differences is reduced, at
  the several per cent level, with the NNLO+PS matching.

  This kind of matching ambiguities should be added to the usual
  renormalization/factorization scale variations and deserves further
  investigation.  An example of such a study of matching uncertainties
  can be found in Ref.~\cite{Bagnaschi:2015bop},
  for the Higgs transverse momentum distribution in
  gluon fusion.

\item 
QCD and EW effects are separately available at first
  perturbative order and have been extensively tested in Section
  \ref{sec:tuned}. The possibility of combining the differential
  K-factors in a factorized ansatz has been shown to be accurate,
  compared to the ${\cal O}(\alpha\alpha_s)$ results available in pole
  approximation at the $W$ ($Z$) resonance, for observables that are
  insensitive to a redistribution of events by QCD radiation, such as
  in the transverse-mass distribution of the $W$ or $Z$ bosons. Naive
  products fail to capture the dominant QCDxEW corrections in
  distributions such as in the transverse momentum
  of the lepton, which is sensitive to QCD initial-state radiation and
  photonic final-state radiation. For the invariant-mass distribution
  of the neutral-current process the naive product approach is
  insufficient as well because of large photonic final-state
  corrections and initial-state QCD corrections which
  depend on the reconstructed invariant mass in a non-trivial way.

\item The \powheg implementation of QCD+EW corrections shares with the
  other codes of the present report the NLO-(QCD+EW) accuracy for the
  total cross section.  On the other hand, it offers one possible
  solution to the matching of fixed- and all-orders results, both in
  QCD and in the EW sectors, and in turn it introduces mixed QCDxEW
  factorizable corrections to all orders.

\item  The interplay between QCD and QED corrections is not trivial, as it
  can be checked in observables like the charged-lepton transverse
  momentum distribution, where one can appreciate the large size of
  mixed \oaas and higher corrections.  The impact, in the same QCD
  framework, of subleading effects due to weak radiative corrections
  and to the exact treatment of real radiation matrix elements is not
  negligible in view of precision EW measurements, e.g. being the
  correction at the several per mill level in the case of the
  lepton-pair transverse mass distribution.

\end{itemize}

\noindent
Higher-order effects and theoretical uncertainties:
\begin{itemize}
\item The estimate of the accuracy available in the prediction of DY
  observables requires the distinction between: 1) higher-order
  corrections which have been computed and are available in at least
  one code and 2) missing higher-order terms which are unknown, whose
  effect can only be estimated.
\item The present report provides, for item 1), guidance to assess the
  size of the corrections which are missing in one code, thanks to the
  analysis of Section \ref{sec:impact}, so that they can be treated as
  a theoretical systematic error, when they are not included in the
  simulation.
\item On the other hand, item 2) requires a detailed, systematic discussion,
  which can start from the results of the present report, but goes
  beyond its scope.  The estimate of the actual size of missing higher
  orders is an observable-dependent statement.  In some specific cases
  the available fixed-order perturbative results may offer a handle to
  estimate the remaining missing corrections.  On the other hand, the
  quantities which require matching of fixed- and all-order
  results are simultaneously affected by several sources of
  uncertainty whose systematic evaluation will require a dedicated
  effort (see, e.g., the discussion in Section~\ref{sec:nnlops}).

\end{itemize}

\section*{Acknowledgements}
Work on this report was partially carried out at the
{\em LHC Run II and the Precision Frontier} workshop at the Kavli Institute for Theoretical Physics (KITP), supported by NSF PHY11-25915, at the
MIAPP workshop {\em Challenges,
    Innovations and Developments in Precision Calculations for the LHC}, supported by the Munich Institute for Astro- and Particle Physics
(MIAPP) of the DFG cluster of excellence ``Origin and Structure of the Universe'',
 and the {\em Prospects and
precision at the Large Hadron Collider at 14 TeV} workshop at the Galileo Galilei Institute for Theoretical Physics (GGI). 
We are grateful to the CERN LPCC and Michelangelo Mangano for 
hosting workshops at the LPCC, which allowed for close interactions of members of this working group with members of the experimental collaborations involved in EW precision analysis. 
We thank the Department of Physics at the University of Milan and INFN for financial support for the kick-off meeting {\em $W$ mass workshop}, and John Campbell and Fermilab for hosting the second $W$ mass workshop.

\clearpage
\appendix
\section{Description of MC codes}
\label{app:codes}

In the following we provide a brief description of the MC codes used
in this study. For details we refer to the relevant references which
are also provided.  

\subsection{DYNNLO}

{\tt DYNNLO}~\cite{Catani:2009sm} is a parton-level Monte Carlo program that computes the cross section for
vector-boson production in $pp$ and $p{\bar p}$ collisions. The calculation is performed up to
NNLO in QCD perturbation theory. The program includes $\gamma-Z$ interference, finite-width
effects, the leptonic decay of the vector boson and the corresponding spin correlations. The
user is allowed to apply arbitrary (though infrared safe) cuts on the final state and to plot
the corresponding distributions in the form of binned histograms. The program is available at
{\tt http://theory.fi.infn.it/grazzini/dy.html }\,, and more details can be found 
in the associated Ref.~\cite{Catani:2009sm}.

\subsection{DYNNLOPS}

DYNNLOPS \cite{Karlberg:2014qua} is a framework for matching fixed order
NNLO-QCD predictions of Drell-Yan production to parton showers. The method is
based on a reweighting procedure which uses events generated with the Bj-MiNLO
generators of the POWHEG-BOX
\cite{Alioli:2010xd,Hamilton:2012np,Hamilton:2012rf} and fixed order NNLO-QCD
predicitons obtained in DYNNLO \cite{Catani:2009sm}. Due to the MiNLO procedure
the Bj generator is fully inclusive and NLO accurate for zero and one-jet phase
space regions. By reweighting the events over the three dimensional phase space
of the massive vector boson we acquire NNLO accurate events, which can be passed
to a parton shower in the same way as usual POWHEG events. The MiNLO Sudakov
form factor is constructed in such a way that the reweighting procedure does not
introduce any spurious terms which could spoil the NNLO accuracy of the
calculation. Although the MiNLO Sudakov form factor is not formally NNLL
accurate, very good numerical agreement has been observed between dedicated
resummation calculations of the vector boson transverse momentum and DYNNLOPS.

To date this procedure has been implemented for Higgs production
\cite{Hamilton:2013fea}, Drell-Yan production and associated Higgs production
\cite{Astill:2016hpa}. In all cases public codes exist and can be obtained
through the POWHEG-BOX Version 2 by first checking out the repositories of the
Zj and Wj generators. The code allows the user to set all relevant input
parameters themself and to apply cuts on the final state leptons and jets. An
example analysis is also provided which the user can modify to their need. The
code is provided with step-by-step instructions and requires only little more
work to run compared to the Bj-MiNLO generators themselves.

\noindent
{\bf Acknowledgements} 
A.~Karlberg is supported by the British Science and Technology Facilities
Council and by the Buckee Scholarship at Merton College.

\subsection{FEWZ}

{\tt FEWZ} calculates the fully differential production of 
dilepton pairs via the neutral-current (intermediate photons and $Z$-bosons) and charged-current
processes. It is designed to make predictions for hadron-collider observables with realistic 
acceptance cuts at NNLO in the strong coupling constant. All
spin correlations and finite-width effects are included. In the neutral-current case 
it allows for the computation of the NLO electroweak corrections
as well. Technical details regarding several aspects of {\tt FEWZ} relevant 
to users of the code are discussed below.
\begin{itemize}
\item All inputs, including cuts on leptons and jets, electroweak couplings, and other 
parameters which control run setting, are set in an external input
file, allowing the user complete flexibility to customize {\tt FEWZ}.
\item Kinematic distributions are produced automatically during a run, with little overhead. 
The user can select which histograms to fill in an external
input file. Most distributions of interest are included in the default version of {\tt FEWZ}.
\item When running with PDF sets that contain error eigenvectors, all eigenvectors are 
calculated automatically for each histogram bin. The resulting output
can be combined using the included scripts to produce a final output file that contains 
the integration error as well as PDF error for both the total cross
section and each histogram bin. {\tt FEWZ} can be run using either LHAPDF, 
or with one of several PDF sets with native support.
\item Shell scripts are provided for farming out the sectors in parallel either locally or 
on \emph{Condor}, and a finishing script which combines the
results of individual sectors. 
In addition to the basic operation of combining the sectors and computing PDF errors, 
the finishing script can perform
operations such as addition, subtraction, multiplication, and division on different runs, 
all while treating the integration and PDF errors consistently.
\item The user can either choose from two hard-coded schemes for the input parameters, 
the $\alpha(M_Z)$ or $G_{\mu} $ scheme, or specify each coupling
manually. However, if the user decides to manually input the coupling parameters, only 
the QED corrections will be included in order to protect gauge
invariance.
\end{itemize}
For more details on the usage or validation of {\tt FEWZ} we refer the user to the publications 
\cite{Gavin:2010az,Gavin:2012sy,Li:2012wna} and the online documentation
at {\tt http://gate.hep.anl.gov/fpetriello/FEWZ.html}.

\subsection{HORACE}

HORACE is a parton-level Monte Carlo generator for precision
simulations of charged-current and neutral-current Drell-Yan processes
$p p \to W \to l \nu_l$ and $p p \to \gamma, Z \to l^+ l^-$, $l = e,
\mu$ at hadron colliders.

It is available at the web site {\tt
  http://www2.pv.infn.it/\~{}hepcomplex/horace.html}. It can be used
to generate both weighted and unweighted events and to obtain
predictions under realistic event selection conditions.

In a nutshell, the program includes the exact NLO electroweak (EW)
radiative corrections matched with a QED Parton Shower (PS) to take
into account higher-order QED leading logarithmic contributions due to
multiple photon emission from any charged legs, according to the
formulation described in detail in
\cite{CarloniCalame:2006zq,CarloniCalame:2007cd}.  Therefore, the
code, on top of the exact EW NLO corrections, includes the leading
effects due to initial and final state multiple photon radiation, as
well as its interference Thanks to the PS approach implemented in the
code, the transverse degree of freedom of the emitted photons beyond
${\cal O}(\alpha)$ are kept under control.  The generator can also run
including only final-state-like QED corrections in a pure PS approach,
as described in \cite{CarloniCalame:2003ux,CarloniCalame:2005vc}.
Fixed-order or PS QCD contributions are not accounted for in the
program.

As different classes of corrections are included in HORACE, it can be
used to provide an estimate of higher-order effects and theoretical
uncertainties, as documented in the report.

In detail, in HORACE the following EW contributions are taken into account
\begin{itemize}

\item Complete NLO EW corrections matched to multiple photon contributions.

\item Leading universal EW effects beyond NLO (running $\alpha$, $\rho$ parameter).

\item Different EW input parameter schemes ($\alpha_{G_\mu}$, $\alpha(0)$, $\alpha(M_Z)$)

\item Photon-induced processes ($\gamma q$ and $\gamma\gamma$ contributions).

\item Pair corrections in the leading logarithmic approximation.

\end{itemize}

\subsection{PHOTOS}

For a long time, the {\tt PHOTOS} Monte Carlo program
\cite{Barberio:1990ms,Barberio:1993qi} was used for the generation of
bremsstrahlung in the decay of particles and resonances. The core of
the algorithm operates on elementary decays. Thanks to carefully
studied properties of QED and investigation of several options for
exact phase space parameterization, an algorithm could be constructed.
With certain probability, {\tt PHOTOS} algorithm replaces the
kinematic configuration of the Born level decay with a new one,
where a bremsstrahlung photon or photons are added and other particle
momenta are modified.  Over the years the program evolved into a high
precision tool \cite{Golonka:2006tw}, for example it was found very
useful in the interpretation of data for the precision measurement of
the $W$ mass by CDF and D0 \cite{Abazov:2012bv,Aaltonen:2012bp}. In
the 2005 program version 2.15 multi-photon radiation was introduced
\cite{Golonka:2005pn}. To gain flexibility of its application, the
     {\tt FORTRAN} implementation is being replaced gradually by {\tt
       C++} and instead of {\tt HEPEVT}, the C++ event structure {\tt
       HepMC} \cite{Dobbs:2001ck} is used as the event
     record. Emission kernel based on complete first order matrix
     elements for QED final state bremsstrahlung was introduced,
     following papers \cite{Golonka:2006tw,Nanava:2009vg} in
     \cite{Davidson:2010ew}.

Here we describe several initializations for {\tt PHOTOS}, which may
be of interest for the study of effects due to final state photonic
bremsstrahlung in $W$ or $Z$ decays. We do not intent a detailed
documentation, but we will rather point to parameters which need to be
changed with respect to defaults and the code documented in
\cite{Davidson:2010ew}.

In practical applications for detector response simulations {\tt
  PHOTOS} in exponentiation mode will be certainly the best choice,
both in case of $Z$ and $W$ decays. In case of C++ applications kernel
featuring first order matrix element is then available as well.  The
initialization methods {\tt Photos::setMeCorrectionWtForW(bool corr)},
{\tt Photos::setMeCorrectionWtForZ(bool corr)} and \\ {\tt
  Photos::setExponentiation(bool expo)} should be all set true.

If matrix elements initialization is set {\tt false} universal process
independent kernel is used. This may be of interest to cross check the
numerical importance of matrix element effect, which was missing for
example in the {\tt FORTRAN} implementation of {\tt PHOTOS}.  From our
study \cite{Arbuzov:2012dx} we conclude that the matrix element was
necessary to improve precision from 0.3\% of the {\tt FORTRAN }
version of {\tt PHOTOS} to 0.2\% precision level now. This uncertainty
is for all QED final state emissions: photons, additional pairs and
interference effect combined.

For the studies of bremsstrahlung systematic on observables relating
$W$ and $Z$ decays one may be interested in degrading emission kernels
to the level when the same formulae are used in $W$ and $Z$ decays. In
case of the $Z$ decays kernel is applied for both outgoing leptons,
but it is then the same as for the photon emission in $W$ decay.  Not
only {\tt Photos::setMeCorrectionWtForW(bool corr)}, {\tt
  Photos::setMeCorrectionWtForZ(bool corr)} should be set to false,
but also \\ {\tt Photos::setInterference(bool interference)} and \\ {\tt
  Photos::setCorrectionWtForW(bool corr)}. The size of this part of
the bremsstrahlung effect, which is distinct for $W$ and $Z$ decays, can
be then studied by comparison.

There are two other modes which are of importance. Single photon
emission mode and double photon emission mode. Both of these modes are
for the studies of theoretical effects.

Single photon mode, activated with {\tt Photos::setExponentiation(bool
  expo)} and {\tt Photos::setDoubleBrem(bool doub)} both set to false,
is suitable to evaluate if definition of what is QED Final State
Radiation (FSR) matrix element is the same in {\tt PHOTOS} as in the
calculation of complete electroweak corrections. This has to be
verified, as we have done in case of studies with {\tt SANC}.  We have
validated that indeed calculation of pure weak effects with
contribution of final state QED bremsstrahlung removed can be used
together with {\tt PHOTOS} because QED bremsstrahlung is defined in
both packages in the same way. The complete calculation resulting from
use of pure weak calculator {\tt SANC} and {\tt PHOTOS} simultaneously
has its systematic error under precise control.  One should keep in
mind that comparisons and studies of separating out pure EW from QED
FSR are not straigtforward. In the single photon mode, the so-called
$k_0$ bias, resulting from the fact that below this threshold real
photons are not generated by {\tt PHOTOS} but their kinematic effect
may be present in the part of QED FSR corrections removed from pure
weak calculation.

Careful definition of separation between QED FSR and pure weak
corrections is specially important in case of $W$, charged and
relatively broad resonance, decay.

In case of of the two photon mode, activated with {\tt
  Photos::setDoubleBrem(bool doub)}, the $k_0$ bias is even stronger
than in the single photon one. The purpose of this mode is to check
how the iterative algorithm of {\tt PHOTOS} works.  Comparisons with
the calculations faring exact double photon emission amplitudes can be
performed that way as it was done in early time with tests using
papers \cite{RichterWas:1993ta,RichterWas:1994ep}, a step in this
direction is documented in \cite{Doan:2013qqa} in context of the
$\phi^*$ observable. General scheme for such studies of particular
terms, such as interference corrections, or effects of second order
QED matrix element embedded in exclusive exponentiation is now
available for predictions for $pp$ collisions as well, see
Ref. \cite{Jadach:2013aha}.

To conclude, the {\tt PHOTOS} Monte Carlo program is suitable now for
applications at the 0.2\% precision level for QED FSR emission and
observables of single $W$ or $Z$ production and decay. This result is
valid for {\tt C++} {\tt HepMC} applications including $\phist$
observable when kernels based on matrix element can be used. Otherwise
precision of 0.3\% should be assumed.  Further improvement on
precision is possible. Better test or implementation of pair emission
is then needed as well as detailed discussion of interferences effect
which may at certain moment need to be implemented as well with the
help of correction weight added into {\tt PHOTOS} and also initial
state emission/parton shower algorithm.  Finally let us point out that
tests of Ref.~\cite{Arbuzov:2012dx} provide interesting technical
tests of {\tt SANC} as well.

\noindent
{\bf Acknowledgements} 
This project is financed in part from funds of
Polish National Science Centre under decisions DEC-2011/03/B/ST2/00220
and \\ DEC-2012/04/M/ST2/00240.
Useful discussions with E. Richter-Was are acknowledged.

\subsection{POWHEG\_BMNNP and POWHEG\_BMNNPV}

Here we describe the simulation of Drell-Yan (DY) processes in the
POWHEG BOX performed by means of the two separate packages: {\tt
  W\_ew-BMNNP}~\cite{Barze:2012tt} for the $p p \to W \to l\nu$
process and {\tt Z\_ew-BMNNPV}~\cite{Barze:2013yca} for $p p \to
Z/\gamma^* \to l^+ l^-$. They are available in the public repository
of the POWHEG BOX~\cite{Alioli:2010xd} (Version 2) at the web site
{\tt http://powhegbox.mib.infn.it}.

The common feature of the two packages is the treatment of the hard
matrix elements with NLO QCD and NLO Electroweak (EW) corrections,
supplemented with QCD and QED higher order contributions within the
POWHEG framework.  The QCD virtual corrections and real radiation
matrix elements are the same as the ones contained in {\tt
  POWHEG\_W(Z)}~\cite{Alioli:2008gx}, while the expressions of the
virtual EW corrections are the ones publicly available in
Ref.~\cite{Dittmaier:2001ay} for the charged-current DY process and in
Ref.~(\cite{Dittmaier:2009cr}) for the neutral-current DY process.
The infrared and collinear singularities of EW origin in the loop
integrals are regulated using a hybrid scheme: the singularities
associated with the colored charged particles and the photon are
regulated with dimensional regularization, while QED mass
singularities are regulated by keeping finite lepton masses.  The soft
and collinear singularities of the real radiation matrix elements are
subtracted using the FKS subtraction scheme~\cite{Frixione:1995ms},
both for QCD radiation as well as for QED radiation described by the
matrix elements associated to one-photon emission off quarks and
leptons $q \bar{q}^{\prime} \to W \to l\nu + \gamma$ and $q \bar{q}
\to Z/\gamma^* \to l^+ l^- + \gamma$.  The singularities associated
with the unstable nature of the $W/Z$ vector bosons circulating in the
loops are treated according to the factorization
scheme~\cite{Dittmaier:2001ay,Dittmaier:2009cr} and the complex mass
scheme~\cite{Denner:2005fg,Denner:2006ic}.  The generation of the
hardest radiation is performed by means of the product of Sudakov form
factors associated with the singular regions and defined in terms of
the QCD and QED real radiation matrix elements. Thus the generation of
a radiative event, {\em i.e.} containing an additional QCD parton or
an additional photon\footnote{In the present version the
  ``photon-induced'' processes are not considered.}, is the result of
a competition between QCD and QED emission.

The NLO QCD and EW corrections are matched with Parton Shower (PS)
contributions, according to the POWHEG method: once the configuration
with the hardest (in transverse momentum) emission has been generated,
the subsequent radiation process is handled by the PS (both for QCD
and QED radiation) ordered in $p_{\rm T}$, applying a veto
technique. The multiple photon emission from external leptons is
included by default by means of the package
PHOTOS~\cite{Golonka:2005pn}, switching off the contribution of QED
radiation from the PS. Alternatively, it can be treated by the PS
itself, and in this case also multiple QED radiation from initial
state partons is simulated.

In summary, the POWHEG DY libraries {\tt W\_ew-BMNNP} and {\tt
  Z\_ew-BMNNPV} share the following features:
\begin{itemize}
\item normalization with QCD + EW corrections at NLO accuracy
\item complete SM NLO corrections matched to a mixed
  QCD$\otimes$QED parton cascade, where the particles present in the
  shower are coloured particles or photons
\item mixed ${\cal O}(\alpha \alpha_s)$ contributions partially taken into account (according to 
a factorized prescription, by construction)
\end{itemize}
The adopted input parameter schemes are the following ones: 
\begin{itemize}
\item charged-current DY: $G_\mu$ scheme as default, where the input 
parameters are $G_\mu$, $M_W$ and $M_Z$. The user can also 
switch to the $\alpha(0)$ scheme (even if not recommended), 
where the input parameters are 
$\alpha(0)$, $M_W$ and $M_Z$. In this scheme the masses 
of the light quarks in the fermionic corrections are taken finite 
and their values are chosen in such a way to reproduce the hadronic 
contribution to the hadronic vacuum polarization; 
\item neutral-current DY: in addition to the above two choices, also 
the scheme $\alpha(M_Z)$ can be switched on, where, 
instead of $\alpha(0)$, the value of $\alpha(M_Z)$ 
is used as input. 
\end{itemize}
For user convenience, the contribution of QCD or EW corrections can be
switched off by a proper flag.

\subsection{POWHEG\_BW}

In {\tt POWHEG\_BW} the full EW ${\cal O}(\alpha)$ radiative
corrections of Ref.~\cite{Baur:1998kt,Baur:2004ig} contained in the
public MC code {\tt WGRAD2} are added to the NLO QCD calculation of
the $pp \to W \to l \nu$ process of {\tt
  POWHEG-W}~\cite{Alioli:2008gx}.  The resulting MC code, called in
the following {\tt POWHEG-W\_EW}, is publicly available at the {\tt
  POWHEG BOX} web page and allows the simultaneous study
of the effects of both QCD and NLO EW corrections and with both {\tt
  Pythia} and {\tt Herwig}.  Note that the effects of photon-induced
processes and of multiple photon radiation are not included and that
QED corrections in {\tt Pythia} need to be switched off to avoid
double counting.

As default, {\tt POWHEG-W\_EW} produces results in the constant-width
scheme and by using the fine structure constant, $\alpha(0)$, in both
the LO and NLO EW calculation. More options can be found in {\tt
  subroutine init\_phys\_EW} but should be used with care and under
the advisement of the authors. Since QED radiation has the dominant
effect on observables relevant to the $W$ mass measurement, there is
the possibility of only including resonant weak corrections by
choosing qnonr=0, i.e. the weak box diagrams are neglected. Their
impact is important in kinematic distributions away from the resonance
region.  The full weak 1-loop corrections are included with
qnonr=1. The full set of QED contributions (QED=4) is included as
default, i.e. initial-state and final-state radiation as well as
interference contributions, but subsets can be studied separately by
choosing the flag 'QED' accordingly. The QED factorization scheme can
either chosen to be the DIS scheme (lfc=1) or the $\overline{\rm MS}$
scheme (lfc=0), and both schemes are defined in analogy to the
corresponding QCD factorization schemes. A description of the QED
factorization scheme as implemented in {\tt POWHEG-W\_EW} can be found
in Ref.~\cite{Baur:1998kt}.
  
Fermion masses only enter to the EW gauge boson self-energies and as
regulators of the collinear singularity.  The mass of the charged
lepton is included in the phase space generation of the final-state
four-momenta and serves as a regulator of the singularity arising from
collinear photon radiation off the charged lepton. Thus, no collinear
cut needs to be applied (collcut=0 in {\tt POWHEG-W\_EW}) on
final-state photon radiation, allowing the study of finite lepton-mass
effects.  Note that the application of a collinear cut on final-state
photon radiation (collcut=1) is only allowed in the electron case and
only when a recombination of the electron and photon momenta is
performed in the collinear region (usually defined by $\Delta
R_{e\gamma}<R_{cut}$, see Ref.~\cite{Baur:1998kt} for a detailed
discussion).

\noindent
{\bf Acknowledgements}
This research was supported by the National Science Foundation under award
  No.~PHY-1118138 and a LHC Theory Initiative Graduate Fellowship, NSF
  Grant No.~PHY-0705682.


\subsection{RADY}
{\tt RADY} is a Monte Carlo program for the calculation of
\underline{RA}diative corrections to
\underline{D}rell--\underline{Y}an processes, i.e.\ $pp/p\bar p\to
W/Z\to l\nu_l/l^+l^-$.  As a flexible Monte Carlo integrator, it
supports all kinds of event definitions (any experimental cuts,
collinear-safe or non-collinear-safe treatment of photons, jet
algorithms, etc.).  A large variety of radiative corrections can be
included in predictions, not only to achieve high precision, but also
to allow for estimates of various scheme dependences and other
theoretical uncertainties.  In detail, RADY supports:
\begin{itemize}
\item Next-to-leading order (NLO) QCD and electroweak (EW) corrections
  within the SM.  For Z production the individually gauge-invariant
  subsets of photonic final-state radiation, initial-state radiation,
  and initial--final interferences, as well as the genuine weak
  corrections can be investigated separately.
\item
NLO QCD and EW corrections within the Minimal Supersymmetric SM (MSSM).
Genuine supersymmetric corrections can be investigated separately.
\item
Multi-photon radiation effects beyond NLO in the collinear approximation
via structure functions.
\item
Leading universal EW effects beyond NLO ($\rho$-parameter, running $\alpha$).
\item
Leading EW Sudakov logarithms beyond NLO.
\item
Corrections induced by initial-state photons
($\gamma q$ and $\gamma\gamma$ collisions).
\item
Different EW input-parameter schemes ($\alpha(0)$, $\alpha(M_{\mathrm{Z}})$, $\alpha_{G_\mu}$).
\item
Different gauge-invariant schemes for treating the W/Z resonances
(complex-mass scheme, factorization scheme, pole scheme).
\item
Different technical treatments of soft and/or collinear photon/gluon
emission (dipole subtraction and two variants of phase-space slicing).
Note that, in particular, the treatment of collinear logarithms of the lepton masses
is very efficient in the extended dipole subtraction scheme.
\item
Mixed NNLO QCD$\times$EW corrections of ${\cal O}(\alpha \alpha_s)$
based on the pole approximation worked out in
Refs.~\cite{Dittmaier:2014qza,Dittmaier:2015rxo}.
\end{itemize}

\subsection{SANC: mcsanc-v1.01 and mcsanc-v1.20}

The SANC system (Support for Analytic and Numeric Calculations for
experiments at colliders)~\cite{Andonov:2004hi} implements
calculations of complete (real and virtual) NLO QCD and EW corrections
for the Drell--Yan CC~\cite{Arbuzov:2005dd} and
NC~\cite{Arbuzov:2007db} processes, associative Higgs and gauge boson
production~\cite{Bardin:2005dp}, single top
production~\cite{Bardin:2010mz,Bardin:2011ti} and several other
processes at the partonic level.  Here we give a brief summary of the
main properties of this framework. For the complete list of SANC
processes see~\cite{Andonov:2008ga}.  All calculations are performed
within the OMS (on-mass-shell) renormalization scheme in the $R_\xi$
gauge, \cite{Bardin:1999ak}, which allows an explicit control of the
gauge invariance by examining the cancellation of the gauge parameters
in analytical expression for matrix element.  The use of the OMS
scheme leads to running ($s$-dependent) width in vector boson
propagators (cf. Eq. (2.10) of Ref.~\cite{Dittmaier:2009cr}).

The list of processes implemented in the \texttt{mcsanc}-v1.01 {\em
  Monte-Carlo integrator},~\cite{Bardin:2012jk,Bondarenko:2013nu}, is
given in the Table~1 and the tree level diagrams are shown in Figure~1
of Ref.~\cite{Bondarenko:2013nu}.

NLO corrections contain terms proportional to logarithms of the quark
masses, $\log(\hat{s}/m_{u,d}^2)$.  They come from the initial state
radiation contributions including hard, soft and virtual photon or
gluon emission.  In the case of hadron collisions these logs have been
already {\em effectively} taken into account in the parton density
functions (PDF) and have to be consistently subtracted.  The
\texttt{mcsanc}-v1.01 supports both \MSbar and DIS subtraction
schemes.  A solution described in \cite{Diener:2005me} allows to avoid
the double counting of the initial quark mass singularities contained
in the results for the corrections to the free quark cross section and
the ones contained in the corresponding PDF.  The latter should also
be taken in the same scheme with the same factorization scale.

For example, the \MSbar QED subtraction to the fixed (leading) order
in $\alpha$ is given by:
\begin{eqnarray}
\label{msbarq}
\bar{q}(x,M^2)& = &q(x,M^2) -
\int_x^1 \frac{\mathrm{d} z}{z} \, q\biggl(\frac{x}{z},M^2\biggr) \,
\frac{\alpha}{2\pi} \, Q_q^2
\biggl[ \frac{1+z^2}{1-z}
\biggl\{\ln\biggl(\frac{M^2}{m_q^2}\biggr)-2\ln(1-z)-1\biggr\} \biggr]_+
\nonumber \\ 
& \equiv& q(x,M^2) - \Delta q,
\nonumber 
\end{eqnarray}
where \(q(x,M^2)\) is the parton density function in the \MSbar scheme
computed using the QED DGLAP evolution.

The differential hadronic cross section for DY processes with one-loop
EW corrections is given by:
\begin{eqnarray}
  \mathrm{d}\sigma^{pp \to \ell\ell'X} = & \sum_{q_{1}q_{2}} \int\limits_0^1 \int\limits_0^1 
  \mathrm{d}x_1 \, \mathrm{d}x_2 \, \bar{q}_1(x_1,M^2) \, \bar{q}_2(x_2,M^2)\,
  \mathrm{d}\hat{\sigma}^{q_1 q_2\to \ell\ell'},
\end{eqnarray}
where \(\bar{q}_1(x_1,M^2), \bar{q}_2(x_2,M^2) \) are the parton density
functions of the incoming quarks modified by the subtraction of the quark mass
singularities and \( \hat{\sigma}^{q_1 q_2\to \ell\ell'} \) is the partonic cross
section of corresponding hard process.  The sum is performed over all
possible quark combinations for a given type of process ($q_1q_2 = u\bar{d},
u\bar{s}, c\bar{d}, c\bar{s}$ for CC and $q_1q_2 = u\bar{u}, d\bar{d},
s\bar{s}, c\bar{c}, b\bar{b}$ for NC). The expressions for other processes are similar.

The effect of applying different EW schemes in the SANC system is
discussed in~\cite{Arbuzov:2007db}.  The SANC system supports
\(\alpha(0), G_{\mu}, \alpha(M_Z)\), of which the
\(G_{\mu}\)-scheme~\cite{Degrassi:1996ps} can be preferable since it
minimizes EW radiative corrections to the inclusive DY cross section.

\begin{figure}[h]
\begin{center}
  \includegraphics[width=0.6\textwidth]{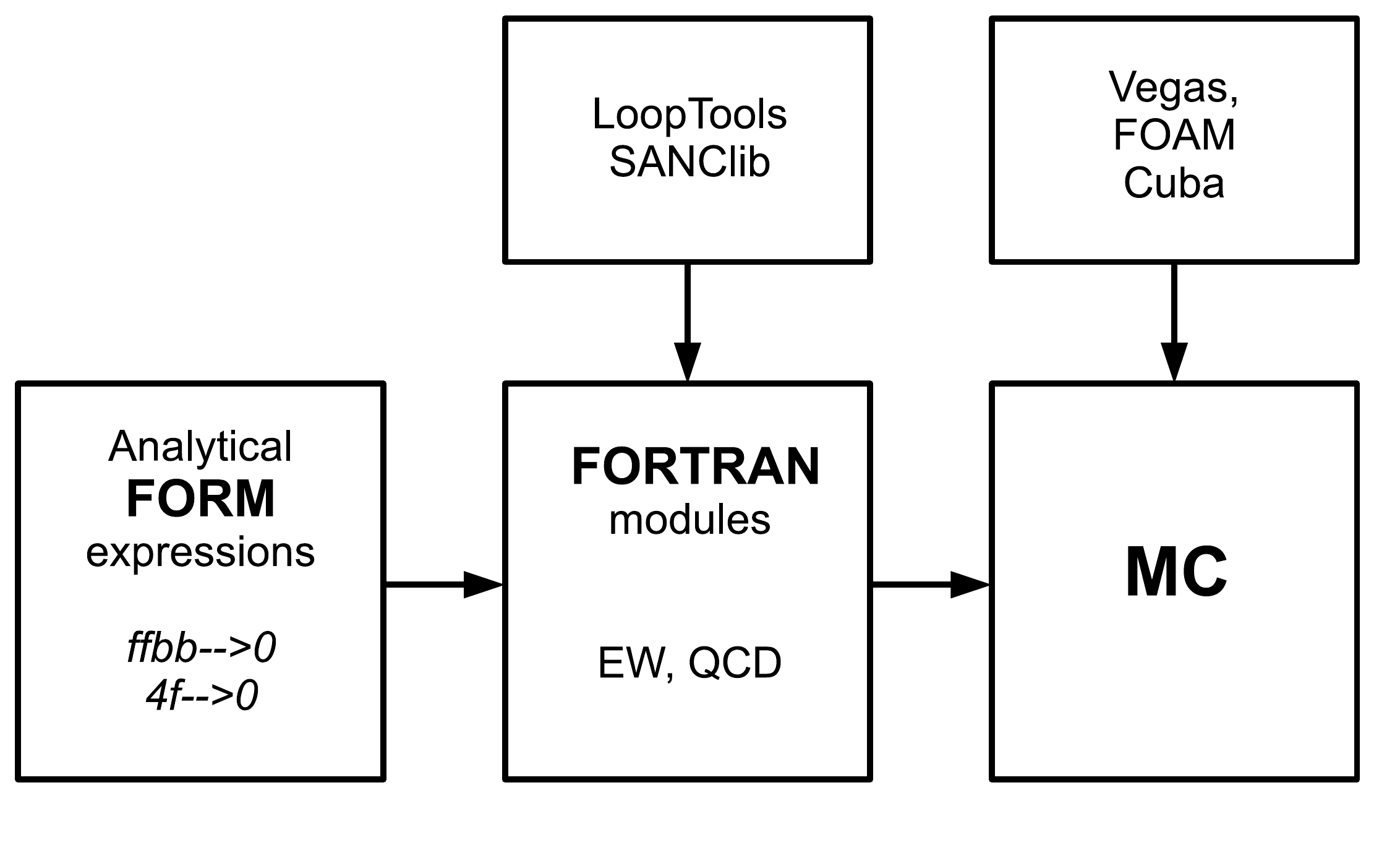}
\end{center}
\caption{The SANC framework scheme.}
\label{fig:sanc-scheme}
\end{figure}

The scheme of the SANC framework is shown on the
Figure~\ref{fig:sanc-scheme}. Analytical expressions are obtained for
the formfactors and amplitudes of generalized processes \(ffbb \to 0\)
and \(4f\to 0\) and stored as the FORM~\cite{Vermaseren:2000nd}
language
expressions~\cite{Andonov:2004hi,Bardin:2005dp,Arbuzov:2007ke,Andonov:2009nn}.
The latter are translated to the Fortran modules~\cite{Andonov:2008ga}
for specific parton level processes with an unified treatment QCD and
EW NLO corrections. The modules are utilising
Looptools~\cite{Hahn:1998yk} and SANClib~\cite{Bardin:2009ix} packages
for loop integrals evaluation.  To build a Monte-Carlo code one
convolutes the partonic cross sections from the modules with the
parton density functions and feeds the result as an integrand to any
Monte-Carlo algorithm implementation, e.g. FOAM~\cite{Jadach:2005ex}
or Cuba~\cite{Hahn:2004fe}.

Depending on the process and type of corrections, we subdivide the
total NLO cross section at the partonic level into several terms: $
\displaystyle{d\sigma = \sum^{id=6}_{id=1}d\sigma_{id}}, $
differential over a generic observable which is a function of the
final state momenta.  The individual terms depend on auxiliary
parameters $\bar{\omega}$ (photon energy which separates phase spaces
associated with the soft and hard photon emission) and $\lambda$
(photon mass which regularizes infrared divergences) which are
introduced in the NLO calculations. They cancel out after summation in
any physically observable differential NLO cross section.  In general,
NLO level hard sub-processes consist of: LO -- leading order (id=0),
\textit{virt} -- virtual (id=2), \textit{real} brems(glue)-strahlung,
$q\bar{q}$-,$\;gq$-channels (id=3--4,6) and \textit{subt} --
subtraction (id=1,5); \textit{real}, in turn, is subdivided into
\textit{soft} (id=3) and \textit{hard} (id=4) contributions by the
soft-hard separator parameter $\bar{\omega}$.  (For description of
id's see Section 2.1 of \cite{Bardin:2012jk}.)  The entire NLO
sub-process cross section is independent of both unphysical parameters
$\bar{\omega}$ and $m_q$.
 
The \texttt{mcsanc}-v1.01 code~\cite{Bondarenko:2013nu} was thoroughly cross checked against
another tools to provide reliable results.  Many numerical comparisons
with the well known MCFM~\cite{Campbell:2010ff} package are presented
in Ref.~\cite{Bardin:2012jk}. The NLO QCD values are in agreement
within statistical errors.  To conclude, we note, that the ``best what
{\texttt{mcsanc}} can do at pure NLO level'' {\it i.e.} the
recommended approximation, is computation of distributions in the
$G_\mu$ EW scheme with running widths.

The new \texttt{mcsanc}-v1.20 version of integrator is published in~\cite{Arbuzov:2015yja}.
The extensions concern implementation of Drell--Yan-like processes and include a systematic
treatment of the photon-induced contribution in proton--proton collisions and electroweak
corrections beyond NLO approximation. There are also technical improvements such as the calculation
of the forward-backward asymmetry for the neutral current Drell--Yan process.
Results were compared to the ones presented in~\cite{Brensing:2007qm,Dittmaier:2009cr}.
The numbers illustrate good agreement within the statistical errors of the Monte Carlo
integration.

\noindent
{\bf Acknowledgements}
This work was supported in part by the RFBR grant 12-02-91526-CERN\_a and by the Dynasty Foundation.


\subsection{WINHAC}
\label{sec:winhac}

\winhac \cite{Placzek:2003zg,Placzek:2009jy,Placzek:2013moa} is a Monte Carlo event generator 
for Drell--Yan (DY) processes in proton--proton, proton--antiproton 
as well as nucleus--nucleus collisions.
It features multiphoton radiation in the charge-current ($W$-boson mediated) 
DY processes within the Yennie--Frautschi--Suura (YFS) exclusive exponentiation
scheme \cite{yfs:1961} and the ${\cal O}(\alpha)$ electroweak (EW) 
radiative corrections with initial-state photon radiation (ISR) subtracted in a gauge invariant way.
The analytical formulae of the ${\cal O}(\alpha)$ virtual and soft-photon 
corrections have been obtained by the SANC group and provided in form of 
a numerical library~\cite{Bardin:2008fn}.
They are implemented in \winhac\  in two versions: (1) as the EW corrections
to $W$-boson decays and (2) as the EW corrections to the full charged-current 
DY process. In the latter case the quark mass singularities of the  
ISR are subtracted  in a gauge-invariant way. Two subtraction methods are implemented
in the current version of \winhac:
(1) the ``YFS-like scheme'' described in \cite{Bardin:2008fn}
and  (2) the ``dipole-subtraction scheme'', similar to a recently
developed method for matching NLO QCD effects with parton showers \cite{Jadach:2015mza}. 
Generation of ISR photons is handed to the parton shower generators,
such as {\tt Pythia} or {\tt Herwig}. Therefore, the predictions of \winhac\  may differ 
slightly from the calculations based on the $\overline{\rm MS}$ or DIS QED 
subtraction schemes. 

The current version, 1.37, of \winhac\ includes the Les Houches Accord (LHA)
interface to parton shower generators, such as {\tt Pythia}, {\tt Herwig}, etc.
This interface allows to write \winhac\  generated events into a disk file
or a named ({\sf FIFO}) pipe, which can then be read in and processed further
by an appropriate generator of QED/QCD parton showers and hadronisation.
Using the {\sf FIFO} pipe instead of an ordinary disk file has some advantages:
programs run faster, one does not have to deal with huge data files, 
very large event statistics can be generated without overloading disk/quota
capacity. We include a demo program in which events from \winhac\  are
sent to {\tt PYTHIA 6.4} for parton showering and hadronisation through one {\sf FIFO} pipe
and then sent back through another {\sf FIFO} pipe to \winhac\  for event analysis.
In addition to the LHA interface, \winhac\  includes also an internal interface
to {\tt PYTHIA 6.4}, in which appropriate {\tt PYTHIA} routines are called directly
from the \winhac\  code. It is less universal but faster in CPU time 
and can be used for some dedicated studies, 
see e.g.\ Refs.~\cite{Krasny:2007cy,Fayette:2008wt, Krasny:2010vd}.
Moreover, it includes options for correcting the {\tt PYTHIA~6} problem  
of wrong charge asymmetries of the DY leptons transverse momenta, 
see Ref.~\cite{Krasny:2012pa}.
 
In addition to unpolarized $W$-boson production, the program provides 
options for generation of polarized $W$-bosons in three different 
reference frames. \winhac\  also includes the neutral-current 
($Z/\gamma$) Drell--Yan process at the Born level and with the FSR QED
corrections generated by \photos \cite{Davidson:2010ew} 
(though a dedicated interface). 
\photos\ can also be used to generate QED FSR in the $W$-boson case, 
which might be useful for some studies. 
  
\winhac\  is interfaced with the {\tt LHAPDF} package and provides
the possibility to compute auxiliary weights corresponding to PDF errors; 
all these weights are calculated in a single MC run.
In the case of nucleus--nucleus collisions, an option for switching on/off
nuclear shadowing effects for PDFs is provided. Nuclear beams are defined
through the input parameters by setting atomic numbers $A$, charge numbers $Z$ 
and energies of two colliding nuclei. This collider option was applied 
to studies presented in Refs.~\cite{Krasny:2005cb,Krasny:2007cy}.

The QED FSR corrections in \winhac\  were compared numerically
with the ones implemented in the MC generator \horace\ and a good agreement 
of the two programs for several observables was
found \cite{CarloniCalame:2004qw}. 
Implementation of the ${\cal O}(\alpha)$ EW corrections was 
successfully cross-checked (to a high precision) with the \sanc\ 
program \cite{Bardin:2008fn}.

Several options and steering parameters available in \winhac\  make it a flexible
Monte Carlo tool for various studies and tests related to the DY processes,
particularly in the context of the Higgs-boson production at the LHC 
\cite{Krasny:2013aca,Krasny:2015dka,Krasny:2016bvx}.
The original \winhac\  program is written in Fortran, however rewriting it
in C$++$ is already quite advanced \cite{Sobol:2011zz}.
A similar event generator for the neutral-current Drell--Yan process, called
{\tt ZINHAC}, is under development. We also work on the QCD NLO parton-shower
algorithm and matching it with the QCD NLO hard process matrix elements,
see \cite{Jadach:2011cr,Jadach:2012vs,Jadach:2015mza}. 

\noindent
{\bf Acknowledgements}
This research is supported in part by the programme of the French--Polish 
co-operation between IN2P3 and COPIN no.\ 05-116
and by the Polish National Centre of Science grant no.\ DEC-2012/04/M/ST2/00240.
We also thank the Galileo Galilei Institute for Theoretical Physics in Florence for the hospitality
and the INFN for partial support during the GGI Workshop ``Prospects and Precision at the Large Hadron Collider at 14 TeV'' where some parts of this work were done.

\subsection{WZGRAD}

{\tt WZGRAD} combines the Monte Carlo programs {\tt
  WGRAD2}~\cite{Baur:1998kt,Baur:2004ig} and {\tt
  ZGRAD2}~\cite{Baur:2001ze}. It is a parton-level Monte Carlo program
that includes the complete ${\cal O}(\alpha)$ electroweak radiative
corrections to $p\,p\hskip-7pt\hbox{$^{^{(\!-\!)}}$} \to
W^\pm\to\ell^\pm\nu X$ ({\tt WGRAD2}) and
$p\,p\hskip-7pt\hbox{$^{^{(\!-\!)}}$} \to\gamma,\, Z\to\ell^+\ell^- X$
($\ell=e,\,\mu$) ({\tt ZGRAD2}). For the numerical evaluation, the
Monte Carlo phase space slicing method for next-to-leading-order (NLO)
calculations described in Ref.~\cite{Baer:1990ra,Harris:2001sx} is
used. Final-state charged lepton mass effects are included in the
following approximation. The lepton mass regularizes the collinear
singularity associated with final state photon radiation. The
associated mass singular logarithms of the form $\ln(\hat
s/m_\ell^2)$, where $\hat s$ is the squared parton center of mass
energy and $m_\ell$ is the charged lepton mass, are included in the
calculation, but the very small terms of ${\cal O}(m_\ell^2/\hat s)$
are neglected.

As a result of the absorption of the universal initial-state mass
singularities by redefined ({\it renormalized})
PDFs~\cite{Baur:1998kt,DeRujula:1979grv}, the cross sections become
dependent on the QED factorization scale $\mu_{\rm QED}$. In order to
treat the ${\cal O}(\alpha)$ initial-state photonic corrections to $W$
and $Z$ production in hadronic collisions in a consistent way, the
parton distribution functions should be used which include QED
corrections such as NNPDF2.3QED~\cite{Ball:2013hta}.  Absorbing the
collinear singularity into the PDFs introduces a QED factorization
scheme dependence. The squared matrix elements for different QED
factorization schemes differ by the finite ${\cal O}(\alpha)$ terms
which are absorbed into the PDFs in addition to the singular
terms. {\sc WZGRAD} can be used both in the QED ${\rm \overline{MS}}$
and DIS schemes, which are defined analogously to the usual ${\rm
  \overline{MS}}$~\cite{Bardeen:1978yd} and DIS~\cite{Owens:1992hd}
schemes used in QCD calculations.

It is recommended that {\sc WZGRAD} is used with a constant width and
the $G_\mu$ input scheme, which correspondents to the EW input scheme
used for producing the benchmark results in this report.  Radiative
corrections beyond ${\cal O}(\alpha)$ are partially implemented as
described in Section~\ref{sec:renormalization-ho-universal}.

\clearpage

\section{Tuned comparison of total cross sections at NLO EW and NLO QCD for $W^\pm$ and $Z$ production with LHCb cuts}
\label{app:lhcb}

\begin{table}[h]
\begin{center}
\begin{tabular}{|c|l|c|c|c|}
\hline
        & LO  & NLO & NLO & NLO  \\ 
code    &     &  QCD  & EW $\mu$  & EW $e$ \\ 
\hline
HORACE  & 841.82(3)   &    $\times$       & 876.93(4) & 862.99(5)\\
\hline
WZGRAD  & 841.820(7)   &  $\times$         & 876.81(2) & 862.86(3) \\
\hline
RADY    & 841.822(3)  & 928.61(9) & 876.92(1) & 862.96(1) \\
\hline
SANC    & 841.818(8)  & 928.8(1)  & 876.61(2) & 862.59(2) \\
\hline
FEWZ    & 841.80(5)   & 928.8(1) & $\times$ & $\times$   \\
\hline
POWHEG-w & 841.7(2)    & 928.67(4)& $\times$  & $\times$ \\
\hline
\end{tabular}
\caption{\label{tab:xsec-wp-lhc8-lhcb}  $p p \to W^+\to l^+ \nu_l$
  cross sections (in pb) at the 8 TeV LHC, with LHCb cuts and  bare leptons.}
\end{center}
\end{table}

\begin{table}[h]
\begin{center}
\begin{tabular}{|c|c|c|c|}
\hline
     & LO   & NLO-EW $\mu$ calo & NLO EW $e$ calo  \\ 
code &         &  &  \\ 
\hline
HORACE   & 841.82(3) & 831.94(4) &  876.33(5) \\
\hline
WZGRAD   & 841.820(7) & 831.57(2) & 875.97(3)  \\
\hline
SANC        &  841.818(8)& 831.36(2) & 875.88(1)   \\
\hline
\end{tabular}
\caption{\label{tab:xsec-wp-lhc8-lhcb-rec}  $p p \to W^+\to l^+ \nu_l$
  cross sections (in pb) at the 8 TeV LHC, with LHCb cuts and  ``calorimetric'' leptons.}

\end{center}
\end{table}

\begin{table}[h]
\begin{center}
\begin{tabular}{|c|l|c|c|c|}
\hline
        & LO  & NLO & NLO  & NLO  \\ 
code    &    &QCD    & EW $\mu$ & EW $e$ \\ 
\hline
HORACE  & 640.36(2)  &  $\times$         & 664.81(3)  & 652.66(3) \\
\hline
WZGRAD  & 640.358(5)  &  $\times$ &  664.79(1)  & 652.68(2)\\
\hline
RADY    & 640.353(2) & 665.93(9) & 664.828(8) & 652.712(8) \\
\hline
SANC    & 640.357(2) & 666.8(1)  & 664.784(6) & 652.630(6)\\
\hline 
FEWZ   &  640.35(2)  & 666.00(8) & $\times$ & $\times$ \\
\hline
POWHEG-w & 640.36(2)  &666.23(6) & $\times$  & $\times$ \\
\hline
\end{tabular}
\caption{\label{tab:xsec-wm-lhc8-lhcb}  $p p \to W^-\to l^- \bar\nu_l$
  cross sections (in pb) at the 8 TeV LHC, with LHCb cuts and
  bare leptons.}
\end{center}
\end{table}

\begin{table}[h]
\begin{center}
\begin{tabular}{|c|c|c|c|}
\hline
     & LO   & NLO-EW $\mu$ calo & NLO EW $e$ calo  \\ 
code &         &  &  \\ 
\hline
HORACE   & 640.36(2) & 630.82(3) & 665.24(4) \\
\hline
WZGRAD   & 640.358(5) & 630.60(1)& 665.16(2)  \\
\hline
SANC        & 640.357(2) & 630.597(6) &665.139(8)   \\
\hline
\end{tabular}
\caption{\label{tab:xsec-wm-lhc8-lhcb-rec}  $p p \to W^-\to l^- \bar\nu_l$
  cross sections (in pb) at the 8 TeV LHC, with LHCb cuts and  ``calorimetric'' leptons.}

\end{center}
\end{table}

\begin{table}[h]
\begin{center}
\begin{tabular}{|c|l|c|c|c|}
\hline
        & LO        & NLO       & NLO          & NLO  \\ 
code    &           & QCD       & EW $\mu$ & EW $e$ \\ 
\hline
HORACE  & 75.009(2) & $\times$          & 77.838(4) & 76.053(4) \\
\hline
WZGRAD  & 75.0090(7) & $\times$          & 77.931(2) & 76.233(3)           \\
\hline
RADY    & 75.0112(3)& 87.173(8) & 77.9518(9)& 76.2523(9)\\
\hline
SANC    & 75.0087(4)& 87.23(1)  & 77.881(4) & 76.182(2) \\
\hline
FEWZ   & 75.001(3) & 87.19(1)  & ($\times$) & ($\times$) \\
\hline
POWHEG-z &  75.04(2) & 87.188(4) & $\times$  & $\times$ \\
\hline
\end{tabular}
\caption{\label{tab:xsec-z-lhc8-lhcb}  $p p \to \gamma,Z \to l^-
  l^+$  cross sections (in pb) at the 8 TeV LHC, with LHCb cuts
  and bare leptons.}
\end{center}
\end{table}

\begin{table}[h]
\begin{center}
\begin{tabular}{|l|l|l|l|}
\hline
     & LO   & NLO-EW $\mu$ calo & NLO EW $e$ calo  \\ 
code &         &  &  \\ 
\hline
HORACE   & 75.009(2) & 67.979(4) & 77.142(4) \\
\hline
WZGRAD   & 75.0090(7) & 67.961(2) &  77.304(3) \\
\hline
SANC        &75.0047(9)  & 67.9821(9) & 77.245(2)  \\
\hline
\end{tabular}
\caption{\label{tab:xsec-z-lhc8-lhcb-rec}  $p p \to \gamma , Z\to l^+ l^-$
  cross sections (in pb) at the 8 TeV LHC, with LHCb cuts and  ``calorimetric'' leptons.}

\end{center}
\end{table}



\clearpage

\newpage
\bibliographystyle{atlasnote}
\bibliography{report-main}

\end{document}